\documentclass[sigconf, screen]{acmart}

\AtBeginDocument{%
  \providecommand\BibTeX{{%
    \normalfont B\kern-0.5em{\scshape i\kern-0.25em b}\kern-0.8em\TeX}}}

\setcopyright{acmlicensed}
\copyrightyear{2026}
\acmYear{2026}
\acmDOI{XXXXXXX.XXXXXXX}

\acmConference[GI '26]{Graphics Interface 2026}{June 09--12,
  2026}{Kitchener - Waterloo, Ontario, Canada}
\acmISBN{979-8-4007-XXXX-X/2026/06}

\usepackage{_macros}

\showrevisions{REVISIONRED}

\begin{document}

\newcommand{\an}[1]{\authorcomment{GOLD}{AN}{#1}}
\newcommand{\jian}[1]{\authorcomment{BLUE}{JZ}{#1}}
\newcommand{\shimon}[1]{\textcolor{GREEN}{[SA: #1]}}
\title{Exploring Above-Neck Unimanual Swipe Gestures for Off-Device Earable Interaction}

\author{Shaikh Shawon Arefin Shimon}
\email{ssarefin@uwaterloo.ca}
\orcid{0000-0001-5007-2828}
\affiliation{%
  \institution{University of Waterloo}
  \city{Waterloo}
  \state{Ontario}
  \country{Canada}
}

\author{Ali Neshati}
\orcid{0000-0002-0405-1169}
\email{ali.neshati@ontariotechu.ca}
\affiliation{%
  \institution{Ontario Tech University}
  \city{Oshawa}
  \state{Ontario}
  \country{Canada}
}

\author{Junwei Sun}
\email{jwsun@idt.eitech.edu.cn}
\orcid{0009-0001-5933-0122}
\affiliation{%
  \institution{Ningbo Institute of Digital Twin}
  \city{Ningbo}
  \country{China}
}

\author{Qiang Xu}
\email{qiang.xu1@huawei.com}
\orcid{0000-0002-0077-1003}
\affiliation{%
  \institution{Huawei Technologies Canada Co., Ltd.}
  \city{Markham}
  \state{Ontario}
  \country{Canada}
}

\author{Jian Zhao}
\orcid{0000-0001-5008-4319}
\email{jianzhao@uwaterloo.ca}
\affiliation{%
  \institution{University of Waterloo}
  \city{Waterloo}
  \state{Ontario}
  \country{Canada}
}

\renewcommand{\shortauthors}{Arefin Shimon \etal{}}

\begin{abstract}

Despite their growing popularity, in-ear \textit{Earable} / \textit{Hearable} devices (\ie{}, ear-mounted wearables) face interaction challenges due to limited input space and compact form factors.
To enhance interaction capabilities, researchers are exploring off-device hand-based input spaces above the neck using midair and onskin gestures.
However, existing literature primarily focuses on axial swipes (\ie{}, horizontal and vertical), leaving nonaxial swipes (\ie{}, unidirectional swipes with varied orientations) and angular swipes (\eg{}, ``L'', ``U'', or ``V'') largely underexplored despite their potential interaction advantages.
To address this gap, we conducted a within-subject gesture motion analysis study with 24 participants, analyzing 5,568 swipes of varying shape, orientation, and complexity.
Our results revealed preferred starting and ending regions for different unidirectional and angular swipe shapes, as well as intuitive swipe shapes within the off-device, above-neck manual interaction space.
We further examine off-device swipe characteristics, discuss the feasibility of recognizing these earable gestures with current sensing technologies, and highlight their potential application in various scenarios.
These findings broaden the understanding of off-device earable gestures and provide design insights for integrating suitable nonaxial and angular swipes alongside traditional axial gestures to enhance interaction with in-ear earable devices.
   
\end{abstract}



\maketitle

\section{Introduction}
\label{section: Introduction}
{

Ear-mounted wearables (\ie{}, \textit{earables}~/~\textit{hearables}) are widely used for their auditory capabilities, and are increasingly explored as head-based sensing platforms \cite{roddiger2022LitReview}. 
Wireless in-ear devices, such as \textit{Apple Airpods}~\cite{apple2025airpodspro3}, are especially popular for their compactness, portability, and convenience~\cite{zhou2020survey}, but their small form factor imposes substantial constraints on direct on-device interaction. 
Simple actions such as \textit{accepting or rejecting calls} and \textit{adjusting volume} can be performed through taps, multi-presses, or vertical swipes on the device body, more complex tasks (\eg{}, \textit{menu navigation} or \textit{application switching}) are challenging due to limited on-device input space.
While \textit{Voice} input can partially mitigates these constraints, it is often less effective than other input methods for head-mounted devices \cite{Kollee2014ExploringHMD}, may be socially inappropriate in certain contexts, \cite{Pandey2021AcceptabilitySpeech}, and is susceptible to environmental noise \cite{Li2019SpeechRecognitionResearch}. These limitations motivate exploring alternative, off-device interaction techniques for in-ear earables.

Researchers have explored various off-device modalities -- such as midair and onskin manual gestures, head motion, full-body movement, and silent commands -- to extend in-ear earable interaction beyond physical, on-device input \cite{lissermann2014earput, xu2020earbuddy, jin2022earcommand, alkiek2022eargest}. 
Among these, unimanual (\ie{}, one-handed) hand-to-face gestures above the neck have emerged as particularly appealing to end-users \cite{roddiger2022LitReview}. 
These gestures occur naturally in daily behavior \cite{nicas2008study, kwok2015face}, and facilitate one-handed interaction with the ear-mounted device while keeping the other hand available for additional tasks.
Prior studies also show that swipe gestures performed in midair and onskin regions above the neck are among the most preferred unimanual off-device gestures for earables \cite{chen2020exploring, rateau2022leveraging}. 
However, direct physical interaction with the outer ear can cause in-ear earables to displace from the ear canal \cite{Iguma2023Input}, making direct touch interactions on the ear itself unsuitable for interacting with these devices. 
This motivates a focus on unimanual hand-to-face swipe gestures performed in the onskin and midair spaces \emph{around} and \emph{above} the outer ear, rather than on the ear surface itself.

Previous work has demonstrated that unimanual gestures can be differentiated by both shape and location in midair and onskin interaction spaces around the ear \cite{xu2020earbuddy, lissermann2014earput, kikuchi2017eartouch, Alkiek2023EarBender}. 
More recently, Shimon \etal{} \cite{arefinshimon2024fingertracker} explored location-based reuse of axial (horizontal and vertical) swipe, tap, and pinch gestures for off-device earable input. 
Despite these advances, most prior studies have focused primarily on axial swipe directions ~\cite{Grossman2015Typing, xu2020earbuddy, roddiger2022LitReview}. 
While swipe or stroke gestures of varying complexity, difficult ranking and number of strokes in the above-neck space have been proposed \cite{Vatavu2022Understanding}, there is limited understanding of how nonaxial unidirectional swipes and angular swipe shapes -- such as ``L'', ``U'', and ``V'' -- perform in off-device earable interaction.
This gap is notable, as previous studies highlight that these non-axial and angular swipes yield above-average user agreement \cite{rateau2022leveraging}. 
These unistroke gestures also exhibit low shape complexity and low execution difficulty \cite{Vatavu2022Understanding, Isokoski2001Model, vatavu2011estimating}. 
Prior work on symbolic and marking menu-based interaction further demonstrates that users naturally employ these swipes resembling checkmark (\faCheck) or caret ($\boldsymbol{\bigwedge}$) for semantically meaningful actions, (\eg{}, confirmation, selection) across various input modalities, including onskin and midair interactions \cite{Vatavu2022Understanding, Wobbrock2007Gestures, Wittorf2016Eliciting}. 
By allowing angular swipe~/~stroke gestures with fixed contortions to be rotated or re-oriented, we can expand the off-device input vocabulary and encode distinct earable input commands while avoiding precision issues associated with small gesture region boundaries suggested in location-based gesture reuse \cite{arefinshimon2024fingertracker} for midair and onskin, off-device earable spaces.
Furthermore, these expressive, angular manual gestures can serve as delimiters to mitigate the \textit{Midas Touch} problem in manual hand-to-face interaction for head-based wearables \cite{Weng2021Facesight} as their explicit and deliberate articulation minimizes the chance of unintentional gesture activation.

{
\begingroup
\setlength{\floatsep}{-1ex} 
\setlength{\intextsep}{-1ex} 
\setlength{\abovecaptionskip}{-1ex}
\setlength{\belowcaptionskip}{-1ex}

\begin{figure}
    \centering
    \includegraphics[width=180pt]{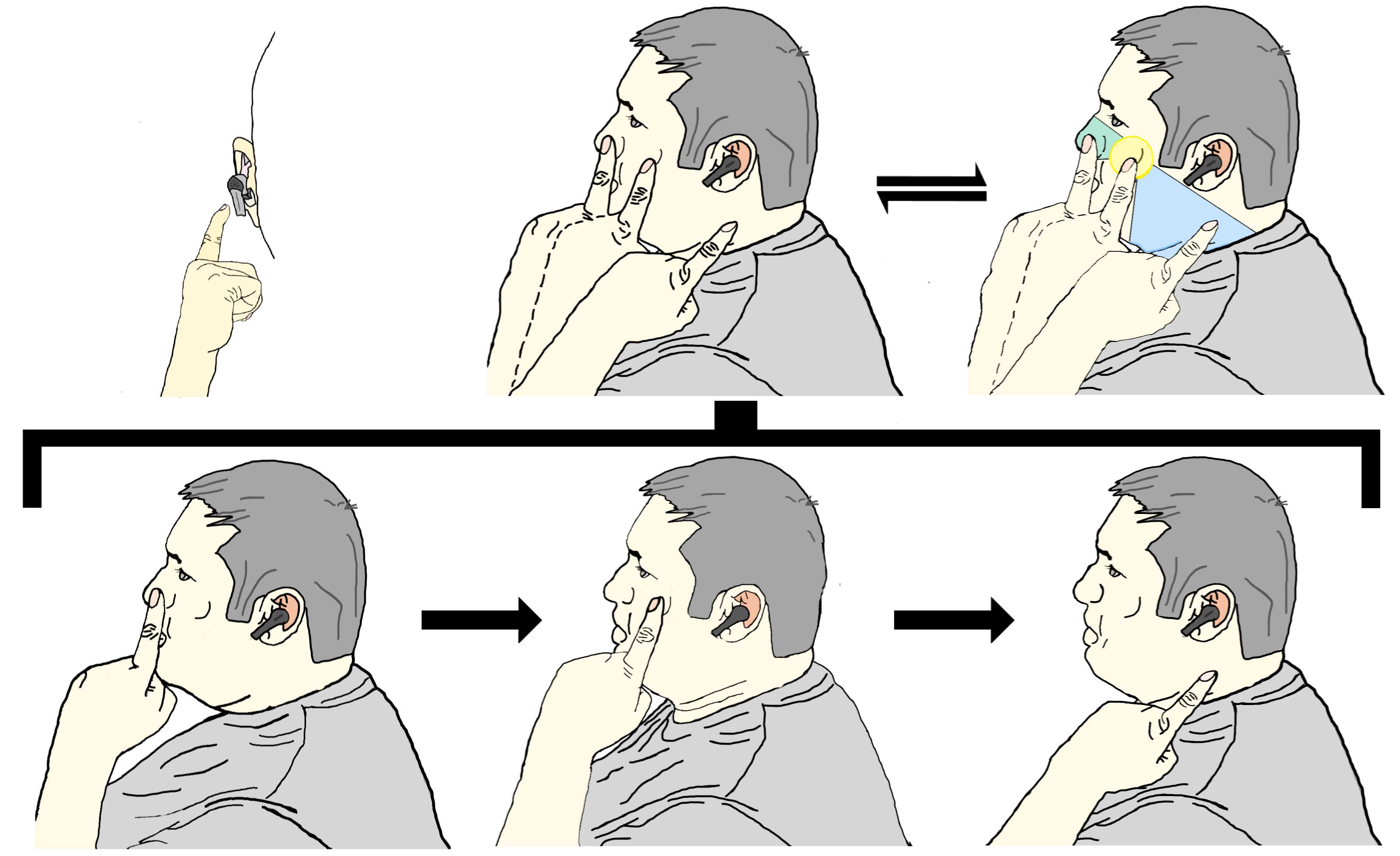}
    \vspace{0.2cm}
    \caption{Unimanual (one-handed) swipe using the left hand from a starting (\textit{Green}) region \textbf{S} through a \textit{swipe focal point} (\textit{Yellow}) to an ending (\textit{Blue}) region \textbf{E} on the left hand side of the face.} 
    \label{fig:R2R-swipe-motion}
    \Description{Six images showing how a right-handed participant performs a unimanual swipe on the left side of their face from a starting region through a swipe focal point to an ending region using their non-dominant (i.e, left) hand while wearing an in-ear earbud in the left ear. Top right image shows the front view of the non-dominant left side of the face, showing left-hand moving around the left ear and the left cheekbone. Top middle figure shows the side view of the motion described in the top-left image. The top-right image expands on the top middle figure, and shows how left index finger moves from the swipe starting region around the nose, marked in green, to the swipe focal point over the cheekbone marked in yellow, to the swipe ending region around the jawline highlighted in blue. Group of 3 images on the bottom expand the motion depicted in top-center image. Bottom left image shows the finger over the nose, bottom middle image shows the finger over the cheekbone, and the bottom right figure show the left index finger over the jawline.}
\end{figure}

\begin{figure}
    \centering
    \begin{minipage}{50pt}
        \centering
        \includegraphics[width=21pt,frame]{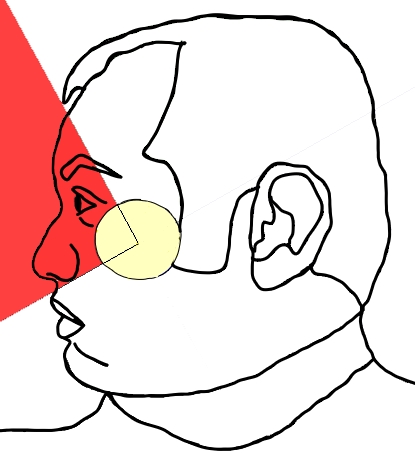}
        \includegraphics[width=21pt,frame]{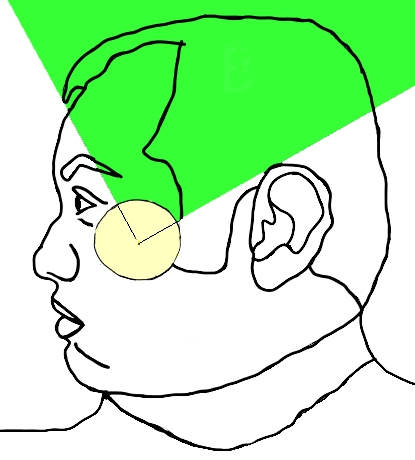}
        \\
        \includegraphics[width=21pt,frame]{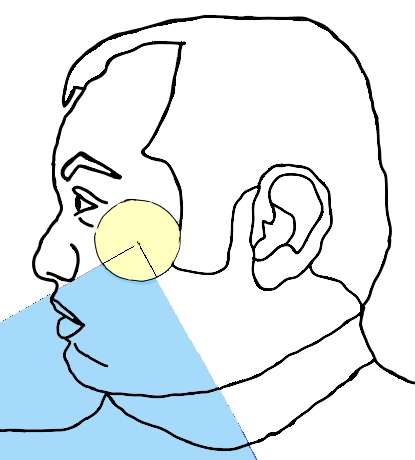}
        \includegraphics[width=21pt,frame]{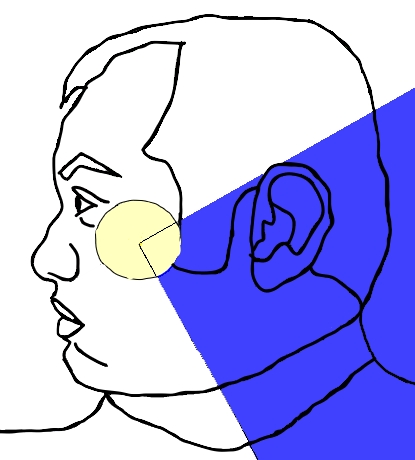}
        \\
        \includegraphics[width=10.0pt]{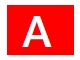}
        \includegraphics[width=10.0pt]{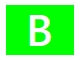}
        \includegraphics[width=10.0pt]{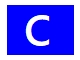}
        \includegraphics[width=10.0pt]{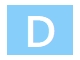}
        \\
        \vspace{-0.15cm}
        \subcaption[figure]{4 Midair}    
        \label{fig:midair_4R2R_in_gesture_region_section}                  
    \end{minipage}%
    \hfill
    \begin{minipage}{10.5pt}\end{minipage}
    \begin{minipage}{75pt}
        \centering
        \includegraphics[width=21pt,frame]{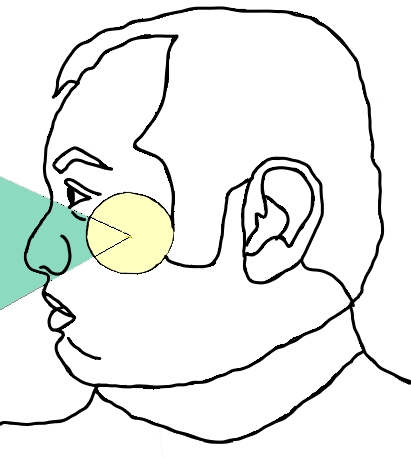}
        \includegraphics[width=21pt,frame]{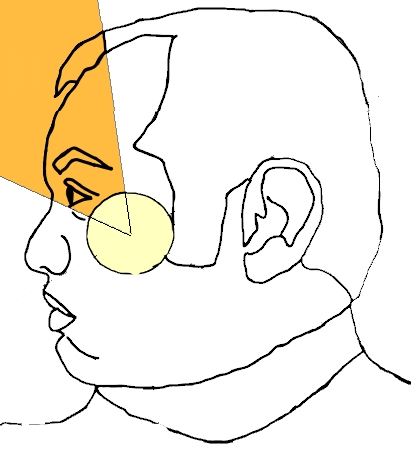}
        \includegraphics[width=21pt,frame]{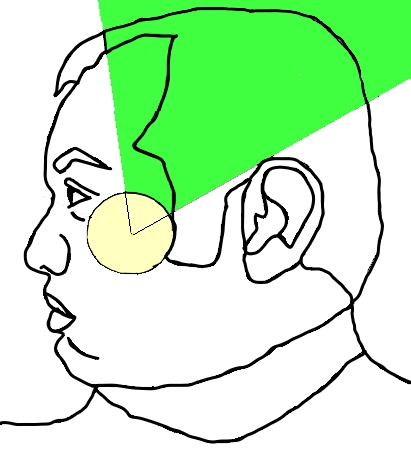}
        \\
        \includegraphics[width=21pt,frame]{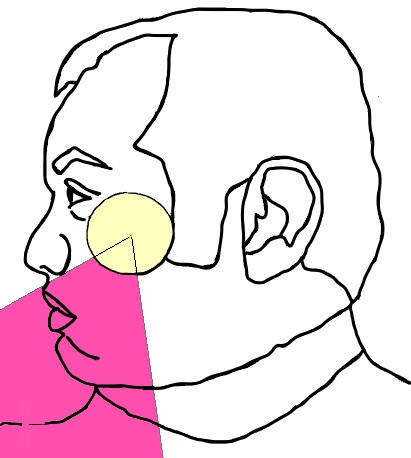}
        \includegraphics[width=21pt,frame]{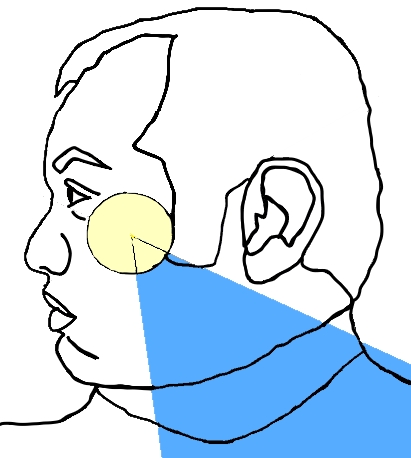}
        \includegraphics[width=21pt,frame]{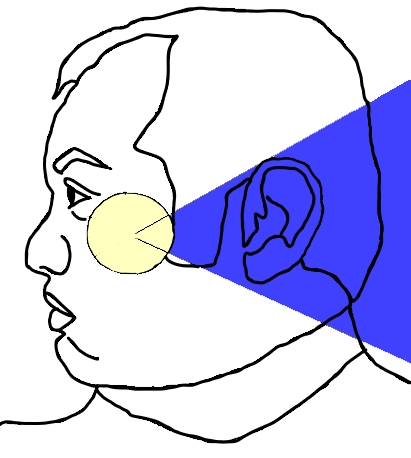}
        \\
        \includegraphics[width=10.0pt]{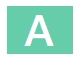}
        \includegraphics[width=10.0pt]{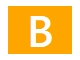}
        \includegraphics[width=10.0pt]{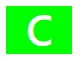}
        \includegraphics[width=10.0pt]{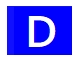}
        \includegraphics[width=10.0pt]{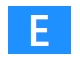}
        \includegraphics[width=10.0pt]{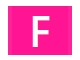}
        \\
        \vspace{-0.15cm}
        \subcaption[figure]{6 Midair}   
        \label{fig:midair_6R2R_in_gesture_region_section}                             
    \end{minipage}%
    \begin{minipage}{10.5pt}\end{minipage}
    \hfill
    \begin{minipage}{100pt}
        \centering
        \includegraphics[width=21pt,frame]{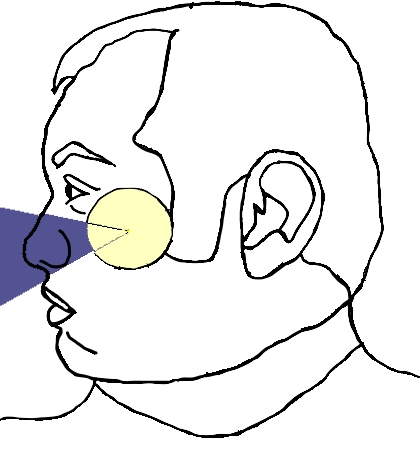}
        \includegraphics[width=21pt,frame]{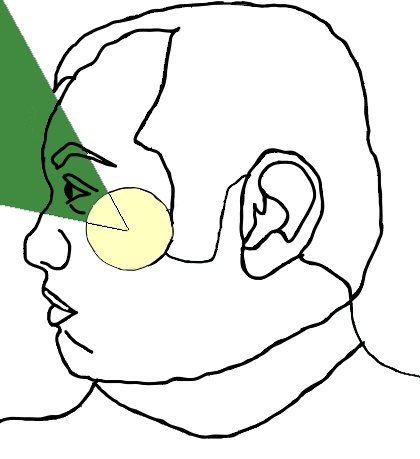}
        \includegraphics[width=21pt,frame]{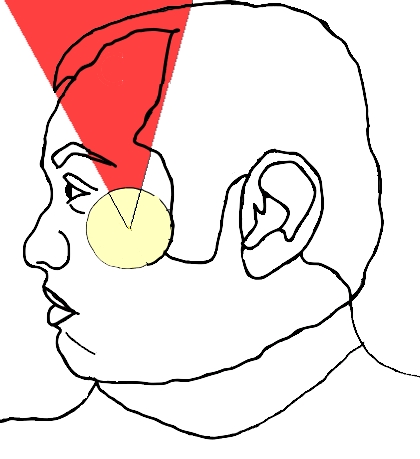}
        \includegraphics[width=21pt,frame]{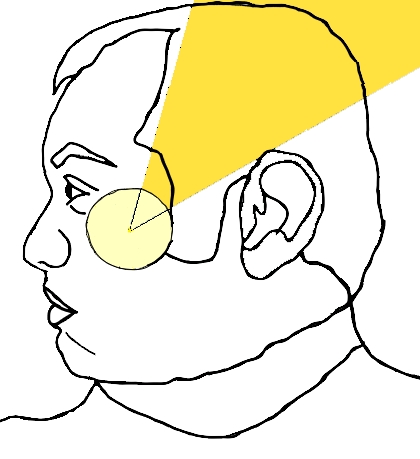}
        \\
        \includegraphics[width=21pt,frame]{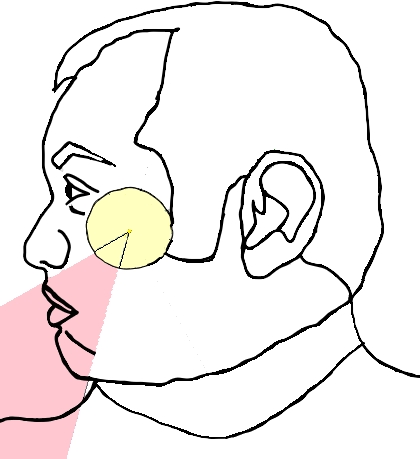}
        \includegraphics[width=21pt,frame]{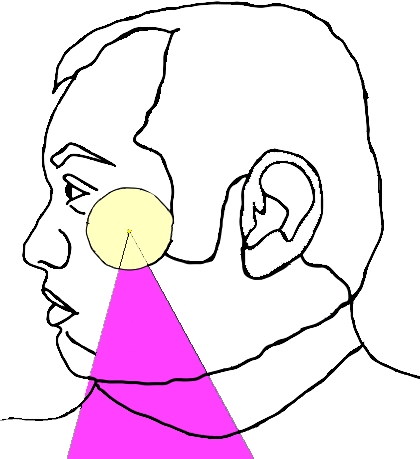}
        \includegraphics[width=21pt,frame]{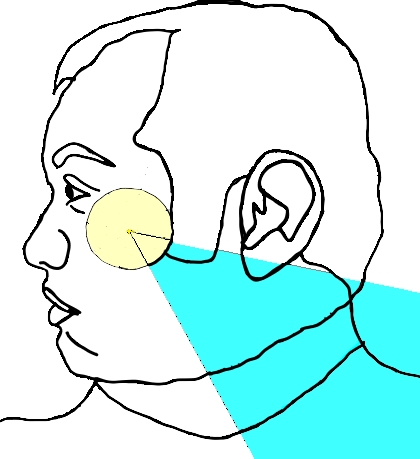}
        \includegraphics[width=21pt,frame]{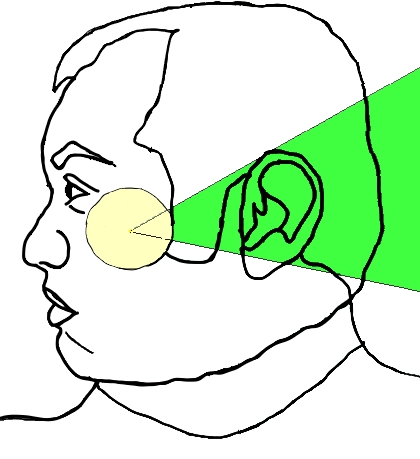}
        \\
        \includegraphics[width=10.0pt]{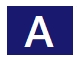}
        \includegraphics[width=10.0pt]{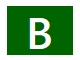}
        \includegraphics[width=10.0pt]{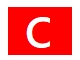}
        \includegraphics[width=10.0pt]{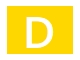}
        \includegraphics[width=10.0pt]{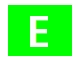}
        \includegraphics[width=10.0pt]{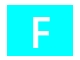}
        \includegraphics[width=10.0pt]{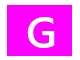}
        \includegraphics[width=10.0pt]{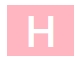}
        \\
        \vspace{-0.15cm}
        \subcaption[figure]{8 Midair}   
        \label{fig:midair_8R2R_in_gesture_region_section}                              
    \end{minipage}%
    \hfill\\
    \vspace{0.1cm} 
    \begin{minipage}{50pt}
        \centering
        \includegraphics[width=21pt,frame]{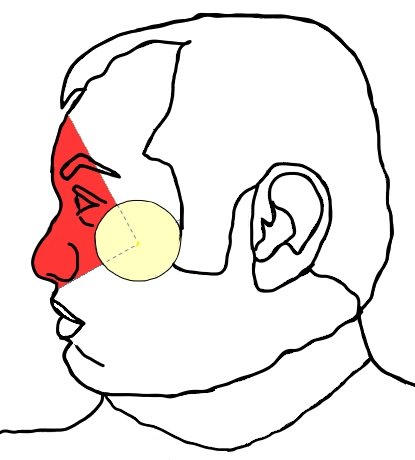}
        \includegraphics[width=21pt,frame]{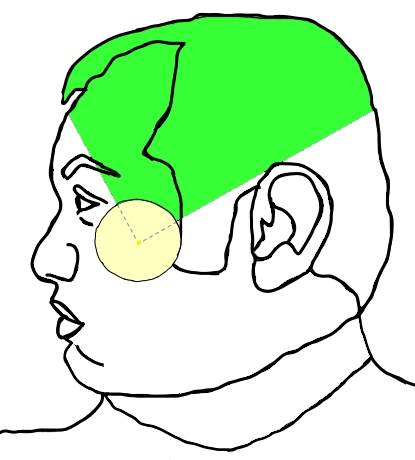}
        \\
        \includegraphics[width=21pt,frame]{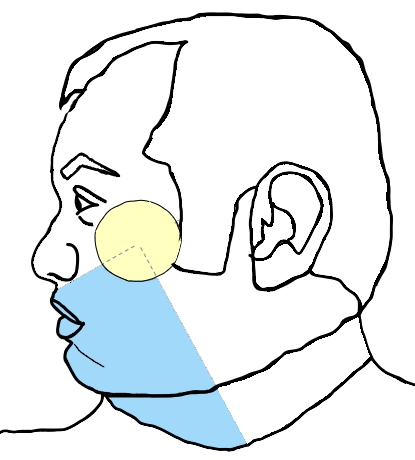}
        \includegraphics[width=21pt,frame]{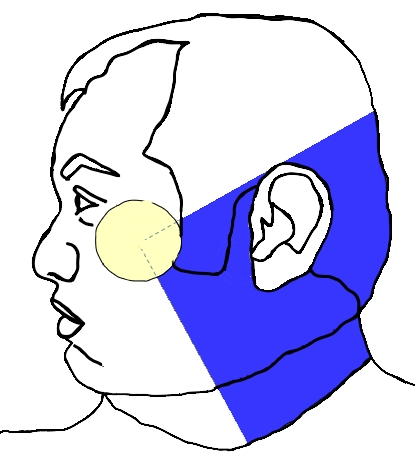}
        \\
        \includegraphics[width=10.0pt]{figures/01_Colored_Segmentation/10_Colors_4R2R_A.jpg}
        \includegraphics[width=10.0pt]{figures/01_Colored_Segmentation/10_Colors_4R2R_B.jpg}
        \includegraphics[width=10.0pt]{figures/01_Colored_Segmentation/10_Colors_4R2R_C.jpg}
        \includegraphics[width=10.0pt]{figures/01_Colored_Segmentation/10_Colors_4R2R_D.jpg}
        \\
        \vspace{-0.15cm}
        \subcaption[figure]{4 Onskin}    
        \label{fig:onskin_4R2R_in_gesture_region_section}                  
    \end{minipage}%
    \hfill
    \begin{minipage}{10.5pt}\end{minipage}
    \begin{minipage}{75pt}
        \centering
        \includegraphics[width=21pt,frame]{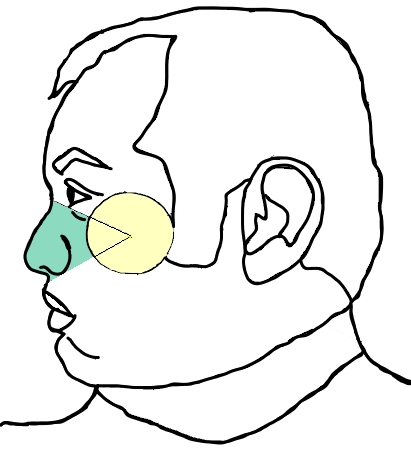}
        \includegraphics[width=21pt,frame]{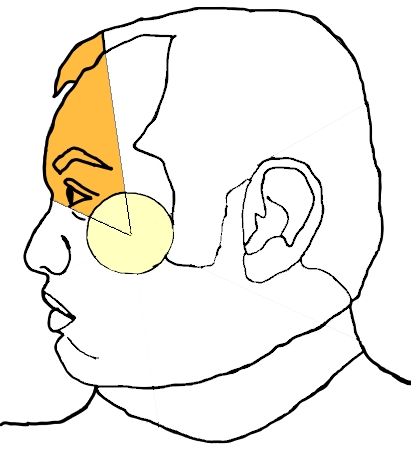}
        \includegraphics[width=21pt,frame]{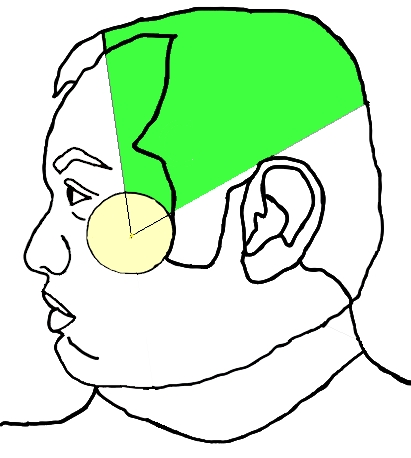}
        \\
        \includegraphics[width=21pt,frame]{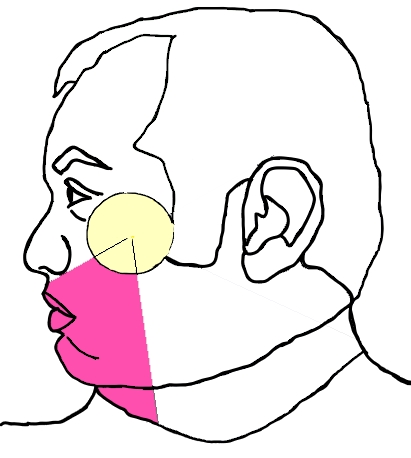}
        \includegraphics[width=21pt,frame]{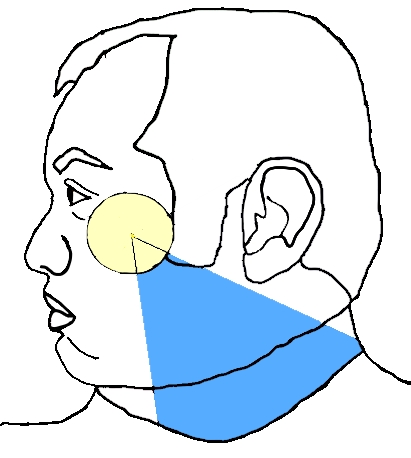}
        \includegraphics[width=21pt,frame]{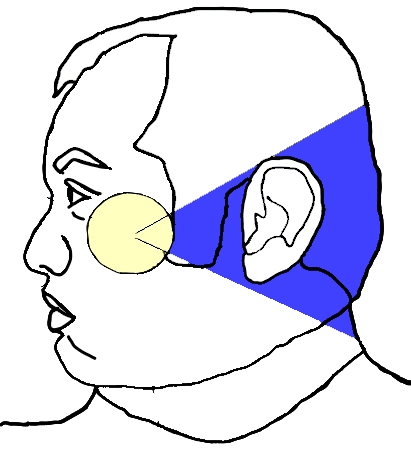}
        \\
        \includegraphics[width=10.0pt]{figures/01_Colored_Segmentation/10_Colors_6R2R_A.jpg}
        \includegraphics[width=10.0pt]{figures/01_Colored_Segmentation/10_Colors_6R2R_B.jpg}
        \includegraphics[width=10.0pt]{figures/01_Colored_Segmentation/10_Colors_6R2R_C.jpg}
        \includegraphics[width=10.0pt]{figures/01_Colored_Segmentation/10_Colors_6R2R_D.jpg}
        \includegraphics[width=10.0pt]{figures/01_Colored_Segmentation/10_Colors_6R2R_E.jpg}
        \includegraphics[width=10.0pt]{figures/01_Colored_Segmentation/10_Colors_6R2R_F.jpg}
        \\
        \vspace{-0.15cm}
        \subcaption[figure]{6 Onskin}   
        \label{fig:onskin_6R2R_in_gesture_region_section}                             
    \end{minipage}%
    \begin{minipage}{10.5pt}\end{minipage}  
    \hfill
    \begin{minipage}{100pt}
        \centering
        \includegraphics[width=21pt,frame]{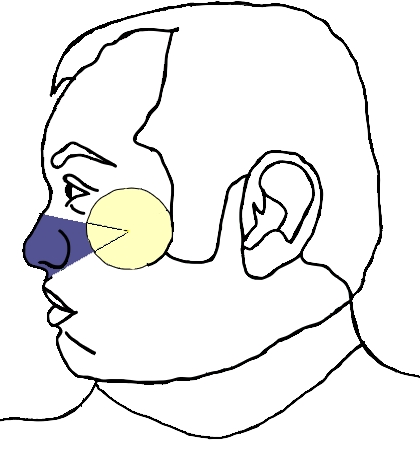}
        \includegraphics[width=21pt,frame]{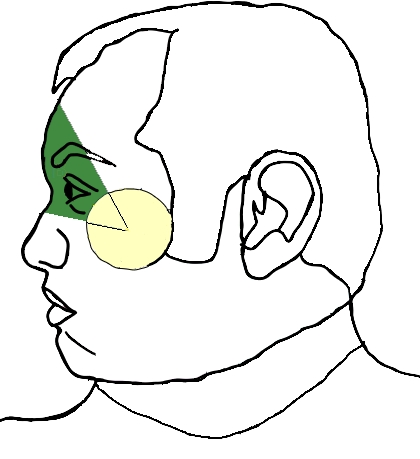}
        \includegraphics[width=21pt,frame]{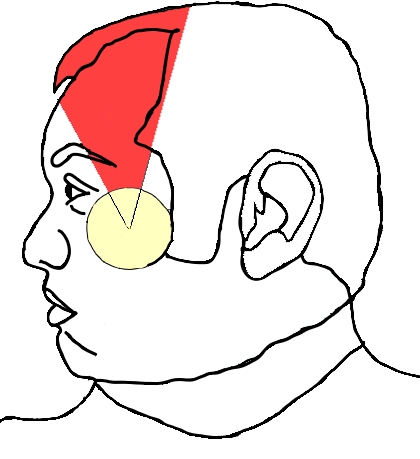}
        \includegraphics[width=21pt,frame]{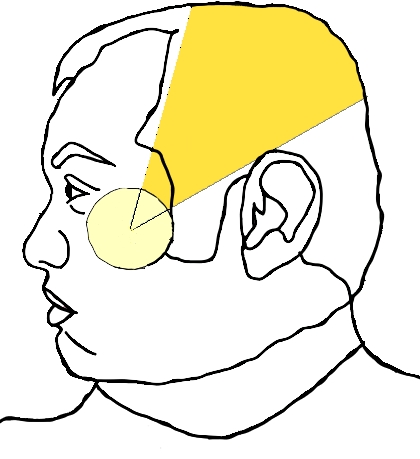}
        \\
        \includegraphics[width=21pt,frame]{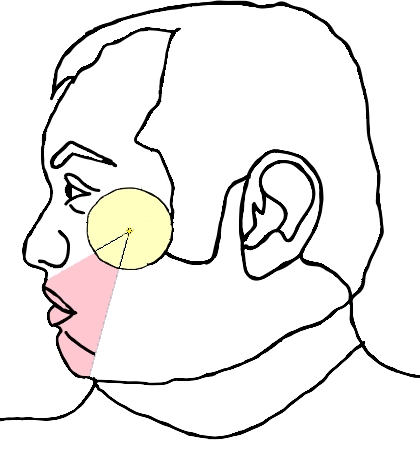}
        \includegraphics[width=21pt,frame]{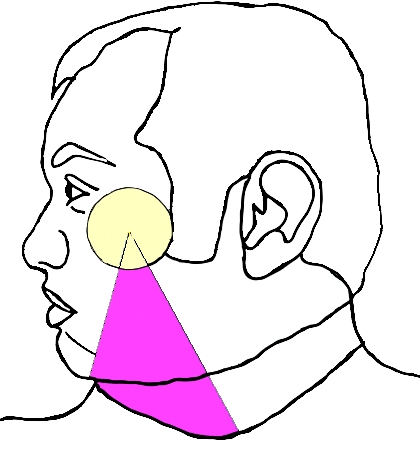}
        \includegraphics[width=21pt,frame]{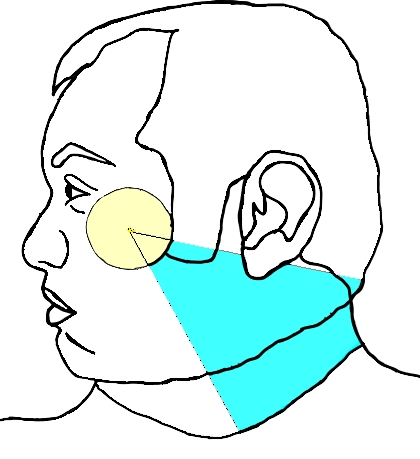}
        \includegraphics[width=21pt,frame]{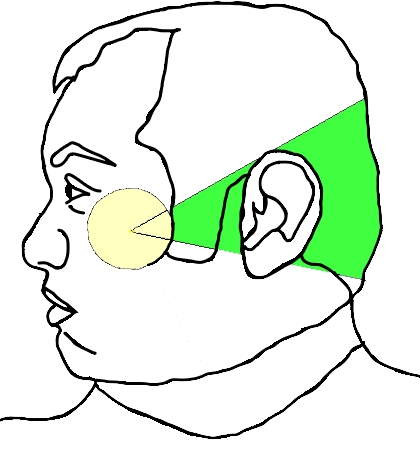}
        \\
        \includegraphics[width=10.0pt]{figures/01_Colored_Segmentation/10_Colors_8R2R_A.jpg}
        \includegraphics[width=10.0pt]{figures/01_Colored_Segmentation/10_Colors_8R2R_B.jpg}
        \includegraphics[width=10.0pt]{figures/01_Colored_Segmentation/10_Colors_8R2R_C.jpg}
        \includegraphics[width=10.0pt]{figures/01_Colored_Segmentation/10_Colors_8R2R_D.jpg}
        \includegraphics[width=10.0pt]{figures/01_Colored_Segmentation/10_Colors_8R2R_E.jpg}
        \includegraphics[width=10.0pt]{figures/01_Colored_Segmentation/10_Colors_8R2R_F.jpg}
        \includegraphics[width=10.0pt]{figures/01_Colored_Segmentation/10_Colors_8R2R_G.jpg}
        \includegraphics[width=10.0pt]{figures/01_Colored_Segmentation/10_Colors_8R2R_H.jpg}
        \\
        \vspace{-0.15cm}
        \subcaption[figure]{8 Onskin}   
        \label{fig:onskin_8R2R_in_gesture_region_section}                              
    \end{minipage}%
    \hfill\\
    \vspace{0.3cm}    
    \caption{4, 6, and 8 region segmentation on (onskin)~/~above (midair) a \textit{swipe focal point}~(\eg{}, cheekbone, marked in Yellow).} 
    \label{fig:-02-r2r-swipe-motion}
    \Description{Six images labeled (a) 4 Midair, (b) 6 Midair, (c) 8 Midair, (d) 4 Onskin, (e) 6 Onskin and (f) 8 Onskin show midair and onskin segmentation of 4, 6 and 8 region distribution using color codes for different regions. The swipe regions are distributed around the swipe focal point on or above the cheekbone, which is marked in yellow. In (a) 4 Midair, 4 midair regions around the cheekbone are highlighted and labeled using 4 colors. In (b) 6 Midair, 6 midair regions around the cheekbone are highlighted and labeled using 6 colors. In (c) 8 Midair, 8 midair regions around the cheekbone are highlighted and labeled using 8 colors. In (d) 4 Onskin, 4 onskin regions around the cheekbone are highlighted and labeled using 4 colors. In (e) 6 Onskin, 6 onskin regions around the cheekbone are highlighted and labeled using 6 colors. In (f) 8 Onskin, 8 onskin regions around the cheekbone are highlighted and labeled using 8 colors.}
\end{figure}

\endgroup
}

To address this research gap, we conducted a controlled within-subject study with 24 participants, analyzing 5{,}568 unimanual hand-to-face swipes performed both in midair above the ear, and on the skin around the ear for off-device interaction with an in-ear earable.
Rather than being limited by current gesture-sensing constraints of such devices, we adopted an  \textit{imaginary interface} approach to examine user-preferred motion independent of sensing limitations \cite{gustafson2010imaginary, Dezfuli_Palm_RC}, and employed an optical motion-capture system~\cite{arefinshimon2024fingertracker, Faleel2021HPUI, Cheng2024BodyTouch} for detailed analysis of unidirectional and angular single-stroke swipe gestures.
We examined axial (horizontal and vertical), non-axial (unidirectional but non-horizontal~/~vertical), and angular (\eg{}, ``L'', ``U'', and ``V'') swipes that start at a defined region, pass through a focal point near a facial landmark, and end at another region (Figure~\ref{fig:R2R-swipe-motion}).
We systematically studied spatial density effects on various swipe shapes by evaluating region segmentations of varying granularity (4, 6, and 8 regions) centered around the cheekbone focal region (Figure~\ref{fig:-02-r2r-swipe-motion}).

Our findings show that although participants preferred onskin swipes at lower region densities, unidirectional and angular midair swipes scaled more effectively as swipe region density increased, outperforming onskin interaction in speed and precision under highly segmented conditions. 
However, across both interaction spaces, usability degraded beyond the 6-region segmentation, with several onskin cheekbone regions exhibiting poor usability even in the 6-region condition. 

Taken together, these findings inform our proposed design guidelines for structuring practical subsets of unidirectional and angular swipe gestures for off-device earable interaction, which can enable more intuitive and efficient interaction within increasingly segmented above-neck input spaces. 
By characterizing the usability and scalability limits of midair and onskin swipe regions, our contribution extends prior research on unimanual off-device earable interaction and provides a foundation for richer, more scalable gesture vocabulary design for future in-ear earable systems.
}

\section{Related Work}
\label{section: Related Work}
{
\subsection{Off-Device Earable Interaction Techniques}
\label{subsection:-Related_Work-OffDeviceInteractionTechnique}
{
Earable interaction researchers have proposed 
\textit{facial expressions} \cite{Verma2021ExpressEar, Choi2022PPGface, song2022facelistener, Zhang2024Earphone}, 
\textit{silent commands}~\cite{digehsara2022user, jin2022earcommand, zeng2023msilent, shariff2022cw},
\textit{jaw motion-based gestures}~\cite{Bedri2015Stick, Khanna2021JawSense, srivastava2022muteit, Sun2021TeethTap},
\textit{head movement}~\cite{ferlini2019head, gashi2021hierarchical, Hu2023HeadTrack, radhakrishnan2021applying} and \textit{body movement-based gestures}~\cite{zhu2023char} as potential non-manual approaches for off-device interaction with earable devices. 
However, facial expression-based gestures have less acceptability in public settings \cite{koelle2020social} due to the possibility of attracting unwarranted, negative attention~\cite{eastwood2003negative}.
Voice or silent command-based interactions require memorization of a large set of unique commands and are not well suited for user interface navigational tasks~\cite{farringdon1999co} in other wearable or interactive devices. 
After repeated usage, such gestures can generate jaw and facial muscle fatigue among end-users.
Large whole-body movement-based interactions have been observed to attract negative attention in public spaces \cite{chen2020exploring} similar to approaches based on facial expression.
Although proposed head-movement interaction models are more subtle than whole-body interaction, whole-body and head-movement-based gestures can evoke fatigue and physical strain to gesture frequency and path length. 

Prior work demonstrates that manual (\ie,{} hand-based) gestures are often unconscious and spontaneous behaviors~\cite{barroso1980self, ekman1972hand, harrigan1987self, mueller2019self}, with hand-to-face motions (\eg,{} face touching) occurring as frequently as 15.7 to 23 times per hour~\cite{nicas2008study}. 
In natural, unconstrained settings, users have shown to prefer subtle above-neck unimanual gestures rather than larger full-body movements, especially when at least one hand is free~\cite{hossain2021exploring}. 
These gestures are also more socially acceptable, as large manual motions attract unwanted attention~\cite{koelle2020social}.
Above-neck gestures align with user expectations for in-ear earables, where one hand interacts with a single earbud, leaving the alternate hand available for other possible tasks.
Hand-to-face onskin or midair gestures also avoid the \textit{Gorilla Arm} fatigue caused by large midair movements away from the body ~\cite{hincapie2014consumed}.
These benefits make unimanual, hand-to-face gestures a popular choice for off-device earable input~\cite{cao2023earace, roddiger2022LitReview, choudhury2021earable}.

Earable systems capable of recognizing midair and onskin unimanual gestures have been proposed in \textit{FreeDigiter}\cite{metzger2004freedigiter}, \textit{SonicASL} \cite{jin2021sonicasl}, \textit{Earbuddy} \cite{xu2020earbuddy}, \textit{FaceSense} \cite{kakaraparthi2021FaceSense}, \textit{OESense} \cite{ma2021oesense}, \textit{EarGest} \cite{alkiek2022eargest}, \textit{EarAce} 
\cite{cao2023earace}, LeakyFeeder \cite{yang2025leakyfeeder}, EarHover \cite{EarHover_Suzuki_2024}, MAF \cite{yang2024maf}, \textit{UiEar} \cite{Zhao2024UiEar}, \textit{TinyssimoRadar} \cite{tinyssimoradar_ronco_2024}.
These works (revisited in detail in Section \ref{discussion: swipeRecognitionFeasibility}) show that it is viable to detect both gesture shape and gesture location for midair and onskin spaces above the neck for unimanual swipes, which sets the foundation of our study.
}

\subsection{Input Space Segmentation for Enabling Gesture Variations}
\label{subsection:-related work-Interaction-Space-Segmentation}
{
While complex gesture shapes and recognition techniques have been widely studied across wearable devices, interaction-space segmentation to enable gesture variation through location, orientation, and contortion remains underexplored for earables. 
Smartwatch research has examined dividing circular bezels into multiple regions to support location-based swipe variation \cite{wong2020exploring, rey2022understanding}, while Dezfuli \etal{}~\cite{Dezfuli_Palm_RC} segmented an imaginary hand~/~palm gesture space to enable onskin swipe and tap input. 
Li \etal{}~\cite{Li2021Depth} explored pie-menus for midair gesture interaction with remote displays using front-of-body manual gestures.
Similarly, Tu \etal{}~\cite{Tu2021ArmMenu} explored radial menu interaction for distance displays with lateral arm movements in front of the body.  
Gil \etal{}~\cite{gil2023thumbair} explored region placement and region count for in-air thumb typing on a commercial head-mounted display (HMD). 
Although these works target other wearables, their evaluation metrics and study designs inform gesture reuse across multiple regions in alternative interaction spaces.

Lissermann \etal{}~\cite{lissermann2014earput} studied region-based reuse of unimanual touches, taps, and grasp on and above the outer ear for earable interaction, but did not examine continuous swipes or gestures on regions beyond the outer ear.
Zhang \etal{}~\cite{zhang2023toward} examined onskin tap recognition across 5 outer-ear regions using a bone-conduction earable device. 
Elicitation studies by Hanae \etal{}~\cite{rateau2022leveraging} and Chen \etal{}~\cite{chen2020exploring} show that users prefer subtle unimanual gestures in onskin and midair areas above the neck, with single-finger horizontal~/~vertical swipes, taps, and two-finger pinches receiving the highest acceptance. 
Building on this work, Shimon \etal{}~\cite{arefinshimon2024fingertracker} extended outer-ear segmentation to midair and onskin spaces around the ear using non-overlapping regions, identifying key boundaries for the highest-agreement gesture primitives reported in \cite{rateau2022leveraging}.

However, the location-based swipe boundaries proposed in \cite{arefinshimon2024fingertracker} constrain natural swipe speed and gesture extent in order to avoid region boundary crossing. 
While prior earable work has largely focused on axial swipes, elicitation study results from \cite{rateau2022leveraging} demonstrate above-average acceptance of non-axial and angular swipes. 
Varying swipe orientation and angular contortion can therefore expand the earable input vocabulary in both midair and onskin spaces without relying on densely segmented, non-overlapping regions. 
Although non-axial and angular swipes have been studied on smartwatch screens and bezels \cite{wong2020exploring, rey2022understanding}, comparable analyses for hand-to-face midair and onskin swipes in off-device earable interaction remain unavailable, which we address in this paper.
}

\subsection{Gesture Analysis using Motion Capture Data}
\label{subsection:-related work-Vicon-based-Gesture-Analysis}
{
Beyond generating ground-truth data for on-device gesture recognition systems in development, motion-capture technologies such as \textit{Vicon}~\cite{Vicon_2023}, \textit{OptiTrack}~\cite{Optitrack_2024}, and \textit{Kinect}~\cite{Kinect_2024} are widely used to study gesture motion and showcase proof-of-concepts for novel interaction techniques~\cite{wang2020ev, qiu2016using}. 
By decoupling interaction design from sensing limitations, this imaginary-interface approach enables early investigation of emerging interaction spaces before robust on-device sensing support matures in mobile and wearable devices.

In the context of VR interaction, Faleel \etal{}~\cite{Faleel2021HPUI} explored \textit{Hand Proximate User Interfaces} (HPUIs) for head-mounted displays using \textit{Vicon} tracking when HMD hand tracking remained unreliable under self-occlusion, while Holfeld \etal{}~\cite{Holfeld2023Evaluating} later used \textit{Vicon} motion tracking to evaluate HPUI design guidelines independent of headset sensing limitations. 
Similarly, \textit{BodyTrack} by Cheng \etal{}~\cite{Cheng2024BodyTouch} used an \textit{OptiTrack}-based motion capture system to explore full-body interaction spaces and identify potential gestures for VR headsets before integrated full-body tracking hardware became widely available. 

Beyond VR systems,
Dezfuli \etal{}~\cite{Dezfuli_Palm_RC} used \textit{OptiTrack}-based motion tracking to explore an imaginary palm-based interactive surface for TV control. 
Shimon \etal{}~\cite{arefinshimon2024fingertracker} investigated location-based gesture reuse for unimanual off-device earable interaction using a \textit{Vicon} system instead of earable hardware, motivated by advances in above-neck gesture sensing for ear-mounted wearables.
Inspired by and grounded in these proof-of-concept approaches, we leverage the high-precision tracking capability of \textit{Vicon}-based motion capture system in this study to extensively investigate off-device earable swipes in both onskin and midair input spaces. 
We believe our collected data and anlaysis could be valuable for interaction space exploration to support the development of capable gesture recognition systems.
}

}

\section{Experiment}
\label{section:experiment}
{

{
\begingroup
\setlength{\floatsep}{-1ex} 
\setlength{\intextsep}{-1ex} 
\setlength{\abovecaptionskip}{-1ex}
\setlength{\belowcaptionskip}{-1ex}
\begin{figure*}[!tb]
    \centering
    \includegraphics[width=480pt]{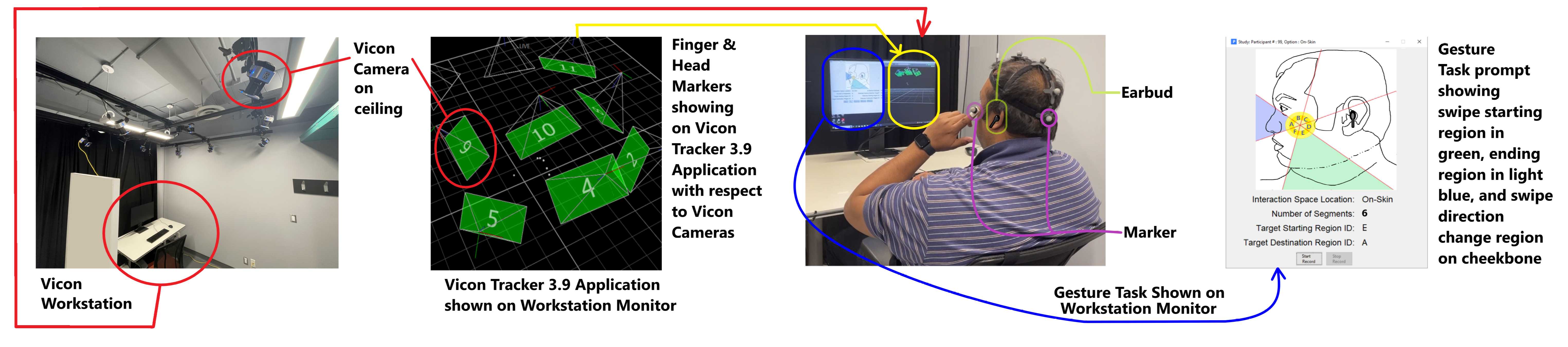}
    \vspace{-0.025cm}
    \caption{Different parts of study area and apparatus.}
    \label{fig:study_area_setup_consolidated}
    \Description{4 Images showing different parts of study area and apparatus. The leftmost figure shows the working area with the ceiling mounted Vicon cameras looking down on the Vicon Workstation area. Center left figure shows the Vicon camera views tracking head and body-mounted markers in Vicon Tracker 3.9 application. Center-right figure shows an user performing around-ear swipe based on gesture task shown on workstation monitor in a split-screen arrangement. The Vicon tracker application occupies the other half of the monitor screen showing tracking of head and finger worn markers. The rightmost figure shows the details of the gesture task shown in the figure to the left. It highlights the swipe starting area on the face in green, and swipe ending area in blue - giving directions on how to perform the unimanual hand-to-face swipe.}
\end{figure*}
\endgroup
}

\subsection{Experiment Overview}
\label{subsection:experiment-overview}
{
This study examines unidirectional and angular hand-to-face swipe gestures for off-device earable interaction using a marker-based motion capture system. 
We focus exclusively on swipe gestures, excluding other unimanual inputs such as taps and pinches due to their distinct motion characteristics and behavioral patterns \cite{Yamanaka2024Behavioral}. 
Gestures were performed in above-neck midair and onskin interaction spaces and analyzed within a head-centered coordinate frame to capture spatial articulation, performance, and workload characteristics. 
The study adopts an observational, user-centered approach to understand how users naturally articulate hand-to-face manual swipe gestures for off-device interaction with in-ear earable devices; and to inform the design of robust and usable gesture sets for such use-case.
}

\subsection{Research Questions}
\label{subsection:experiment-research-questions}
{


This work investigates how interaction space, region density, and spatial region location influence the articulation and usability of above-neck hand-to-face swipe gestures for off-device interaction with in-ear earables, motivating the following research questions:

\begin{itemize}
    \item \textbf{RQ1:} How do swipe performance and perceived workload differ between midair and onskin interaction spaces when the number of gesture regions is held constant?
    \item \textbf{RQ2:} How does increasing the number of available starting and ending regions affect swipe performance and perceived workload within a given interaction space?
    \item \textbf{RQ3:} Within a fixed gesture block (\ie{}, fixed interaction space and region density), how do swipe performance and subjective region preference vary across different starting and ending region locations?
\end{itemize}

Section \ref{subsection:experiment-measures} explains the metrics used to examine these research questions, and the rationale behind the metric choices.
}

\subsection{Study Design}
\label{subsection:experiment-study-design}
{
We employed a within-subject $2 \times 3$ factorial design with interaction space (midair vs.\ onskin) and region density (4, 6, and 8 regions) as independent variables. 
Each participant completed one gesture block for each factor combination, yielding 6 blocks in total. 
For each interaction space, participants performed 116 unique region-to-region swipe conditions, covering all possible starting and ending region pairings across region densities ($4\times4=16$, $6\times6=36$, $8\times8=64$). 
To control for order effects, the study was divided into two counterbalanced halves by interaction space, each comprising three gesture blocks, with swipe conditions presented in randomized order within each block.
}

{
\begingroup
\setlength{\floatsep}{-1ex} 
\setlength{\intextsep}{-1ex} 
\setlength{\abovecaptionskip}{-1ex}
\setlength{\belowcaptionskip}{-1ex}
\begin{figure*}[!tb]
    \centering
    \begin{subfigure}[t]{0.245\linewidth}
        \centering
        \captionsetup{justification=centering}        
        \includegraphics[height=70pt]{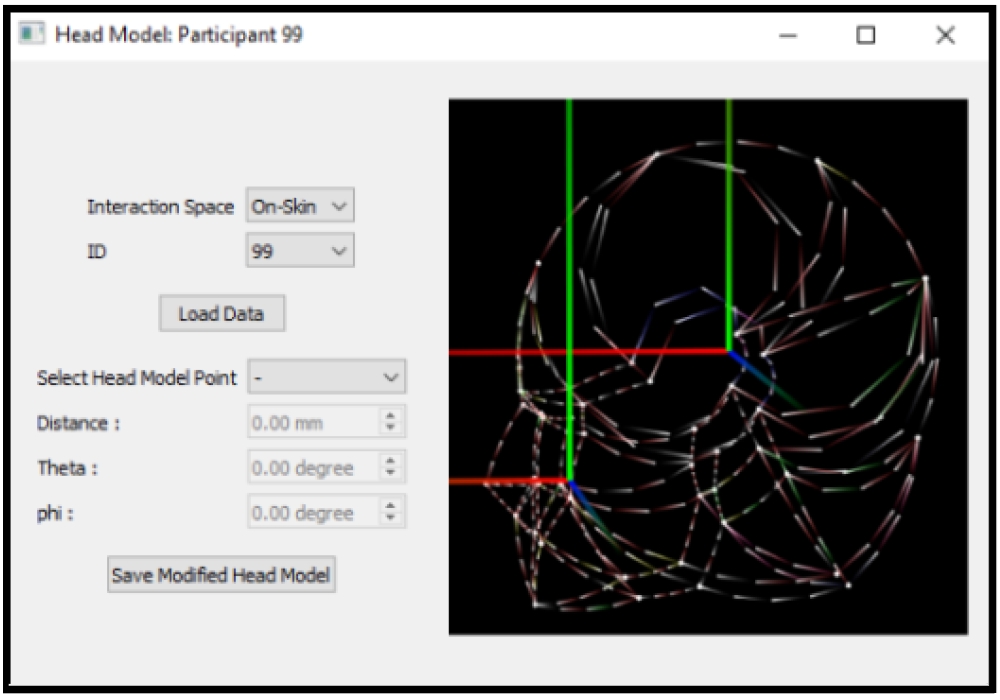}
        \vspace{-0.15cm}
        \caption{Head model\\adjustment window}
    \label{fig:different_views_custom_app_02}
    \end{subfigure}
    \hfill
    \begin{subfigure}[t]{0.245\linewidth}
        \centering
        \captionsetup{justification=centering}
        \includegraphics[height=70pt]{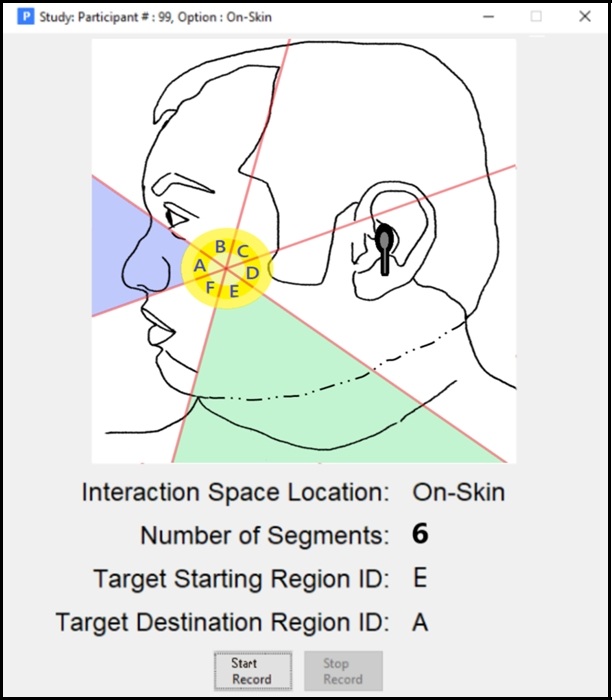}
        \vspace{-0.15cm}
        \caption{Task window}
    \label{fig:different_views_custom_app_04}
    \end{subfigure}
    \hfill
    \begin{subfigure}[t]{0.245\linewidth}
        \centering
        \captionsetup{justification=centering}
        \includegraphics[height=70pt]{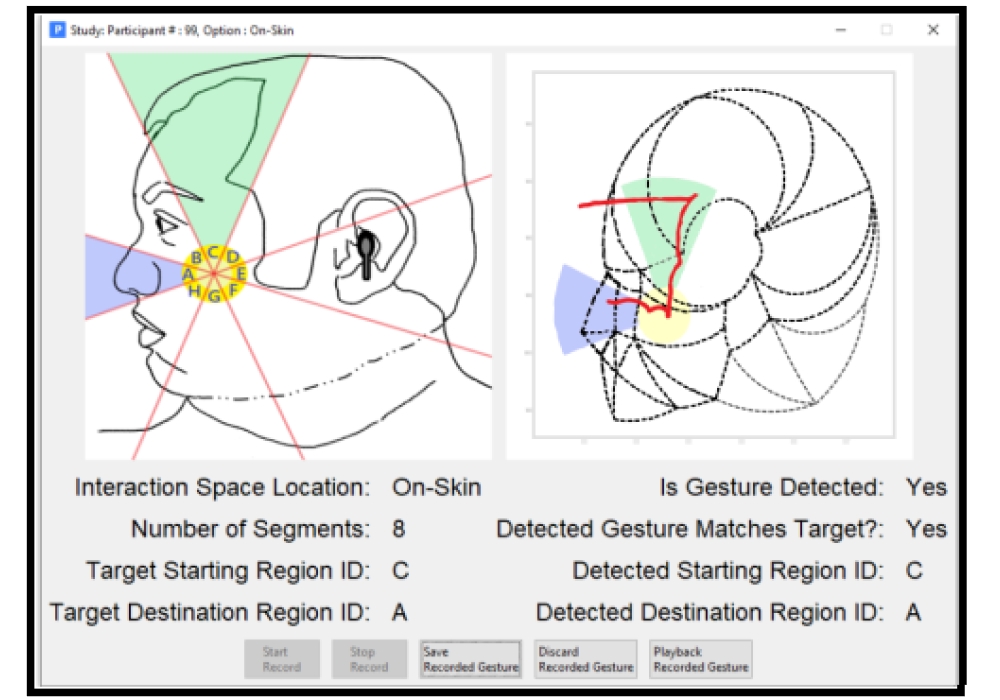}
        \vspace{-0.15cm}
        \caption{Task window\\with feedback}
    \label{fig:different_views_custom_app_01}
    \end{subfigure}
    \hfill
    \begin{subfigure}[t]{0.245\linewidth}
        \centering
        \captionsetup{justification=centering}   
        \includegraphics[height=70pt]{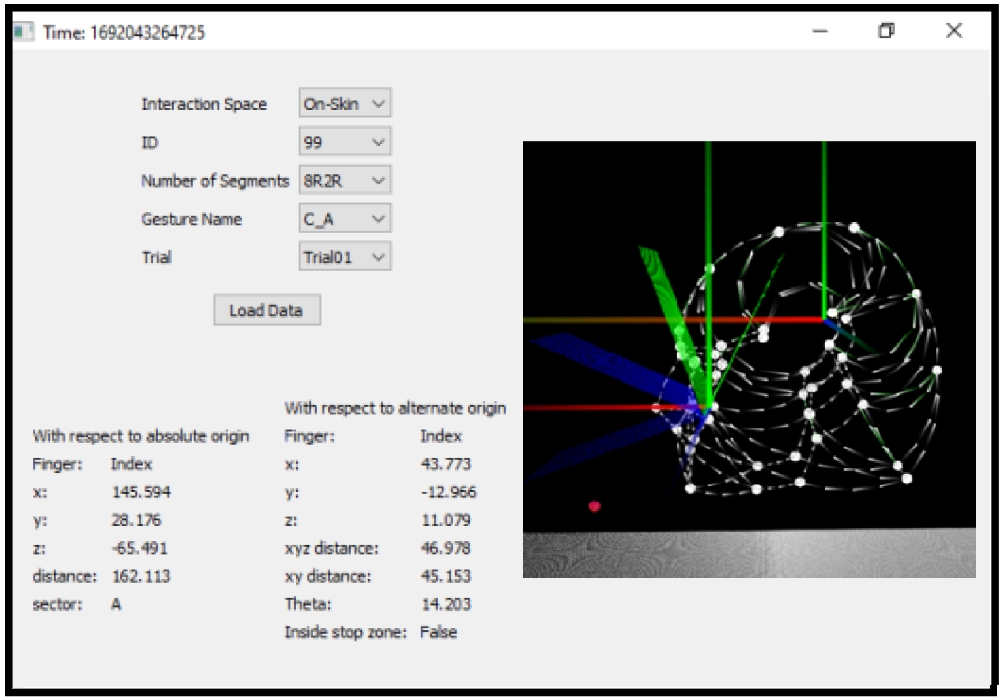}
        \vspace{-0.15cm}
        \caption{3D gesture\\playback window}
        \label{fig:different_views_custom_app_03}
    \end{subfigure}
    \vspace{0.3cm}
    \caption{Different views in custom gesture analysis application.}
    \label{fig:different_views_custom_app}
    \Description{Three images showing different views in custom gesture analysis application. The left image shows the task window with a red swipe trajectory overlay on the user's head model after a hand-to-face swipe.
    The center image features a head model adjustment window for participants to customize their models prior to data collection.
    The right image displays a playback window for recorded swipes, with a red dot indicating the swipe location on the head model during the trajectory.}    
\end{figure*}
\endgroup
}

\subsection{Participants}
\label{subsection:experiment-participants}
{
We recruited 24 participants (15 male, 9 female; age 18--45, $M=29.33$, $SD=7.20$) from the university and surrounding community. 
All participants had at least a high school diploma, with 22 (90\%) holding a bachelor’s degree or higher, including 11 (45.8\%) with a graduate degree.
All participants used their non-dominant hand for gesture interaction; 21 (87.5\%) were right-handed, and 3 (12.5\%) were left-handed.
Although 12 participants (50\%) reported regularly wearing eyewear, 21 participants completed the study without eyewear.
18 participants (75\%) owned wireless earbuds (0--20 hours~/~week of use, $M=11.58$, $SD=8.37$). 
None reported any hand, arm, or finger joint issues that could affect gesture performance.

This study, approved by the \textit{Institutional Review Board (IRB)}, was conducted over two 90-minute sessions on separate days to reduce fatigue, with participants receiving \$50 compensation.
In total, 5{,}568 gesture motions were collected across all participants, interaction spaces, and swipe conditions.
}

\subsection{Apparatus and Sensing Setup}
\label{subsection:experiment-apparatus}
{
The study used a ceiling-mounted $3\times3$ \textit{Vicon} motion-capture camera grid connected to a workstation (16-core AMD Ryzen~9 CPU, 32\,GB RAM) running Windows~10~Pro (Figure~\ref{fig:study_area_setup_consolidated}). 
The workstation ran \textit{Vicon Tracker~3.9} software alongside a custom Python/ PyQT-based gesture analysis application (Figure \ref{fig:different_views_custom_app}), with task instructions displayed on a 24-inch UHD monitor (Figure~\ref{fig:study_area_setup_consolidated}).
Participants were seated in front of the workstation, and fitted with a head strap holding five \textit{Vicon} markers along the sagittal plane~\cite{vella2008anatomy}, a marker on the index finger of the non-dominant hand, and a disconnected in-ear wireless earbud~\cite{aukey_ep_t27_official} on the non-dominant side as a prop for imagined earable-based gesture sensing.
Swipe gestures were performed unimanually with the non-dominant index finger, while the dominant hand controlled a keyboard and mouse, reflecting common non-dominant hand use for secondary input~\cite{dobbelstein2017effects, lee2010hand, arefinshimon2024fingertracker}.
A marked resting position on the table indicated the start and end points of each trial.

Marker trajectories were analyzed in a head-centered coordinate frame defined by the intersection of the sagittal plane and an eye-level horizontal plane, a reference frame used in analyze hand-to-face and near-head gestures \cite{arefinshimon2024fingertracker}. 
Analysis was limited to motion within a 30 cm-radius spherical volume centered on the head, corresponding to the close and mid distance ranges of around-device gesture spaces \cite{ahlstrom2014you} and capturing the spatial extent of above-neck manual interaction \cite{arefinshimon2024fingertracker, rateau2022leveraging}.
Motion outside this region was excluded to reduce confounding from large arm movements and to ensure consistent measurement of off-device gesture timing.
}

\subsection{Gesture Layout and Swipe Focal Point}
\label{subsection:experiment-gesture-layout}
{
The above-neck midair and onskin off-device interaction spaces were segmented into fan-shaped regions arranged around a shared \textit{swipe focal point} (Figure \ref{fig:-02-r2r-swipe-motion}), defined as a facial landmark at which swipe direction transitions occur.
Swipe gestures were modeled as rapid unimanual index-finger motions originating in a starting region, passing through the focal point, and terminating in an ending region (Figure \ref{fig:R2R-swipe-motion}).
The swipe layout was informed by prior hand-to-face interaction literature and a co-design session with (6) participants. 
A 4-region configuration was first adopted to support horizontal (anterior--posterior) and vertical (inferior--superior) swipes, identified as fundamental hand-to-face swipe primitives in prior work \cite{serrano2014exploring}.
Denser configurations with 6-regions and 8-regions were subsequently evaluated to enable angular (rotational) swipes and finer-grained input.

Prior research highlights a directional bias toward horizontal panning along the posterior–anterior axis between the ear and nose in face-centric gesture spaces~\cite{serrano2014exploring}. 
Building on these findings and related work on hand-to-face and earable interactions~\cite{arefinshimon2024fingertracker, xu2020earbuddy, sato_exploring_2024}, we identified the cheekbone as the optimal focal point for onskin swipes. 
Its high haptic salience, visual accessibility for tracking direction changes, and balanced spacing relative to facial landmarks—such as the eye corner, nostril, nasal bridge, and jawline—support the creation of evenly distributed region boundaries. 
In contrast, placing the focal point above the ear for midair swipes increased muscle strain during non-axial movements, making it unsuitable for angular swipes (e.g., “L”, “U”, and “V” shapes). 
Therefore, a shared focal region above the cheekbone was adopted for both onskin and midair gestures. 

Guided by these observations, swipe regions were oriented along the posterior–anterior axis defined between the nostril and the top of the outer ear helix for all region densities. 
To maintain swipe reliability, region density was capped at eight regions to avoid spans that were too narrow.
}

\subsection{Tasks}
\label{subsection:experiment-tasks}
{
Participants completed a series of single unimanual, unistroke swipe tasks between specified start and end regions around the cheekbone (Figure \ref{fig:different_views_custom_app_04}). 
Tasks included both unidirectional straight swipes and angular swipes requiring a directional change at the focal point. 
Each trial began and ended with the hand at the designated resting position on the workstation: the finger entered the head-centered gesture space (Section \ref{subsection:experiment-apparatus}), passed through the focal point, and returned to rest. 
Across gesture blocks, all valid region pairings were systematically tested.
}

\subsection{Procedure}
\label{subsection:experiment-procedure}
{
Each session followed a fixed procedure for administering the unidirectional and angular, above-neck swipe tasks described above. 
The study was conducted over two 90-minute sessions on separate days, each corresponding to one interaction space. 
The first session began with a study briefing, informed consent, and a demographic survey. 
Each session then proceeded to a task instruction phase, and an environment walkthrough specific to that session's interaction space.
Participants were instructed to avoid touching the in-ear earable, or the outer ear to prevent the ear-mounted device from being displaced.

Before data collection, participants completed a calibration phase to generate an individualized head model for swipe analysis, with optional adjustment to ensure accurate alignment with each participant’s facial features (Figure~\ref{fig:different_views_custom_app_02}). 
Participants then completed counterbalanced gesture blocks presented by the custom application (Figure~\ref{fig:different_views_custom_app_04}). 
Each trial began and ended with the non-dominant hand resting at a designated table location, while recording was initiated and terminated using the dominant hand.

Immediate visual feedback of the recorded swipe trajectory was presented after each trial (Figures~\ref{fig:different_views_custom_app_01} and~\ref{fig:different_views_custom_app_03}), allowing participants to accept the trial, or initiate a retrial. 
Participants were instructed to base retrial decisions exclusively on the geometric correspondence between the trajectory and the individualized head model, rather than on any system-level gesture classification or recognition output. 
After each trial ended, feedback was displayed as a gesture trajectory visualization; however, no performance-related metrics were provided. 
This feedback mechanism functioned as a data quality control step, enabling detection and correction of occasional head-model misalignment or tracking artifacts. As a result, sensing- and calibration-related confounds were reduced while the influence on gesture articulation was minimized.

Trials lasted approximately 15--25 seconds, with short breaks between gesture blocks and additional rest permitted as needed.
At the end of each gesture block, participants completed a brief questionnaire and participated in a short Q\&A session. 
Each session concluded with a semi-structured interview focusing on swipe starting and ending regions, swipe shape characteristics, and overall gesture preferences. 
Following data collection, swipe trajectories were labeled using an integrated baseline recognizer within our custom gesture analysis application, informed by \textit{Vicon} tracking data, with labels subsequently verified through manual review prior to accuracy analysis.
}

}

\section{Measures and Analysis}
\label{section:measures-analysis}
{

{
\begingroup
\setlength{\floatsep}{-1ex} 
\setlength{\intextsep}{-1ex} 
\setlength{\abovecaptionskip}{-1ex}
\setlength{\belowcaptionskip}{-1ex}
\begin{figure}[!tb]
    \centering
    \includegraphics[keepaspectratio, width=240pt]{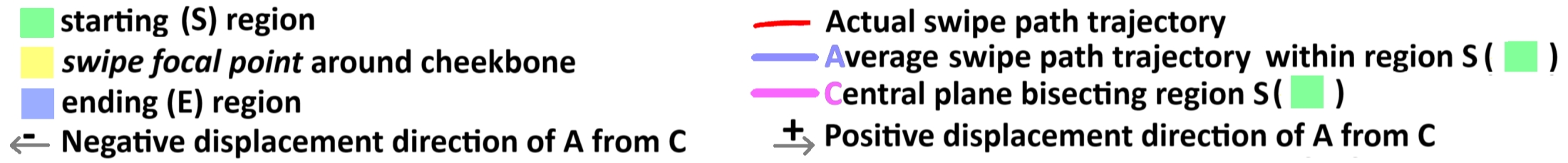}
    \begin{subfigure}{0.25\linewidth}
        \centering
        \captionsetup{justification=centering}    
        \includegraphics[width=0.9\linewidth]{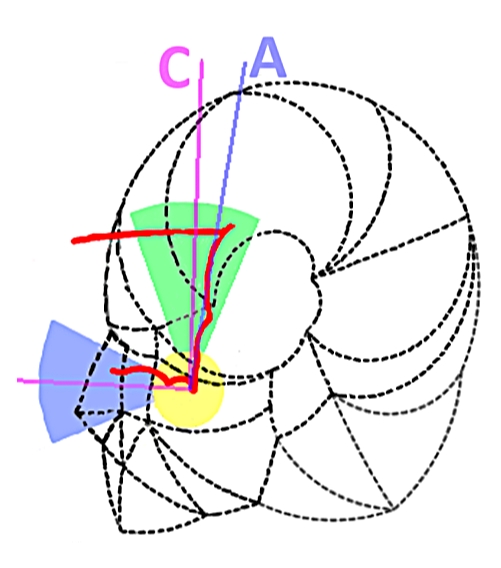}
        \vspace{-0.15cm}
        \caption{}
        \label{fig:Face-AngularDisplacement-a}
    \end{subfigure}
    \hfill
    \begin{subfigure}{0.25\linewidth}
        \centering
        \captionsetup{justification=centering}    
        \includegraphics[width=0.9\linewidth]{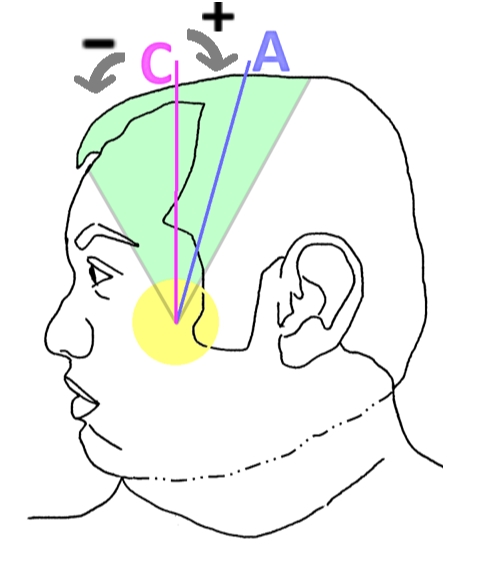}
        \vspace{-0.15cm}
        \caption{}
        \label{fig:Face-AngularDisplacement-b}
    \end{subfigure}
    \hfill
    \begin{subfigure}{0.4\linewidth}
        \centering
        \captionsetup{justification=centering}    
        \includegraphics[width=0.9\linewidth]{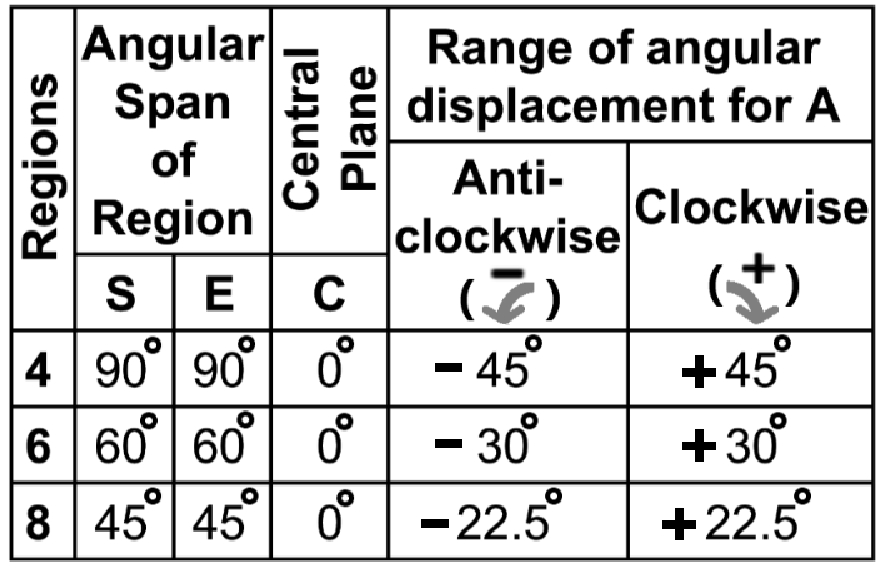}   
        \vspace{-0.15cm}
        \caption{}
        \label{fig:Face-AngularDisplacement-c}        
    \end{subfigure}
    \vspace{0.3cm}
    \caption{
    Swipe interaction regions and angular displacement on the left side of the face. 
    (a) A unimanual swipe trajectory (red line) passing through the starting region (\textit{Green}), focal region (\textit{Yellow}), and ending region (\textit{Blue}). 
    (b) Central plane (\textbf{C}) and average trajectory (\textbf{A}) of the swipe segment within the starting region, with angular displacement defined relative to the central plane. 
    (c) Region configurations for 4-, 6-, and 8-region distributions showing the angular span of each region and the corresponding range of angular displacement of A from the central plane C of the starting or ending region. 
    Greater angular displacement of \textbf{A} from \textbf{C} increases the likelihood that a swipe crosses into an incorrect starting or ending region.
    }
    \label{fig:Face-AngularDisplacement}
    \Description{Figure illustrating swipe interaction regions and angular displacement on the left side of a person's face. 
    (a) A left-side view of a head showing a unimanual swipe trajectory. The starting region is shown in green, the focal transition region in yellow, and the ending region in blue. A red line traces the swipe path from the starting region, through the focal region, to the ending region. 
    (b) The starting region with two reference lines: the central plane (C), shown as a pink vertical line, and the average swipe trajectory (A), shown as a blue line. Grey arrows indicate positive and negative angular displacement of the trajectory relative to the central plane. 
    (c) A table summarizing region configurations for 4-, 6-, and 8-region distributions. For each configuration, the table lists the angular span of each region and the corresponding range of angular displacement from the central plane of the starting or ending region.
    }
\end{figure}
\endgroup
}

\subsection{Measures}
\label{subsection:experiment-measures}
{
Following prior HCI research on swipe and stroke gesture performance, we evaluated gesture behavior using a combination of objective performance metrics and subjective workload and preference measures. 
We adopted \textit{accuracy} \cite{anthony2010lightweight, Vatavu2012Gestures, Wobbrock2007Gestures}, \textit{gesture articulation time} \cite{Cao2007Modeling, Wobbrock2007Gestures, remizova2022midair, Hafizi2023Invehicle}, and \textit{gesture path length} \cite{Leiva2021HowWeSwipe} as core performance metrics, complemented by perceived workload (NASA-TLX) \cite{xu2020earbuddy, Havlucu2017ItMade} and self-reported region preferences \cite{vatavu2011estimating, Wobbrock2009User}.  
To capture spatial deviations in swipe execution, we additionally introduced \textit{angular displacement}, inspired by the \textit{bending error (BE)} metric for stroke gestures proposed by Vatavu \etal{}~\cite{Vatavu2013Relative}, which quantifies the magnitude and direction of swipe skew relative to region geometry.

\begin{itemize}
    \item \textbf{DV1 – Accuracy:} Percentage (\%) of gestures correctly initiated and terminated within the intended starting and ending regions, defined onskin or in midair.
    
    \item \textbf{DV2 – Gesture time:} Total duration (ms) from the hand entering the head-centered gesture space (Section~\ref{subsection:experiment-apparatus}) to completing the swipe and exiting the space.
    
    \item \textbf{DV3 – Gesture path length:} Cumulative spatial length (mm) of the swipe trajectory traced by the index finger during gesture time (DV2).
    
    \item \textbf{DV4 – Angular displacement:} Angular deviation ($^\circ$) of the swipe trajectory (\ie{}, drift) within the starting (\textbf{S}) or ending (\textbf{E}) region relative to the region’s central reference plane (\textbf{C}) (Figure~\ref{fig:Face-AngularDisplacement}).
\end{itemize}

After each gesture block, perceived task workload was assessed using the NASA Task Load Index (NASA-TLX) on a 1–20 scale. 
Participants also reported preferences for swipe starting and ending regions using a single region-level rating (\eg{}., 1–4 for the 4-region segmentation), with higher values indicating stronger preference. 
RQ1 and RQ2 were evaluated using performance measures (DV1–DV4) together with NASA-TLX workload scores, while RQ3 examined region-level effects using accuracy (DV1), angular displacement (DV4), and subjective region ratings. 
Time- and distance-based measures (DV2, DV3) were excluded from RQ3 because differences in reach distance and movement direction across regions confound direct region-level comparison under a fixed head posture.
}

\subsection{Data Analysis}
\label{subsection:experiment-analysis}
{
Statistical analyses employed repeated-measures techniques appropriate for the within-subject study design. Normality assumptions were assessed using the \textit{Shapiro--Wilk} test. 
For two-level comparisons, paired \textit{t}-tests were used when normality was satisfied, and Wilcoxon signed-rank tests were used otherwise. 
For comparisons involving three levels, \textit{repeated-measures ANOVA} was applied to normally distributed data, while \textit{Friedman tests} were used when normality assumptions were violated. 
Where applicable, post-hoc pairwise comparisons were conducted using paired \textit{t}-tests or \textit{Wilcoxon signed-rank} tests with \textit{bonferroni correction} for multiple comparisons. 
A significance threshold of $\alpha = .05$ was used throughout.
}

}

\section{Results}
\label{section - Results}
{

{
Section~\ref{subsection:Results_RQ1_RQ2} presents quantitative analyses examining the effects of midair vs.\ onskin interaction space (\textbf{RQ1}) and region density (\textbf{RQ2}) on swipe performance and perceived workload. 
Building on these findings, Section~\ref{subsection:RQ3_QualitativeResults} outlines how swipe starting and ending regions, as well as individual swipe shapes, influence swipe interaction through combined quantitative and qualitative analyses (\textbf{RQ3}), focusing on region layouts identified as effective in the preceding analyses. 
Together, these results motivate a refined subset of swipe interactions and a modified 5-region layout proposed in Section~\ref{discussion: merging regions} to better support hand-to-face off-device interaction with earable devices.
}

\subsection{Effect of Interaction Space (RQ1) and Region Density (RQ2) on Swipe Gestures}
\label{subsection:Results_RQ1_RQ2}
{

{
{
\begingroup
\setlength{\floatsep}{-1ex} 
\setlength{\intextsep}{-1ex} 
\setlength{\abovecaptionskip}{-1ex}
\setlength{\belowcaptionskip}{-1ex}
\setlength{\tabcolsep}{2pt}

\begin{figure*}[!tb]    
    \centering
    \includegraphics[width=350pt]{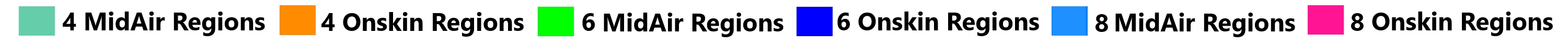}
    \includegraphics[width=500pt]{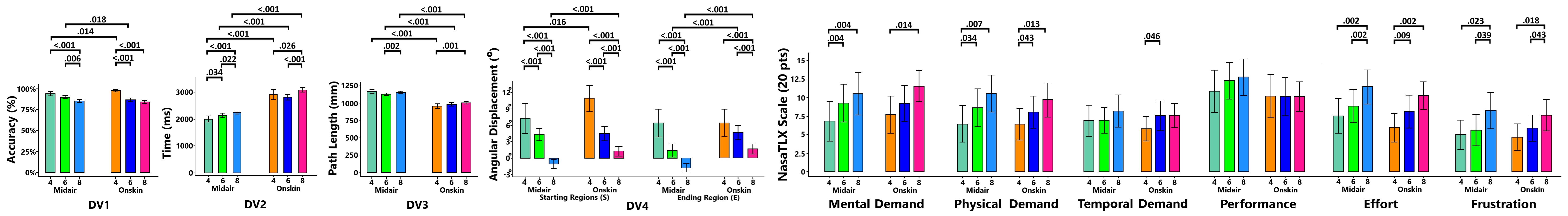}
    \vspace{0.2cm}
    \captionof{figure}{Gesture metrics (DV1--DV4) and NasaTLX workload for RQ1 and RQ2.}
    \label{fig:AllGestures_RQ1_RQ2}  
    \Description{Various bar graphs showing accuracy (DV1), time (DV2), path length (DV3), angular displacement (DV4) for the starting and ending regions on the top row, and various NasaTLX metrics such as mental demand, physical demand, temporal demand, performance, effort, frustration on the bottom row for six gesture blocks to address research questions RQ1 and RQ2. The statistically significant differences between blocks are highlighted using black and blue overhead lines above the bars. The 6 gesture blocks are color coded.}
\end{figure*}

\endgroup
}

In the 4-region configuration, onskin axial and L-shaped swipes through the cheekbone area significantly outperformed midair swipe counterparts in both accuracy and user preference.
Participants attributed this advantage to tactile feedback and clearer spatial constraints, which helped reduce accidental or unintentional contact with sensitive facial areas.
While midair swipes enabled faster and larger movements, they lacked stabilizing feedback.
However, as region density increased, individual region spans decreased for both interaction spaces. 
Onskin regions became increasingly constrained and placement-sensitive (\eg{}, near the eye corner), whereas midair swipes benefited from a larger available motion envelope. 
Accordingly, midair swipes became comparatively more effective at higher densities, reducing the relative benefit of tactile feedback (Figure~\ref{fig:AllGestures_RQ1_RQ2}).
Performance metrics, workload ratings, and participant feedback all indicate that accuracy decreases and effort and frustration increase when more than 6 regions are used in either interaction space, establishing a practical upper bound on region density.
We therefore recommend limiting swipe layouts to a maximum of 6 regions around the focal interaction area (Figure~\ref{fig:-02-r2r-swipe-motion}) to ensure reliable unidirectional and angular swipes.

The remainder of this subsection elaborates on these trends, with detailed statistical test results, including normality test results reported in Tables~\ref{tab:RQ1_Metrics} (RQ1) and \ref{tab:RQ2_GestureMetics} (RQ2) in appendix \ref{appendix:Detailed Statistical Results-RQ1-RQ2}.
Significant post hoc differences across interaction spaces and region densities have been summarized in Figure~\ref{fig:AllGestures_RQ1_RQ2}.
}

\subsubsection{Gesture Metrics}
\label{subsection:Results_RQ1_RQ2_Gesture_Metrics}
{

Accuracy \textbf{(DV1)} was significantly affected by region density for both interaction spaces (Friedman test; midair $\chi^2(2)=57.90$, onskin $\chi^2(2)=164.76$, both $p<.001$).
Post-hoc Wilcoxon signed-rank tests with Bonferroni correction showed accuracy decreases for onskin (4$\leftrightarrow$6; $p<.001$) and midair swipes (6$\leftrightarrow$8; $p=.006$), with overall declines from 4 to 8 regions in both spaces ($p<.001$).
Between interaction spaces, onskin swipes were more accurate at 4 regions (Wilcoxon $W=148.5$, $p=.014$), midair swipes at 6 regions ($W=5922.0$, $p=.018$), with negligible differences at 8 regions.
Accuracy declined with increasing region density for both interaction spaces, with a steeper reduction for onskin swipes.

Swipe time \textbf{(DV2)} was significantly affected by region density for both midair (Friedman $\chi^2(2)=55.20$, $p<.001$) and onskin swipes (Friedman $\chi^2(2)=20.33$, $p<.001$). 
For onskin swipes, time increased only at higher densities (4$\leftrightarrow$8: $p<.001$; 6$\leftrightarrow$8: $p=.026$), with no significant change from 4 to 6 regions, whereas midair swipe time increased at all density steps (4$\leftrightarrow$6: $p=.034$; 6$\leftrightarrow$8: $p=.022$; 4$\leftrightarrow$8: $p<.001$). 
Across all region densities, onskin swipes were consistently slower than midair swipes (Wilcoxon signed-rank, all $p<.001$).

Path length \textbf{(DV3)} differed significantly between interaction spaces, with midair swipes producing longer paths than onskin swipes across all region densities (Wilcoxon signed-rank, all $p<.001$). 
Region density also had a significant effect on path length for both midair (Friedman $\chi^2(2)=39.47$, $p<.001$) and onskin gestures (Friedman $\chi^2(2)=16.35$, $p<.001$). 
Post-hoc comparisons indicated that midair path length increased only from 6 to 8 regions ($p=.002$), whereas onskin path length increased only from 4 to 8 regions ($p<.001$).

Angular displacement \textbf{(DV4)} varied significantly with region density for both starting and ending regions in midair and onskin interaction (Friedman test; midair start $\chi^2(2)=148.49$, onskin start $\chi^2(2)=148.29$, midair end $\chi^2(2)=118.21$, onskin end $\chi^2(2)=46.41$; all $p<.001$).
Angular displacement decreased with increasing region density, with significant reductions across most comparisons ($p<.001$), except for the onskin ending region between 4 and 6 regions.
At swipe initiation, midair swipes were more closely aligned with the central plane of the region than onskin swipes in the 4-region ($W=27268$, $p=.016$) and 8-region ($W=439705$, $p<.001$) layouts. 
At swipe termination, onskin swipes deviated further from the central plane of the ending region than midair swipes in the 6-region ($W=120336$, $p<.001$) and 8-region ($W=411172$, $p<.001$) layouts. 
Participants reported that skewing during both initiation and termination was manageable up to 6 regions but overly constrained in the 8-region configuration due to interference from facial features such as the eye corner.
}

\subsubsection{NASA-TLX Perceived Workload}
\label{subsection:Results_RQ1_RQ2_NasaTLX}
{

\textbf{Mental Demand} increased significantly with region density for both interaction spaces. 
For midair swipes, a significant effect was observed (Friedman $\chi^2(2)=14.395$, $p<.001$), with increases from 4 to 6 and from 4 to 8 regions (both $p=.004$). 
For onskin swipes, mental demand was also significantly affected (Friedman $\chi^2(2)=11.341$, $p<.003$), driven primarily by the increase from 4 to 8 regions ($p=.014$).

\textbf{Physical Demand} significantly increased with increase in region density for both midair (Friedman $\chi^2(2)=13.300$, $p<.001$) and onskin swipes (Friedman $\chi^2(2)=10.173$, $p<.001$). 
Physical demand increased from 4 to 6 regions (midair: $p=.034$; onskin: $p=.043$) and from 4 to 8 regions (midair: $p=.007$; onskin: $p=.013$), with no additional increase from 6 to 8 regions.

\textbf{Temporal Demand} remained largely stable across region densities, with a significant increase  observed only for onskin swipes (Friedman $\chi^2(2)=10.203$, $p<.006$) between 4 and 6 regions ($p=.046$).

Self-reported \textbf{Performance} ratings remained stable for onskin swipes, while midair performance declined with increasing region density; however, these differences were not statistically significant.

Self-reported \textbf{Effort} increased significantly with region density for both interaction spaces (Friedman; midair $\chi^2(2)=17.82$, onskin $\chi^2(2)=20.00$, both $p<.001$). 
Effort increased from 4 to 6 regions for onskin swipes ($p=.009$) and from 6 to 8 regions for midair swipes ($p=.002$), with no significant change at the other step. 
Overall, effort increased from 4 to 8 regions for both interaction spaces (both $p=.002$).

Similarly, \textbf{Frustration} increased significantly with region density for both midair (Friedman $\chi^2(2)=10.265$, $p=.006$) and onskin interaction (Friedman $\chi^2(2)=13.618$, $p=.001$). 
Frustration did not increase significantly from 4 to 6 regions, but increased significantly from 6 to 8 regions (midair: $p=.039$; onskin: $p=.043$) and from 4 to 8 regions (midair: $p=.023$; onskin: $p=.018$). 

Overall, perceived workload remained manageable up to 6 regions, beyond which interaction became increasingly effortful and frustrating.
}

\subsection{Effect of Starting (or Ending) Position (RQ3) Choice on Swipe Gestures}
\label{subsection:RQ3_QualitativeResults}
{

This subsection summarizes how the choice of swipe starting and ending regions influences performance in midair and onskin interaction spaces, focusing on the 4- and 6-region segmentations identified as effective in Section~\ref{subsection:Results_RQ1_RQ2}. 
Given the large volume of 4- and 6-region result, this section presents key findings for each region configuration, while detailed statistical analyses and swipe-level breakdowns are provided in Appendix~\ref{appendix:Detailed Results RQ3}.

\subsubsection{4-Region Segmentation}
\label{subsection:RQ3_QualitativeResults-4Region}
{
{
\begingroup
\setlength{\floatsep}{-1ex} 
\setlength{\intextsep}{-1ex} 
\setlength{\abovecaptionskip}{-1ex}
\setlength{\belowcaptionskip}{-1ex}
\setlength{\tabcolsep}{2pt}
\begin{figure}[!tb]
    \centering
    \includegraphics[width = 0.99\linewidth]{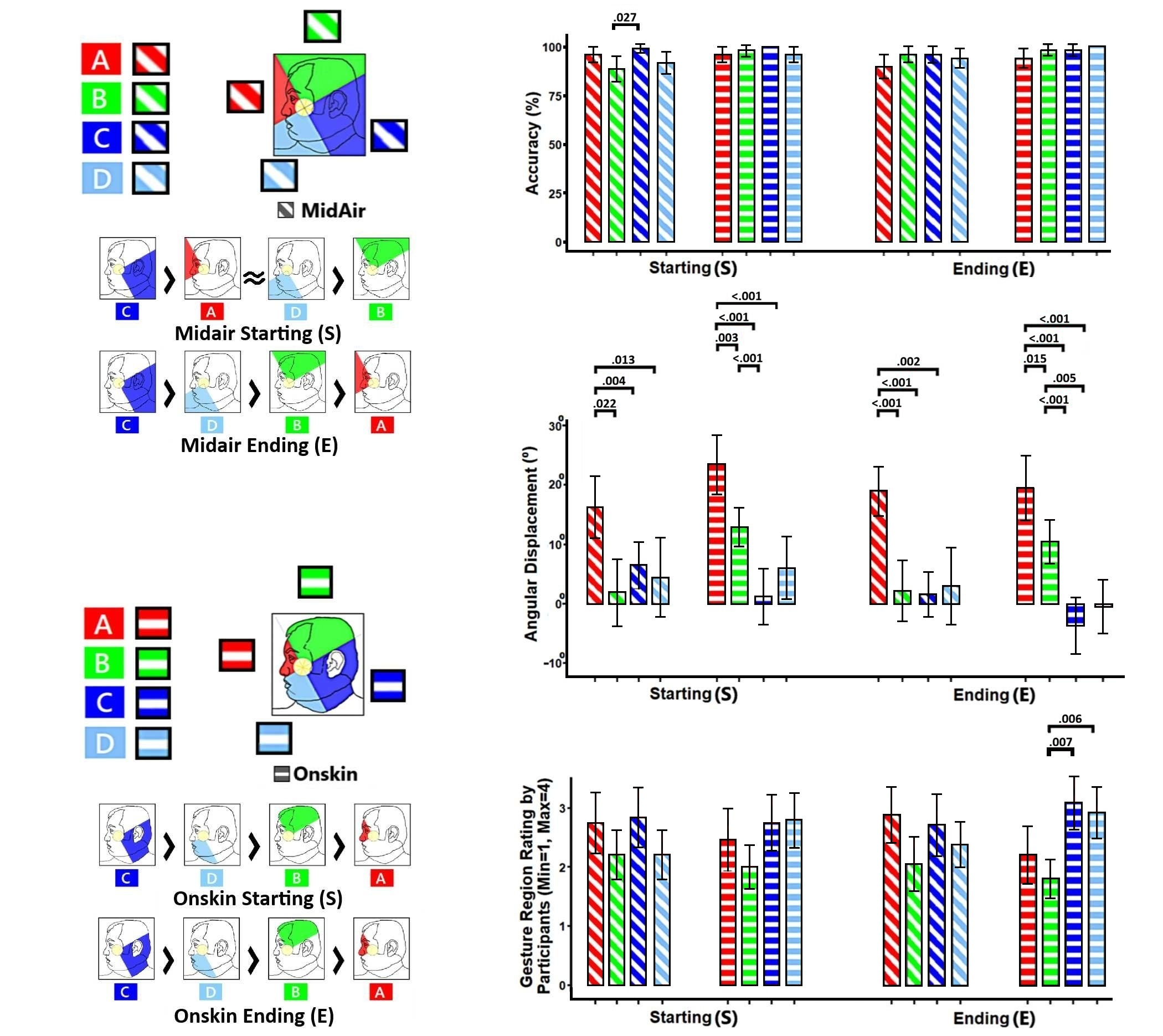}          
    \vspace{0.25cm} 
    \caption{Gesture region layout, metrics and ranking for 4 regions.}  
    \Description{Gesture region layout, metrics and ranking for 4 regions.}  
    \label{fig:GestureMetrics_4R2R_RQ3_all_Comparisons}   
\end{figure}

\begin{table}[!tb]
    \centering
    \renewcommand{\arraystretch}{1.1}
    \resizebox{240pt}{!}{
    \begin{tabular}{c|ccp{7cm}}
        \toprule
        \textbf{Space} & \textbf{Position} & \textbf{Ranked Regions} & \textbf{Key Evidence} \\
        \midrule
        
        \multirow{4}{*}{Midair}
        & \multirow{2}{*}{Start (S)}
        & \multirow{2}{*}{C $>$ A $\approx$ D $>$ B }
        & \multirow{2}{7cm}{C: $\uparrow$Accuracy, $\downarrow$Displacement; A: $\uparrow$displacement; good accuracy and ratings; within visual range} \\
        & & & \\ \cline{2-4}
        
        & \multirow{2}{*}{End (E)}
        & \multirow{2}{*}{C $>$ D $>$ B $>$ A}
        & \multirow{2}{7cm}{C~/~D: lowest displacement \\ 
        A: highest displacement} \\
        & & & \\
        
        \midrule
        \midrule
        
        \multirow{4}{*}{Onskin}
        & \multirow{2}{*}{Start (S)}
        & \multirow{2}{*}{C $>$ D $>$ B $>$ A}
        & \multirow{2}{7cm}{C~/~D: $\downarrow$Displacement; C~/~D: $\uparrow$Ratings\\
        A: low rating, accuracy, high displacement} \\
        & & & \\ \cline{2-4}
        
        & \multirow{2}{*}{End (E)}
        & \multirow{2}{*}{C $>$ D $>$ B $>$ A}
        & \multirow{2}{7cm}{C~/~D: $\downarrow$Displacement; C~/~D: $\uparrow$Ratings\\
        A: low rating, accuracy, high displacement} \\
        & & & \\
    \bottomrule
    \end{tabular}
    }
    \vspace{0.25cm}
    \caption{Relative performance of 4 swipe regions across interaction spaces. 
    Arrows indicate metric trends contributing to the ranking (Accuracy~/~Ranking $\uparrow$ higher \& Displacement $\downarrow$ lower is better). 
    Accuracy = DV1, Angular Displacement = DV4, Ratings = Subjective swipe region ratings.}
    \Description{Relative performance of 4 swipe regions across interaction spaces. 
    Arrows indicate metric trends contributing to the ranking ($\uparrow$ higher is better, $\downarrow$ lower is better). 
    Accuracy = DV1, Angular Displacement = DV4, Ratings = subjective swipe region ratings.}
    \label{tab:4region_performance_summary}
\end{table}
\endgroup
}

{
\begingroup
\setlength{\floatsep}{-1ex} 
\setlength{\intextsep}{-1ex} 
\setlength{\abovecaptionskip}{-1ex}
\setlength{\belowcaptionskip}{-1ex}
\setlength{\tabcolsep}{2pt}

\begin{figure}[t]
    \centering
    \includegraphics[width = 0.95\linewidth]{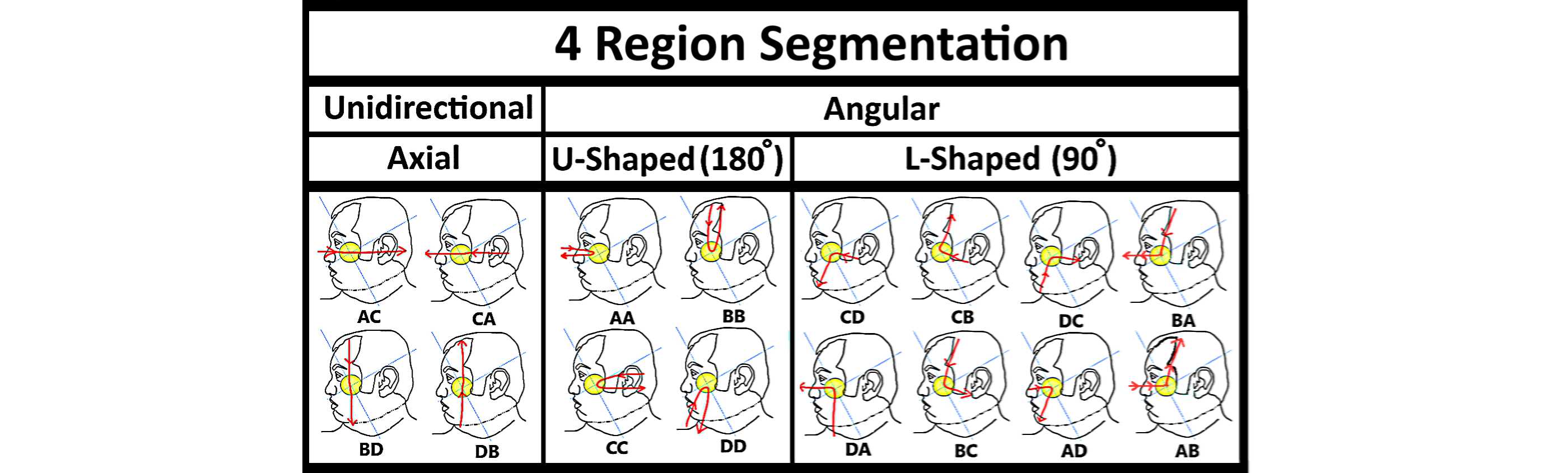}  
    \vspace{0.25cm} 
    \caption{Visual references for individual swipe shapes in 4-region segmentation.}  
    \Description{Visual references for individual swipe shapes in 4-region segmentation.} \label{fig:GestureMetrics_4R2R_RQ3_MainPaper_Visual_Swipe_References}     
\end{figure}

\begin{table}[t]
    \centering
    \renewcommand{\arraystretch}{1.1}
    \begin{subtable}{1.0\linewidth}
        \centering
        \resizebox{240pt}{!}{
        {
        \begin{tabular}{c c c c c}
        \toprule
        \textbf{Swipe Group} & 
        \textbf{Stable} & 
        \textbf{Moderate} & 
        \textbf{Unstable} & 
        \textbf{Observation} \\
        \midrule
        
        \begin{tabular}[c]{@{}c@{}}Axial\end{tabular} &
        \begin{tabular}[c]{@{}c@{}}AC, BD\\DB\end{tabular} &
        \begin{tabular}[c]{@{}c@{}}CA\end{tabular} &
        \begin{tabular}[c]{@{}c@{}}--\end{tabular} &
        \begin{tabular}[c]{@{}c@{}}Axial swipes show high stability.\end{tabular} \\\hline
        
        \begin{tabular}[c]{@{}c@{}}U-shaped\\($180^\circ$)\end{tabular} &
        \begin{tabular}[c]{@{}c@{}}BB, CC\\DD\end{tabular} &
        \begin{tabular}[c]{@{}c@{}}AA\end{tabular} &
        \begin{tabular}[c]{@{}c@{}}--\end{tabular} &
        \begin{tabular}[c]{@{}c@{}}Loops reliable near\\
        ear, temple and chin.\end{tabular} \\\hline
        
        \begin{tabular}[c]{@{}c@{}}L-shaped\\($90^\circ$)\end{tabular} &
        \begin{tabular}[c]{@{}c@{}}CB\end{tabular} &
        \begin{tabular}[c]{@{}c@{}}BC, DC\\CD\end{tabular} &
        \begin{tabular}[c]{@{}c@{}}AB, AD\\BA, DA\end{tabular} &
        \begin{tabular}[c]{@{}c@{}}Angular swipes show higher variability;\\
        especially when terminating around eye.\end{tabular} \\\hline

        \bottomrule
        \end{tabular}
        }
        }
        \vspace{0.05cm}
        \caption{Midair}
        \label{tab:midair_4region_swipe_shape_stability}
    \end{subtable}

    \begin{subtable}{1.0\linewidth}
        \centering
        \resizebox{240pt}{!}{
        {
        \begin{tabular}{c c c c c}
        \toprule
        \textbf{Swipe Group} & 
        \textbf{Stable} & 
        \textbf{Moderate} & 
        \textbf{Unstable} & 
        \textbf{Observation} \\
        \midrule
        
        \begin{tabular}[c]{@{}c@{}}Axial\end{tabular} &
        \begin{tabular}[c]{@{}c@{}}AC, CA\\BD, DB\end{tabular} &
        \begin{tabular}[c]{@{}c@{}}--\end{tabular} &
        \begin{tabular}[c]{@{}c@{}}--\end{tabular} &
        \begin{tabular}[c]{@{}c@{}}Axial swipes show high stability.\end{tabular} \\\hline
        
        \begin{tabular}[c]{@{}c@{}}U-shaped\\($180^\circ$)\end{tabular} &
        \begin{tabular}[c]{@{}c@{}}BB, CC\\DD\end{tabular} &
        \begin{tabular}[c]{@{}c@{}}AA\end{tabular} &
        \begin{tabular}[c]{@{}c@{}}--\end{tabular} &
        \begin{tabular}[c]{@{}c@{}}U-shaped Loops reliable near ear and chin\\
        Nose loop shows larger but managable drift.\end{tabular} \\\hline
        
        \begin{tabular}[c]{@{}c@{}}L-shaped\\($90^\circ$)\end{tabular} &
        \begin{tabular}[c]{@{}c@{}}--\end{tabular} &
        \begin{tabular}[c]{@{}c@{}}AD, DC, CD\\BC, CB\end{tabular} &
        \begin{tabular}[c]{@{}c@{}}AB, BA\\DA\end{tabular} &
        \begin{tabular}[c]{@{}c@{}}Angular swipes involving\\
        region \textbf{A} showed weakest stability.\end{tabular} \\\hline        
        \bottomrule
        \end{tabular}
        }
        }
        \vspace{0.05cm}
        \caption{Onskin}
        \label{tab:onskin_4region_swipe_shape_stability}
    \end{subtable}    
    \vspace{-0.4cm}
    \caption{Swipe shape (Figure~\ref{fig:GestureMetrics_4R2R_RQ3_MainPaper_Visual_Swipe_References}) stability under 4-region segmentation derived from accuracy (DV1), angular displacement (DV4), and subjective ratings.}
    \Description{Swipe shape (Figure~\ref{fig:GestureMetrics_4R2R_RQ3_MainPaper_Visual_Swipe_References}) stability under 4-region segmentation derived from accuracy (DV1), angular displacement (DV4), and subjective ratings.}
    \label{tab:4region_swipe_shape_stability}
\end{table}

\endgroup
}
The analysis of the \textbf{4-region} layout highlights distinct spatial interaction patterns for midair and onskin swipes above~/~around the cheekbone. 
As shown in Figure~\ref{fig:GestureMetrics_4R2R_RQ3_all_Comparisons}, the regions correspond to the anatomical contour: \textbf{A} near the nose–eye boundary, \textbf{B} at the temple, \textbf{C} around the ear and nape, and \textbf{D} along the cheek–chin area. 
Trends for swipe initiation and termination are summarized in Table~\ref{tab:4region_performance_summary}, with swipe-shape stability results summarized in Tables~\ref{tab:midair_4region_swipe_shape_stability} \&~\ref{tab:onskin_4region_swipe_shape_stability}, and shape trajectories illustrated in Figure~\ref{fig:GestureMetrics_4R2R_RQ3_MainPaper_Visual_Swipe_References} for visual reference.  
Detailed statistical results are discussed in Appendix~\ref{section:Results_RQ3_4R2R_Midair} \& ~\ref{section:Results_RQ3_4R2R_Onskin}.

Both region and swipe-level analyses indicate that \textbf{axial swipes} (\eg{}, \textbf{AC}, \textbf{CA}, \textbf{BD}, and \textbf{DB}) exhibit consistent performance across interaction spaces (Figure \ref{fig:GestureMetrics_4R2R_RQ3_MainPaper_Visual_Swipe_References}). 
\textbf{U-shaped swipes} around the ear and cheek regions (\textbf{BB}, \textbf{CC}, and \textbf{DD}) demonstrate strong stability and user preference. 
Conversely, U-shaped swipes near the nose (\textbf{AA}) showed increased angular drift, due to users' reported aversion to the sensitive ocular area. 
Despite this variation, participants generally found the 4-region layout manageable, especially for midair swipes.

Variability was more pronounced for \textbf{angular L-shaped swipes} involving the nose (\eg{}, \textbf{AB}, \textbf{BA}, and \textbf{DA}), with users having difficulty maintaining consistent trajectories near the eye corners, particularly when swipes reached the corner of the eye. 
Many participants felt more comfortable executing these swipes in midair, as this approach provided a greater spatial buffer from the eyes compared to onskin swipes.

Notably, subjective ratings in the swipe-level analysis indicated low preference for certain \textbf{L-shaped swipes beyond the field of view} (\ie{}, between \textbf{B–C} and \textbf{C–D}), despite their low drift and high recognition accuracy. 
Participants expressed discomfort when onskin swipes crossed the hair region during \textbf{B–C} swipes. 
In midair interactions, lower ratings were primarily linked to increased wrist bending and the necessity to complete significant portions of the swipe outside the line of sight.
}

\subsubsection{6-Region Segmentation}
\label{subsection:RQ3_QualitativeResults-6Region}
{

{
\begingroup
\setlength{\floatsep}{-1ex} 
\setlength{\intextsep}{-1ex} 
\setlength{\abovecaptionskip}{-1ex}
\setlength{\belowcaptionskip}{-1ex}
\setlength{\tabcolsep}{3pt}
\begin{figure}[t]
    \centering
    \includegraphics[width = 0.99\linewidth]{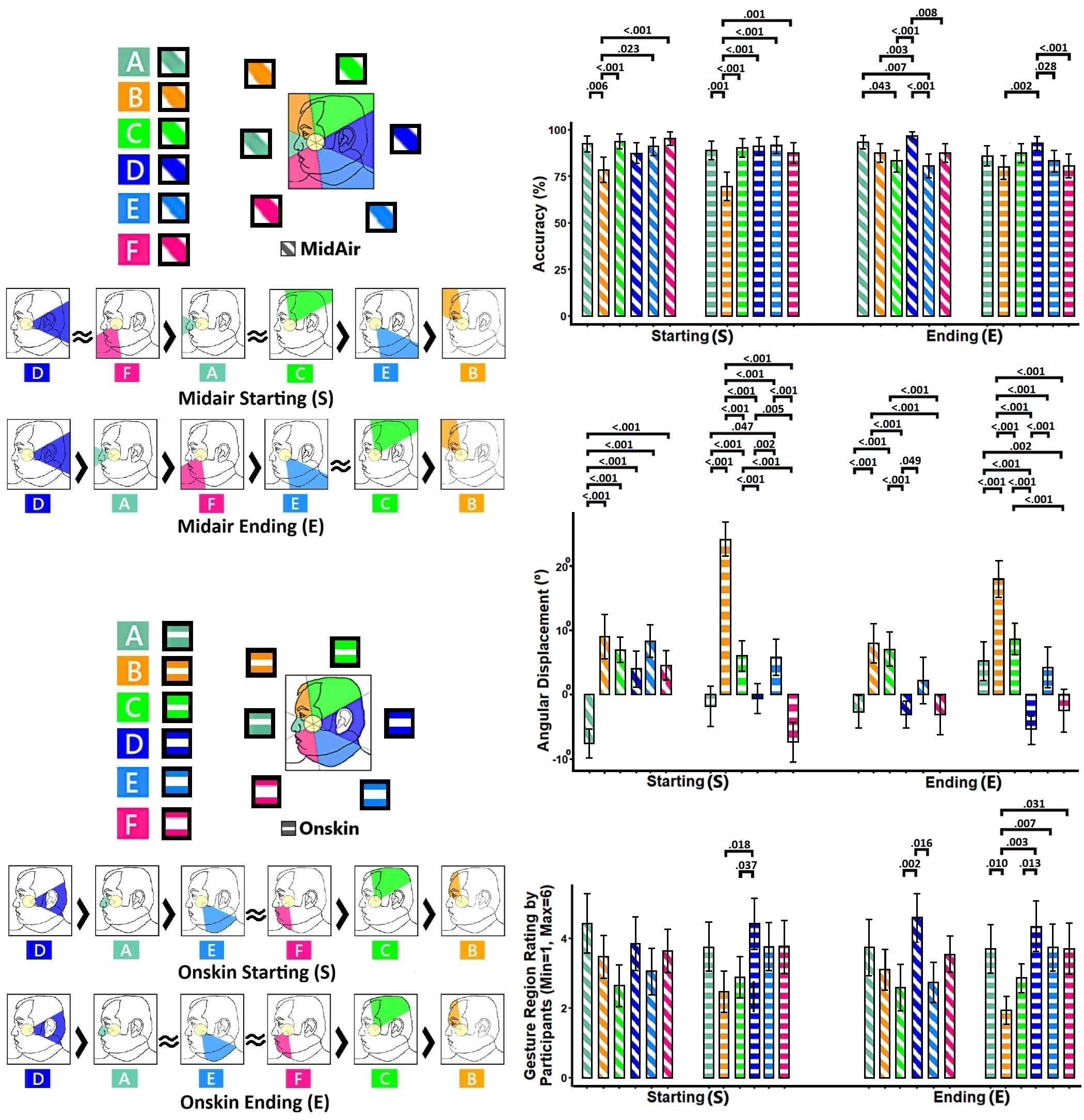}          
    \vspace{0.25cm} 
    \caption{Gesture region layout, metrics and ranking for 6 regions.}  
    \Description{Gesture region layout, metrics and ranking for 6 regions.}  
    \label{fig:GestureMetrics_6R2R_RQ3_all_Comparisons}   
\end{figure}
\begin{table}[t]
    \centering
    \renewcommand{\arraystretch}{1.1}
    \resizebox{240pt}{!}{
    \begin{tabular}{c|c c l}
    \toprule
    \textbf{Space} & \textbf{Position} & \textbf{Ranked Regions} & \textbf{Key Evidence} \\
    \midrule
    
    \multirow{4}{*}{Midair}
    & Start (S)
    & D $\approx$ F $>$ A $\approx$ C $>$ E $>$ B
    & \begin{tabular}[c]{@{}l@{}}
    D~/~F: $\uparrow$Accuracy, $\downarrow$Displacement; stable initiation\\
    A~/~C: good accuracy within FOV, mild directional bias\\
    E: limited visual feedback\\B: ocular avoidance, lowest accuracy
    \end{tabular} \\\cline{2-4}
    
    & End (E)
    & D $>$ A $>$ F $>$ C $\approx$ E $>$ B
    & \begin{tabular}[c]{@{}l@{}}
    D: $\uparrow$Accuracy, $\downarrow$Displacement; stable termination\\
    A: strong visual guidance; F: moderate stability\\
    C~/~E: reduced separability;\\
    B: highest angular skew
    \end{tabular} \\
    
    \midrule
    \midrule
    
    \multirow{4}{*}{Onskin}
    & Start (S)
    & D $>$ A $>$ E $\approx$ F $>$ C $>$ B
    & \begin{tabular}[c]{@{}l@{}}
    D: $\uparrow$Ratings, $\downarrow$Displacement; most reliable tactile start\\
    A~/~F: combined visual + haptic feedback\\
    C: moderate stability\\
    B: ocular constraint, highest displacement
    \end{tabular} \\\cline{2-4}
    
    & End (E)
    & D $>$ A $\approx$ E $\approx$ F $>$ C $>$ B
    & \begin{tabular}[c]{@{}l@{}}
    D~/~A: lowest displacement and highest accuracy\\
    E~/~F: intermediate endpoints\\
    C: reduced separability\\
    B: highest skew and lowest ratings
    \end{tabular} \\
    
    \bottomrule
    \end{tabular}
    }
    \vspace{0.25cm}
    \caption{Relative performance of 6 swipe regions across interaction spaces. 
    Arrows indicate metric trends contributing to the ranking (Accuracy~/~ranking $\uparrow$ higher \& Displacement $\downarrow$ lower is better). Accuracy = DV1, Angular Displacement = DV4, Ratings = subjective swipe region ratings.}
    \Description{Relative performance of 6 swipe regions across interaction spaces. 
    Arrows indicate metric trends contributing to the ranking (Accuracy~/~ranking $\uparrow$ higher \& Displacement $\downarrow$ lower is better). Accuracy = DV1, Angular Displacement = DV4, Ratings = subjective swipe region ratings.}
    \label{tab:6region_performance_summary}
\end{table}
\endgroup
}
{
\begingroup
\setlength{\floatsep}{-1ex} 
\setlength{\intextsep}{-1ex} 
\setlength{\abovecaptionskip}{-1ex}
\setlength{\belowcaptionskip}{-1ex}
\setlength{\tabcolsep}{3pt}
\begin{figure}[t]
    \centering
    \includegraphics[width = 0.95\linewidth]{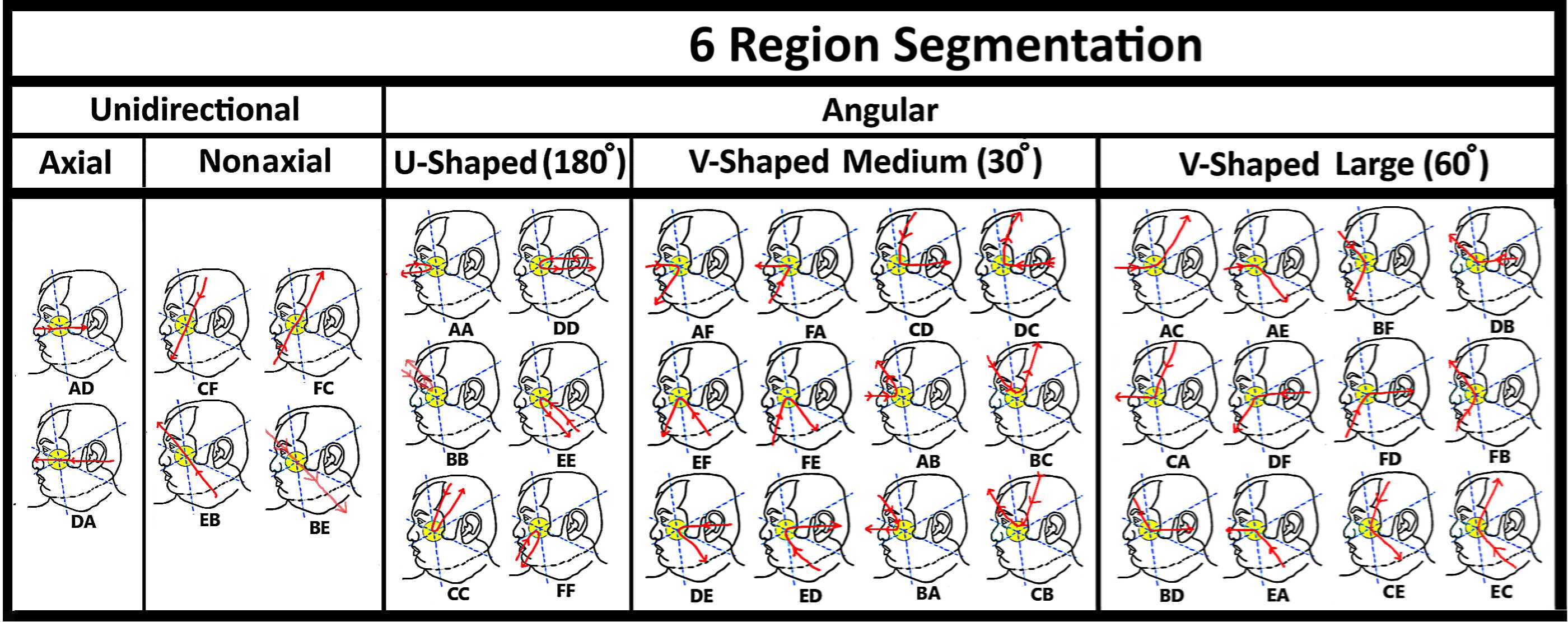}  
    \vspace{0.25cm} 
    \caption{Visual references for individual swipe shapes in 6-region segmentation.}  
    \Description{Visual references for individual swipe shapes in 6-region segmentation.}   
    \label{fig:GestureMetrics_6R2R_RQ3_MainPaper_Visual_Swipe_References}   
\end{figure}
\vfill
\begin{table}[t]
    \centering
    \renewcommand{\arraystretch}{1.1}
    \begin{subtable}{1.0\linewidth}
        \centering
        \resizebox{240pt}{!}
        {
        {
        \begin{tabular}{c c c c c}
        \toprule
        \textbf{Swipe Group} & 
        \textbf{Stable} & 
        \textbf{Moderate} & 
        \textbf{Unstable} & 
        \textbf{Observation} \\
        \midrule
        
        \begin{tabular}[c]{@{}c@{}}Axial\end{tabular} &
        \begin{tabular}[c]{@{}c@{}}AD\\DA\end{tabular} &
        \begin{tabular}[c]{@{}c@{}}--\end{tabular} &
        \begin{tabular}[c]{@{}c@{}}--\end{tabular} &
        \begin{tabular}[c]{@{}c@{}}Stable on cheek--chin axis;\\
        lowest deviation\end{tabular} \\\hline
        
        \begin{tabular}[c]{@{}c@{}}Non-axial\end{tabular} &
        \begin{tabular}[c]{@{}c@{}}CF\\FC\end{tabular} &
        \begin{tabular}[c]{@{}c@{}}EB\end{tabular} &
        \begin{tabular}[c]{@{}c@{}}BE\end{tabular} &
        \begin{tabular}[c]{@{}c@{}}Stable toward F;\\
        B, E weaker\end{tabular} \\\hline
        
        \begin{tabular}[c]{@{}c@{}}U-shaped\\($180^\circ$)\end{tabular} &
        \begin{tabular}[c]{@{}c@{}}AA, DD\\EE, FF\end{tabular} &
        \begin{tabular}[c]{@{}c@{}}BB\\CC\end{tabular} &
        \begin{tabular}[c]{@{}c@{}}--\end{tabular} &
        \begin{tabular}[c]{@{}c@{}}U-shapes mostly stable;\\
        B~/~C less consistent\end{tabular} \\\hline
        
        \begin{tabular}[c]{@{}c@{}}V-shaped\\Medium ($30^\circ$)\end{tabular} &
        \begin{tabular}[c]{@{}c@{}}CD, DC\\FA\end{tabular} &
        \begin{tabular}[c]{@{}c@{}}AF, FE, EF\\DE, ED\end{tabular} &
        \begin{tabular}[c]{@{}c@{}}AB, BA\\BC, CB\end{tabular} &
        \begin{tabular}[c]{@{}c@{}}Stable near D;\\
        Upper-face angular swipes weaken\end{tabular} \\\hline
        
        \begin{tabular}[c]{@{}c@{}}V-shaped\\Large ($60^\circ$)\end{tabular} &
        \begin{tabular}[c]{@{}c@{}}CA, DF\\BD, EA\end{tabular} &
        \begin{tabular}[c]{@{}c@{}}AC, AE, FD\\ FB, EC\end{tabular} &
        \begin{tabular}[c]{@{}c@{}}BF, DB\\CE\end{tabular} &
        \begin{tabular}[c]{@{}c@{}}Stable on D--F axis;\\
        long B/C/E crossings degrade.\end{tabular} \\
        
        \bottomrule
        \end{tabular}
        }
        }
        \vspace{0.05cm}
        \caption{Midair}
        \Description{Midair}
        \label{tab:midair_6region_swipe_shape_stability}
    \end{subtable}

    \begin{subtable}{1.0\linewidth}
        \centering
        \resizebox{240pt}{!}{
        {
        \begin{tabular}{c c c c c}
        \toprule
        \textbf{Swipe Group} & 
        \textbf{Stable} & 
        \textbf{Moderate} & 
        \textbf{Unstable} & 
        \textbf{Observation} \\
        \midrule
        
        \begin{tabular}[c]{@{}c@{}}Axial\end{tabular} &
        \begin{tabular}[c]{@{}c@{}}AD\\DA\end{tabular} &
        \begin{tabular}[c]{@{}c@{}}--\end{tabular} &
        \begin{tabular}[c]{@{}c@{}}--\end{tabular} &
        \begin{tabular}[c]{@{}c@{}}Stable on cheek--chin axis;\\
        lowest deviation\end{tabular} \\\hline
        
        \begin{tabular}[c]{@{}c@{}}Non-axial\end{tabular} &
        \begin{tabular}[c]{@{}c@{}}CF\end{tabular} &
        \begin{tabular}[c]{@{}c@{}}FC\end{tabular} &
        \begin{tabular}[c]{@{}c@{}}EB\\BE\end{tabular} &
        \begin{tabular}[c]{@{}c@{}}Stable toward F;\\
        swipes intersecting B unstable\end{tabular} \\\hline
        
        \begin{tabular}[c]{@{}c@{}}U-shaped\\($180^\circ$)\end{tabular} &
        \begin{tabular}[c]{@{}c@{}}AA, CC\\DD, EE\end{tabular} &
        \begin{tabular}[c]{@{}c@{}}FF\end{tabular} &
        \begin{tabular}[c]{@{}c@{}}BB\end{tabular} &
        \begin{tabular}[c]{@{}c@{}}Stable at A/D/E/F anchors;\\
        loops at B unstable\end{tabular} \\\hline
        
        \begin{tabular}[c]{@{}c@{}}V-shaped\\Medium ($30^\circ$)\end{tabular} &
        \begin{tabular}[c]{@{}c@{}}CD, DC, ED\\EF, FE\end{tabular} &
        \begin{tabular}[c]{@{}c@{}}FA, AF\\DE\end{tabular} &
        \begin{tabular}[c]{@{}c@{}}AB, BA\\BC, CB\end{tabular} &
        \begin{tabular}[c]{@{}c@{}}Stable near D;\\
        swipes crossing B unstable\end{tabular} \\\hline
        
        \begin{tabular}[c]{@{}c@{}}V-shaped\\Large ($60^\circ$)\end{tabular} &
        \begin{tabular}[c]{@{}c@{}}DF\\FD\end{tabular} &
        \begin{tabular}[c]{@{}c@{}}AC, AE, CA\\EA, CE, EC\end{tabular} &
        \begin{tabular}[c]{@{}c@{}}BD, BF\\DB, FB\end{tabular} &
        \begin{tabular}[c]{@{}c@{}}Stable on D--F axis;\\
        eye-crossing swipes unstable\end{tabular} \\
        
        \bottomrule
        \end{tabular}
        }
        }
        \vspace{0.05cm}
        \caption{Onskin}
        \Description{Midair}
        \label{tab:onskin_6region_swipe_shape_stability}
    \end{subtable}    
    \vspace{-0.4cm}
    \caption{Swipe shape (Figure~\ref{fig:GestureMetrics_6R2R_RQ3_MainPaper_Visual_Swipe_References}) stability under 6-region segmentation derived from accuracy (DV1), angular displacement (DV4), and subjective ratings.}
    \Description{Swipe shape (Figure~\ref{fig:GestureMetrics_6R2R_RQ3_MainPaper_Visual_Swipe_References}) stability under 6-region segmentation derived from accuracy (DV1), angular displacement (DV4), and subjective ratings.}
    \label{tab:6region_swipe_shape_stability}
\end{table}

\endgroup
}

\textbf{6-region} layout analysis reveals finer spatial distinctions beyond the coarse patterns observed in the 4-region layout. 
As shown in Figure~\ref{fig:GestureMetrics_6R2R_RQ3_all_Comparisons}, swipe regions are distributed around the cheekbone focal area: \textbf{A} above the nose, \textbf{B} near the outer eye corner, \textbf{C} above the temple, \textbf{D} above the ear, \textbf{E} near the mandibular angle, and \textbf{F} near the chin. 
Table~\ref{tab:6region_performance_summary} summarizes region-level performance for swipe initiation and termination, while Tables~\ref{tab:midair_6region_swipe_shape_stability} \&~\ref{tab:onskin_6region_swipe_shape_stability} summarize swipe-shape stability across interaction spaces. 
Figure~\ref{fig:GestureMetrics_6R2R_RQ3_MainPaper_Visual_Swipe_References} provides representative swipe trajectories, with detailed statistical comparisons reported in Appendix~\ref{section:Results_RQ3_6R2R_Midair} \& ~\ref{section:Results_RQ3_6R2R_Onskin}.

Taken together, the region- and swipe-level analyses reveal that the ear-adjacent region \textbf{D} and the chin-adjacent regions \textbf{F} and \textbf{A} (Figure \ref{fig:GestureMetrics_6R2R_RQ3_all_Comparisons}) produced the most stable trajectories across interaction spaces. 
In contrast, the eye-adjacent region \textbf{B} exhibited the lowest accuracy and user preference, primarily due to strong avoidance of the eye corner, particularly during onskin interactions.

Drift and avoidance patterns in regions \textbf{B} and \textbf{C} (Figure \ref{fig:GestureMetrics_6R2R_RQ3_all_Comparisons} and Table \ref{tab:6region_performance_summary}) highlight challenges in consistently distinguishing these upper-face areas midair. 
Their limited spatial extent and the presence of hair complicate onskin swipes, suggesting that these regions might be merged for interaction purposes.

In contrast, the lower-face regions along the cheekbone (\textbf{A}, \textbf{D}, \textbf{E}, and \textbf{F}) offer better spatial anchoring for onskin swipes. 
Thus, effective gesture sets should primarily focus on the lower half of the anterior–posterior axis between the nose and ear. Within this region, the metrics support treating \textbf{E} and \textbf{F} as distinct areas, especially for onskin interactions.

}
}

}

}

\section{Discussion}
\label{section: discussion}

{
\begingroup
\setlength{\floatsep}{-1ex} 
\setlength{\intextsep}{-1ex} 
\setlength{\abovecaptionskip}{-1ex}
\setlength{\belowcaptionskip}{-1ex}
\begin{figure}[!tb]
    \centering  
    \includegraphics[width=240pt]{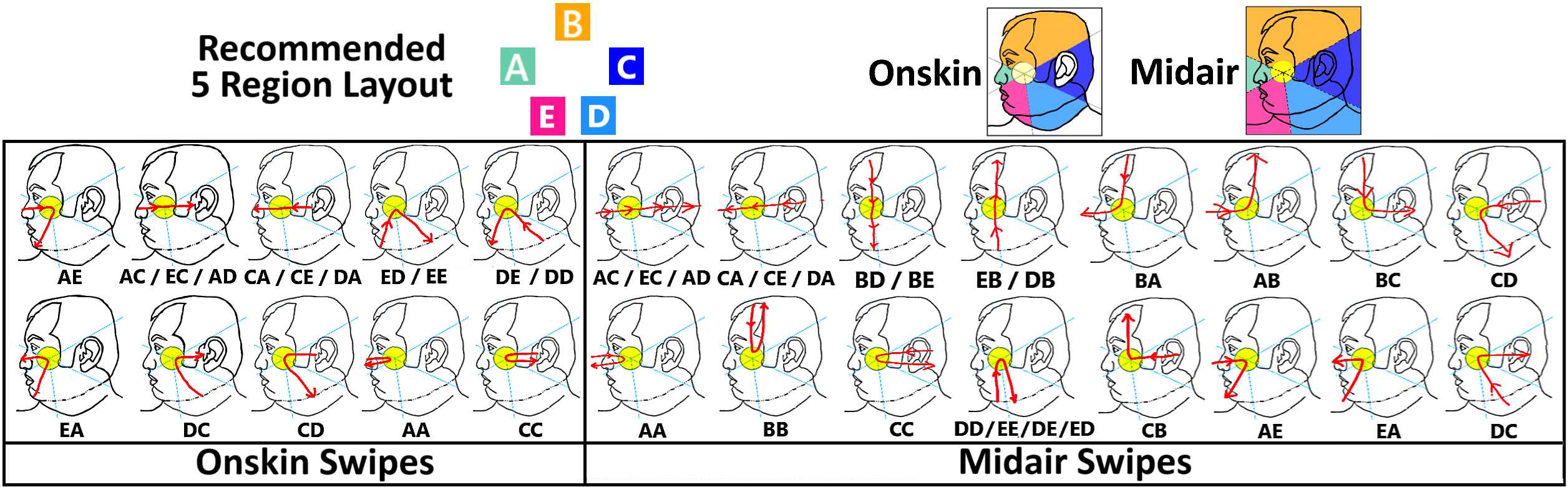}
    \vspace{-0.15cm}
    \caption{Proposed 5 region pattern, and a set of 26 onskin and midair swipes (10 onskin, 16 midair) after merging regions from 6-region pattern.}
    \Description{Proposed five region pattern, and a set of ten onskin and sixteen midair swipes - a total of twenty six swipe shapes after merging regions from six-region pattern.}
    \label{fig:proposed_5_regions}  
\end{figure}
\endgroup
}

\subsection{Proposed 5-Region Pattern For Midair and Onskin Swipes}
\label{discussion: merging regions}
{
Drawing from consistent patterns in region preference, swipe performance, and gesture articulation across 4- and 6-region layouts, we propose a candidate 5-region pattern as a concise summary of the above-neck, off-device earable swipe design space.
Although not yet experimentally validated, this evidence-informed hypothesis reflects converging trends across region densities and interaction modalities. 
Addressing the density threshold observed in the 6-region layout and the recurring overlap between eye- and temple-adjacent regions, we merge the \textit{eye} (\textbf{B}) and \textit{temple} (\textbf{C}), while retaining the \textit{nose} (\textbf{A}), \textit{outer ear} (\textbf{D}), \textit{mandible angle} (\textbf{E}), and \textit{chin} (\textbf{F}) for both midair and onskin interaction. 
This layout (Figure~\ref{fig:proposed_5_regions}) reduces region density, preserves spatial distinctiveness around key facial landmarks, and prevents over-segmentation near the eyes.

Consistent with prior work~\cite{ma2021oesense, cao2023earace} and our qualitative observations (Section~\ref{subsection:RQ3_QualitativeResults}), the mandible angle and chin remain separate due to their motor and perceptual distinctiveness. 
We avoid recommending onskin interaction within the merged temple region due to proximity to the eyes and potential interference from eyewear, while noting that midair interaction remains more feasible.
Finally, angular midair swipes—especially around the mandible and chin—showed consistent directional bias and low discriminability, suggesting that users did not reliably articulate or perceive some of these swipes as distinct.
We therefore consolidate them into higher-level gesture abstractions aligned with dominant motion axes.
Overall, the proposed 5-region pattern and set of proposed swipes (Figure \ref{fig:proposed_5_regions}) captures observed articulation tendencies while reducing gesture complexity.
}

\subsection{Toward Sensing Proposed Midair and Onskin Swipes}
\label{discussion: swipeRecognitionFeasibility}
{
Prior work demonstrates that both onskin and midair hand-to-face gestures can be sensed using ear-worn and in-ear platforms across diverse sensing modalities, including microphones and acoustic leakage, inertial sensing, wireless signal variation, cameras, atmospheric pressure, electric-field sensing, bio-impedance, and mmWave radar.
As summarized in Section~\ref{subsection:-Related_Work-OffDeviceInteractionTechnique}, multiple earable systems have already recognized taps, swipes, and near-face motions on or around facial regions relevant to our study. 
For example, \textit{EarBuddy}~\cite{xu2020earbuddy}, \textit{EarAcE}~\cite{cao2023earace}, and \textit{OESense}~\cite{ma2021oesense} recognize onskin interactions on regions such as the cheekbone, temple, mandible angle, and chin, while \textit{FaceSense}~\cite{kakaraparthi2021FaceSense} explores camera placement near the earlobe to capture hand-to-face interaction. 
Similarly, midair sensing around the face has been demonstrated using BLE signal strength (\textit{EarGest}~\cite{alkiek2022eargest}), acoustic leakage (\textit{LeakyFeeder}~\cite{yang2025leakyfeeder}, \textit{EarHover}~\cite{EarHover_Suzuki_2024}, \textit{MAF}~\cite{yang2024maf}), IMUs (\textit{UiEar}~\cite{Zhao2024UiEar}), atmospheric pressure~\cite{Iguma2023Input}, and in-ear mmWave radar (\textit{TinyssimoRadar}~\cite{tinyssimoradar_ronco_2024}). 
Recent and forthcoming industry efforts further suggest the practical viability of near-face midair sensing on consumer earables~\cite{jackson_wearable_2023, airpod_camera_updated_apple_2026}.

Our candidate 5-region layout (Section~\ref{discussion: merging regions}) was derived from observed articulation and region-level tendencies, emphasizing facial areas that already have precedent in the sensing literature (\eg{}, nose~/~cheek-adjacent regions, temple, tragus~/~outer ear, mandible angle, and chin). 
Because the 5-region configuration itself was not evaluated as an experimental condition, we treat it as an evidence-informed design hypothesis rather than a validated sensing layout; consequently, the sensing implications of heterogeneous region sizes and boundaries remain to be empirically examined.

Nevertheless, the proposed gesture vocabulary is compatible with recognition strategies used in prior systems. 
For onskin interaction, many proposed unidirectional and angular swipes can be modeled as trajectories across a small set of touch-relevant landmarks. 
For example, a long onskin swipe (\eg{}, \textbf{EA} in Figure~\ref{fig:proposed_5_regions}) can be interpreted as a sequence of contacts and short transitions (chin $\rightarrow$ cheekbone $\rightarrow$ nose), where each subregion has independently demonstrated discriminative sensing characteristics in prior work. 
Similarly, for midair interaction, several systems already recognize directional swipes and near-face motion primitives; angular swipes may therefore be decomposed into short axial segments (\eg{}, a vertical segment followed by a horizontal segment), enabling recognition pipelines that compose primitives into higher-level gestures.

Taken together, these findings and prior sensing systems suggest a plausible path toward recognizing the 26 proposed midair and onskin swipes discussed in Section~\ref{discussion: merging regions}, while also motivating follow-up sensing studies to validate region layouts, shape-level constraints, and end-to-end robustness on practical earable hardware. 
Consistent with prior work on imaginary and off-device interfaces~\cite{Dezfuli_Palm_RC}, our contributions ground candidate gesture designs in empirically observed articulation patterns and sensing capabilities already demonstrated in the earable interaction literature.
}
\section{Limitation and Future Work}
\label{section:limitation_and_futurework}
{
While this work synthesizes a candidate 5-region layout and a corresponding set of 26 unidirectional and angular midair and onskin swipes for off-device earable interaction, the configuration itself was not evaluated as an experimental condition and therefore remains an evidence-informed design hypothesis. 
Additionally, our study was conducted under seated, controlled conditions to isolate spatial effects around the face, limiting generalizability to mobile everyday contexts where motor variability, comfort, and social acceptability may differ.

As follow-up work, we plan to empirically examine how region size, shape, and heterogeneity affect performance, learnability, recognition accuracy, and robustness across static and mobile settings. 
We further plan to validate key swipe-shape design decisions (\eg{}, consolidating angular midair swipes and separating midair and onskin interaction) through targeted swipe-level evaluations informed by prior findings on location-specific tap, swipe, and pinch gestures in above-neck earable interaction~\cite{arefinshimon2024fingertracker}. 
Additionally, while emerging in-ear sensing platforms show promise~\cite{tinyssimoradar_ronco_2024, sato_exploring_2024}, we plan to examine recognition accuracy, latency, and robustness to articulation variability for the proposed gestures under practical sensing conditions.

Beyond validating the proposed interaction space, we also aim to explore hybrid hand-to-face techniques, similar to \textit{Air+Touch}~\cite{Chen2014AirAndtouch}, that combine stable onskin anchors (\eg{}, the cheekbone focal region) with midair motion for off-device earable interaction. 
From an application perspective, we also plan to investigate how off-device earable swipe gestures integrate with XR systems, remote display interaction, and earable voice interfaces such as \textit{Siri} on \textit{Apple AirPods}~\cite{Chen2024Enabling}, where hand-to-face gestures may complement speech and lightweight manual input to support multimodal interaction without sustained visual attention~\cite{Wu2024Body}.
}

\section{Conclusion}
\label{section - Conclusion}
{
This work investigates unimanual hand-to-face unidirectional and angular swipe gestures as an off-device input modality for in-ear earables. 
Through controlled analysis across interaction spaces, region densities, and region locations, we characterize how users naturally articulate these single-stroke swipes, and how these behaviors relate to performance and region preference for swipe gestures. 
Based on these empirical observations, we synthesize a candidate 5-region pattern and a reduced set of midair and onskin unidirectional and angular swipes to structure and reason about the design space of off-device manual earable interaction.
This synthesis represents an evidence-informed design hypothesis rather than a validated gesture configuration.
Overall, this work advances understanding of hand-to-face swipe behavior for off-device interaction with in-ear earable device like \textit{Apple Airpod}, and provides a grounded foundation for future research on validating region layouts, refining shape-aware gesture mappings, and exploring sensing and application-level implications.
Although grounded in earable interaction, the contributions of this work generalize to a wider class of above-neck, contact-free gestural interaction techniques for head-worn systems. By characterizing how users perform and structure hand-to-face swipes, this work provides actionable insight for the design of off-device gestural input across earables, head-mounted displays, and other head-worn wearables.
}

\bibliographystyle{ACM-Reference-Format}

\bibliography{references.bib}

@inproceedings{Vatavu2013Relative,
    author = {Vatavu, Radu-Daniel and Anthony, Lisa and Wobbrock, Jacob O.},
    title = {Relative Accuracy Measures for Stroke Gestures},
    booktitle = {Proceedings of the 15th ACM on International Conference on Multimodal Interaction},
    series = {ICMI '13},
    location = {Sydney, Australia},
    publisher = {ACM},
    address = {New York, NY, USA},
    year = {2013},
    pages = {279--286},
    doi = {10.1145/2522848.2522875},
}

@inproceedings{Isokoski2001Model,
    author = {Isokoski, Poika},
    title = {Model for Unistroke Writing Time},
    booktitle = {Proceedings of the 2001 SIGCHI Conference on Human Factors in Computing Systems},
    series = {CHI '01},
    year = {2001},
    location = {Seattle, WA, USA},
    pages = {357–364},
	numpages = {8},
    url = {https://doi.org/10.1145/365024.365299},
	doi = {10.1145/365024.365299},
    publisher = {ACM},
    address = {New York, NY, USA},
	isbn = {1581133278},
}

@inproceedings{Vatavu2022Understanding,
    author = {Vatavu, Radu-Daniel and Ungurean, Ovidiu-Ciprian},
    title = {Understanding Gesture Input Articulation with Upper-Body Wearables for Users with Upper-Body Motor Impairments},
    booktitle = {Proceedings of the 2022 SIGCHI Conference on Human Factors in Computing Systems},
    series = {CHI '22},
    year = {2022},
    location = {New Orleans, LA, USA},
    articleno = {2},
	pages = {1--16},
    numpages = {16},
    url = {https://doi.org/10.1145/3491102.3501964},
	doi = {10.1145/3491102.3501964},
    publisher = {ACM},
    address = {New York, NY, USA},
	isbn = {9781450391573},
}

@inproceedings{Wittorf2016Eliciting,
    author = {Wittorf, Markus L. and Jakobsen, Mikkel R.},
	title = {Eliciting Mid-Air Gestures for Wall-Display Interaction},
    booktitle = {Proceedings of the 9th Nordic Conference on Human-Computer Interaction},
    series = {NordiCHI '16},
    year = {2016},
    location = {Gothenburg, Sweden},
    articleno = {3},
	pages = {1--4},
    numpages = {4},
    url = {https://doi.org/10.1145/2971485.2971503},
	doi = {10.1145/2971485.2971503},
    publisher = {ACM},
    address = {New York, NY, USA},
}

@inproceedings{Yamanaka2024Behavioral,
    author = {Yamanaka, Shota and Usuba, Hiroki and Sato, Junichi},
    title = {Behavioral Differences between Tap and Swipe: Observations on Time, Error, Touch-point Distribution, and Trajectory for Tap-and-swipe Enabled Targets},
    booktitle = {Proceedings of the 2024 SIGCHI Conference on Human Factors in Computing Systems},
    series = {CHI '24},
    year = {2024},
    location = {Honolulu, HI, USA},
    articleno = {549},
    pages = {1--12},
	numpages = {12},
	url = {https://doi.org/10.1145/3613904.3642272},
	doi = {10.1145/3613904.3642272},
    publisher = {ACM},
    address = {New York, NY, USA},    
    isbn = {9798400703300},  
}

@inproceedings{Hafizi2023Invehicle,
    author = {Hafizi, Arman and Henderson, Jay and Neshati, Ali and Zhou, Wei and Lank, Edward and Vogel, Daniel},
    title = {In-vehicle Performance and Distraction for Midair and Touch Directional Gestures},
    booktitle = {Proceedings of the 2023 SIGCHI Conference on Human Factors in Computing Systems},
    series = {CHI '23},
    year = {2023},
    location = {Hamburg, Germany},
    articleno = {316},
    pages = {1--13},
    numpages = {13},
    url = {https://doi.org/10.1145/3544548.3581335},
	doi = {10.1145/3544548.3581335},
    publisher = {ACM},
    address = {New York, NY, USA},
	isbn = {9781450394215},
}

@inproceedings{Weng2021Facesight,
    author = {Weng, Yueting and Yu, Chun and Shi, Yingtian and Zhao, Yuhang and Yan, Yukang and Shi, Yuanchun},
    title = {FaceSight: Enabling Hand-to-Face Gesture Interaction on AR Glasses with a Downward-Facing Camera Vision},
    booktitle = {Proceedings of the 2021 SIGCHI Conference on Human Factors in Computing Systems},
    series = {CHI '21},
    year = {2021},
    location = {Yokohama, Japan},
    articleno = {10},
	pages = {1--14},
    numpages = {14},
    url = {https://doi.org/10.1145/3411764.3445484},
	doi = {10.1145/3411764.3445484},
    publisher = {ACM},
    address = {New York, NY, USA},
	isbn = {9781450380966},
}

@inproceedings{Wu2024Body,
    author = {Wu, Liwei and Lafreniere, Ben and Grossman, Tovi and White, Thomas and Santosa, Stephanie},
    title = {Body Language for VUIs: Exploring Gestures to Enhance Interactions with Voice User Interfaces},
    booktitle = {Proceedings of the 2024 ACM Designing Interactive Systems Conference},
    series = {DIS '24},   
	year = {2024},
    location = {Copenhagen, Denmark},   
    pages = {133–150},
	numpages = {18},
    url = {https://doi.org/10.1145/3643834.3660691},
	doi = {10.1145/3643834.3660691},
	publisher = {ACM},
    address = {New York, NY, USA},
    isbn = {9798400705830},
}

@inproceedings{Chen2024Enabling,
    author = {Chen, Tao and Yang, Yongjie and Qiu, Chonghao and Fan, Xiaoran and Guo, Xiuzhen and Shangguan, Longfei},
    title = {Enabling Hands-Free Voice Assistant Activation on Earphones},
    booktitle = {Proceedings of the 22nd Annual International Conference on Mobile Systems, Applications and Services},
    series = {MOBISYS '24},
    year = {2024},
    location = {Minato-ku, Tokyo, Japan},
    pages = {155--168},
	numpages = {14},
    url = {https://doi.org/10.1145/3643832.3661890},
    doi = {10.1145/3643832.3661890},
    publisher = {ACM},
    address = {New York, NY, USA},
}

@inproceedings{Havlucu2017ItMade,
    author = {Havlucu, Hayati and Ergin, Mehmet Yark{\i}n and Bostan, {\.{I}}dil and Buruk, O{\u{g}}uz Turan and G{\"o}ksun, Tilbe and {\"O}zcan, O{\u{g}}uzhan},
    title = {It Made More Sense: Comparison of User-Elicited On-skin Touch and Freehand Gesture Sets},
    booktitle = {Proceedings of the 5th International Conference on Distributed, Ambient and Pervasive Interactions, held as part of HCI International 2017},
    series = {DAPI '17},
    year = {2017},
    location = {Vancouver, BC, Canada},
    pages = {159--171},
    numpages = {13},
    url = {https://doi.org/10.1007/978-3-319-58697-7_11},
    doi = {10.1007/978-3-319-58697-7_11},
    publisher = {Springer},
    address = {Cham, Switzerland},
	isbn= {978-3-319-58697-7},
}

@inproceedings{Leiva2021HowWeSwipe,
	author = {Leiva, Luis A. and Kim, Sunjun and Cui, Wenzhe and Bi, Xiaojun and Oulasvirta, Antti},
    title = {How We Swipe: A Large-scale Shape-writing Dataset and Empirical Findings},
    booktitle = {Proceedings of the 23rd International Conference on Mobile Human-Computer Interaction},
    series = {MobileHCI '21},
    year = {2021},
    location = {Toulouse \& Virtual, France},
    articleno = {11},
    pages = {1--13},
	numpages = {13},
    url = {https://doi.org/10.1145/3447526.3472059},
    doi = {10.1145/3447526.3472059},
    publisher = {ACM},
    address = {New York, NY, USA},
	isbn = {9781450383288},
}

@inproceedings{vatavu2011estimating,
    author={Vatavu, Radu-Daniel and Vogel, Daniel and Casiez, G{\'e}ry and Grisoni, Laurent},
    title = {Estimating the Perceived Difficulty of Pen Gestures},
    booktitle= {Proceedings of the 13th IFIP TC 13 International Conference, Part 1},
    series = {Interact '11},
    year = {2011},
    location = {Lisbon, Portugal},
    pages = {89--106},
    numpages = {18},
	url = {https://doi.org/10.1007/978-3-642-23771-3_9}, 
    doi = {10.1007/978-3-642-23771-3_9},    
	publisher = {Springer},
	address = {Berlin-Heidelberg, Germany},
	isbn = {9783642237713},
}

@inproceedings{anthony2010lightweight,
    author = {Anthony, Lisa and Wobbrock, Jacob O.},
    title = {A Lightweight Multistroke Recognizer for User Interface Prototypes},    
    booktitle = {Proceedings of Graphics Interface 2010},
    series = {GI '10},
    year = {2010},
    location = {Ottawa, ON, Canada},
    pages = {245--252},
    numpages = {8},
    url = {https://dl.acm.org/doi/10.5555/1839214.1839258},
    doi = {10.5555/1839214.1839258},
    publisher = {CIPS},    
    address = {Toronto, ON, Canada},
}

@inproceedings{Vatavu2012Gestures,
    author = {Vatavu, Radu-Daniel and Anthony, Lisa and Wobbrock, Jacob O.},
    title = {Gestures as Point Clouds: A {\$P} Recognizer for User Interface Prototypes},    
    booktitle = {Proceedings of the 14th ACM International Conference on Multimodal Interaction},
    series = {ICMI '12},
    year = {2012},
    location = {Santa Monica, CA, USA},
    pages = {273--280},
    numpages = {8},
    url = {https://doi.org/10.1145/2388676.2388732},
	doi = {10.1145/2388676.2388732},
    publisher = {ACM},
    address = {New York, NY, USA},
	isbn = {9781450314671},
}

@inproceedings{Wobbrock2007Gestures,
    author = {Wobbrock, Jacob O. and Wilson, Andrew D. and Li, Yang},
    title = {Gestures without libraries, toolkits or training: a {\$1} recognizer for user interface prototypes}, 
    booktitle = {Proceedings of the 20th Annual ACM Symposium on User Interface Software and Technology},
    series = {UIST '07},
	year = {2007},
    location = {Newport, RI, USA},
    pages = {159--168},
    numpages = {10},
    url = {https://doi.org/10.1145/1294211.1294238},
	doi = {10.1145/1294211.1294238},
	publisher = {ACM},
    address = {New York, NY, USA},
    isbn = {9781595936790},
}

@inproceedings{wang2020ev,
    author = {Wang, Lichen and Sun, Bin and Robinson, Joseph and Jing, Taotao and Fu, Yun},
    title = {EV-Action: Electromyography-Vision Multi-Modal Action Dataset}, 
    booktitle = {Proceedings of the 2020 15th IEEE International Conference on Automatic Face and Gesture Recognition}, 
    series = {FG '20},
    year = {2020},
    location = {Buenos Aires, Argentina},
    pages = {160--167},
	numpages = {8},
	url = {https://doi.org/10.1109/FG47880.2020.00018},
    doi = {10.1109/FG47880.2020.00018},
    publisher = {IEEE},
    address = {Piscataway, NJ, USA},
	isbn = {9781728130798},
}

@inproceedings{Cao2007Modeling,
    author = {Cao, Xiang and Zhai, Shumin},
    title = {Modeling Human Performance of Pen Stroke Gestures},    
    booktitle = {Proceedings of the 2007 SIGCHI Conference on Human Factors in Computing Systems},
    series = {CHI '07},
    year = {2007},
    location = {San Jose, CA, USA},
    pages = {1495--1504},
	numpages = {10},
	url = {https://doi.org/10.1145/1240624.1240850},
    doi = {10.1145/1240624.1240850},
    publisher = {ACM},
    address = {New York, NY, USA},
}

@inproceedings{Wobbrock2009User,
    author = {Wobbrock, Jacob O. and Morris, Meredith Ringel and Wilson, Andrew D.},
    title = {User-Defined Gestures for Surface Computing},    
    booktitle = {Proceedings of the 2009 SIGCHI Conference on Human Factors in Computing Systems},
    series = {CHI '09},
    year = {2009},	
    location = {Boston, MA, USA},
    pages = {1083--1092},
    numpages = {10},
    url = {https://doi.org/10.1145/1518701.1518866},
	doi = {10.1145/1518701.1518866},
    publisher = {ACM},
    address = {New York, NY, USA},
	isbn = {9781605582467},
}

@inproceedings{yang2025leakyfeeder,
    author = {Yang, Yongjie and Chen, Tao and An, Zhenlin and Cao, Shirui and Fan, Xiaoran and Shangguan, Longfei},
    title = {LeakyFeeder: In-Air Gesture Control Through Leaky Acoustic Waves},
    booktitle = {Proceedings of the 23rd ACM Conference on Embedded Networked Sensor Systems},
    series = {SenSys '25},
    year = {2025},
    location = {Irvine, CA, USA},
    pages = {144--157},
	numpages = {14},
    url = {https://doi.org/10.1145/3715014.3722054},
	doi = {10.1145/3715014.3722054},
    publisher = {ACM},
    address = {New York, NY, USA},
	isbn = {9798400714795},
}

@inproceedings{yang2024maf,
    author = {Yang, Yongjie and Chen, Tao and Huang, Yujing and Guo, Xiuzhen and Shangguan, Longfei},
	title = {MAF: Exploring Mobile Acoustic Field for Hand-to-Face Gesture Interactions},
    booktitle = {Proceedings of the 2024 SIGCHI Conference on Human Factors in Computing Systems},
    series = {CHI '24},
    year = {2024},
    location = {Honolulu, HI, USA},
    articleno = {638},
	pages = {1--20},
    numpages = {20},
    url = {https://doi.org/10.1145/3613904.3642437},
	doi = {10.1145/3613904.3642437},
	publisher = {ACM},
    address = {New York, NY, USA},
	isbn = {9798400703300},
}

@inproceedings{metzger2004freedigiter,
    author = {Metzger, C. and Anderson, M. and Starner, T.},
    title = {{FreeDigiter}: A Contact-Free Device for Gesture Control},
    booktitle = {Proceedings of the 2004 8th IEEE International Symposium on Wearable Computers}, 
    series = {ISWC '04},
    year = {2004},
    location = {Arlington, VA, USA},
    volume = {1},
    pages = {18--21},
	numpages = {4},	
	url = {https://doi.org/10.1109/ISWC.2004.23},
	doi = {10.1109/ISWC.2004.23},
    publisher={IEEE},
    address={Piscataway, NJ, USA},    
}

@inproceedings{choudhury2021earable,
    author = {Choudhury, Romit Roy},
    title = {Earable Computing: A New Area to Think About},
    booktitle = {Proceedings of the 22nd International Workshop on Mobile Computing Systems and Applications},
    series = {HotMobile '21},
    year = {2021},
    location = {Virtual Event, UK},
    pages = {147--153},
	numpages = {7},
	url = {https://doi.org/10.1145/3446382.3450216},
	doi = {10.1145/3446382.3450216},
	isbn = {9781450383233},
    publisher = {ACM},
    address = {New York, NY, USA},
}

@inproceedings{ahlstrom2014you,
    author = {Ahlstr\"{o}m, David and Hasan, Khalad and Irani, Pourang},
    title = {Are You Comfortable Doing That?: Acceptance Studies of Around-Device Gestures in and for Public Settings},    
    booktitle = {Proceedings of the 16th International Conference on Human-Computer Interaction with Mobile Devices \& Services},
    series = {MobileHCI '14},
    location = {Toronto, ON, Canada},
    publisher = {ACM},
    address = {New York, NY, USA},    
    year = {2014},
    pages = {193--202},
    doi = {10.1145/2628363.2628381},
}

@inproceedings{dobbelstein2017effects,
    author = {Dobbelstein, David and Haas, Gabriel and Rukzio, Enrico},
    title = {The Effects of Mobility, Encumbrance, and (Non-)dominant Hand on Interaction with Smartwatches},    
    booktitle = {Proceedings of the 2017 ACM International Symposium on Wearable Computers},
    series = {ISWC '17},
    year = {2017},
    location = {Maui, HI, USA},
    pages = {90--93},
    numpages = {4},
	url = {https://doi.org/10.1145/3123021.3123033},	
    doi = {10.1145/3123021.3123033},
	isbn = {9781450351881},
    publisher = {ACM},
    address = {New York, NY, USA},
}

@inproceedings{Iguma2023Input,
    author = {Iguma, Koki and Murao, Kazuya and Watanabe, Hiroki},
    title = {Input Interface with Touch and Non-touch Interactions using Atmospheric Pressure for Hearable Devices},
    booktitle = {Proceedings of the 2023 ACM International Symposium on Wearable Computers},
    series = {ISWC '23},
    year = {2023},
    location = {Cancun, Quintana Roo, Mexico},
    pages = {1–5},
	numpages = {5},
	url = {https://doi.org/10.1145/3594738.3611354},
    doi = {10.1145/3594738.3611354},
	publisher = {ACM},
    address = {New York, NY, USA},
	isbn = {9798400701993},
}

@inproceedings{zhang2023toward,
    author = {Zhang, Yuke and Takaki, Ken and Murakami, Hiroaki and Sasatani, Takuya and Kawahara, Yoshihiro},
    title = {Toward Continuous Finger Positioning on Ear Using Bone Conduction Speaker},
    booktitle = {Proceedings of the 20th ACM Conference on Embedded Networked Sensor Systems},
    series = {SenSys '22},
    year = {2023},
	location = {Boston, MA, USA},
	pages = {847--848},
    numpages = {2},
	url = {https://doi.org/10.1145/3560905.3568075},
	doi = {10.1145/3560905.3568075},
	publisher = {ACM},
    address = {New York, NY, USA},
    isbn = {9781450398862},
}

@inproceedings{kikuchi2017eartouch,
	author = {Kikuchi, Takashi and Sugiura, Yuta and Masai, Katsutoshi and Sugimoto, Maki and Thomas, Bruce H.},    
	title = {{EarTouch}: Turning the Ear into an Input Surface},
    booktitle = {Proceedings of the 19th International Conference on Human-Computer Interaction with Mobile Devices and Services},
    series = {MobileHCI '17},
    year = {2017},
    location = {Vienna, Austria},
    articleno = {27},
    pages = {1--6},
    numpages = {6},
	url = {https://doi.org/10.1145/3098279.3098538},
	doi = {10.1145/3098279.3098538},
    publisher = {ACM},
    address = {New York, NY, USA},
	isbn = {9781450350754},
}

@inproceedings{xu2020earbuddy,
    author = {Xu, Xuhai and Shi, Haitian and Yi, Xin and Liu, WenJia and Yan, Yukang and Shi, Yuanchun and Mariakakis, Alex and Mankoff, Jennifer and Dey, Anind K.},
    title = {EarBuddy: Enabling On-Face Interaction via Wireless Earbuds},
    booktitle = {Proceedings of the 2020 SIGCHI Conference on Human Factors in Computing Systems},
    series = {CHI '20},
    year = {2020},
	location = {Honolulu, HI, USA},   
    pages = {1--14},
	numpages = {14},
	url = {https://doi.org/10.1145/3313831.3376836},
	doi = {10.1145/3313831.3376836},	
    publisher = {ACM},
    address = {New York, NY, USA},   
    isbn = {9781450367080},
}

@inproceedings{ma2021oesense,
	author = {Ma, Dong and Ferlini, Andrea and Mascolo, Cecilia},
    title = {OESense: Employing Occlusion Effect for in-Ear Human Sensing},
    booktitle = {Proceedings of the 19th Annual International Conference on Mobile Systems, Applications, and Services},
    series = {MobiSys '21},
    year = {2021},
    location = {Virtual Event, WI, USA},
    pages = {175--187},
	url = {https://doi.org/10.1145/3458864.3467680},
	doi = {10.1145/3458864.3467680},
    publisher = {ACM},
    address = {New York, NY, USA},
	isbn = {9781450384438},
}

@inproceedings{song2022facelistener,
    author = {Song, Xingzhe and Huang, Kai and Gao, Wei},
    title = {FaceListener: Recognizing Human Facial Expressions via Acoustic Sensing on Commodity Headphones}, 
    booktitle = {Proceedings of the 2022 21st ACM/IEEE International Conference on Information Processing in Sensor Networks}, 
    series = {IPSN '22},
    year = {2022},
    location = {Milan, Italy},
    volume = {},
    number = {},
    pages = {145-157},
    numpages = {13},
	url = {https://doi.org/10.1109/IPSN54338.2022.00019},
    doi = {10.1109/IPSN54338.2022.00019},
    publisher = {IEEE},
    address = {Piscataway, NJ, USA},
	isbn = {9781665496247},
}

@inproceedings{lissermann2014earput,
    author = {Lissermann, Roman and Huber, Jochen and Hadjakos, Aristotelis and Nanayakkara, Suranga and M\"{u}hlh\"{a}user, Max},
    title = {{EarPut}: Augmenting Ear-Worn Devices for Ear-Based Interaction},
    booktitle = {Proceedings of the 26th Australian Computer-Human Interaction Conference on Designing Futures: The Future of Design},
    series = {OzCHI '14},
    year = {2014},
    location = {Sydney, New South Wales, Australia},
    pages = {300--307},
    numpages = {8},	
	url = {https://doi.org/10.1145/2686612.2686655},
	doi = {10.1145/2686612.2686655},
	publisher = {ACM},
    address = {New York, NY, USA},
    isbn = {9781450306539},
}

@inproceedings{alkiek2022eargest,
    author = {Alkiek, Khaled and Harras, Khaled A. and Youssef, Moustafa},
    title = {EarGest: Hand Gesture Recognition with Earables},
    booktitle = {Proceedings of the 2022 19th Annual IEEE International Conference on Sensing, Communication, and Networking},
    publisher = {IEEE},
    address = {Piscataway, NJ, USA},   
    series = {SECON '22},
    location = {Stockholm, Sweden},
    year = {2022},
    pages = {91--99},
    doi = {10.1109/SECON55815.2022.9918622},
}

@inproceedings{gustafson2010imaginary,
    author = {Gustafson, Sean and Bierwirth, Daniel and Baudisch, Patrick},
    title = {Imaginary Interfaces: Spatial Interaction with Empty Hands and without Visual Feedback},
    booktitle = {Proceedings of the 23nd Annual ACM Symposium on User Interface Software and Technology},
    series = {UIST '10},
    year = {2010},
    location = {New York, NY, USA},
    pages = {3--12},
    numpages = {10},
	url = {https://doi.org/10.1145/1866029.1866033},
	doi = {10.1145/1866029.1866033},
	publisher = {ACM},
    address = {New York, NY, USA},	
	isbn = {9781450302715},
}

@inproceedings{ferlini2019head,
    author = {Ferlini, Andrea and Montanari, Alessandro and Mascolo, Cecilia and Harle, Robert},
    title = {Head Motion Tracking Through in-Ear Wearables},
    booktitle = {Proceedings of the 1st International Workshop on Earable Computing},
    series = {EarComp '19},
    year = {2020},
    location = {London, UK},
    pages = {8--13},
	numpages = {6},
    url = {https://doi.org/10.1145/3345615.3361131},
	doi = {10.1145/3345615.3361131},
    publisher = {ACM},
    address = {New York, NY, USA},
	isbn = {9781450369022},
}

@inproceedings{radhakrishnan2021applying,
    author={Radhakrishnan, Meera and Misra, Kushaan and Ravichandran, V.},
    title={Applying “Earable” Inertial Sensing for Real-time Head Posture Detection}, 
    booktitle={Proceedings of the 2021 IEEE International Conference on Pervasive Computing and Communications Workshops and other Affiliated Events}, 
    series = {PerCom Workshops '21},
    year={2021},
    location = {Virtual Event, Germany},
    pages={176-181},
	numpages = {6},
	url = {https://doi.org/10.1109/PerComWorkshops51409.2021.9430988},
    doi = {10.1109/PerComWorkshops51409.2021.9430988},
    publisher = {IEEE},
    address = {Piscataway, NJ, USA},
	isbn = {9781665404242},
}

@inproceedings{serrano2014exploring,
    author = {Serrano, Marcos and Ens, Barrett M. and Irani, Pourang P.},
    title = {Exploring the Use of Hand-to-face Input for Interacting with Head-worn Displays},
    booktitle = {Proceedings of the 2014 SIGCHI Conference on Human Factors in Computing Systems},
    series = {CHI '14},
    year = {2014},
    location = {Toronto, ON, Canada},
    volume = {},
    number = {},
    pages = {3181--3190},
	numpages = {10},
	url = {https://doi.org/10.1145/2556288.2556984},
	doi = {10.1145/2556288.2556984},	
    publisher = {ACM},
    address = {New York, NY, USA},
}

@inproceedings{gashi2021hierarchical,
    author = {Gashi, Shkurta and Saeed, Aaqib and Vicini, Alessandra and Di Lascio, Elena and Santini, Silvia},
    title = {Hierarchical Classification and Transfer Learning to Recognize Head Gestures and Facial Expressions Using Earbuds},
    booktitle = {Proceedings of the 2021 International Conference on Multimodal Interaction},
    series = {ICMI '21},
    year = {2021},
    location = {Montr\'{e}al, QC, Canada},   
    pages = {168–176},
	numpages = {9},
    url = {https://doi.org/10.1145/3462244.3479921},
	doi = {10.1145/3462244.3479921},
    publisher = {ACM},
    address = {New York, NY, USA},
	isbn = {9781450384810},
}

@inproceedings{zhu2023char,
    author = {Zhu, Peizhao and Zou, Yongpan and Li, Wenyuan and Wu, Kaishun},
    title = {CHAR: Composite Head-body Activities Recognition with A Single Earable Device}, 
    booktitle = {Proceedings of the 2023 IEEE International Conference on Pervasive Computing and Communications}, 
    series = {PerCom '23},
    year = {2023},
	location = {Atlanta, GA, USA},     
    pages = {212-221},
	numpages = {10},
	url = {https://doi.org/10.1109/PERCOM56429.2023.10099218},
    doi = {10.1109/PERCOM56429.2023.10099218},
    publisher = {IEEE},
    address = {Piscataway, NJ, USA},
    isbn = {9781665453783},
}

@inproceedings{zhou2020survey,
    author = {Zhou, Quan and Fang, Bin and Shan, Jianhua and Sun, Fuchun and Guo, Di},
    title = {A Survey of the Development of Wearable Devices}, 
    booktitle = {Proceedings of the 2020 5th International Conference on Advanced Robotics and Mechatronics}, 
    series = {ICARM '20},
    year = {2020},
    location = {Shenzhen, China},
    pages = {198-203},
	numpages = {6},
	url = {https://doi.org/ICARM49381.2020.9195351},
    doi = {10.1109/ICARM49381.2020.9195351},
    publisher = {IEEE},
    address = {Piscataway, NJ, USA},
	isbn = {9781728164793},
}

@inproceedings{hincapie2014consumed,
    author = {Hincapi\'{e}-Ramos, Juan David and Guo, Xiang and Moghadasian, Paymahn and Irani, Pourang},
    title = {Consumed Endurance: A Metric to Quantify Arm Fatigue of Mid-Air Interactions},
    booktitle = {Proceedings of the 2014 SIGCHI Conference on Human Factors in Computing Systems},
    series = {CHI '14},
    location = {Toronto, ON, Canada},
    publisher = {ACM},
    address = {New York, NY, USA},
    year = {2014},
    pages = {1063--1072},
    doi = {10.1145/2556288.2557130},
}

@inproceedings{wong2020exploring,
    author = {Wong, Pui Chung and Zhu, Kening and Yang, Xing-Dong and Fu, Hongbo},
    title = {Exploring Eyes-free Bezel-initiated Swipe on Round Smartwatches},
    booktitle = {Proceedings of the 2020 SIGCHI Conference on Human Factors in Computing Systems},
    series = {CHI '20},
    year = {2020},
    location = {Honolulu, HI, USA},
    pages = {1--11},
	numpages = {11},
    url = {https://doi.org/10.1145/3313831.3376393},
	doi = {10.1145/3313831.3376393},	
	publisher = {ACM},
    address = {New York, NY, USA},
	isbn = {9781450367080},    
}

@inproceedings{Chen2014AirAndtouch,
    author = {Chen, Xiang 'Anthony' and Schwarz, Julia and Harrison, Chris and Mankoff, Jennifer and Hudson, Scott E.},
    title = {Air+touch: Interweaving Touch \& In-air Gestures},
    booktitle = {Proceedings of the 27th Annual ACM Symposium on User Interface Software and Technology},
    series = {UIST '14},
    year = {2014},
	isbn = {9781450330695},
    location = {Honolulu, HI, USA},
    pages = {519--525},
	numpages = {7},
	url = {https://doi.org/10.1145/2642918.2647392},
    doi = {10.1145/2642918.2647392},
    publisher = {ACM},
    address = {New York, NY, USA},
}

@inproceedings{Alkiek2023EarBender,
    author = {Alkiek, Khaled and Youssef, Moustafa and Harras, Khaled A.},
    title = {EarBender: Enabling Rich IMU-based Natural Hand-to-Ear Interaction in Commodity Earables},
    booktitle = {Adjunct Proceedings of the 2023 ACM International Joint Conference on Pervasive and Ubiquitous Computing \& the 2023 ACM International Symposium on Wearable Computing},
    publisher = {ACM},
    address = {New York, NY, USA},
    series = {UbiComp/ISWC '23 Adjunct},
    location = {Cancun, Quintana Roo, Mexico},
    year = {2023},
    pages = {333--338},
    doi = {10.1145/3594739.3610671},
}

@inproceedings{Kollee2014ExploringHMD,
	author = {Kollee, Barry and Kratz, Sven and Dunnigan, Anthony},
    title = {Exploring Gestural Interaction in Smart Spaces Using Head-Mounted Devices with Ego-centric Sensing},
    booktitle = {Proceedings of the 2nd ACM Symposium on Spatial User Interaction},
    series = {SUI '14},
    year = {2014},
    location = {Honolulu, HI, USA},
    pages = {40--49},
	numpages = {10},
    url = {https://doi.org/10.1145/2659766.2659781},
	doi = {10.1145/2659766.2659781},
    publisher = {ACM},
    address = {New York, NY, USA},
	isbn = {9781450328203},
}

@inproceedings{Li2019SpeechRecognitionResearch,
    author = {Li, Ning and Zhou, Tuoyang and Zhou, Yingwei and Guo, Chen and Fu, Deqiang and Li, Xiaoqing and Guo, Zijing},
    title = {Research on Human-Computer Interaction Mode of Speech Recognition Based on Environment Elements of Command and Control System}, 
    booktitle = {Proceedings of the 2019 5th International Conference on Big Data and Information Analytics}, 
    series = {BigDIA '19},
    year = {2019},
    location = {Kunming, China},
	volume = {},
    number = {},
    pages = {170-175},
	numpages = {6},	
	url = {https://doi.org/10.1109/BigDIA.2019.8802812},
    doi = {10.1109/BigDIA.2019.8802812},
    publisher = {IEEE},
    address = {Piscataway, NJ, USA},
	isbn = {9781728139333},
}

@inproceedings{Pandey2021AcceptabilitySpeech,
	author = {Pandey, Laxmi and Hasan, Khalad and Arif, Ahmed Sabbir},    
	title = {Acceptability of Speech and Silent Speech Input Methods in Private and Public},
    booktitle = {Proceedings of the 2021 SIGCHI Conference on Human Factors in Computing Systems},
    series = {CHI '21},
    year = {2021},
    location = {Yokohama, Japan},
    articleno = {251},
    pages = {1--19},
    numpages = {13},
    url = {https://doi.org/10.1145/3411764.3445430},
	doi = {10.1145/3411764.3445430},
    publisher = {ACM},
    address = {New York, NY, USA},
	isbn = {9781450380966},
}

@inproceedings{Sun2021TeethTap,
    author = {Sun, Wei and Li, Franklin Mingzhe and Steeper, Benjamin and Xu, Songlin and Tian, Feng and Zhang, Cheng},
    title = {TeethTap: Recognizing Discrete Teeth Gestures Using Motion and Acoustic Sensing on an Earpiece},
    booktitle = {Proceedings of the 26th International Conference on Intelligent User Interfaces},
    series = {IUI '21},
    year = {2021},
    location = {College Station, TX, USA},
	volume = {},
    number = {},
    pages = {161--169},
	numpages = {9},
	url = {https://doi.org/10.1145/3397481.3450645},
	doi = {10.1145/3397481.3450645},
    publisher = {ACM},
    address = {New York, NY, USA},
	isbn = {9781450380171},	
}

@inproceedings{Khanna2021JawSense,
    author = {Khanna, Prerna and Srivastava, Tanmay and Pan, Shijia and Jain, Shubham and Nguyen, Phuc},
	title = {JawSense: Recognizing Unvoiced Sound using a Low-cost Ear-worn System},
    booktitle = {Proceedings of the 22nd International Workshop on Mobile Computing Systems and Applications},
    series = {HotMobile '21},
    year = {2021},
    location = {Virtual Event, UK},
    pages = {44--49},
	numpages = {6},
	url = {https://doi.org/10.1145/3446382.3448363},
	doi = {10.1145/3446382.3448363},
    publisher = {ACM},
    address = {New York, NY, USA},
	isbn = {9781450383233},
}

@inproceedings{shariff2022cw,
    author = {Mohd Shariff, Khairul Khaizi and Nadiah Yusni, Auni and Md Ali, Mohd Adli and Syahirul Amin Megat Ali, Megat and Megat Tajuddin, Megat Zuhairy and Younis, M A A},
    title = {CW Radar Based Silent Speech Interface Using CNN},
    booktitle = {Proceedings of the 2022 IEEE Symposium on Wireless Technology \& Applications},
    series = {ISWTA '22},
	year = {2022},
    location = {Kuala Lumpur, Malaysia},
    volume = {},
    number = {},
    pages = {76--81},
    numpages = {6},
	url = {https://doi.org/10.1109/ISWTA55313.2022.9942730},	
    doi = {10.1109/ISWTA55313.2022.9942730},	
	publisher = {IEEE},
    address = {Piscataway, NJ, USA},
}

@inproceedings{Bedri2015Stick,
    author = {Bedri, Abdelkareem and Byrd, David and Presti, Peter and Sahni, Himanshu and Gue, Zehua and Starner, Thad},
    title = {Stick It in Your Ear: Building an In-Ear Jaw Movement Sensor},
    booktitle = {Adjunct Proceedings of the 2015 ACM International Joint Conference on Pervasive and Ubiquitous Computing and Proceedings of the 2015 ACM International Symposium on Wearable Computers},
    series = {UbiComp/ISWC'15 Adjunct},
    year = {2015},
    location = {Osaka, Japan},
    pages = {1333--1338},
	numpages = {6},
	url = {https://doi.org/10.1145/2800835.2807933},
    doi = {10.1145/2800835.2807933},
    publisher = {ACM},
    address = {New York, NY, USA},
}

@inproceedings{koelle2020social,
    author = {Koelle, Marion and Ananthanarayan, Swamy and Boll, Susanne},
    title = {Social Acceptability in HCI: A Survey of Methods, Measures, and Design Strategies},
    booktitle = {Proceedings of the 2020 SIGCHI Conference on Human Factors in Computing Systems},
    series = {CHI '20},
    year = {2020},
    location = {Honolulu, HI, USA},
    pages = {1–19},
	numpages = {19},
    url = {https://doi.org/10.1145/3313831.3376162},
	doi = {10.1145/3313831.3376162},
    publisher = {ACM},
    address = {New York, NY, USA},
	isbn = {9781450367080},
}

@inproceedings{farringdon1999co,
    author = {Farringdon, Jonny and Oni, Vanessa and Kan, Chi Ming and Poll, Leo},
    title = {Co-Modal Browser-An Interface for Wearable Computers},
    booktitle = {Digest of Papers. Third International Symposium on Wearable Computers},
    series = {ISWC '99},
    year = {1999},
    location  = {San Francisco, CA, USA},
    pages = {45-51},
	numpages = {7},
	url = {https://doi.org/10.1109/ISWC.1999.806644},	
    doi = {10.1109/ISWC.1999.806644},
    publisher = {IEEE},
    address = {Piscataway, NJ, USA},
}

@inproceedings{hossain2021exploring,
    author = {Hossain, Tamzid and Islam, Md. Fahimul and Delamare, William and Chowdhury, Farida and Hasan, Khalad},
    title = {Exploring Social Acceptability and Users’ Preferences of Head-and Eye-Based Interaction with Mobile Devices},
    booktitle = {Proceedings of the 20th International Conference on Mobile and Ubiquitous Multimedia},
    series = {MUM '21},
    year = {2022},
    location = {Leuven, Belgium},
    pages = {12--23},
    numpages = {10},
    url = {https://doi.org/10.1145/3490632.3490636},
	doi = {10.1145/3490632.3490636},
    publisher = {ACM},
    address = {New York, NY, USA},
	isbn = {9781450386432},
}

@inproceedings{Dezfuli_Palm_RC,
    author = {Dezfuli, Niloofar and Khalilbeigi, Mohammadreza and Huber, Jochen and M\"{u}ller, Florian and M\"{u}hlh\"{a}user, Max},
    title = {PalmRC: Imaginary Palm-Based Remote Control for Eyes-Free Television Interaction},
    booktitle = {Proceedings of the 10th European Conference on Interactive TV and Video},
    series = {EuroITV '12},
    year = {2012},	
    location = {Berlin, Germany},
    pages = {27--34},
	numpages = {8},
	url = {https://doi.org/10.1145/2325616.2325623},	
    doi = {10.1145/2325616.2325623},
	isbn = {9781450311076},
    publisher = {ACM},
    address = {New York, NY, USA},
}

@inproceedings{Grossman2015Typing,
    author = {Grossman, Tovi and Chen, Xiang Anthony and Fitzmaurice, George},
    title = {Typing on Glasses: Adapting Text Entry to Smart Eyewear},
    booktitle = {Proceedings of the 17th International Conference on Human-Computer Interaction with Mobile Devices and Services},
    series = {MobileHCI '15},
    year = {2015},
    location = {Copenhagen, Denmark},
    pages = {144--152},
	numpages = {9},
    url = {https://doi.org/10.1145/2785830.2785867},
	doi = {10.1145/2785830.2785867},
    publisher = {ACM},
    address = {New York, NY, USA},
	isbn = {9781450336529},
}

@inproceedings{Holfeld2023Evaluating,
    author = {Perella-Holfeld, Francisco and Faleel, Shariff AM and Irani, Pourang},
    title = {Evaluating Design Guidelines for Hand-Proximate User Interfaces},
    booktitle = {Proceedings of the 2023 ACM Designing Interactive Systems Conference},
    series = {DIS '23},
    year = {2023},
    location = {Pittsburgh, PA, USA},
    pages = {1159--1173},
	numpages = {15},
	url = {https://doi.org/10.1145/3563657.3596117},
	doi = {10.1145/3563657.3596117},
    publisher = {ACM},
    address = {New York, NY, USA},
	isbn = {9781450398930},
}

@inproceedings{EarHover_Suzuki_2024,
    author = {Suzuki, Shunta and Amesaka, Takashi and Watanabe, Hiroki and Shizuki, Buntarou and Sugiura, Yuta},
    title = {EarHover: Mid-Air Gesture Recognition for Hearables Using Sound Leakage Signals},
    booktitle = {Proceedings of the 37th Annual ACM Symposium on User Interface Software and Technology},
    series = {UIST '24},
    year = {2024},
    location = {Pittsburgh, PA, USA},
    articleno = {129},
	pages = {1--13},
    numpages = {13},
    url = {https://doi.org/10.1145/3654777.3676367},
	doi = {10.1145/3654777.3676367},
    publisher = {ACM},
    address = {New York, NY, USA},
	isbn = {9798400706288},
}

@inproceedings{tinyssimoradar_ronco_2024,
    author = {Ronco, Andrea and Schilk, Philipp and Magno, Michele},
    title = {TinyssimoRadar: In-Ear Hand Gesture Recognition with Ultra-Low Power mmWave Radars}, 
    booktitle = {Proceedings of the 2024 IEEE/ACM Ninth International Conference on Internet-of-Things Design and Implementation}, 
    series = {IoTDI '24},
    year = {2024},
    location = {Hong Kong, Hong Kong SAR},
    volume = {},
    number = {},
    pages = {192-202},
	numpages = {11},
	url = {https://doi.org/10.1109/IoTDI61053.2024.00021},
    doi = {10.1109/IoTDI61053.2024.00021},
	publisher = {IEEE},
    address = {Piscataway, NJ, USA},
}

@article{remizova2022midair,
    title = {Midair Gestural Techniques for Translation Tasks in Large-Display Interaction},
    author = {Remizova, Vera and Gizatdinova, Yulia and Surakka, Veikko},
    journal = {Adv. Hum.-Comput. Interact.},
	issue_date = {July 2022},
    volume = {2022},
    number = {1},
	month = jul,
	year = {2022},
	articleno = {9362916},
    pages = {1--13},
	numpages = {13},
	url = {https://doi.org/10.1155/2022/9362916},
    doi = {10.1155/2022/9362916},
    publisher = {Wiley Online Library},
    address = {Hoboken, NJ, USA},
}

@article{Li2021Depth,
    author = {Li, Wenmin and Wan, Xueyi and Shi, Yanwei and Yao, Nailang and Wang, Ci and Gao, Zaifeng},
    title = {Depth and Breadth of Pie Menus for Mid-air Gesture Interaction},
    journal = {Int. J. Hum.-Comput. Interact.},
	issue_date = {August 2020},
    volume = {37},
    number = {2},
	month = aug,
    year = {2021},
    pages = {131--140},
	numpages = {10},
	url = {https://doi.org/10.1080/10447318.2020.1809245},
    doi = {10.1080/10447318.2020.1809245},
    publisher = {Taylor \& Francis},
    address = {Oxfordshire, UK},
}

@article{Tu2021ArmMenu,
    author = {Huawei Tu and Weiyang Huan and Xingdong Yang and Xiangshi Ren and Feng Tian},
    title = {{ArmMenu}: Command Input on Distant Displays with Proprioception-Based Lateral Arm Movements},
    journal = {Behav. Inf. Technol.},
	issue_date = {April 2020},
    volume = {40},
    number = {13},
    year = {2021},
    pages = {1428--1447},
    numpages = {20},
	url = {https://doi.org/10.1080/0144929X.2020.1757760},
    doi = {10.1080/0144929X.2020.1757760},
    publisher = {Taylor \& Francis},
    address = {Oxfordshire, UK},
}

@article{jin2021sonicasl,
    author = {Jin, Yincheng and Gao, Yang and Zhu, Yanjun and Wang, Wei and Li, Jiyang and Choi, Seokmin and Li, Zhangyu and Chauhan, Jagmohan and Dey, Anind K. and Jin, Zhanpeng},
	title = {SonicASL: An Acoustic-based Sign Language Gesture Recognizer Using Earphones},
    journal = {Proc. ACM Interact. Mob. Wearable Ubiquitous Technol.},
	issue_date = {June 2021},
    volume = {5},
    number = {2},
    month = jun,
    year = {2021},
    articleno = {67},
    pages = {1-30},
    numpages = {30},
	url = {https://doi.org/10.1145/3463519},
	doi = {10.1145/3463519},
    publisher = {ACM},
    address = {New York, NY, USA},
}

@article{roddiger2022LitReview,
    author = {R\"{o}ddiger, Tobias and Clarke, Christopher and Breitling, Paula and Schneegans, Tim and Zhao, Haibin and Gellersen, Hans and Beigl, Michael},
    title = {Sensing with Earables: A Systematic Literature Review and Taxonomy of Phenomena},
    journal = {Proc. ACM Interact. Mob. Wearable Ubiquitous Technol.},
	issue_date = {September 2022},
    volume = {6},
    number = {3},
	month = sep,
    year = {2022},
    pages = {1--57},
	numpages = {57},
    url = {https://doi.org/10.1145/3550314},
	doi = {10.1145/3550314},
    publisher = {ACM},
    address = {New York, NY, USA},
}

@article{chen2020exploring,
    author = {Chen, Yu-Chun and Liao, Chia-Ying and Hsu, Shuo-wen and Huang, Da-Yuan and Chen, Bing-Yu},
    title = {Exploring User Defined Gestures for Ear-Based Interactions},
    journal = {Proc. ACM Hum.-Comput. Interact.},
	issue_date = {November 2020},
    volume = {4},
    number = {ISS},
    month = nov,
    year = {2020},
    articleno = {186},
    numpages = {20},
	url = {https://doi.org/10.1145/3427314},
    doi = {10.1145/3427314},
    publisher = {ACM},
    address = {New York, NY, USA},
}

@article{rateau2022leveraging,
	author = {Rateau, Hanae and Lank, Edward and Liu, Zhe},
	title = {Leveraging Smartwatch and Earbuds Gesture Capture to Support Wearable Interaction},
    journal = {Proc. ACM Hum.-Comput. Interact.},
	issue_date = {December 2022},
	volume = {6},
	number = {ISS},
	month = nov,
	year = {2022},
	articleno = {557},
    pages = {31--50},
	numpages = {20},
	url = {https://doi.org/10.1145/3567710},
	doi = {10.1145/3567710},
    publisher = {ACM},
    address = {New York, NY, USA},    
}

@article{cao2023earace,
    author = {Cao, Yetong and Cai, Chao and Yu, Anbo and Li, Fan and Luo, Jun},
    title = {EarAcE: Empowering Versatile Acoustic Sensing via Earable Active Noise Cancellation Platform},
    journal = {Proc. ACM Interact. Mob. Wearable Ubiquitous Technol.},
	issue_date = {June 2023},
    volume = {7},
    number = {2},
    month = jun,
    year = {2023},
    articleno = {47},
    numpages = {23},
    pages = {1--23},
	url = {https://doi.org/10.1145/3596242},
    doi = {10.1145/3596242},
    publisher = {ACM},
    address = {New York, NY, USA},
}

@article{verma2021expressear,
    author = {Verma, Dhruv and Bhalla, Sejal and Sahnan, Dhruv and Shukla, Jainendra and Parnami, Aman},
    title = {ExpressEar: Sensing Fine-Grained Facial Expressions with Earables},
    journal = {Proc. ACM Interact. Mob. Wearable Ubiquitous Technol.},
	issue_date = {Sept 2021},
    volume = {5},
    number = {3},
    month = sep,
    year = {2021},
    articleno = {129},
    pages = {1--28},
    numpages = {28},
	url = {https://doi.org/10.1145/3478085},
	doi = {10.1145/3478085},
    publisher = {ACM},
    address = {New York, NY, USA},
}

@article{jin2022earcommand,
    author = {Jin, Yincheng and Gao, Yang and Xu, Xuhai and Choi, Seokmin and Li, Jiyang and Liu, Feng and Li, Zhengxiong and Jin, Zhanpeng},
	title = {EarCommand: "Hearing" Your Silent Speech Commands In Ear},
    journal = {Proc. ACM Interact. Mob. Wearable Ubiquitous Technol.},
	issue_date = {June 2022},
    volume = {6},
    number = {2},
    month = jul,
    year = {2022},
    articleno = {57},
    pages = {1-28},
    numpages = {28},
	url = {https://doi.org/10.1145/3534613},
	doi = {10.1145/3534613},
    publisher = {ACM},
    address = {New York, NY, USA},
}

@article{srivastava2022muteit,
    author = {Srivastava, Tanmay and Khanna, Prerna and Pan, Shijia and Nguyen, Phuc and Jain, Shubham},
    title = {MuteIt: Jaw Motion Based Unvoiced Command Recognition Using Earable},
    journal = {Proc. ACM Interact. Mob. Wearable Ubiquitous Technol.},
	issue_date = {September 2022},
    volume = {6},
    number = {3},
    month = sep,
    year = {2022},
    articleno = {140},
    pages = {1--26},
    numpages = {26},
	url = {https://doi.org/10.1145/3550281},
	doi = {10.1145/3550281},
    publisher = {ACM},
    address = {New York, NY, USA},    
}

@article{rey2022understanding,
    author = {Rey, Bradley and Zhu, Kening and Perrault, Simon Tangi and Bardot, Sandra and Neshati, Ali and Irani, Pourang},
    title = {Understanding and Adapting Bezel-to-Bezel Interactions for Circular Smartwatches in Mobile and Encumbered Scenarios},
    journal = {Proc. ACM Hum.-Comput. Interact.},
	issue_date = {September 2022},
    volume = {6},
    number = {MHCI},
    month = sep,
    year = {2022},
    articleno = {201},
	pages = {1--28},
    numpages = {28},
    url = {https://doi.org/10.1145/3546736},
	doi = {10.1145/3546736},
	publisher = {ACM},
    address = {New York, NY, USA},
}

@article{barroso1980self,
    author = {Barroso, Felix and Freedman, Norbert and Grand, Stanley},
    title = {Self-Touching, Performance, and Attentional Processes},    
    journal = {Percept. Mot. Skills},
	issue_date = {June 1980},
    volume = {50},
    number = {3\_suppl},
    month = jun,
    year = {1980},
    numpages = {7},
    pages = {1083--1089},
	url = {https://doi.org/10.2466/pms.1980.50.3c.1083},
    doi = {10.2466/pms.1980.50.3c.1083},
    publisher = {SAGE Publications},
    address = {Los Angeles, CA, USA},
}

@article{kwok2015face,
    author = {Kwok, Yen Lee Angela and Gralton, Jan and McLaws, Mary-Louise},
    title = {Face Touching: A Frequent Habit That Has Implications for Hand Hygiene},    
    journal = {Am. J. Infect. Control},
	issue_date = {February 2015},
    volume = {43},
    number = {2},
	month = feb,
    year = {2015},
    pages = {112--114},
	numpages = {3},
    url = {https://www.sciencedirect.com/science/article/pii/S0196655314012814},
	doi = {10.1016/j.ajic.2014.10.015},
    publisher = {Elsevier},
    address = {Amsterdam, Netherlands},
	issn = {0196-6553},
}

@article{ekman1972hand,
    author = {Ekman, Paul and Friesen, Wallace V.},
    title = {Hand Movements},
    journal = {J. Commun.},
	issue_date = {December 1972},
    volume = {22},
    number = {4},
    month = dec,	
    year = {1972},    
    pages = {353--374},
	numpages = {22},
	url = {https://doi.org/10.1111/j.1460-2466.1972.tb00163.x},
    doi = {10.1111/j.1460-2466.1972.tb00163.x},
    publisher = {Oxford University Press},
    address = {Oxford, UK},
}

@article{harrigan1987self,
    author = {Harrigan, Jinni A and Kues, John R and Steffen, John J and Rosenthal, Robert},
    title = {Self-Touching and Impressions of Others},    
    journal = {Pers. Soc. Psychol. Bull.},
	issue_date = {December 1987},
    volume = {13},
    number = {4},
	month = dec,
    year = {1987},
    pages = {497--512},
    numpages = {16},
	url = {https://doi.org/10.1177/0146167287134007},
    doi = {10.1177/0146167287134007},
    publisher = {Sage Publications},
    location = {Los Angeles, CA, USA},
}

@article{nicas2008study,
    author = {Nicas, Mark and Best, Daniel},
    title = {A Study Quantifying the Hand-to-Face Contact Rate and Its Potential Application to Predicting Respiratory Tract Infection},
    journal = {J. Gen. Psychol.},
	issue_date = {April 2008},
    volume = {5},
    number = {6},
	month = apr,
    year = {2008},
    pages = {347--352},
    numpages = {6},
	url = {https://doi.org/10.1080/15459620802003896},
    doi = {10.1080/15459620802003896},
    publisher = {Taylor \& Francis},
    address = {Oxfordshire, UK},
}

@article{mueller2019self,
    author = {Mueller, Stephanie Margarete and Martin, Sven and Grunwald, Martin},
    title = {Self-Touch: Contact Durations and Point of Touch of Spontaneous Facial Self-Touches Differ Depending on Cognitive and Emotional Load},    
    journal = {PLOS ONE},
	issue_date = {March 2019},
    volume = {14},
    number = {3},
	month = mar,
    year = {2019},
    articleno = {e0213677},
	pages = {1--19},
    numpages = {19},
	url = {https://doi.org/10.1371/journal.pone.0213677},
    doi = {10.1371/journal.pone.0213677},
    publisher = {Public Library of Science},
    address = {San Francisco, CA USA},
}

@article{Faleel2021HPUI,
    author = {Faleel, Shariff AM and Gammon, Michael and Fan, Kevin and Huang, Da-Yuan and Li, Wei and Irani, Pourang},
    title = {{HPUI}: Hand Proximate User Interfaces for One-Handed Interactions on Head Mounted Displays}, 
    journal = {IEEE Trans. Vis. Comput. Graph.},
	issue_date = {August 2021},
    volume = {27},
    number = {11},
	month = aug,
    year = {2021},
    pages = {4215-4225},
	numpages = {11},
	url = {https://doi.org/10.1109/TVCG.2021.3106493},
    doi = {10.1109/TVCG.2021.3106493},
    publisher = {IEEE},
    address = {Piscataway, NJ, USA},
}

@article{Hu2023HeadTrack,
    author = {Hu, Jingyang and Jiang, Hongbo and Xiao, Zhu and Chen, Siyu and Dustdar, Schahram and Liu, Jiangchuan},
    title = {{HeadTrack}: Real-Time Human-Computer Interaction via Wireless Earphones},    
    journal = {IEEE J. Sel. Areas Commun.},
	issue_date = {December 2023},
    volume = {42},
    number = {4},
	month = apr,
    year = {2024},
    pages = {990-1002},
    numpages = {13},
	url = {https://doi.org/10.1109/JSAC.2023.3345381},
    doi = {10.1109/JSAC.2023.3345381},
    publisher = {IEEE},
    address = {Piscataway, NJ, USA},
	ISSN = {1558-0008},
}

@article{Zhang2024Earphone,
    author = {Zhang, Shijia and Lu, Taiting and Zhou, Hao and Liu, Yilin and Liu, Runze and Gowda, Mahanth},
    title = {I Am an Earphone and I Can Hear My User’s Face: Facial Landmark Tracking Using Smart Earphones},
    journal = {ACM Trans. Internet Things},
	issue_date = {December 2023},
    volume = {5},
    number = {1},
    month = dec,
    year = {2023},
    articleno = {1},
	pages = {1--29},
    numpages = {29},
	url = {https://doi.org/10.1145/3614438},
	doi = {10.1145/3614438},
    publisher = {ACM},
    address = {New York, NY, USA},
}

@article{Choi2022PPGface,
    author = {Choi, Seokmin and Gao, Yang and Jin, Yincheng and Kim, Se jun and Li, Jiyang and Xu, Wenyao and Jin, Zhanpeng},
    title = {PPGface: Like What You Are Watching? Earphones Can "Feel" Your Facial Expressions},
    journal = {Proc. ACM Interact. Mob. Wearable Ubiquitous Technol.},
	issue_date = {June 2022},
    volume = {6},
    number = {2},
    month = jul,
    year = {2022},
    articleno = {48},
    numpages = {32},
	url = {https://doi.org/10.1145/3534597},
    doi = {10.1145/3534597},
    publisher = {ACM},
    address = {New York, NY, USA},
}

@article{digehsara2022user,
    author = {Digehsara, Pouriya Amini and de Menezes, Joao V{\i}tor Possamai and Wagner, Christoph and B{\"a}rhold, Michael and Schaffer, Petr and Plettemeier, Dirk and Birkholz, Peter},
    title = {A User-Friendly Headset for Radar-Based Silent Speech Recognition},    
    journal = {Proc. Interspeech},
	issue_date = {September 2022},
    volume = {7},
    number = {21},
    month = sep,
    year = {2022},
    pages = {4835--4839},
	numpages = {5},
	url = {https://doi.org/10.21437/Interspeech.2022-10090},
    doi = {10.21437/Interspeech.2022-10090},
    publisher = {ISCA},
    address = {Grenoble, France},
}

@article{arefinshimon2024fingertracker,
    author = {Arefin Shimon, Shaikh Shawon and Neshati, Ali and Sun, Junwei and Xu, Qiang and Zhao, Jian},
    title = {Exploring Uni-manual Around Ear Off-Device Gestures for Earables},
    journal = {Proc. ACM Interact. Mob. Wearable Ubiquitous Technol.},
	issue_date = {March 2024},
    volume = {8},
    number = {1},
    month = mar,
    year = {2024},
    articleno = {3},
    numpages = {29},
	pages = {1--29},
	url = {https://doi.org/10.1145/3643513},
	doi = {10.1145/3643513},
    publisher = {ACM},
    address = {New York, NY, USA},
}

@article{zeng2023msilent,
    author = {Zeng, Shang and Wan, Haoran and Shi, Shuyu and Wang, Wei},
    title = {MSilent: Towards General Corpus Silent Speech Recognition Using COTS MmWave Radar},
    journal = {Proc. ACM Interact. Mob. Wearable Ubiquitous Technol.},
	issue_date = {March 2023},
    volume = {7},
    number = {1},
    month = mar,
    year = {2023},
    articleno = {39},
	pages = {1--28},
    numpages = {28},
	url = {https://doi.org/10.1145/3580838},
	doi = {10.1145/3580838},
    publisher = {ACM},
    address = {New York, NY, USA},
}

@article{eastwood2003negative,
    author = {Eastwood, John D and Smilek, Daniel and Merikle, Philip M},
    title = {Negative Facial Expression Captures Attention and Disrupts Performance},
    journal = {Percept. Psychophys.},
	issue_date = {April 2003},
    volume = {65},
    number = {3},
    month = apr,
    year = {2003},    
    pages = {352--358},
	numpages = {7},
	url = {https://doi.org/10.3758/BF03194566},
    doi = {10.3758/BF03194566},
	publisher = {Springer},
	address = {Berlin-Heidelberg, Germany},
}

@article{kakaraparthi2021FaceSense,
    author = {Kakaraparthi, Vimal and Shao, Qijia and Carver, Charles J. and Pham, Tien and Bui, Nam and Nguyen, Phuc and Zhou, Xia and Vu, Tam},
	title = {FaceSense: Sensing Face Touch with an Ear-worn System},
    journal = {Proc. ACM Interact. Mob. Wearable Ubiquitous Technol.},
	issue_date = {September 2021},
    volume = {5},
    number = {3},
    month = sep,
    year = {2021},
    articleno = {110},
    pages = {1-27},
    numpages = {27},
	url = {https://doi.org/10.1145/3478129},
	doi = {10.1145/3478129},	
    publisher = {ACM},
    address = {New York, NY, USA},
}

@article{gil2023thumbair,
    author = {Gil, Hyunjae and Oakley, Ian},
    title = {ThumbAir: In-Air Typing for Head Mounted Displays},
    journal = {Proc. ACM Interact. Mob. Wearable Ubiquitous Technol.},
	issue_date = {December 2022},
    volume = {6},
    number = {4},
	month = jan,
    year = {2023},
    articleno = {164},
    pages = {1-30},
    numpages = {30},
    url = {https://doi.org/10.1145/3569474},
	doi = {10.1145/3569474},
    publisher = {ACM},
    address = {New York, NY, USA},
}

@article{Cheng2024BodyTouch,
    author = {Cheng, Wen-Wei and Chan, Liwei},
    title = {BodyTouch: Investigating Eye-Free, On-Body and Near-Body Touch Interactions with HMDs},
    journal = {Proc. ACM Interact. Mob. Wearable Ubiquitous Technol.},
	issue_date = {December 2023},
    volume = {7},
    number = {4},
    month = jan,
    year = {2024},
    articleno = {152},
    numpages = {22},
	url = {https://doi.org/10.1145/3631426},
    doi = {10.1145/3631426},
    publisher = {ACM},
    address = {New York, NY, USA},
}

@article{qiu2016using,
    author = {Qiu, Sen and Wang, Zhelong and Zhao, Hongyu and Hu, Huosheng},
    title = {Using Distributed Wearable Sensors to Measure and Evaluate Human Lower Limb Motions},
    journal = {IEEE Trans. Instrum. Meas.},
	issue_date = {April 2016},
    volume = {65},
    number = {4},
	month = apr,
    year = {2016},
    pages = {939--950},
    numpages = {12},
	url = {https://doi.org/10.1109/TIM.2015.2504078},
    doi = {10.1109/TIM.2015.2504078},
    publisher = {IEEE},
    address = {Piscataway, NJ, USA},
	ISSN = {1557-9662},
}

@article{lee2010hand,
    author = {Lee, Jae Yeol and Rhee, Gue Won and Seo, Dong Woo},
    title = {Hand Gesture-based Tangible Interactions for Manipulating Virtual Objects in a Mixed Reality Environment},
    journal = {Int. J. Adv. Manuf. Technol.},
	issue_date = {April 2010},
    volume = {51},
	month = apr,
    year = {2010},
    pages = {1069--1082},
	numpages = {14},
	url = {https://link.springer.com/article/10.1007/s00170-010-2671-x},
    doi = {10.1007/s00170-010-2671-x},
    publisher = {Springer},
    address = {Cham, Switzerland},
}

@article{sato_exploring_2024,
    author = {Sato, Yukina and Amesaka, Takashi and Yamamoto, Takumi and Watanabe, Hiroki and Sugiura, Yuta},
    title = {Exploring User-Defined Gestures as Input for Hearables and Recognizing Ear-Level Gestures with {IMUs}},   
    journal = {Proc. ACM Hum.-Comput. Interact.},
	issue_date = {September 2024},
    volume = {8},
    number = {MHCI},
    month = sep,
    year = {2024},
    articleno = {258},
    pages = {1--23},
    numpages = {23},
	url = {https://doi.org/10.1145/3676503},
	doi = {10.1145/3676503},
    publisher = {ACM},
    address = {New York, NY, USA},
}

@article{Zhao2024UiEar,
	author={Zhao, Guangrong and Shen, Yiran and Li, Feng and Liu, Lei and Cui, Lizhen and Wen, Hongkai},
    title = {{Ui-Ear}: On-Face Gesture Recognition Through On-Ear Vibration Sensing},    
    journal = {IEEE Trans. Mob. Comput.}, 
	issue_date = {March 2025},
    volume = {24},
    number = {3},
    year = {2025},
    pages = {1482--1495},
	numpages = {14},
	url = {https://doi.org/10.1109/TMC.2024.3480216},
    doi = {10.1109/TMC.2024.3480216},
    publisher = {IEEE},
    address = {Piscataway, NJ, USA},
}

@misc{aukey_ep_t27_official,
    author = {AUKEY Official},
    title = {AUKEY EP-T27 Soundstream Wireless Earbuds Noise Cancelling IPX7 Waterproof Black},
    year={2025},
    url = {https://www.aukey.com/products/aukey-soundstream-wireless-earbuds-noise-cancelling-ipx7-waterproof-black},
}

@misc{Vicon_2023, 
    title={Vicon: What is Motion Capture - Intelligence in Motion},
    author={Vicon Motion Systems Ltd UK},
    year={2023}, 
    url={https://www.vicon.com/about-us/what-is-motion-capture/}, 
}

@misc{Optitrack_2024, 
    title = {Optitrack: The Best Just Got Better},
    author = {NaturalPoint{,} Inc. DBA OptiTrack},
    year = {2024}, 
    url = {https://www.optitrack.com/about/}, 
}

@misc{Kinect_2024, 
    title = {Azure Kinect DK},
    author = {Microsoft},
    year = {2024}, 
    url = {https://azure.microsoft.com/en-us/products/kinect-dk}, 
}

@misc{airpod_camera_updated_apple_2026,
	title = {Beyond Audio: Why the AirPods Ultra’s New Infrared Sensors Are a Game-Changer},
	author = {Hutchinson, Roland},
	language = {en},
	urldate = {2026-05-13},
	journal = {Geeky Gadgets},
	month = may,
	year = {2026},
	lastaccessed = {May 19, 2026},
	url = {https://www.geeky-gadgets.com/apple-airpods-ultra-cameras-ai/},
}

@online{apple2025airpodspro3,
    title        = {Introducing AirPods Pro 3, the Ultimate Audio Experience},
    author       = {{Apple Inc.}},
    organization = {Apple Inc.},
    year         = {2025},
    month        = {May},
    day          = {13},
    lastaccessed = {May 18, 2026},
    url          = {https://www.apple.com/newsroom/2025/09/introducing-airpods-pro-3-the-ultimate-audio-experience/},
}

@misc{jackson_wearable_2023,
    title = {Wearable Interactive Audio Device},    
    author = {Jackson, Benjamin G. and Baugh, Brenton A. and Bloom, David H. and Roper, Gemma A. and Bark, Karlin Y. and Hulbert, Thomas S.},
    year = {2023}, 
    month = {Aug},
    note = {Patent No. US11743623B2, Filed July 13th., 2020, Issued Aug. 29th., 2023},
    url = {https://patents.google.com/patent/US11743623B2},
	lastaccessed = {},
}

@book{vella2008anatomy,
    author = {Vella, M.},
    title = {Anatomy for Strength and Fitness Training},
    year = {2008},
    isbn = {9781847731531},
    edition = {1st},
    publisher = {New Holland},
    address   = {London, UK},
    url = {https://books.google.ca/books?id=Q3_MvAEACAAJ},
}

\clearpage 

\appendix
\label{section: appendix}
\begin{appendices}

\setcounter{table}{0}
\renewcommand{\thetable}{\Alph{section}.\arabic{table}}
\setcounter{figure}{0}
\renewcommand{\thefigure}{\Alph{section}.\arabic{figure}}

\section{Detailed Statistical Results for Section \ref{subsection:Results_RQ1_RQ2}}
\label{appendix:Detailed Statistical Results-RQ1-RQ2}
{

\subsection{RQ1}
\label{appendix:Detailed Statistical Results-RQ1}
{

{
\begingroup
\setlength{\floatsep}{-1ex} 
\setlength{\intextsep}{-1ex} 
\setlength{\abovecaptionskip}{-1ex}
\setlength{\belowcaptionskip}{-1ex}
\setlength{\tabcolsep}{2pt}
\begin{table}[!h]
    \centering
    \vspace{.25cm}
    \renewcommand{\arraystretch}{1.05}
    \resizebox{235pt}{!}
    {
    \begin{tabular}{|cc||c||cccc||cccc||ccc|}
    \hline
    \multicolumn{2}{|c||}{}                                                                                                                                                                                    &                                   & \multicolumn{4}{c||}{}                                                                                                                                                           & \multicolumn{4}{c||}{}                                                                                                                                                              			& \multicolumn{3}{c|}{}                                                                                                                                                                         \\
    \multicolumn{2}{|c||}{}                                                                                                                                                                                    &                                   & \multicolumn{4}{c||}{\multirow{-2}{*}{\textbf{\begin{tabular}[c]{@{}c@{}}Shapiro-Wilk \\ Normality Test\end{tabular}}}}                                                          & \multicolumn{4}{c||}{}                                                                                                                                                              			& \multicolumn{3}{c|}{\multirow{-2}{*}{\textbf{\begin{tabular}[c]{@{}c@{}}Paired\\ t-test\end{tabular}}}}                                                                                       \\ \cline{4-7} \cline{12-14} 
    \multicolumn{2}{|c||}{}                                                                                                                                                                                    &                                   & \multicolumn{2}{c|}{\textbf{Midair}}                                                              & \multicolumn{2}{c||}{\textbf{Onskin}}                                        & \multicolumn{4}{c||}{\multirow{-3}{*}{\textbf{\begin{tabular}[c]{@{}c@{}}Wilcoxon Signed \\ Rank Test\\ (Paired Data)\end{tabular}}}}                                               			& \multicolumn{1}{c|}{}                                  & \multicolumn{1}{c|}{}                              & \multicolumn{1}{c|}{}                                                           \\ \cline{4-11}
    \multicolumn{2}{|c||}{\multirow{-4}{*}{\Large{\textbf{\begin{tabular}[c]{@{}c@{}}Metric\end{tabular}}}}}                                                                      & \multirow{-4}{*}{\rotatebox[origin=c]{90}{\textbf{Regions}}}   & \multicolumn{1}{c|}{\textbf{W}}  & \multicolumn{1}{c|}{\textbf{p}}                                & \multicolumn{1}{c|}{\textbf{W}}  & \textbf{p}                                & \multicolumn{1}{c|}{\textbf{W}}                   		& \multicolumn{1}{c|}{\textbf{Z}}                     & \multicolumn{1}{c|}{\textbf{p}}                                & \textbf{r}     & \multicolumn{1}{c|}{\multirow{-2}{*}{\textbf{t(23)}}}  & \multicolumn{1}{c|}{\multirow{-2}{*}{\textbf{p}}}  & \multirow{-2}{*}{\textbf{\begin{tabular}[c]{@{}c@{}}Cohen's\\ d\end{tabular}}}  \\ \hline
    \multicolumn{2}{|c||}{}                                                                                                                                                                                    & 4                                 & \multicolumn{1}{c|}{$.258$}      & \multicolumn{1}{c|}{{\color[HTML]{0000FF} \boldmath{$<.001$}}} & \multicolumn{1}{c|}{$.144$}      & {\color[HTML]{0000FF} \boldmath{$<.001$}} & \multicolumn{1}{c|}{$148.500$}                          & \multicolumn{1}{c|}{$-2.475$}                       & \multicolumn{1}{c|}{{\color[HTML]{0000FF} \textbf{.014}}}      & {$.089$}       & \multicolumn{1}{c|}{\cellcolor[HTML]{C0C0C0}{-}}       & \multicolumn{1}{c|}{\cellcolor[HTML]{C0C0C0}{-}}   & \multicolumn{1}{c|}{\cellcolor[HTML]{C0C0C0}{-}}                                \\ \cline{3-14} 
    \multicolumn{2}{|c||}{}                                                                                                                                                                                    & 6                                 & \multicolumn{1}{c|}{$.348$}      & \multicolumn{1}{c|}{{\color[HTML]{0000FF} \boldmath{$<.001$}}} & \multicolumn{1}{c|}{$.404$}      & {\color[HTML]{0000FF} \boldmath{$<.001$}} & \multicolumn{1}{c|}{$5922.000$}                         & \multicolumn{1}{c|}{$2.366$}                        & \multicolumn{1}{c|}{{\color[HTML]{0000FF} \textbf{.018}}}      & {$.056$}       & \multicolumn{1}{c|}{\cellcolor[HTML]{C0C0C0}{-}}       & \multicolumn{1}{c|}{\cellcolor[HTML]{C0C0C0}{-}}   & \multicolumn{1}{c|}{\cellcolor[HTML]{C0C0C0}{-}}                                \\ \cline{3-14} 
    \multicolumn{2}{|c||}{\multirow{-3}{*}{\textbf{\begin{tabular}[c]{@{}c@{}}Accuracy\\ (DV1)\end{tabular}}}}                                                                                                 & 8                                 & \multicolumn{1}{c|}{$.423$}      & \multicolumn{1}{c|}{{\color[HTML]{0000FF} \boldmath{$<.001$}}} & \multicolumn{1}{c|}{$.439$}      & {\color[HTML]{0000FF} \boldmath{$<.001$}} & \multicolumn{1}{c|}{$37233.000$}                        & \multicolumn{1}{c|}{$.875$}                         & \multicolumn{1}{c|}{$.381$}                                    & {$.015$}       & \multicolumn{1}{c|}{\cellcolor[HTML]{C0C0C0}{-}}       & \multicolumn{1}{c|}{\cellcolor[HTML]{C0C0C0}{-}}   & \multicolumn{1}{c|}{\cellcolor[HTML]{C0C0C0}{-}}                                \\ \hline
    \multicolumn{2}{|c||}{}                                                                                                                                                                                    & 4                                 & \multicolumn{1}{c|}{$.865$}      & \multicolumn{1}{c|}{{\color[HTML]{0000FF} \boldmath{$<.001$}}} & \multicolumn{1}{c|}{$.924$}      & {\color[HTML]{0000FF} \boldmath{$<.001$}} & \multicolumn{1}{c|}{$13924.000$}                        & \multicolumn{1}{c|}{$-10.584$}                      & \multicolumn{1}{c|}{{\color[HTML]{0000FF} \boldmath{$<.001$}}} & {$.382$}       & \multicolumn{1}{c|}{\cellcolor[HTML]{C0C0C0}{-}}       & \multicolumn{1}{c|}{\cellcolor[HTML]{C0C0C0}{-}}   & \multicolumn{1}{c|}{\cellcolor[HTML]{C0C0C0}{-}}                                \\ \cline{3-14} 
    \multicolumn{2}{|c||}{}                                                                                                                                                                                    & 6                                 & \multicolumn{1}{c|}{$.925$}      & \multicolumn{1}{c|}{{\color[HTML]{0000FF} \boldmath{$<.001$}}} & \multicolumn{1}{c|}{$.951$}      & {\color[HTML]{0000FF} \boldmath{$<.001$}} & \multicolumn{1}{c|}{$79484.000$}                        & \multicolumn{1}{c|}{$-14.631$}                      & \multicolumn{1}{c|}{{\color[HTML]{0000FF} \boldmath{$<.001$}}} & {$.352$}       & \multicolumn{1}{c|}{\cellcolor[HTML]{C0C0C0}{-}}       & \multicolumn{1}{c|}{\cellcolor[HTML]{C0C0C0}{-}}   & \multicolumn{1}{c|}{\cellcolor[HTML]{C0C0C0}{-}}                                \\ \cline{3-14} 
    \multicolumn{2}{|c||}{\multirow{-3}{*}{\textbf{\begin{tabular}[c]{@{}c@{}}Time\\ (DV2)\end{tabular}}}}                                                                                                     & 8                                 & \multicolumn{1}{c|}{$.944$}      & \multicolumn{1}{c|}{{\color[HTML]{0000FF} \boldmath{$<.001$}}} & \multicolumn{1}{c|}{$.970$}      & {\color[HTML]{0000FF} \boldmath{$<.001$}} & \multicolumn{1}{c|}{$218413.000$}                       & \multicolumn{1}{c|}{$-21.360$}                      & \multicolumn{1}{c|}{{\color[HTML]{0000FF} \boldmath{$<.001$}}} & {$.385$}       & \multicolumn{1}{c|}{\cellcolor[HTML]{C0C0C0}{-}}       & \multicolumn{1}{c|}{\cellcolor[HTML]{C0C0C0}{-}}   & \multicolumn{1}{c|}{\cellcolor[HTML]{C0C0C0}{-}}                                \\ \hline
    \multicolumn{2}{|c||}{}                                                                                                                                                                                    & 4                                 & \multicolumn{1}{c|}{$.945$}      & \multicolumn{1}{c|}{{\color[HTML]{0000FF} \boldmath{$<.001$}}} & \multicolumn{1}{c|}{$.891$}      & {\color[HTML]{0000FF} \boldmath{$<.001$}} & \multicolumn{1}{c|}{$61402.000$}                        & \multicolumn{1}{c|}{$11.230$}                       & \multicolumn{1}{c|}{{\color[HTML]{0000FF} \boldmath{$<.001$}}} & {$.405$}       & \multicolumn{1}{c|}{\cellcolor[HTML]{C0C0C0}{-}}       & \multicolumn{1}{c|}{\cellcolor[HTML]{C0C0C0}{-}}   & \multicolumn{1}{c|}{\cellcolor[HTML]{C0C0C0}{-}}                                \\ \cline{3-14} 
    \multicolumn{2}{|c||}{}                                                                                                                                                                                    & 6                                 & \multicolumn{1}{c|}{$.961$}      & \multicolumn{1}{c|}{{\color[HTML]{0000FF} \boldmath{$<.001$}}} & \multicolumn{1}{c|}{$.934$}      & {\color[HTML]{0000FF} \boldmath{$<.001$}} & \multicolumn{1}{c|}{$280603.000$}                       & \multicolumn{1}{c|}{$12.778$}                       & \multicolumn{1}{c|}{{\color[HTML]{0000FF} \boldmath{$<.001$}}} & {$.307$}       & \multicolumn{1}{c|}{\cellcolor[HTML]{C0C0C0}{-}}       & \multicolumn{1}{c|}{\cellcolor[HTML]{C0C0C0}{-}}   & \multicolumn{1}{c|}{\cellcolor[HTML]{C0C0C0}{-}}                                \\ \cline{3-14} 
    \multicolumn{2}{|c||}{\multirow{-3}{*}{\textbf{\begin{tabular}[c]{@{}c@{}}Path\\ Length\\ (DV3)\end{tabular}}}}                                                                                            & 8                                 & \multicolumn{1}{c|}{$.968$}      & \multicolumn{1}{c|}{{\color[HTML]{0000FF} \boldmath{$<.001$}}} & \multicolumn{1}{c|}{$.893$}      & {\color[HTML]{0000FF} \boldmath{$<.001$}} & \multicolumn{1}{c|}{$919425.000$}                       & \multicolumn{1}{c|}{$18.935$}                       & \multicolumn{1}{c|}{{\color[HTML]{0000FF} \boldmath{$<.001$}}} & {$.342$}       & \multicolumn{1}{c|}{\cellcolor[HTML]{C0C0C0}{-}}       & \multicolumn{1}{c|}{\cellcolor[HTML]{C0C0C0}{-}}   & \multicolumn{1}{c|}{\cellcolor[HTML]{C0C0C0}{-}}                                \\ \hline
    \multicolumn{1}{|c|}{}                                                                                                                            & \multicolumn{1}{c||}{}                                 & 4                                 & \multicolumn{1}{c|}{$.946$}      & \multicolumn{1}{c|}{{\color[HTML]{0000FF} \boldmath{$<.001$}}} & \multicolumn{1}{c|}{$.956$}      & {\color[HTML]{0000FF} \boldmath{$<.001$}} & \multicolumn{1}{c|}{$27268.000$}                        & \multicolumn{1}{c|}{$-2.446$}                       & \multicolumn{1}{c|}{{\color[HTML]{0000FF} \boldmath{$.016$}}}  & {$.088$}       & \multicolumn{1}{c|}{\cellcolor[HTML]{C0C0C0}{-}}       & \multicolumn{1}{c|}{\cellcolor[HTML]{C0C0C0}{-}}   & \multicolumn{1}{c|}{\cellcolor[HTML]{C0C0C0}{-}}                                \\ \cline{3-14} 
    \multicolumn{1}{|c|}{}                                                                                                                            & \multicolumn{1}{c||}{}                                 & 6                                 & \multicolumn{1}{c|}{$.979$}      & \multicolumn{1}{c|}{{\color[HTML]{0000FF} \boldmath{$<.001$}}} & \multicolumn{1}{c|}{$.977$}      & {\color[HTML]{0000FF} \boldmath{$<.001$}} & \multicolumn{1}{c|}{$148493.000$}                       & \multicolumn{1}{c|}{$.320$}                         & \multicolumn{1}{c|}{$.841$}                                    & {$.007$}       & \multicolumn{1}{c|}{\cellcolor[HTML]{C0C0C0}{-}}       & \multicolumn{1}{c|}{\cellcolor[HTML]{C0C0C0}{-}}   & \multicolumn{1}{c|}{\cellcolor[HTML]{C0C0C0}{-}}                                \\ \cline{3-14} 
    \multicolumn{1}{|c|}{}                                                                                                                            & \multicolumn{1}{c||}{\multirow{-3}{*}{\textbf{S}}}     & 8                                 & \multicolumn{1}{c|}{$.952$}      & \multicolumn{1}{c|}{{\color[HTML]{0000FF} \boldmath{$<.001$}}} & \multicolumn{1}{c|}{$.895$}      & {\color[HTML]{0000FF} \boldmath{$<.001$}} & \multicolumn{1}{c|}{$439705.000$}                       & \multicolumn{1}{c|}{$4.543$}                        & \multicolumn{1}{c|}{{\color[HTML]{0000FF} \boldmath{$<.001$}}} & {$.082$}       & \multicolumn{1}{c|}{\cellcolor[HTML]{C0C0C0}{-}}       & \multicolumn{1}{c|}{\cellcolor[HTML]{C0C0C0}{-}}   & \multicolumn{1}{c|}{\cellcolor[HTML]{C0C0C0}{-}}                                \\ \cline{2-14} 
    \multicolumn{1}{|c|}{}                                                                                                                            & \multicolumn{1}{c||}{}                                 & 4                                 & \multicolumn{1}{c|}{$.961$}      & \multicolumn{1}{c|}{{\color[HTML]{0000FF} \boldmath{$<.001$}}} & \multicolumn{1}{c|}{$.966$}      & {\color[HTML]{0000FF} \boldmath{$<.001$}} & \multicolumn{1}{c|}{$30875.000$}                        & \multicolumn{1}{c|}{$-.247$}                        & \multicolumn{1}{c|}{$.849$}                                    & {$.009$}       & \multicolumn{1}{c|}{\cellcolor[HTML]{C0C0C0}{-}}       & \multicolumn{1}{c|}{\cellcolor[HTML]{C0C0C0}{-}}   & \multicolumn{1}{c|}{\cellcolor[HTML]{C0C0C0}{-}}                                \\ \cline{3-14} 
    \multicolumn{1}{|c|}{}                                                                                                                            & \multicolumn{1}{c||}{}                                 & 6                                 & \multicolumn{1}{c|}{$.976$}      & \multicolumn{1}{c|}{{\color[HTML]{0000FF} \boldmath{$<.001$}}} & \multicolumn{1}{c|}{$.975$}      & {\color[HTML]{0000FF} \boldmath{$<.001$}} & \multicolumn{1}{c|}{$120336.000$}                       & \multicolumn{1}{c|}{$-4.068$}                       & \multicolumn{1}{c|}{{\color[HTML]{0000FF} \boldmath{$<.001$}}} & {$.097$}       & \multicolumn{1}{c|}{\cellcolor[HTML]{C0C0C0}{-}}       & \multicolumn{1}{c|}{\cellcolor[HTML]{C0C0C0}{-}}   & \multicolumn{1}{c|}{\cellcolor[HTML]{C0C0C0}{-}}                                \\ \cline{3-14} 
    \multicolumn{1}{|c|}{\multirow{-6}{*}{\rotatebox[origin=c]{90}{\textbf{\begin{tabular}[c]{@{}c@{}}Angular\\ Displacement\\ (DV4)\end{tabular}}}}} & \multicolumn{1}{c||}{\multirow{-3}{*}{\textbf{E}}}     & 8                                 & \multicolumn{1}{c|}{$.980$}      & \multicolumn{1}{c|}{{\color[HTML]{0000FF} \boldmath{$<.001$}}} & \multicolumn{1}{c|}{$.902$}      & {\color[HTML]{0000FF} \boldmath{$<.001$}} & \multicolumn{1}{c|}{$411172.000$}                       & \multicolumn{1}{c|}{$-7.317$}                       & \multicolumn{1}{c|}{{\color[HTML]{0000FF} \boldmath{$<.001$}}} & {$.132$}       & \multicolumn{1}{c|}{\cellcolor[HTML]{C0C0C0}{-}}       & \multicolumn{1}{c|}{\cellcolor[HTML]{C0C0C0}{-}}   & \multicolumn{1}{c|}{\cellcolor[HTML]{C0C0C0}{-}}                                \\ \hline\hline
    \multicolumn{2}{|c||}{}                                                                                                                                                                                    & 4                                 & \multicolumn{1}{c|}{$.815$}      & \multicolumn{1}{c|}{{\color[HTML]{0000FF} \boldmath{$<.001$}}} & \multicolumn{1}{c|}{$.901$}      & {\color[HTML]{0000FF} \boldmath{$.023$}}  & \multicolumn{1}{c|}{$62.500$}                           & \multicolumn{1}{c|}{$-1.585$}                       & \multicolumn{1}{c|}{$.195$}                                    & {$.229$}       & \multicolumn{1}{c|}{\cellcolor[HTML]{C0C0C0}{-}}       & \multicolumn{1}{c|}{\cellcolor[HTML]{C0C0C0}{-}}   & \multicolumn{1}{c|}{\cellcolor[HTML]{C0C0C0}{-}}                                \\ \cline{3-14} 
    \multicolumn{2}{|c||}{}                                                                                                                                                                                    & 6                                 & \multicolumn{1}{c|}{$.914$}      & \multicolumn{1}{c|}{{\color[HTML]{0000FF} \boldmath{$.043$}}}  & \multicolumn{1}{c|}{$.924$}      & {$.073$}                                  & \multicolumn{1}{c|}{$105.500$}                          & \multicolumn{1}{c|}{$-.100$}                        & \multicolumn{1}{c|}{$1.000$}                                   & {$.014$}       & \multicolumn{1}{c|}{\cellcolor[HTML]{C0C0C0}{-}}       & \multicolumn{1}{c|}{\cellcolor[HTML]{C0C0C0}{-}}   & \multicolumn{1}{c|}{\cellcolor[HTML]{C0C0C0}{-}}                                \\ \cline{3-14} 
    \multicolumn{2}{|c||}{\multirow{-3}{*}{\rotatebox[origin=c]{20}{\textbf{\begin{tabular}[c]{@{}c@{}}Mental\\ Demand\end{tabular}}}}}                                                                        & 8                                 & \multicolumn{1}{c|}{$.929$}      & \multicolumn{1}{c|}{{\color[HTML]{0000FF} \boldmath{$.006$}}}  & \multicolumn{1}{c|}{$.905$}      & {\color[HTML]{0000FF} \boldmath{$.028$}}  & \multicolumn{1}{c|}{$117.000$}                          & \multicolumn{1}{c|}{$-.672$}                        & \multicolumn{1}{c|}{$.532$}                                    & {$.097$}       & \multicolumn{1}{c|}{\cellcolor[HTML]{C0C0C0}{-}}       & \multicolumn{1}{c|}{\cellcolor[HTML]{C0C0C0}{-}}   & \multicolumn{1}{c|}{\cellcolor[HTML]{C0C0C0}{-}}                                \\ \hline
    \multicolumn{2}{|c||}{}                                                                                                                                                                                    & 4                                 & \multicolumn{1}{c|}{$.801$}      & \multicolumn{1}{c|}{{\color[HTML]{0000FF} \boldmath{$<.001$}}} & \multicolumn{1}{c|}{$.848$}      & {\color[HTML]{0000FF} \boldmath{$<.001$}} & \multicolumn{1}{c|}{$79.500$}                           & \multicolumn{1}{c|}{$.392$}                         & \multicolumn{1}{c|}{$.905$}                                    & {$.056$}       & \multicolumn{1}{c|}{\cellcolor[HTML]{C0C0C0}{-}}       & \multicolumn{1}{c|}{\cellcolor[HTML]{C0C0C0}{-}}   & \multicolumn{1}{c|}{\cellcolor[HTML]{C0C0C0}{-}}                                \\ \cline{3-14} 
    \multicolumn{2}{|c||}{}                                                                                                                                                                                    & 6                                 & \multicolumn{1}{c|}{$.914$}      & \multicolumn{1}{c|}{{\color[HTML]{0000FF} \boldmath{$.043$}}}  & \multicolumn{1}{c|}{$.950$}      & {$.276$}                                  & \multicolumn{1}{c|}{$128.500$}                          & \multicolumn{1}{c|}{$1.019$}                        & \multicolumn{1}{c|}{$.389$}                                    & {$.147$}       & \multicolumn{1}{c|}{\cellcolor[HTML]{C0C0C0}{-}}       & \multicolumn{1}{c|}{\cellcolor[HTML]{C0C0C0}{-}}   & \multicolumn{1}{c|}{\cellcolor[HTML]{C0C0C0}{-}}                                \\ \cline{3-14} 
    \multicolumn{2}{|c||}{\multirow{-3}{*}{\rotatebox[origin=c]{20}{\textbf{\begin{tabular}[c]{@{}c@{}}Physical\\ Demand\end{tabular}}}}}                                                                      & 8                                 & \multicolumn{1}{c|}{$.934$}      & \multicolumn{1}{c|}{$.117$}                                    & \multicolumn{1}{c|}{$.955$}      & {$.348$}                                  & \multicolumn{4}{c|}{\cellcolor[HTML]{C0C0C0}{-}}                                                                                                                                                & \multicolumn{1}{c|}{$.671$}                            & \multicolumn{1}{c|}{$.508$}                        & \multicolumn{1}{c|}{$.137$}                                                     \\ \hline
    \multicolumn{2}{|c||}{}                                                                                                                                                                                    & 4                                 & \multicolumn{1}{c|}{$.921$}      & \multicolumn{1}{c|}{$.060$}                                    & \multicolumn{1}{c|}{$.899$}      & {\color[HTML]{0000FF} \boldmath{$.020$}}  & \multicolumn{1}{c|}{$131.500$}                          & \multicolumn{1}{c|}{$1.106$}                        & \multicolumn{1}{c|}{$.330$}                                    & {$.160$}       & \multicolumn{1}{c|}{\cellcolor[HTML]{C0C0C0}{-}}       & \multicolumn{1}{c|}{\cellcolor[HTML]{C0C0C0}{-}}   & \multicolumn{1}{c|}{\cellcolor[HTML]{C0C0C0}{-}}                                \\ \cline{3-14} 
    \multicolumn{2}{|c||}{}                                                                                                                                                                                    & 6                                 & \multicolumn{1}{c|}{$.958$}      & \multicolumn{1}{c|}{$.408$}                                    & \multicolumn{1}{c|}{$.958$}      & {$.240$}                                  & \multicolumn{4}{c|}{\cellcolor[HTML]{C0C0C0}{-}}                                                                                                                                                & \multicolumn{1}{c|}{$.650$}                            & \multicolumn{1}{c|}{$.522$}                        & \multicolumn{1}{c|}{$.133$}                                                     \\ \cline{3-14} 
    \multicolumn{2}{|c||}{\multirow{-3}{*}{\rotatebox[origin=c]{20}{\textbf{\begin{tabular}[c]{@{}c@{}}Temporal\\Demand\end{tabular}}}}}                                                                       & 8                                 & \multicolumn{1}{c|}{$.955$}      & \multicolumn{1}{c|}{$.343$}                                    & \multicolumn{1}{c|}{$.964$}      & {$.515$}                                  & \multicolumn{4}{c|}{\cellcolor[HTML]{C0C0C0}{-}}                                                                                                                                                & \multicolumn{1}{c|}{$.487$}                            & \multicolumn{1}{c|}{$.631$}                        & \multicolumn{1}{c|}{$.010$}                                                     \\ \hline
    \multicolumn{2}{|c||}{}                                                                                                                                                                                    & 4                                 & \multicolumn{1}{c|}{$.874$}      & \multicolumn{1}{c|}{{\color[HTML]{0000FF} \boldmath{$.006$}}}  & \multicolumn{1}{c|}{$.870$}      & {\color[HTML]{0000FF} \boldmath{$.005$}}  & \multicolumn{1}{c|}{$154.000$}                          & \multicolumn{1}{c|}{$1.018$}                        & \multicolumn{1}{c|}{$.379$}                                    & {$.147$}       & \multicolumn{1}{c|}{\cellcolor[HTML]{C0C0C0}{-}}       & \multicolumn{1}{c|}{\cellcolor[HTML]{C0C0C0}{-}}   & \multicolumn{1}{c|}{\cellcolor[HTML]{C0C0C0}{-}}                                \\ \cline{3-14} 
    \multicolumn{2}{|c||}{}                                                                                                                                                                                    & 6                                 & \multicolumn{1}{c|}{$.900$}      & \multicolumn{1}{c|}{{\color[HTML]{0000FF} \boldmath{$.021$}}}  & \multicolumn{1}{c|}{$.918$}      & {$.052$}                                  & \multicolumn{1}{c|}{$144.000$}                          & \multicolumn{1}{c|}{$1.349$}                        & \multicolumn{1}{c|}{$.150$}                                    & {$.195$}       & \multicolumn{1}{c|}{\cellcolor[HTML]{C0C0C0}{-}}       & \multicolumn{1}{c|}{\cellcolor[HTML]{C0C0C0}{-}}   & \multicolumn{1}{c|}{\cellcolor[HTML]{C0C0C0}{-}}                                \\ \cline{3-14} 
    \multicolumn{2}{|c||}{\multirow{-3}{*}{\rotatebox[origin=c]{20}{\textbf{Performance}}}}                                                                                                                    & 8                                 & \multicolumn{1}{c|}{$.914$}      & \multicolumn{1}{c|}{{\color[HTML]{0000FF} \boldmath{$.042$}}}  & \multicolumn{1}{c|}{$.969$}      & {$.648$}                                  & \multicolumn{1}{c|}{$156.500$}                          & \multicolumn{1}{c|}{$1.389$}                        & \multicolumn{1}{c|}{$.159$}                                    & {$.200$}       & \multicolumn{1}{c|}{\cellcolor[HTML]{C0C0C0}{-}}       & \multicolumn{1}{c|}{\cellcolor[HTML]{C0C0C0}{-}}   & \multicolumn{1}{c|}{\cellcolor[HTML]{C0C0C0}{-}}                                \\ \hline
    \multicolumn{2}{|c||}{}                                                                                                                                                                                    & 4                                 & \multicolumn{1}{c|}{$.918$}      & \multicolumn{1}{c|}{$.052$}                                    & \multicolumn{1}{c|}{$.887$}      & {\color[HTML]{0000FF} \boldmath{$.012$}}  & \multicolumn{1}{c|}{$152.500$}                          & \multicolumn{1}{c|}{$1.285$}                        & \multicolumn{1}{c|}{$.199$}                                    & {$.225$}       & \multicolumn{1}{c|}{\cellcolor[HTML]{C0C0C0}{-}}       & \multicolumn{1}{c|}{\cellcolor[HTML]{C0C0C0}{-}}   & \multicolumn{1}{c|}{\cellcolor[HTML]{C0C0C0}{-}}                                \\ \cline{3-14} 
    \multicolumn{2}{|c||}{}                                                                                                                                                                                    & 6                                 & \multicolumn{1}{c|}{$.914$}      & \multicolumn{1}{c|}{{\color[HTML]{0000FF} \boldmath{$.044$}}}  & \multicolumn{1}{c|}{$.920$}      & {$.058$}                                  & \multicolumn{1}{c|}{$131.000$}                          & \multicolumn{1}{c|}{$1.093$}                        & \multicolumn{1}{c|}{$.338$}                                    & {$.158$}       & \multicolumn{1}{c|}{\cellcolor[HTML]{C0C0C0}{-}}       & \multicolumn{1}{c|}{\cellcolor[HTML]{C0C0C0}{-}}   & \multicolumn{1}{c|}{\cellcolor[HTML]{C0C0C0}{-}}                                \\ \cline{3-14} 
    \multicolumn{2}{|c||}{\multirow{-3}{*}{\rotatebox[origin=c]{20}{\textbf{Effort}}}}                                                                                                                         & 8                                 & \multicolumn{1}{c|}{$.909$}      & \multicolumn{1}{c|}{{\color[HTML]{0000FF} \boldmath{$.034$}}}  & \multicolumn{1}{c|}{$.926$}      & {\color[HTML]{0000FF} \boldmath{$.027$}}  & \multicolumn{1}{c|}{$144.000$}                          & \multicolumn{1}{c|}{$1.234$}                        & \multicolumn{1}{c|}{$.150$}                                    & {$.178$}       & \multicolumn{1}{c|}{\cellcolor[HTML]{C0C0C0}{-}}       & \multicolumn{1}{c|}{\cellcolor[HTML]{C0C0C0}{-}}   & \multicolumn{1}{c|}{\cellcolor[HTML]{C0C0C0}{-}}                                \\ \hline
    \multicolumn{2}{|c||}{}                                                                                                                                                                                    & 4                                 & \multicolumn{1}{c|}{$.799$}      & \multicolumn{1}{c|}{{\color[HTML]{0000FF} \boldmath{$<.001$}}} & \multicolumn{1}{c|}{$.806$}      & {\color[HTML]{0000FF} \boldmath{$<.001$}} & \multicolumn{1}{c|}{$63.000$}                           & \multicolumn{1}{c|}{$.221$}                         & \multicolumn{1}{c|}{$.886$}                                    & {$.031$}       & \multicolumn{1}{c|}{\cellcolor[HTML]{C0C0C0}{-}}       & \multicolumn{1}{c|}{\cellcolor[HTML]{C0C0C0}{-}}   & \multicolumn{1}{c|}{\cellcolor[HTML]{C0C0C0}{-}}                                \\ \cline{3-14} 
    \multicolumn{2}{|c||}{}                                                                                                                                                                                    & 6                                 & \multicolumn{1}{c|}{$.817$}      & \multicolumn{1}{c|}{{\color[HTML]{0000FF} \boldmath{$<.001$}}} & \multicolumn{1}{c|}{$.920$}      & {$.059$}                                  & \multicolumn{1}{c|}{$86.500$}                           & \multicolumn{1}{c|}{$.460$}                         & \multicolumn{1}{c|}{$.746$}                                    & {$.066$}       & \multicolumn{1}{c|}{\cellcolor[HTML]{C0C0C0}{-}}       & \multicolumn{1}{c|}{\cellcolor[HTML]{C0C0C0}{-}}   & \multicolumn{1}{c|}{\cellcolor[HTML]{C0C0C0}{-}}                                \\ \cline{3-14} 
    \multicolumn{2}{|c||}{\multirow{-3}{*}{\rotatebox[origin=c]{20}{\textbf{Frustration}}}}                                                                                                                    & 8                                 & \multicolumn{1}{c|}{$.931$}      & \multicolumn{1}{c|}{$.102$}                                    & \multicolumn{1}{c|}{$.913$}      & {\color[HTML]{0000FF} \boldmath{$.040$}}  & \multicolumn{1}{c|}{$89.500$}                           & \multicolumn{1}{c|}{$.682$}                         & \multicolumn{1}{c|}{$.553$}                                    & {$.098$}       & \multicolumn{1}{c|}{\cellcolor[HTML]{C0C0C0}{-}}       & \multicolumn{1}{c|}{\cellcolor[HTML]{C0C0C0}{-}}   & \multicolumn{1}{c|}{\cellcolor[HTML]{C0C0C0}{-}}                                \\ \hline
    \end{tabular}%
    }
    \vspace{.15cm}
    \caption{Statistical comparison of swipe performance metrics across  midair and onskin spaces for different region segmentation patterns (RQ1). Figure~\ref{fig:AllGestures_RQ1_RQ2} highlights statistically significant pairwise differences.}
    \Description{Statistical comparison of swipe performance metrics across  midair and onskin spaces for different region segmentation patterns (RQ1). Figure~\ref{fig:AllGestures_RQ1_RQ2} highlights statistically significant pairwise differences.\\}
    \label{tab:RQ1_Metrics}    
\end{table}
\endgroup
}

}

\vfill

\subsection{RQ2}
\label{appendix:Detailed Statistical Results-RQ2}
{

{
\begingroup
\setlength{\floatsep}{-1ex} 
\setlength{\intextsep}{-1ex} 
\setlength{\abovecaptionskip}{-1ex}
\setlength{\belowcaptionskip}{-1ex}
\setlength{\tabcolsep}{2pt}
\begin{table*}[!b]
    \centering
    \renewcommand{\arraystretch}{1.05}
    \resizebox{350pt}{!}
    {
    \begin{tabular}{|cc||c||cccccc||cc||ccc||ccc||ccc|}
    \hline
    \multicolumn{2}{|c||}{}                                                                                                                                                         &                                                                                                                  & \multicolumn{6}{c||}{}                                                                                                                                                                                                                                                      		   & \multicolumn{2}{c||}{}                                                                                                        		& \multicolumn{3}{c||}{}                                                                                                                                                    	 & \multicolumn{3}{c||}{}                                                                                                                                                 & \multicolumn{3}{c|}{}                                                                                                                                                                \\ 
    \multicolumn{2}{|c||}{}                                                                                                                                                         &                                                                                                                  & \multicolumn{6}{c||}{}                                                                                                                                                                                                                                                      		   & \multicolumn{2}{c||}{}                                                                                                        		& \multicolumn{3}{c||}{}                                                                                                                                                    	 & \multicolumn{3}{c||}{}                                                                                                                                                 & \multicolumn{3}{c|}{}                                                                                                                                                                \\ 
    \multicolumn{2}{|c||}{}                                                                                                                                                         &                                                                                                                  & \multicolumn{6}{c||}{\multirow{-3}{*}{\textbf{\begin{tabular}[c]{@{}c@{}}Shapiro-Wilk\\ Normality Test\end{tabular}}}}                                                                                                                                                      		   & \multicolumn{2}{c||}{}                                                                                                        		& \multicolumn{3}{c||}{}                                                                                                                                                    	 & \multicolumn{3}{c||}{}                                                                                                                                                 & \multicolumn{3}{c|}{}                                                                                                                                                                \\  \cline{4-9}
    \multicolumn{2}{|c||}{}                                                                                                                                                         &                                                                                                                  & \multicolumn{2}{c|}{\textbf{4 Regions}}                                                           & \multicolumn{2}{c|}{\textbf{6 Regions}}                                                           & \multicolumn{2}{c||}{\textbf{8 Regions}}                                      & \multicolumn{2}{c||}{\multirow{-4}{*}{\textbf{\begin{tabular}[c]{@{}c@{}}Friedman \\ Test\\ (Matched\\ Group)\end{tabular}}}} 		& \multicolumn{3}{c||}{\multirow{-4}{*}{\textbf{\begin{tabular}[c]{@{}c@{}}Post-Hoc Wilcoxon \\ Test with Bonferroni\\ Correction For\\ Pairwise Comparison\end{tabular}}}} 	 & \multicolumn{3}{c||}{\multirow{-4}{*}{\textbf{\begin{tabular}[c]{@{}c@{}}One-way Repeated \\ Measures Anova \\ for Pairwise \\ Comparison\end{tabular}}}}              & \multicolumn{3}{c|}{\multirow{-4}{*}{\textbf{\begin{tabular}[c]{@{}c@{}}Post-Hoc \\Pairwise t-test\\ with Bonferroni\\  Correction\end{tabular}}}}                                   \\  \cline{4-20} 
    \multicolumn{2}{|c||}{\multirow{-5}{*}{\Large{\rotatebox[origin=c]{0}{\textbf{\begin{tabular}[c]{@{}c@{}}Metric\end{tabular}}}}}}                                               & \multirow{-5}{*}{\rotatebox[origin=c]{90}{\textbf{\begin{tabular}[c]{@{}c@{}}Interaction\\ Space\end{tabular}}}} & \multicolumn{1}{c|}{\textbf{W}}  & \multicolumn{1}{c|}{\textbf{p}}                                & \multicolumn{1}{c|}{\textbf{W}}  & \multicolumn{1}{c|}{\textbf{p}}                                & \multicolumn{1}{c|}{\textbf{W}}  & \textbf{p}                                 & \multicolumn{1}{c|}{\textbf{$\chi^2$}}                         & \textbf{p}                                                       	& \multicolumn{1}{c|}{\textbf{$p_{4-6}$}}                         & \multicolumn{1}{c|}{\textbf{$p_{6-8}$}}                        & \textbf{$p_{4-8}$}                       	 & \multicolumn{1}{c|}{\textbf{$F(2, 46)$}}                      & \multicolumn{1}{c|}{\textbf{p}}                     & \textbf{$\eta^2$}                                & \multicolumn{1}{c|}{\textbf{$p_{4-6}$}}                         & \multicolumn{1}{c|}{\textbf{$p_{6-8}$}}                         & \textbf{$p_{4-8}$}                               \\  \hline
    \multicolumn{2}{|c||}{}                                                                                                                                                         & \textbf{Midair}                                                                                                  & \multicolumn{1}{c|}{$.258$}      & \multicolumn{1}{c|}{{\color[HTML]{0000FF} \boldmath{$<.001$}}} & \multicolumn{1}{c|}{$.348$}      & \multicolumn{1}{c|}{{\color[HTML]{0000FF} \boldmath{$<.001$}}} & \multicolumn{1}{c|}{$.423$}      & {\color[HTML]{0000FF} \boldmath{$<.001$}}  & \multicolumn{1}{c|}{$57.901$}                                  & {\color[HTML]{0000FF} \boldmath{$<.001$}}                        	& \multicolumn{1}{c|}{$.064$}                                     & \multicolumn{1}{c|}{{\color[HTML]{0000FF} \boldmath{$.006$}}}  & {\color[HTML]{0000FF} \boldmath{$<.001$}}   & \multicolumn{1}{c|}{\cellcolor[HTML]{D9D9D9}{-}}              & \multicolumn{1}{c|}{\cellcolor[HTML]{D9D9D9}{-}}    & {\cellcolor[HTML]{D9D9D9}{-}}                    & \multicolumn{1}{c|}{\cellcolor[HTML]{D9D9D9}{-}}                & \multicolumn{1}{c|}{\cellcolor[HTML]{D9D9D9}{-}}                & \multicolumn{1}{c|}{\cellcolor[HTML]{D9D9D9}{-}} \\  \cline{3-20} 
    \multicolumn{2}{|c||}{\multirow{-2}{*}{\rotatebox[origin=c]{0}{\textbf{\begin{tabular}[c]{@{}c@{}}Accuracy\\(DV1)\end{tabular}}}}}                                              & \textbf{Onskin}                                                                                                  & \multicolumn{1}{c|}{$.144$}      & \multicolumn{1}{c|}{{\color[HTML]{0000FF} \boldmath{$<.001$}}} & \multicolumn{1}{c|}{$.404$}      & \multicolumn{1}{c|}{{\color[HTML]{0000FF} \boldmath{$<.001$}}} & \multicolumn{1}{c|}{$.439$}      & {\color[HTML]{0000FF} \boldmath{$<.001$}}  & \multicolumn{1}{c|}{$164.760$}                                 & {\color[HTML]{0000FF} \boldmath{$<.001$}}                        	& \multicolumn{1}{c|}{{\color[HTML]{0000FF} \boldmath{$<.001$}}}  & \multicolumn{1}{c|}{$.370$}                                    & {\color[HTML]{0000FF} \boldmath{$<.001$}}   & \multicolumn{1}{c|}{\cellcolor[HTML]{D9D9D9}{-}}              & \multicolumn{1}{c|}{\cellcolor[HTML]{D9D9D9}{-}}    & {\cellcolor[HTML]{D9D9D9}{-}}                    & \multicolumn{1}{c|}{\cellcolor[HTML]{D9D9D9}{-}}                & \multicolumn{1}{c|}{\cellcolor[HTML]{D9D9D9}{-}}                & \multicolumn{1}{c|}{\cellcolor[HTML]{D9D9D9}{-}} \\  \hline
    \multicolumn{2}{|c||}{}                                                                                                                                                         & \textbf{Midair}                                                                                                  & \multicolumn{1}{c|}{$.865$}      & \multicolumn{1}{c|}{{\color[HTML]{0000FF} \boldmath{$<.001$}}} & \multicolumn{1}{c|}{$.925$}      & \multicolumn{1}{c|}{{\color[HTML]{0000FF} \boldmath{$<.001$}}} & \multicolumn{1}{c|}{$.944$}      & {\color[HTML]{0000FF} \boldmath{$<.001$}}  & \multicolumn{1}{c|}{$55.202$}                                  & {\color[HTML]{0000FF} \boldmath{$<.001$}}                        	& \multicolumn{1}{c|}{{\color[HTML]{0000FF} \boldmath{$.034$}}}   & \multicolumn{1}{c|}{{\color[HTML]{0000FF} \boldmath{$.022$}}}  & {\color[HTML]{0000FF} \boldmath{$<.001$}}   & \multicolumn{1}{c|}{\cellcolor[HTML]{D9D9D9}{-}}              & \multicolumn{1}{c|}{\cellcolor[HTML]{D9D9D9}{-}}    & {\cellcolor[HTML]{D9D9D9}{-}}                    & \multicolumn{1}{c|}{\cellcolor[HTML]{D9D9D9}{-}}                & \multicolumn{1}{c|}{\cellcolor[HTML]{D9D9D9}{-}}                & \multicolumn{1}{c|}{\cellcolor[HTML]{D9D9D9}{-}} \\  \cline{3-20} 
    \multicolumn{2}{|c||}{\multirow{-2}{*}{\rotatebox[origin=c]{0}{\textbf{\begin{tabular}[c]{@{}c@{}}Time\\(DV2)\end{tabular}}}}}                                                  & \textbf{Onskin}                                                                                                  & \multicolumn{1}{c|}{$.924$}      & \multicolumn{1}{c|}{{\color[HTML]{0000FF} \boldmath{$<.001$}}} & \multicolumn{1}{c|}{$.951$}      & \multicolumn{1}{c|}{{\color[HTML]{0000FF} \boldmath{$<.001$}}} & \multicolumn{1}{c|}{$.970$}      & {\color[HTML]{0000FF} \boldmath{$<.001$}}  & \multicolumn{1}{c|}{$20.333$}                                  & {\color[HTML]{0000FF} \boldmath{$<.001$}}                        	& \multicolumn{1}{c|}{$1.000$}                                    & \multicolumn{1}{c|}{{\color[HTML]{0000FF} \boldmath{$<.001$}}} & {\color[HTML]{0000FF} \boldmath{$.026$}}    & \multicolumn{1}{c|}{\cellcolor[HTML]{D9D9D9}{-}}              & \multicolumn{1}{c|}{\cellcolor[HTML]{D9D9D9}{-}}    & {\cellcolor[HTML]{D9D9D9}{-}}                    & \multicolumn{1}{c|}{\cellcolor[HTML]{D9D9D9}{-}}                & \multicolumn{1}{c|}{\cellcolor[HTML]{D9D9D9}{-}}                & \multicolumn{1}{c|}{\cellcolor[HTML]{D9D9D9}{-}} \\  \hline
    \multicolumn{2}{|c||}{}                                                                                                                                                         & \textbf{Midair}                                                                                                  & \multicolumn{1}{c|}{$.945$}      & \multicolumn{1}{c|}{{\color[HTML]{0000FF} \boldmath{$<.001$}}} & \multicolumn{1}{c|}{$.961$}      & \multicolumn{1}{c|}{{\color[HTML]{0000FF} \boldmath{$<.001$}}} & \multicolumn{1}{c|}{$.968$}      & {\color[HTML]{0000FF} \boldmath{$<.001$}}  & \multicolumn{1}{c|}{$39.470$}                                  & {\color[HTML]{0000FF} \boldmath{$<.001$}}                        	& \multicolumn{1}{c|}{$.080$}                                     & \multicolumn{1}{c|}{{\color[HTML]{0000FF} \boldmath{$.002$}}}  & {$1.000$}                                   & \multicolumn{1}{c|}{\cellcolor[HTML]{D9D9D9}{-}}              & \multicolumn{1}{c|}{\cellcolor[HTML]{D9D9D9}{-}}    & {\cellcolor[HTML]{D9D9D9}{-}}                    & \multicolumn{1}{c|}{\cellcolor[HTML]{D9D9D9}{-}}                & \multicolumn{1}{c|}{\cellcolor[HTML]{D9D9D9}{-}}                & \multicolumn{1}{c|}{\cellcolor[HTML]{D9D9D9}{-}} \\  \cline{3-20} 
    \multicolumn{2}{|c||}{\multirow{-2}{*}{\rotatebox[origin=c]{0}{\textbf{\begin{tabular}[c]{@{}c@{}}Path Length\\(DV3)\end{tabular}}}}}                                           & \textbf{Onskin}                                                                                                  & \multicolumn{1}{c|}{$.891$}      & \multicolumn{1}{c|}{{\color[HTML]{0000FF} \boldmath{$<.001$}}} & \multicolumn{1}{c|}{$.934$}      & \multicolumn{1}{c|}{{\color[HTML]{0000FF} \boldmath{$<.001$}}} & \multicolumn{1}{c|}{$.893$}      & {\color[HTML]{0000FF} \boldmath{$<.001$}}  & \multicolumn{1}{c|}{$16.353$}                                  & {\color[HTML]{0000FF} \boldmath{$<.001$}}                        	& \multicolumn{1}{c|}{$.238$}                                     & \multicolumn{1}{c|}{$.161$}                                    & {\color[HTML]{0000FF} \boldmath{$.001$}}    & \multicolumn{1}{c|}{\cellcolor[HTML]{D9D9D9}{-}}              & \multicolumn{1}{c|}{\cellcolor[HTML]{D9D9D9}{-}}    & {\cellcolor[HTML]{D9D9D9}{-}}                    & \multicolumn{1}{c|}{\cellcolor[HTML]{D9D9D9}{-}}                & \multicolumn{1}{c|}{\cellcolor[HTML]{D9D9D9}{-}}                & \multicolumn{1}{c|}{\cellcolor[HTML]{D9D9D9}{-}} \\  \hline
    \multicolumn{1}{|c|}{}                                                                                                  						&                               & \textbf{Midair}                                                                                                  & \multicolumn{1}{c|}{$.946$}      & \multicolumn{1}{c|}{{\color[HTML]{0000FF} \boldmath{$<.001$}}} & \multicolumn{1}{c|}{$.979$}      & \multicolumn{1}{c|}{{\color[HTML]{0000FF} \boldmath{$<.001$}}} & \multicolumn{1}{c|}{$.952$}      & {\color[HTML]{0000FF} \boldmath{$<.001$}}  & \multicolumn{1}{c|}{$148.490$}                                 & {\color[HTML]{0000FF} \boldmath{$<.001$}}                        	& \multicolumn{1}{c|}{{\color[HTML]{0000FF} \boldmath{$<.001$}}}  & \multicolumn{1}{c|}{{\color[HTML]{0000FF} \boldmath{$<.001$}}} & {\color[HTML]{0000FF} \boldmath{$<.001$}}   & \multicolumn{1}{c|}{\cellcolor[HTML]{D9D9D9}{-}}              & \multicolumn{1}{c|}{\cellcolor[HTML]{D9D9D9}{-}}    & {\cellcolor[HTML]{D9D9D9}{-}}                    & \multicolumn{1}{c|}{\cellcolor[HTML]{D9D9D9}{-}}                & \multicolumn{1}{c|}{\cellcolor[HTML]{D9D9D9}{-}}                & \multicolumn{1}{c|}{\cellcolor[HTML]{D9D9D9}{-}} \\  \cline{3-20} 
    \multicolumn{1}{|c|}{}                                                                                                  						& \multirow{-2}{*}{\textbf{S}}  & \textbf{Onskin}                                                                                                  & \multicolumn{1}{c|}{$.956$}      & \multicolumn{1}{c|}{{\color[HTML]{0000FF} \boldmath{$<.001$}}} & \multicolumn{1}{c|}{$.977$}      & \multicolumn{1}{c|}{{\color[HTML]{0000FF} \boldmath{$<.001$}}} & \multicolumn{1}{c|}{$.895$}      & {\color[HTML]{0000FF} \boldmath{$<.001$}}  & \multicolumn{1}{c|}{$148.240$}                                 & {\color[HTML]{0000FF} \boldmath{$<.001$}}                        	& \multicolumn{1}{c|}{{\color[HTML]{0000FF} \boldmath{$<.001$}}}  & \multicolumn{1}{c|}{{\color[HTML]{0000FF} \boldmath{$<.001$}}} & {\color[HTML]{0000FF} \boldmath{$<.001$}}   & \multicolumn{1}{c|}{\cellcolor[HTML]{D9D9D9}{-}}              & \multicolumn{1}{c|}{\cellcolor[HTML]{D9D9D9}{-}}    & {\cellcolor[HTML]{D9D9D9}{-}}                    & \multicolumn{1}{c|}{\cellcolor[HTML]{D9D9D9}{-}}                & \multicolumn{1}{c|}{\cellcolor[HTML]{D9D9D9}{-}}                & \multicolumn{1}{c|}{\cellcolor[HTML]{D9D9D9}{-}} \\  \cline{2-20} 
    \multicolumn{1}{|c|}{}                                                                                                  						&                               & \textbf{Midair}                                                                                                  & \multicolumn{1}{c|}{$.961$}      & \multicolumn{1}{c|}{{\color[HTML]{0000FF} \boldmath{$<.001$}}} & \multicolumn{1}{c|}{$.976$}      & \multicolumn{1}{c|}{{\color[HTML]{0000FF} \boldmath{$<.001$}}} & \multicolumn{1}{c|}{$.980$}      & {\color[HTML]{0000FF} \boldmath{$<.001$}}  & \multicolumn{1}{c|}{$118.210$}                                 & {\color[HTML]{0000FF} \boldmath{$<.001$}}                        	& \multicolumn{1}{c|}{{\color[HTML]{0000FF} \boldmath{$<.001$}}}  & \multicolumn{1}{c|}{{\color[HTML]{0000FF} \boldmath{$<.001$}}} & {\color[HTML]{0000FF} \boldmath{$<.001$}}   & \multicolumn{1}{c|}{\cellcolor[HTML]{D9D9D9}{-}}              & \multicolumn{1}{c|}{\cellcolor[HTML]{D9D9D9}{-}}    & {\cellcolor[HTML]{D9D9D9}{-}}                    & \multicolumn{1}{c|}{\cellcolor[HTML]{D9D9D9}{-}}                & \multicolumn{1}{c|}{\cellcolor[HTML]{D9D9D9}{-}}                & \multicolumn{1}{c|}{\cellcolor[HTML]{D9D9D9}{-}} \\  \cline{3-20} 
    \multicolumn{1}{|c|}{\multirow{-4}{*}{\rotatebox[origin=c]{0}{\textbf{\begin{tabular}[c]{@{}c@{}}Angular\\Displacement\\(DV4)\end{tabular}}}}}  & \multirow{-2}{*}{\textbf{E}}  & \textbf{Onskin}                                                                                                  & \multicolumn{1}{c|}{$.966$}      & \multicolumn{1}{c|}{{\color[HTML]{0000FF} \boldmath{$<.001$}}} & \multicolumn{1}{c|}{$.975$}      & \multicolumn{1}{c|}{{\color[HTML]{0000FF} \boldmath{$<.001$}}} & \multicolumn{1}{c|}{$.902$}      & {\color[HTML]{0000FF} \boldmath{$<.001$}}  & \multicolumn{1}{c|}{$46.406$}                                  & {\color[HTML]{0000FF} \boldmath{$<.001$}}                        	& \multicolumn{1}{c|}{$.242$}                                     & \multicolumn{1}{c|}{{\color[HTML]{0000FF} \boldmath{$<.001$}}} & {\color[HTML]{0000FF} \boldmath{$<.001$}}   & \multicolumn{1}{c|}{\cellcolor[HTML]{D9D9D9}{-}}              & \multicolumn{1}{c|}{\cellcolor[HTML]{D9D9D9}{-}}    & {\cellcolor[HTML]{D9D9D9}{-}}                    & \multicolumn{1}{c|}{\cellcolor[HTML]{D9D9D9}{-}}                & \multicolumn{1}{c|}{\cellcolor[HTML]{D9D9D9}{-}}                & \multicolumn{1}{c|}{\cellcolor[HTML]{D9D9D9}{-}} \\  \hline\hline
    \multicolumn{2}{|c||}{}                                                                                                                                                         & \textbf{Midair}                                                                                                  & \multicolumn{1}{c|}{$.815$}      & \multicolumn{1}{c|}{{\color[HTML]{0000FF} \boldmath{$<.001$}}} & \multicolumn{1}{c|}{$.914$}      & \multicolumn{1}{c|}{{\color[HTML]{0000FF} \boldmath{$.043$}}}  & \multicolumn{1}{c|}{$.929$}      & {\color[HTML]{0000FF} \boldmath{$.006$}}   & \multicolumn{1}{c|}{$14.395$}                                  & {\color[HTML]{0000FF} \boldmath{$<.001$}}                        	& \multicolumn{1}{c|}{{\color[HTML]{0000FF} \boldmath{$.004$}}}   & \multicolumn{1}{c|}{$.401$}                                    & {\color[HTML]{0000FF} \boldmath{$.004$}}    & \multicolumn{1}{c|}{\cellcolor[HTML]{D9D9D9}{-}}              & \multicolumn{1}{c|}{\cellcolor[HTML]{D9D9D9}{-}}    & {\cellcolor[HTML]{D9D9D9}{-}}                    & \multicolumn{1}{c|}{\cellcolor[HTML]{D9D9D9}{-}}                & \multicolumn{1}{c|}{\cellcolor[HTML]{D9D9D9}{-}}                & \multicolumn{1}{c|}{\cellcolor[HTML]{D9D9D9}{-}} \\  \cline{3-20} 
    \multicolumn{2}{|c||}{\multirow{-2}{*}{\rotatebox[origin=c]{0}{\textbf{\begin{tabular}[c]{@{}c@{}}Mental\\Demand\end{tabular}}}}}                                               & \textbf{Onskin}                                                                                                  & \multicolumn{1}{c|}{$.901$}      & \multicolumn{1}{c|}{{\color[HTML]{0000FF} \boldmath{$.023$}}}  & \multicolumn{1}{c|}{$.924$}      & \multicolumn{1}{c|}{$.073$}                                    & \multicolumn{1}{c|}{$.905$}      & {\color[HTML]{0000FF} \boldmath{$.028$}}   & \multicolumn{1}{c|}{$11.341$}                                  & {\color[HTML]{0000FF} \boldmath{$.003$}}                          & \multicolumn{1}{c|}{$.158$}                                     & \multicolumn{1}{c|}{$.066$}                                    & {\color[HTML]{0000FF} \boldmath{$.014$}}    & \multicolumn{1}{c|}{\cellcolor[HTML]{D9D9D9}{-}}              & \multicolumn{1}{c|}{\cellcolor[HTML]{D9D9D9}{-}}    & {\cellcolor[HTML]{D9D9D9}{-}}                    & \multicolumn{1}{c|}{\cellcolor[HTML]{D9D9D9}{-}}                & \multicolumn{1}{c|}{\cellcolor[HTML]{D9D9D9}{-}}                & \multicolumn{1}{c|}{\cellcolor[HTML]{D9D9D9}{-}} \\  \hline
    \multicolumn{2}{|c||}{}                                                                                                                                                         & \textbf{Midair}                                                                                                  & \multicolumn{1}{c|}{$.801$}      & \multicolumn{1}{c|}{{\color[HTML]{0000FF} \boldmath{$<.001$}}} & \multicolumn{1}{c|}{$.914$}      & \multicolumn{1}{c|}{{\color[HTML]{0000FF} \boldmath{$.043$}}}  & \multicolumn{1}{c|}{$.934$}      & {$.117$}                                   & \multicolumn{1}{c|}{$13.300$}                                  & {\color[HTML]{0000FF} \boldmath{$.001$}}                          & \multicolumn{1}{c|}{{\color[HTML]{0000FF} \boldmath{$.034$}}}   & \multicolumn{1}{c|}{$.082$}                                    & {\color[HTML]{0000FF} \boldmath{$.007$}}    & \multicolumn{1}{c|}{\cellcolor[HTML]{D9D9D9}{-}}              & \multicolumn{1}{c|}{\cellcolor[HTML]{D9D9D9}{-}}    & {\cellcolor[HTML]{D9D9D9}{-}}                    & \multicolumn{1}{c|}{\cellcolor[HTML]{D9D9D9}{-}}                & \multicolumn{1}{c|}{\cellcolor[HTML]{D9D9D9}{-}}                & \multicolumn{1}{c|}{\cellcolor[HTML]{D9D9D9}{-}} \\  \cline{3-20} 
    \multicolumn{2}{|c||}{\multirow{-2}{*}{\rotatebox[origin=c]{0}{\textbf{\begin{tabular}[c]{@{}c@{}}Physical\\Demand\end{tabular}}}}}                                             & \textbf{Onskin}                                                                                                  & \multicolumn{1}{c|}{$.848$}      & \multicolumn{1}{c|}{{\color[HTML]{0000FF} \boldmath{$<.001$}}} & \multicolumn{1}{c|}{$.950$}      & \multicolumn{1}{c|}{$.276$}                                    & \multicolumn{1}{c|}{$.955$}      & {$.348$}                                   & \multicolumn{1}{c|}{$10.173$}                                  & {\color[HTML]{0000FF} \boldmath{$.006$}}                          & \multicolumn{1}{c|}{{\color[HTML]{0000FF} \boldmath{$.043$}}}   & \multicolumn{1}{c|}{$.200$}                                    & {\color[HTML]{0000FF} \boldmath{$.013$}}    & \multicolumn{1}{c|}{\cellcolor[HTML]{D9D9D9}{-}}              & \multicolumn{1}{c|}{\cellcolor[HTML]{D9D9D9}{-}}    & {\cellcolor[HTML]{D9D9D9}{-}}                    & \multicolumn{1}{c|}{\cellcolor[HTML]{D9D9D9}{-}}                & \multicolumn{1}{c|}{\cellcolor[HTML]{D9D9D9}{-}}                & \multicolumn{1}{c|}{\cellcolor[HTML]{D9D9D9}{-}} \\  \hline
    \multicolumn{2}{|c||}{}                                                                                                                                                         & \textbf{Midair}                                                                                                  & \multicolumn{1}{c|}{$.921$}      & \multicolumn{1}{c|}{$.060$}                                    & \multicolumn{1}{c|}{$.958$}      & \multicolumn{1}{c|}{$.408$}                                    & \multicolumn{1}{c|}{$.955$}      & {$.343$}                                   & \multicolumn{1}{c|}{\cellcolor[HTML]{D9D9D9}}                  & {\cellcolor[HTML]{D9D9D9}{-}}                                  	& \multicolumn{1}{c|}{\cellcolor[HTML]{D9D9D9}{-}}                & \multicolumn{1}{c|}{\cellcolor[HTML]{D9D9D9}{-}}               & {\cellcolor[HTML]{D9D9D9}{-}}               & \multicolumn{1}{c|}{$2.952$}                                  & \multicolumn{1}{c|}{$.062$}                         & {$.016$}                                         & \multicolumn{1}{c|}{\cellcolor[HTML]{D9D9D9}{-}}                & \multicolumn{1}{c|}{\cellcolor[HTML]{D9D9D9}{-}}                & \multicolumn{1}{c|}{\cellcolor[HTML]{D9D9D9}{-}} \\  \cline{3-20} 
    \multicolumn{2}{|c||}{\multirow{-2}{*}{\rotatebox[origin=c]{0}{\textbf{\begin{tabular}[c]{@{}c@{}}Temporal\\ Demand\end{tabular}}}}}                                            & \textbf{Onskin}                                                                                                  & \multicolumn{1}{c|}{$.899$}      & \multicolumn{1}{c|}{{\color[HTML]{0000FF} \boldmath{$.020$}}}  & \multicolumn{1}{c|}{$.958$}      & \multicolumn{1}{c|}{$.240$}                                    & \multicolumn{1}{c|}{$.964$}      & {$.515$}                                   & \multicolumn{1}{c|}{$10.203$}                                  & {\color[HTML]{0000FF} \boldmath{$.006$}}                          & \multicolumn{1}{c|}{{\color[HTML]{0000FF} \boldmath{$.046$}}}   & \multicolumn{1}{c|}{$1.000$}                                   & {$.058$}                                    & \multicolumn{1}{c|}{\cellcolor[HTML]{D9D9D9}{-}}              & \multicolumn{1}{c|}{\cellcolor[HTML]{D9D9D9}{-}}    & {\cellcolor[HTML]{D9D9D9}{-}}                    & \multicolumn{1}{c|}{\cellcolor[HTML]{D9D9D9}{-}}                & \multicolumn{1}{c|}{\cellcolor[HTML]{D9D9D9}{-}}                & \multicolumn{1}{c|}{\cellcolor[HTML]{D9D9D9}{-}} \\  \hline
    \multicolumn{2}{|c||}{}                                                                                                                                                         & \textbf{Midair}                                                                                                  & \multicolumn{1}{c|}{$.874$}      & \multicolumn{1}{c|}{{\color[HTML]{0000FF} \boldmath{$.006$}}}  & \multicolumn{1}{c|}{$.900$}      & \multicolumn{1}{c|}{{\color[HTML]{0000FF} \boldmath{$.021$}}}  & \multicolumn{1}{c|}{$.914$}      & {\color[HTML]{0000FF} \boldmath{$.042$}}   & \multicolumn{1}{c|}{$1.859$}                                   & {$.395$}                                                        	& \multicolumn{1}{c|}{\cellcolor[HTML]{D9D9D9}}                   & \multicolumn{1}{c|}{\cellcolor[HTML]{D9D9D9}}                  & {\cellcolor[HTML]{D9D9D9}}                  & \multicolumn{1}{c|}{\cellcolor[HTML]{D9D9D9}{-}}              & \multicolumn{1}{c|}{\cellcolor[HTML]{D9D9D9}{-}}    & {\cellcolor[HTML]{D9D9D9}{-}}                    & \multicolumn{1}{c|}{\cellcolor[HTML]{D9D9D9}{-}}                & \multicolumn{1}{c|}{\cellcolor[HTML]{D9D9D9}{-}}                & \multicolumn{1}{c|}{\cellcolor[HTML]{D9D9D9}{-}} \\  \cline{3-20} 
    \multicolumn{2}{|c||}{\multirow{-2}{*}{\rotatebox[origin=c]{0}{\textbf{Performance}}}}                                                                                          & \textbf{Onskin}                                                                                                  & \multicolumn{1}{c|}{$.870$}      & \multicolumn{1}{c|}{{\color[HTML]{0000FF} \boldmath{$.005$}}}  & \multicolumn{1}{c|}{$.918$}      & \multicolumn{1}{c|}{$.052$}                                    & \multicolumn{1}{c|}{$.969$}      & {$.648$}                                   & \multicolumn{1}{c|}{$1.126$}                                   & {$.569$}                                                        	& \multicolumn{1}{c|}{\cellcolor[HTML]{D9D9D9}}                   & \multicolumn{1}{c|}{\cellcolor[HTML]{D9D9D9}}                  & {\cellcolor[HTML]{D9D9D9}}               	 & \multicolumn{1}{c|}{\cellcolor[HTML]{D9D9D9}{-}}              & \multicolumn{1}{c|}{\cellcolor[HTML]{D9D9D9}{-}}    & {\cellcolor[HTML]{D9D9D9}{-}}                    & \multicolumn{1}{c|}{\cellcolor[HTML]{D9D9D9}{-}}                & \multicolumn{1}{c|}{\cellcolor[HTML]{D9D9D9}{-}}                & \multicolumn{1}{c|}{\cellcolor[HTML]{D9D9D9}{-}} \\  \hline
    \multicolumn{2}{|c||}{}                                                                                                                                                         & \textbf{Midair}                                                                                                  & \multicolumn{1}{c|}{$.918$}      & \multicolumn{1}{c|}{$.052$}                                    & \multicolumn{1}{c|}{$.914$}      & \multicolumn{1}{c|}{{\color[HTML]{0000FF} \boldmath{$.044$}}}  & \multicolumn{1}{c|}{$.909$}      & {\color[HTML]{0000FF} \boldmath{$.034$}}   & \multicolumn{1}{c|}{$17.816$}                                  & {\color[HTML]{0000FF} \boldmath{$<.001$}}                        	& \multicolumn{1}{c|}{$.183$}                                     & \multicolumn{1}{c|}{{\color[HTML]{0000FF} \boldmath{$.002$}}}  & {\color[HTML]{0000FF} \boldmath{$.002$}}    & \multicolumn{1}{c|}{\cellcolor[HTML]{D9D9D9}{-}}              & \multicolumn{1}{c|}{\cellcolor[HTML]{D9D9D9}{-}}    & {\cellcolor[HTML]{D9D9D9}{-}}                    & \multicolumn{1}{c|}{\cellcolor[HTML]{D9D9D9}{-}}                & \multicolumn{1}{c|}{\cellcolor[HTML]{D9D9D9}{-}}                & \multicolumn{1}{c|}{\cellcolor[HTML]{D9D9D9}{-}} \\  \cline{3-20} 
    \multicolumn{2}{|c||}{\multirow{-2}{*}{\rotatebox[origin=c]{0}{\textbf{Effort}}}}                                                                                               & \textbf{Onskin}                                                                                                  & \multicolumn{1}{c|}{$.887$}      & \multicolumn{1}{c|}{{\color[HTML]{0000FF} \boldmath{$.012$}}}  & \multicolumn{1}{c|}{$.920$}      & \multicolumn{1}{c|}{$.058$}                                    & \multicolumn{1}{c|}{$.926$}      & {\color[HTML]{0000FF} \boldmath{$.027$}}   & \multicolumn{1}{c|}{$20.000$}                                  & {\color[HTML]{0000FF} \boldmath{$<.001$}}                        	& \multicolumn{1}{c|}{{\color[HTML]{0000FF} \boldmath{$.009$}}}   & \multicolumn{1}{c|}{$.091$}                                    & {\color[HTML]{0000FF} \boldmath{$.002$}}    & \multicolumn{1}{c|}{\cellcolor[HTML]{D9D9D9}{-}}              & \multicolumn{1}{c|}{\cellcolor[HTML]{D9D9D9}{-}}    & {\cellcolor[HTML]{D9D9D9}{-}}                    & \multicolumn{1}{c|}{\cellcolor[HTML]{D9D9D9}{-}}                & \multicolumn{1}{c|}{\cellcolor[HTML]{D9D9D9}{-}}                & \multicolumn{1}{c|}{\cellcolor[HTML]{D9D9D9}{-}} \\  \hline
    \multicolumn{2}{|c||}{}                                                                                                                                                         & \textbf{Midair}                                                                                                  & \multicolumn{1}{c|}{$.799$}      & \multicolumn{1}{c|}{{\color[HTML]{0000FF} \boldmath{$<.001$}}} & \multicolumn{1}{c|}{$.817$}      & \multicolumn{1}{c|}{{\color[HTML]{0000FF} \boldmath{$<.001$}}} & \multicolumn{1}{c|}{$.931$}      & {$.102$}                                   & \multicolumn{1}{c|}{$10.265$}                                  & {\color[HTML]{0000FF} \boldmath{$.006$}}                          & \multicolumn{1}{c|}{$1.000$}                                    & \multicolumn{1}{c|}{{\color[HTML]{0000FF} \boldmath{$.039$}}}  & {\color[HTML]{0000FF} \boldmath{$.023$}}    & \multicolumn{1}{c|}{\cellcolor[HTML]{D9D9D9}{-}}              & \multicolumn{1}{c|}{\cellcolor[HTML]{D9D9D9}{-}}    & {\cellcolor[HTML]{D9D9D9}{-}}                    & \multicolumn{1}{c|}{\cellcolor[HTML]{D9D9D9}{-}}                & \multicolumn{1}{c|}{\cellcolor[HTML]{D9D9D9}{-}}                & \multicolumn{1}{c|}{\cellcolor[HTML]{D9D9D9}{-}} \\  \cline{3-20} 
    \multicolumn{2}{|c||}{\multirow{-2}{*}{\rotatebox[origin=c]{0}{\textbf{Frustration}}}}                                                                                          & \textbf{Onskin}                                                                                                  & \multicolumn{1}{c|}{$.806$}      & \multicolumn{1}{c|}{{\color[HTML]{0000FF} \boldmath{$<.001$}}} & \multicolumn{1}{c|}{$.920$}      & \multicolumn{1}{c|}{$.059$}                                    & \multicolumn{1}{c|}{$.913$}      & {\color[HTML]{0000FF} \boldmath{$.040$}}   & \multicolumn{1}{c|}{$13.618$}                                  & {\color[HTML]{0000FF} \boldmath{$.001$}}                          & \multicolumn{1}{c|}{$.113$}                                     & \multicolumn{1}{c|}{{\color[HTML]{0000FF} \boldmath{$.043$}}}  & {\color[HTML]{0000FF} \boldmath{$.018$}}    & \multicolumn{1}{c|}{\cellcolor[HTML]{D9D9D9}{-}}              & \multicolumn{1}{c|}{\cellcolor[HTML]{D9D9D9}{-}}    & {\cellcolor[HTML]{D9D9D9}{-}}                    & \multicolumn{1}{c|}{\cellcolor[HTML]{D9D9D9}{-}}                & \multicolumn{1}{c|}{\cellcolor[HTML]{D9D9D9}{-}}                & \multicolumn{1}{c|}{\cellcolor[HTML]{D9D9D9}{-}} \\  \hline
    \end{tabular}%
    }
    \vspace{0.15cm}
    \caption{Statistical comparison of swipe performance metrics across different region segmentation pattern for midair and onskin spaces (RQ2). Figure~\ref{fig:AllGestures_RQ1_RQ2} highlights statistically significant pairwise differences.} 
    \Description{Statistical comparison of swipe performance metrics across different region segmentation pattern for midair and onskin spaces (RQ2). Figure~\ref{fig:AllGestures_RQ1_RQ2} highlights statistically significant pairwise differences.}
    \label{tab:RQ2_GestureMetics}        
\end{table*}
\endgroup
}

}

}

\clearpage
\newpage

\section{Detailed Results for Section \ref{subsection:RQ3_QualitativeResults}}
\label{appendix:Detailed Results RQ3}
{

{
\begingroup
\setlength{\floatsep}{-1ex} 
\setlength{\intextsep}{-1ex} 
\setlength{\abovecaptionskip}{-1ex}
\setlength{\belowcaptionskip}{-1ex}
\setlength{\tabcolsep}{2pt}
\begin{figure*}[!b]
    \centering
    \includegraphics[width=500pt]{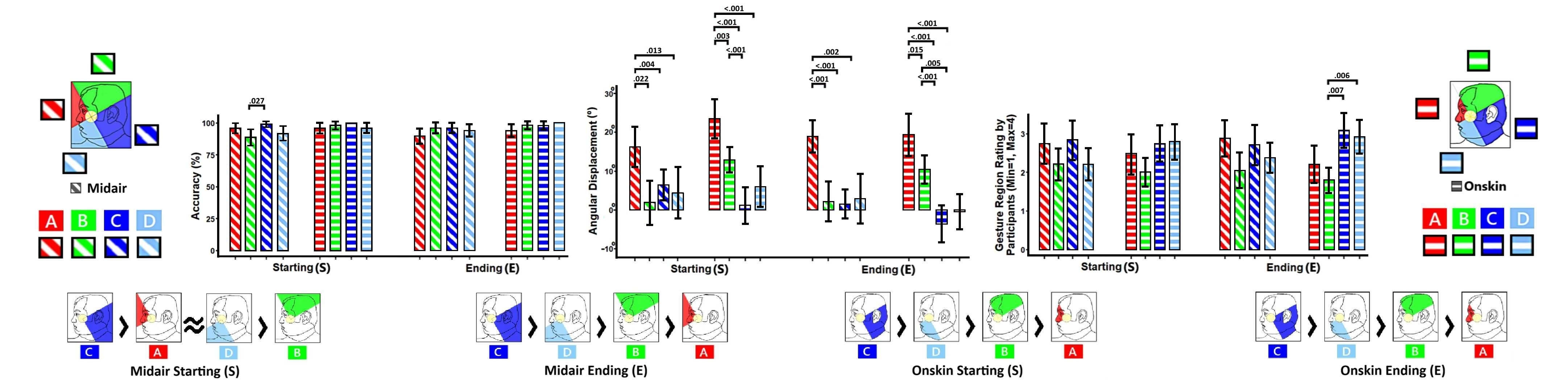}   
    \vspace{.25cm} 
    \caption{Accuracy (DV1), average angular displacement (DV4) \& location preference ratings for staring and ending regions in 4-region segmentation.}  
    \Description{Three vertical bar charts illustrating accuracy, angular displacement, and relative region ratings across various starting and ending regions in a 4-region distribution around the cheekbone. Region A is around the nose, B around the temple, C near the ear and jaw, and D around the mouth and chin, all color-coded against a side image of the face. In the barcharts, slanted colored lines represent midair regions, while horizontal colored lines indicate onskin regions.}
    \label{fig:GestureMetrics_4R2R_RQ3}   
\end{figure*}

\begin{figure*}[!b]
    \centering
    \includegraphics[width=500pt]{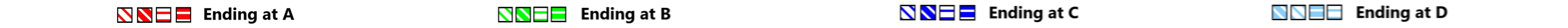}   
    \begin{subfigure}[t]{.355\linewidth}     
        \centering
        \captionsetup{justification=centering}      
        \includegraphics[width = .99\linewidth]{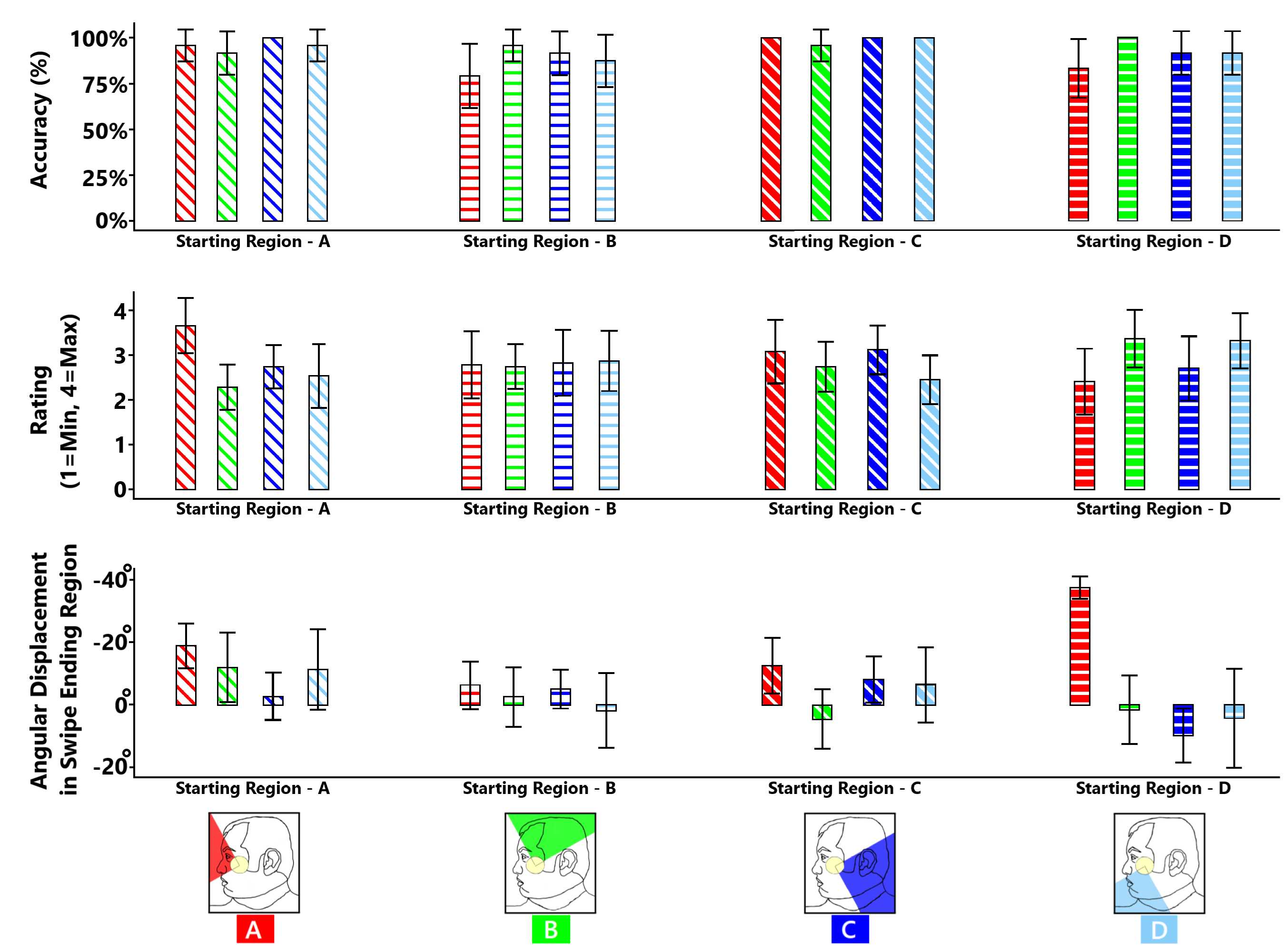}  
        \vspace{-0.5cm} 
        \caption{Midair swipes}  
        \label{fig:GestureMetrics_4R2R_RQ3_Individual_MidairSwipes}   
    \end{subfigure}
    \hfill
    \begin{subfigure}[t]{.274\linewidth}     
        \centering
        \captionsetup{justification=centering}      
        \includegraphics[width = .9\linewidth]{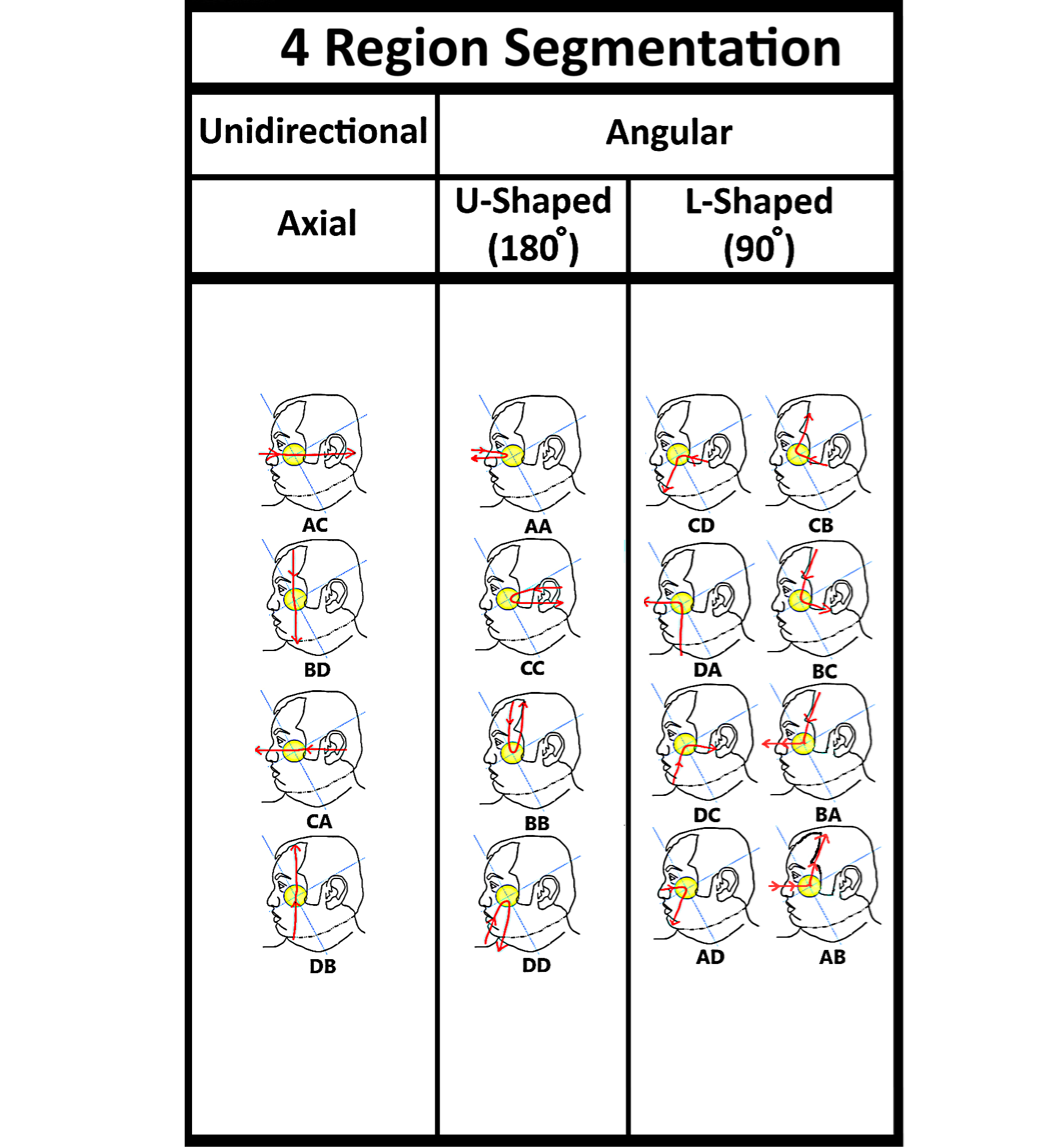}  
        \vspace{-0.15cm} 
        \caption{Visual swipe references}  
        \label{fig:GestureMetrics_4R2R_RQ3_Visual_Swipe_References}   
    \end{subfigure}
    \hfill
    \begin{subfigure}[t]{.355\linewidth}     
        \centering
        \captionsetup{justification=centering}      
        \includegraphics[width = .99\linewidth]{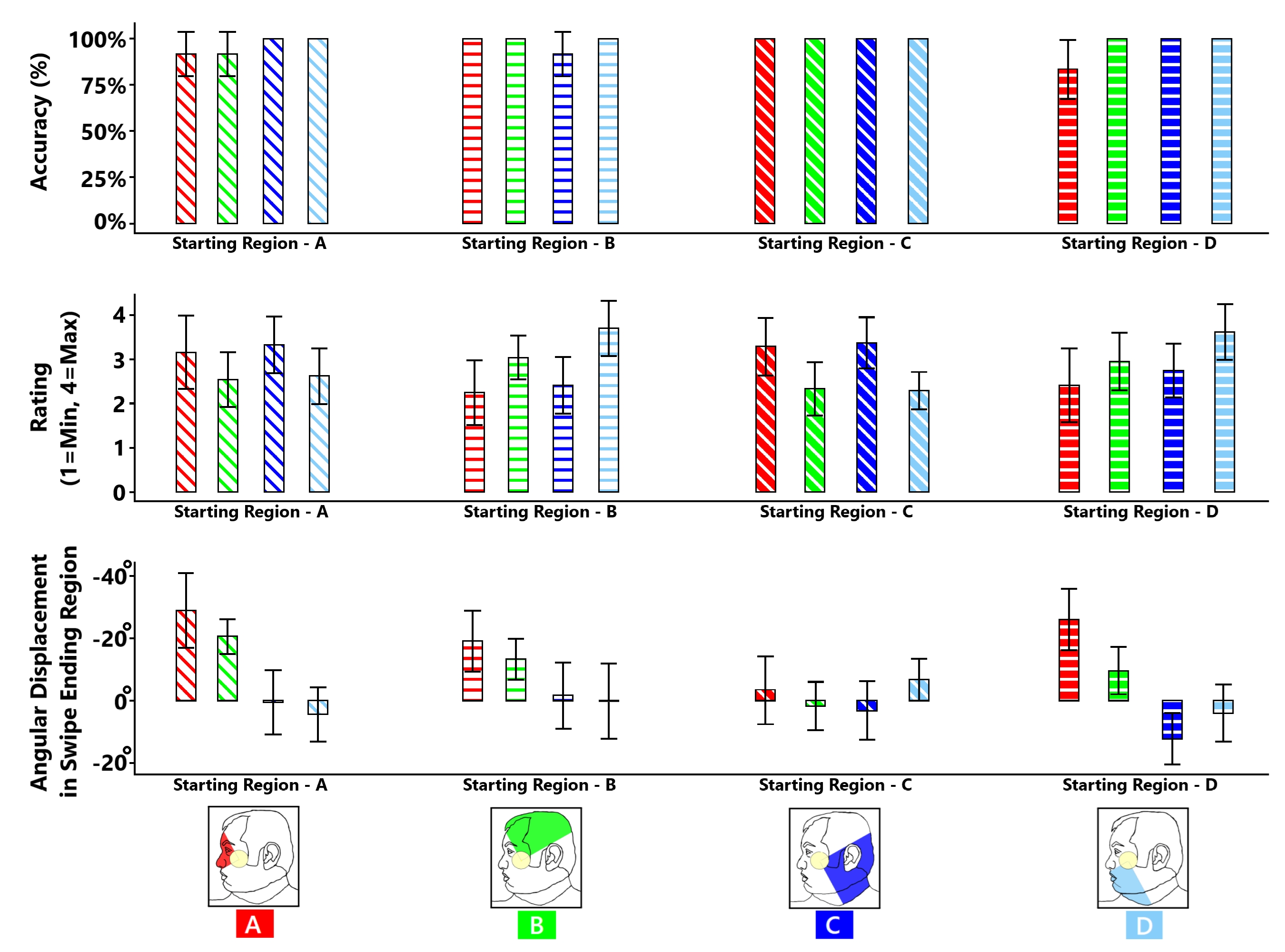}  
        \vspace{-0.5cm} 
        \caption{Onskin swipes}  
        \label{fig:GestureMetrics_4R2R_RQ3_Individual_OnskinSwipes}   
    \end{subfigure}
    \vspace{.35cm} 
    \caption{Accuracy (DV1), average angular displacement (DV4) \& location preference ratings for individual swipe shapes in 4-Region Segmentation.}  
    \Description{Three vertical bar charts illustrating accuracy, angular displacement, and relative region ratings across various starting and ending regions in a 4-region distribution around the cheekbone. Region A is around the nose, B around the temple, C near the ear and jaw, and D around the mouth and chin, all color-coded against a side image of the face. In the barcharts, slanted colored lines represent midair regions, while horizontal colored lines indicate onskin regions.}
    \label{fig:GestureMetrics_4R2R_RQ3_Individual_Swipes}   
\end{figure*}

\begin{table}[!tb]
    \centering
    \resizebox{240pt}{!}
    {%
    \begin{tabular}{|c|c|c||cccccccc||cc||cccc|}
    \hline
                                                                                                                     &                                                                                        					&                                                                 & \multicolumn{8}{c||}{}                                                  																																																																																																																																																																							& \multicolumn{2}{c||}{}                                                                                                                 								& \multicolumn{4}{c|}{}                                                                                                                                                                                							\\
                                                                                                                     &                                                                                        					&                                                                 & \multicolumn{8}{c||}{\multirow{-2}{*}{\textbf{\begin{tabular}[c]{@{}c@{}}Shapiro-Wilk Normality Test\end{tabular}}}}                                                                                                                                                                                                                                                                                                                                                                                                                                                                                                                                                                  																				& \multicolumn{2}{c||}{}                                                                                                                 								& \multicolumn{4}{c|}{}                                                                                                                                                                                							\\ \cline{4-11}
                                                                                                                     &                                                                                        					&                                                                 & \multicolumn{2}{c||}{}                                                                                                                                        								& \multicolumn{2}{c||}{}                                                                                                                                      									& \multicolumn{2}{c||}{}                                                                                                                                      								& \multicolumn{2}{c||}{}                                                                                                                									& \multicolumn{2}{c||}{}                                                                                                                 								& \multicolumn{4}{c|}{}                                                                                                                                                                                							\\
                                                                                                                     &                                                                                        					&                                                                 & \multicolumn{2}{c||}{}                                                                                                     																	& \multicolumn{2}{c||}{}                                                                                                 																		& \multicolumn{2}{c||}{}                                                                                                 																	& \multicolumn{2}{c||}{}                                                                           																			& \multicolumn{2}{c||}{}                                        																										& \multicolumn{4}{c|}{}                                                                                                                                         																\\ 
                                                                                                                     &                                                                                        					&                                                                 & \multicolumn{2}{c||}{\multirow{-3}{*}{\textbf{\begin{tabular}[c]{@{}c@{}}Region\\A\end{tabular}}}}                                                                                          & \multicolumn{2}{c||}{\multirow{-3}{*}{\textbf{\begin{tabular}[c]{@{}c@{}}Region\\B\end{tabular}}}}                                                                                            & \multicolumn{2}{c||}{\multirow{-3}{*}{\textbf{\begin{tabular}[c]{@{}c@{}}Region\\C\end{tabular}}}}                                                                                        & \multicolumn{2}{c||}{\multirow{-3}{*}{\textbf{\begin{tabular}[c]{@{}c@{}}Region\\D\end{tabular}}}}                                                                        & \multicolumn{2}{c||}{\multirow{-5}{*}{\textbf{\begin{tabular}[c]{@{}c@{}}Friedman \\Test for \\Within\\ Subject\\ Factor\end{tabular}}}}                              & \multicolumn{4}{c|}{\multirow{-5}{*}{\textbf{\begin{tabular}[c]{@{}c@{}}Pairwise \\ Comparison\\ using Wilcoxon\\ Rank-Sum Test\end{tabular}}}}                                                      							\\ \cline{4-17} 
    \multirow{-6}{*}{\rotatebox{90}{\textbf{\begin{tabular}[c]{@{}c@{}}Gesture\\ Metrics\end{tabular}}}}             & \multirow{-6}{*}{\rotatebox{90}{\textbf{\begin{tabular}[c]{@{}c@{}}Interaction\\ Space\end{tabular}}}} 	& \multirow{-6}{*}{\textbf{{\rotatebox[origin=c]{90}{Position}}}} & \multicolumn{1}{c|}{\multirow{-1}{*}{\textbf{W}}}                          	& \multicolumn{1}{c||}{\multirow{-1}{*}{\textbf{p}}}                             								& \multicolumn{1}{c|}{\multirow{-1}{*}{\textbf{W}}}                          	& \multicolumn{1}{c||}{\multirow{-1}{*}{\textbf{p}}}                           									& \multicolumn{1}{c|}{\multirow{-1}{*}{\textbf{W}}}                          & \multicolumn{1}{c||}{\multirow{-1}{*}{\textbf{p}}}                            								& \multicolumn{1}{c|}{\multirow{-1}{*}{\textbf{W}}}                          	& \multirow{-1}{*}{\textbf{p}}                            									& \multicolumn{1}{c|}{\multirow{-1}{*}{\boldmath{$\chi^2(3)$}}}               & \multirow{-1}{*}{\textbf{p}}                            								& \multicolumn{1}{c|}{\textbf{}}  & \multicolumn{1}{c|}{\textbf{A}}                                		& \multicolumn{1}{c|}{\textbf{B}}                                	& \textbf{C}                       					\\ \hline
                                                                                                                     &                                                                                        					&                                                                 & \multicolumn{1}{c|}{}                                                      	& \multicolumn{1}{c|}{}                                                          								& \multicolumn{1}{c|}{}                                                      	& \multicolumn{1}{c|}{}                                                        									& \multicolumn{1}{c|}{}                                                      & \multicolumn{1}{c|}{}                                                        								& \multicolumn{1}{c|}{}                                                      	&                                                         									& \multicolumn{1}{c|}{}                                                       &                                                         								& \multicolumn{1}{c|}{\textbf{B}} & \multicolumn{1}{c|}{$.455$}                                     	& \multicolumn{2}{c|}{\cellcolor[HTML]{D9D9D9}}																			\\ \cline{14-16} 
                                                                                                                     &                                                                                        					&                                                                 & \multicolumn{1}{c|}{}                                                      	& \multicolumn{1}{c|}{}                                                          								& \multicolumn{1}{c|}{}                                                      	& \multicolumn{1}{c|}{}                                                        									& \multicolumn{1}{c|}{}                                                      & \multicolumn{1}{c|}{}                                                        								& \multicolumn{1}{c|}{}                                                      	&                                                         									& \multicolumn{1}{c|}{}                                                       &                                                         								& \multicolumn{1}{c|}{\textbf{C}} & \multicolumn{1}{c|}{$1.000$}                                     	& \multicolumn{1}{c|}{{\color[HTML]{0000FF} \boldmath{$.027$}}}     & \multicolumn{1}{c|}{\cellcolor[HTML]{D9D9D9}}		\\ \cline{14-17} 
                                                                                                                     &                                                                                        					& \multirow{-3}{*}{\textbf{S}}                                    & \multicolumn{1}{c|}{\multirow{-3}{*}{{\rotatebox[origin=c]{75}{$.200$}}}} 	& \multicolumn{1}{c|}{\multirow{-3}{*}{{\rotatebox[origin=c]{75}{\color[HTML]{0000FF}{\boldmath{$<.001$}}}}}} 	& \multicolumn{1}{c|}{\multirow{-3}{*}{{\rotatebox[origin=c]{75}{$.200$}}}} 	& \multicolumn{1}{c|}{\multirow{-3}{*}{{\rotatebox[origin=c]{75}{\color[HTML]{0000FF}{\boldmath{$<.001$}}}}}} 	& \multicolumn{1}{c|}{\multirow{-3}{*}{{\rotatebox[origin=c]{75}{$.200$}}}} & \multicolumn{1}{c|}{\multirow{-3}{*}{{\rotatebox[origin=c]{75}{\color[HTML]{0000FF}{\boldmath{$<.001$}}}}}} 	& \multicolumn{1}{c|}{\multirow{-3}{*}{{\rotatebox[origin=c]{75}{$.200$}}}} 	& \multirow{-3}{*}{{\rotatebox[origin=c]{75}{\color[HTML]{0000FF}{\boldmath{$<.001$}}}}} 	& \multicolumn{1}{c|}{\multirow{-3}{*}{{\rotatebox[origin=c]{75}{$10.235$}}}} & \multirow{-3}{*}{{\rotatebox[origin=c]{75}{\color[HTML]{0000FF}{\boldmath{$.017$}}}}}   & \multicolumn{1}{c|}{\textbf{D}} & \multicolumn{1}{c|}{$1.000$}                                     	& \multicolumn{1}{c|}{$1.000$}                                     	& \multicolumn{1}{c|}{$.140$}       				\\ \cline{3-17} 
                                                                                                                     &                                                                                        					&                                                                 & \multicolumn{1}{c|}{}                                                      	& \multicolumn{1}{c|}{}                                                          								& \multicolumn{1}{c|}{}                                                      	& \multicolumn{1}{c|}{}                                                        									& \multicolumn{1}{c|}{}                                                      & \multicolumn{1}{c|}{}                                                        								& \multicolumn{1}{c|}{}                                                      	&                                                         									& \multicolumn{1}{c|}{}                                                       &                                                         								& \multicolumn{1}{c|}{\textbf{B}} & \multicolumn{3}{c|}{\cellcolor[HTML]{D9D9D9}}                                                                                                                      							\\ \cline{14-14}
                                                                                                                     &                                                                                        					&                                                                 & \multicolumn{1}{c|}{}                                                      	& \multicolumn{1}{c|}{}                                                          								& \multicolumn{1}{c|}{}                                                      	& \multicolumn{1}{c|}{}                                                        									& \multicolumn{1}{c|}{}                                                      & \multicolumn{1}{c|}{}                                                        								& \multicolumn{1}{c|}{}                                                      	&                                                         									& \multicolumn{1}{c|}{}                                                       &                                                         								& \multicolumn{1}{c|}{\textbf{C}} & \multicolumn{3}{c|}{\cellcolor[HTML]{D9D9D9}}                                                                                                                      							\\ \cline{14-14}
                                                                                                                     & \multirow{-6}{*}{\textbf{{\rotatebox[origin=c]{90}{Midair}}}}                          					& \multirow{-3}{*}{\textbf{E}}                                    & \multicolumn{1}{c|}{\multirow{-3}{*}{{\rotatebox[origin=c]{75}{$.350$}}}} 	& \multicolumn{1}{c|}{\multirow{-3}{*}{{\rotatebox[origin=c]{75}{\color[HTML]{0000FF}{\boldmath{$<.001$}}}}}}   & \multicolumn{1}{c|}{\multirow{-3}{*}{{\rotatebox[origin=c]{75}{$.200$}}}} 	& \multicolumn{1}{c|}{\multirow{-3}{*}{{\rotatebox[origin=c]{75}{\color[HTML]{0000FF}{\boldmath{$<.001$}}}}}} 	& \multicolumn{1}{c|}{\multirow{-3}{*}{{\rotatebox[origin=c]{75}{$.200$}}}} & \multicolumn{1}{c|}{\multirow{-3}{*}{{\rotatebox[origin=c]{75}{\color[HTML]{0000FF}{\boldmath{$<.001$}}}}}} 	& \multicolumn{1}{c|}{\multirow{-3}{*}{{\rotatebox[origin=c]{75}{$.259$}}}} 	& \multirow{-3}{*}{{\rotatebox[origin=c]{75}{\color[HTML]{0000FF}{\boldmath{$<.001$}}}}} 	& \multicolumn{1}{c|}{\multirow{-3}{*}{{\rotatebox[origin=c]{75}{$4.645$}}}}  & \multirow{-3}{*}{{\rotatebox[origin=c]{75}{$.200$}}}   									& \multicolumn{1}{c|}{\textbf{D}} & \multicolumn{3}{c|}{\cellcolor[HTML]{D9D9D9}}                                                                                                    											\\ \cline{2-14} 
                                                                                                                     &                                                                                        					&                                                                 & \multicolumn{1}{c|}{}                                                      	& \multicolumn{1}{c|}{}                                                          								& \multicolumn{1}{c|}{}                                                      	& \multicolumn{1}{c|}{}                                                        									& \multicolumn{1}{c|}{}                                                      & \multicolumn{1}{c|}{}                                                        								& \multicolumn{1}{c|}{}                                                      	&                                                         									& \multicolumn{1}{c|}{}                                                       &                                                         								& \multicolumn{1}{c|}{\textbf{B}} & \multicolumn{3}{c|}{\cellcolor[HTML]{D9D9D9}}                                                                                                                      							\\ \cline{14-14}
                                                                                                                     &                                                                                        					&                                                                 & \multicolumn{1}{c|}{}                                                      	& \multicolumn{1}{c|}{}                                                          								& \multicolumn{1}{c|}{}                                                      	& \multicolumn{1}{c|}{}                                                        									& \multicolumn{1}{c|}{}                                                      & \multicolumn{1}{c|}{}                                                        								& \multicolumn{1}{c|}{}                                                      	&                                                         									& \multicolumn{1}{c|}{}                                                       &                                                         								& \multicolumn{1}{c|}{\textbf{C}} & \multicolumn{3}{c|}{\cellcolor[HTML]{D9D9D9}}                                                                                                                      							\\ \cline{14-14}
                                                                                                                     &                                                                                        					& \multirow{-3}{*}{\textbf{S}}                                    & \multicolumn{1}{c|}{\multirow{-3}{*}{{\rotatebox[origin=c]{75}{$.144$}}}} 	& \multicolumn{1}{c|}{\multirow{-3}{*}{{\rotatebox[origin=c]{75}{\color[HTML]{0000FF}{\boldmath{$<.001$}}}}}}   & \multicolumn{1}{c|}{\multirow{-3}{*}{{\rotatebox[origin=c]{75}{$.200$}}}} 	& \multicolumn{1}{c|}{\multirow{-3}{*}{{\rotatebox[origin=c]{75}{\color[HTML]{0000FF}{\boldmath{$<.001$}}}}}} 	& \multicolumn{1}{c|}{\multirow{-3}{*}{{\rotatebox[origin=c]{75}{$.127$}}}} & \multicolumn{1}{c|}{\multirow{-3}{*}{{\rotatebox[origin=c]{75}{\color[HTML]{0000FF}{\boldmath{$<.001$}}}}}} 	& \multicolumn{1}{c|}{\multirow{-3}{*}{{\rotatebox[origin=c]{75}{$.200$}}}} 	& \multirow{-3}{*}{{\rotatebox[origin=c]{75}{\color[HTML]{0000FF}{\boldmath{$<.001$}}}}} 	& \multicolumn{1}{c|}{\multirow{-3}{*}{{\rotatebox[origin=c]{75}{$4.400$}}}}  & \multirow{-3}{*}{{\rotatebox[origin=c]{75}{$.221$}}}   									& \multicolumn{1}{c|}{\textbf{D}} & \multicolumn{3}{c|}{\multirow{-6}{*}{\cellcolor[HTML]{D9D9D9}}}                                                                                                    							\\ \cline{3-17} 
                                                                                                                     &                                                                                        					&                                                                 & \multicolumn{1}{c|}{}                                                      	& \multicolumn{1}{c|}{}                                                          								& \multicolumn{1}{c|}{}                                                      	& \multicolumn{1}{c|}{}                                                        									& \multicolumn{1}{c|}{}                                                      & \multicolumn{1}{c|}{}                                                        								& \multicolumn{1}{c|}{}                                                      	&                                                         									& \multicolumn{1}{c|}{}                                                       &                                                         								& \multicolumn{1}{c|}{\textbf{B}} & \multicolumn{1}{c|}{$.430$}                                     	& \multicolumn{2}{c|}{\cellcolor[HTML]{D9D9D9}}																			\\ \cline{14-16} 
                                                                                                                     &                                                                                        					&                                                                 & \multicolumn{1}{c|}{}                                                      	& \multicolumn{1}{c|}{}                                                          								& \multicolumn{1}{c|}{}                                                      	& \multicolumn{1}{c|}{}                                                        									& \multicolumn{1}{c|}{}                                                      & \multicolumn{1}{c|}{}                                                        								& \multicolumn{1}{c|}{}                                                      	&                                                         									& \multicolumn{1}{c|}{}                                                       &                                                         								& \multicolumn{1}{c|}{\textbf{C}} & \multicolumn{1}{c|}{$1.000$}                                     	& \multicolumn{1}{c|}{$1.000$}                                     	& \multicolumn{1}{c|}{\cellcolor[HTML]{D9D9D9}}		\\ \cline{14-17} 
    \multirow{-12}{*}{\rotatebox{90}{\textbf{\begin{tabular}[c]{@{}c@{}}Accuracy\\ (DV1)\end{tabular}}}}             & \multirow{-6}{*}{\textbf{{\rotatebox[origin=c]{90}{Onskin}}}}                          					& \multirow{-3}{*}{\textbf{E}}                                    & \multicolumn{1}{c|}{\multirow{-3}{*}{{\rotatebox[origin=c]{75}{$.259$}}}} 	& \multicolumn{1}{c|}{\multirow{-3}{*}{{\rotatebox[origin=c]{75}{\color[HTML]{0000FF}{\boldmath{$<.001$}}}}}}   & \multicolumn{1}{c|}{\multirow{-3}{*}{{\rotatebox[origin=c]{75}{$.126$}}}} 	& \multicolumn{1}{c|}{\multirow{-3}{*}{{\rotatebox[origin=c]{75}{\color[HTML]{0000FF}{\boldmath{$<.001$}}}}}} 	& \multicolumn{1}{c|}{\multirow{-3}{*}{{\rotatebox[origin=c]{75}{$.126$}}}} & \multicolumn{1}{c|}{\multirow{-3}{*}{{\rotatebox[origin=c]{75}{\color[HTML]{0000FF}{\boldmath{$<.001$}}}}}} 	& \multicolumn{1}{c|}{\multirow{-3}{*}{{\rotatebox[origin=c]{75}{$.144$}}}} 	& \multirow{-3}{*}{{\rotatebox[origin=c]{75}{\color[HTML]{0000FF}{\boldmath{$<.001$}}}}} 	& \multicolumn{1}{c|}{\multirow{-3}{*}{{\rotatebox[origin=c]{75}{$8.769$}}}}  & \multirow{-3}{*}{{\rotatebox[origin=c]{75}{\color[HTML]{0000FF}{\boldmath{$.032$}}}}}   & \multicolumn{1}{c|}{\textbf{D}} & \multicolumn{1}{c|}{$.120$}                                     	& \multicolumn{1}{c|}{$1.000$}                                     	& \multicolumn{1}{c|}{$1.000$}     					\\ \hline
                                                                                                                     &                                                                                        					&                                                                 & \multicolumn{1}{c|}{}                                                      	& \multicolumn{1}{c|}{}                                                          								& \multicolumn{1}{c|}{}                                                      	& \multicolumn{1}{c|}{}                                                        									& \multicolumn{1}{c|}{}                                                      & \multicolumn{1}{c|}{}                                                        								& \multicolumn{1}{c|}{}                                                      	&                                                         									& \multicolumn{1}{c|}{}                                                       &                                                         								& \multicolumn{1}{c|}{\textbf{B}} & \multicolumn{1}{c|}{{\color[HTML]{0000FF} \boldmath{$.022$}}}     	& \multicolumn{2}{c|}{\cellcolor[HTML]{D9D9D9}}																			\\ \cline{14-16} 
                                                                                                                     &                                                                                        					&                                                                 & \multicolumn{1}{c|}{}                                                      	& \multicolumn{1}{c|}{}                                                          								& \multicolumn{1}{c|}{}                                                      	& \multicolumn{1}{c|}{}                                                        									& \multicolumn{1}{c|}{}                                                      & \multicolumn{1}{c|}{}                                                        								& \multicolumn{1}{c|}{}                                                      	&                                                         									& \multicolumn{1}{c|}{}                                                       &                                                         								& \multicolumn{1}{c|}{\textbf{C}} & \multicolumn{1}{c|}{{\color[HTML]{0000FF} \boldmath{$.004$}}}    	& \multicolumn{1}{c|}{$1.000$}                                     	& \multicolumn{1}{c|}{\cellcolor[HTML]{D9D9D9}}		\\ \cline{14-17} 
                                                                                                                     &                                                                                        					& \multirow{-3}{*}{\textbf{S}}                                    & \multicolumn{1}{c|}{\multirow{-3}{*}{{\rotatebox[origin=c]{75}{$.830$}}}} 	& \multicolumn{1}{c|}{\multirow{-3}{*}{{\rotatebox[origin=c]{75}{\color[HTML]{0000FF}{\boldmath{$<.001$}}}}}}   & \multicolumn{1}{c|}{\multirow{-3}{*}{{\rotatebox[origin=c]{75}{$.936$}}}} 	& \multicolumn{1}{c|}{\multirow{-3}{*}{{\rotatebox[origin=c]{75}{\color[HTML]{0000FF}{\boldmath{$<.001$}}}}}} 	& \multicolumn{1}{c|}{\multirow{-3}{*}{{\rotatebox[origin=c]{75}{$.973$}}}} & \multicolumn{1}{c|}{\multirow{-3}{*}{{\rotatebox[origin=c]{75}{\color[HTML]{0000FF}{\boldmath{$<.001$}}}}}} 	& \multicolumn{1}{c|}{\multirow{-3}{*}{{\rotatebox[origin=c]{75}{$.835$}}}} 	& \multirow{-3}{*}{{\rotatebox[origin=c]{75}{\color[HTML]{0000FF}{\boldmath{$<.001$}}}}} 	& \multicolumn{1}{c|}{\multirow{-3}{*}{{\rotatebox[origin=c]{75}{$10.012$}}}} & \multirow{-3}{*}{{\rotatebox[origin=c]{75}{\color[HTML]{0000FF}{\boldmath{$.018$}}}}}   & \multicolumn{1}{c|}{\textbf{D}} & \multicolumn{1}{c|}{{\color[HTML]{0000FF} \boldmath{$.013$}}}     	& \multicolumn{1}{c|}{$1.000$}                                     	& \multicolumn{1}{c|}{$1.000$}     					\\ \cline{3-17} 
                                                                                                                     &                                                                                        					&                                                                 & \multicolumn{1}{c|}{}                                                      	& \multicolumn{1}{c|}{}                                                          								& \multicolumn{1}{c|}{}                                                      	& \multicolumn{1}{c|}{}                                                        									& \multicolumn{1}{c|}{}                                                      & \multicolumn{1}{c|}{}                                                        								& \multicolumn{1}{c|}{}                                                      	&                                                         									& \multicolumn{1}{c|}{}                                                       &                                                         								& \multicolumn{1}{c|}{\textbf{B}} & \multicolumn{1}{c|}{{\color[HTML]{0000FF} \boldmath{$<.001$}}} 		& \multicolumn{2}{c|}{\cellcolor[HTML]{D9D9D9}}																			\\ \cline{14-16} 
                                                                                                                     &                                                                                        					&                                                                 & \multicolumn{1}{c|}{}                                                      	& \multicolumn{1}{c|}{}                                                          								& \multicolumn{1}{c|}{}                                                      	& \multicolumn{1}{c|}{}                                                        									& \multicolumn{1}{c|}{}                                                      & \multicolumn{1}{c|}{}                                                        								& \multicolumn{1}{c|}{}                                                      	&                                                         									& \multicolumn{1}{c|}{}                                                       &                                                         								& \multicolumn{1}{c|}{\textbf{C}} & \multicolumn{1}{c|}{{\color[HTML]{0000FF} \boldmath{$<.001$}}} 		& \multicolumn{1}{c|}{$1.000$}                                     	& \multicolumn{1}{c|}{\cellcolor[HTML]{D9D9D9}}		\\ \cline{14-17} 
                                                                                                                     & \multirow{-6}{*}{\textbf{{\rotatebox[origin=c]{90}{Midair}}}}                          					& \multirow{-3}{*}{\textbf{E}}                                    & \multicolumn{1}{c|}{\multirow{-3}{*}{{\rotatebox[origin=c]{75}{$.928$}}}} 	& \multicolumn{1}{c|}{\multirow{-3}{*}{{\rotatebox[origin=c]{75}{\color[HTML]{0000FF}{\boldmath{$<.001$}}}}}}   & \multicolumn{1}{c|}{\multirow{-3}{*}{{\rotatebox[origin=c]{75}{$.954$}}}} 	& \multicolumn{1}{c|}{\multirow{-3}{*}{{\rotatebox[origin=c]{75}{\color[HTML]{0000FF}{\boldmath{$.001$}}}}}}   	& \multicolumn{1}{c|}{\multirow{-3}{*}{{\rotatebox[origin=c]{75}{$.969$}}}} & \multicolumn{1}{c|}{\multirow{-3}{*}{{\rotatebox[origin=c]{75}{{$.022$}}}}}   									& \multicolumn{1}{c|}{\multirow{-3}{*}{{\rotatebox[origin=c]{75}{$.887$}}}} 	& \multirow{-3}{*}{{\rotatebox[origin=c]{75}{\color[HTML]{0000FF}{\boldmath{$<.001$}}}}} 	& \multicolumn{1}{c|}{\multirow{-3}{*}{{\rotatebox[origin=c]{75}{$19.312$}}}} & \multirow{-3}{*}{{\rotatebox[origin=c]{75}{\color[HTML]{0000FF}{\boldmath{$<.001$}}}}} 	& \multicolumn{1}{c|}{\textbf{D}} & \multicolumn{1}{c|}{{\color[HTML]{0000FF} \boldmath{$.002$}}}   & \multicolumn{1}{c|}{$1.000$}                                  	& \multicolumn{1}{c|}{$1.000$}     					\\ \cline{2-17} 
                                                                                                                     &                                                                                        					&                                                                 & \multicolumn{1}{c|}{}                                                      	& \multicolumn{1}{c|}{}                                                          								& \multicolumn{1}{c|}{}                                                      	& \multicolumn{1}{c|}{}                                                        									& \multicolumn{1}{c|}{}                                                      & \multicolumn{1}{c|}{}                                                        								& \multicolumn{1}{c|}{}                                                      	&                                                         									& \multicolumn{1}{c|}{}                                                       &                                                         								& \multicolumn{1}{c|}{\textbf{B}} & \multicolumn{1}{c|}{{\color[HTML]{0000FF} \boldmath{$.003$}}}     	& \multicolumn{2}{c|}{\cellcolor[HTML]{D9D9D9}}																			\\ \cline{14-16} 
                                                                                                                     &                                                                                        					&                                                                 & \multicolumn{1}{c|}{}                                                      	& \multicolumn{1}{c|}{}                                                          								& \multicolumn{1}{c|}{}                                                      	& \multicolumn{1}{c|}{}                                                        									& \multicolumn{1}{c|}{}                                                      & \multicolumn{1}{c|}{}                                                        								& \multicolumn{1}{c|}{}                                                      	&                                                         									& \multicolumn{1}{c|}{}                                                       &                                                         								& \multicolumn{1}{c|}{\textbf{C}} & \multicolumn{1}{c|}{{\color[HTML]{0000FF} \boldmath{$<.001$}}} 		& \multicolumn{1}{c|}{{\color[HTML]{0000FF} \boldmath{$<.001$}}} 	& \multicolumn{1}{c|}{\cellcolor[HTML]{D9D9D9}}		\\ \cline{14-17} 
                                                                                                                     &                                                                                        					& \multirow{-3}{*}{\textbf{S}}                                    & \multicolumn{1}{c|}{\multirow{-3}{*}{{\rotatebox[origin=c]{75}{$.867$}}}} 	& \multicolumn{1}{c|}{\multirow{-3}{*}{{\rotatebox[origin=c]{75}{\color[HTML]{0000FF}{\boldmath{$<.001$}}}}}}   & \multicolumn{1}{c|}{\multirow{-3}{*}{{\rotatebox[origin=c]{75}{$.959$}}}} 	& \multicolumn{1}{c|}{\multirow{-3}{*}{{\rotatebox[origin=c]{75}{\color[HTML]{0000FF}{\boldmath{$<.001$}}}}}} 	& \multicolumn{1}{c|}{\multirow{-3}{*}{{\rotatebox[origin=c]{75}{$.939$}}}} & \multicolumn{1}{c|}{\multirow{-3}{*}{{\rotatebox[origin=c]{75}{\color[HTML]{0000FF}{\boldmath{$<.001$}}}}}} 	& \multicolumn{1}{c|}{\multirow{-3}{*}{{\rotatebox[origin=c]{75}{$.917$}}}} 	& \multirow{-3}{*}{{\rotatebox[origin=c]{75}{\color[HTML]{0000FF}{\boldmath{$<.001$}}}}} 	& \multicolumn{1}{c|}{\multirow{-3}{*}{{\rotatebox[origin=c]{75}{$50.703$}}}} & \multirow{-3}{*}{{\rotatebox[origin=c]{75}{\color[HTML]{0000FF}{\boldmath{$<.001$}}}}} 	& \multicolumn{1}{c|}{\textbf{D}} & \multicolumn{1}{c|}{{\color[HTML]{0000FF} \boldmath{$<.001$}}} 		& \multicolumn{1}{c|}{$.277$}                                     	& \multicolumn{1}{c|}{$1.000$}     					\\ \cline{3-17} 
                                                                                                                     &                                                                                        					&                                                                 & \multicolumn{1}{c|}{}                                                      	& \multicolumn{1}{c|}{}                                                          								& \multicolumn{1}{c|}{}                                                      	& \multicolumn{1}{c|}{}                                                        									& \multicolumn{1}{c|}{}                                                      & \multicolumn{1}{c|}{}                                                        								& \multicolumn{1}{c|}{}                                                      	&                                                         									& \multicolumn{1}{c|}{}                                                       &                                                         								& \multicolumn{1}{c|}{\textbf{B}} & \multicolumn{1}{c|}{{\color[HTML]{0000FF} \boldmath{$.015$}}}     	& \multicolumn{2}{c|}{\cellcolor[HTML]{D9D9D9}}																			\\ \cline{14-16} 
                                                                                                                     &                                                                                        					&                                                                 & \multicolumn{1}{c|}{}                                                      	& \multicolumn{1}{c|}{}                                                          								& \multicolumn{1}{c|}{}                                                      	& \multicolumn{1}{c|}{}                                                        									& \multicolumn{1}{c|}{}                                                      & \multicolumn{1}{c|}{}                                                        								& \multicolumn{1}{c|}{}                                                      	&                                                         									& \multicolumn{1}{c|}{}                                                       &                                                         								& \multicolumn{1}{c|}{\textbf{C}} & \multicolumn{1}{c|}{{\color[HTML]{0000FF} \boldmath{$<.001$}}} 		& \multicolumn{1}{c|}{{\color[HTML]{0000FF} \boldmath{$<.001$}}} 	& \multicolumn{1}{c|}{\cellcolor[HTML]{D9D9D9}}		\\ \cline{14-17} 
    \multirow{-12}{*}{\rotatebox{90}{\textbf{\begin{tabular}[c]{@{}c@{}}Angular Displacement\\ (DV4)\end{tabular}}}} & \multirow{-6}{*}{\textbf{{\rotatebox[origin=c]{90}{Onskin}}}}                          					& \multirow{-3}{*}{\textbf{E}}                                    & \multicolumn{1}{c|}{\multirow{-3}{*}{{\rotatebox[origin=c]{75}{$.896$}}}} 	& \multicolumn{1}{c|}{\multirow{-3}{*}{{\rotatebox[origin=c]{75}{\color[HTML]{0000FF}{\boldmath{$<.001$}}}}}}   & \multicolumn{1}{c|}{\multirow{-3}{*}{{\rotatebox[origin=c]{75}{$.964$}}}} 	& \multicolumn{1}{c|}{\multirow{-3}{*}{{\rotatebox[origin=c]{75}{\color[HTML]{0000FF}{\boldmath{$.010$}}}}}}   	& \multicolumn{1}{c|}{\multirow{-3}{*}{{\rotatebox[origin=c]{75}{$.923$}}}} & \multicolumn{1}{c|}{\multirow{-3}{*}{{\rotatebox[origin=c]{75}{\color[HTML]{0000FF}{\boldmath{$<.001$}}}}}} 	& \multicolumn{1}{c|}{\multirow{-3}{*}{{\rotatebox[origin=c]{75}{$.930$}}}} 	& \multirow{-3}{*}{{\rotatebox[origin=c]{75}{\color[HTML]{0000FF}{\boldmath{$<.001$}}}}} 	& \multicolumn{1}{c|}{\multirow{-3}{*}{{\rotatebox[origin=c]{75}{$52.088$}}}} & \multirow{-3}{*}{{\rotatebox[origin=c]{75}{\color[HTML]{0000FF}{\boldmath{$<.001$}}}}} 	& \multicolumn{1}{c|}{\textbf{D}} & \multicolumn{1}{c|}{{\color[HTML]{0000FF} \boldmath{$<.001$}}} 		& \multicolumn{1}{c|}{{\color[HTML]{0000FF} \boldmath{$.005$}}}     & \multicolumn{1}{c|}{$1.000$}     					\\ \hline
                                                                                                                     &                                                                                        					&                                                                 & \multicolumn{1}{c|}{}                                                      	& \multicolumn{1}{c|}{}                                                          								& \multicolumn{1}{c|}{}                                                      	& \multicolumn{1}{c|}{}                                                        									& \multicolumn{1}{c|}{}                                                      & \multicolumn{1}{c|}{}                                                        								& \multicolumn{1}{c|}{}                                                      	&                                                         									& \multicolumn{1}{c|}{}                                                       &                                                         								& \multicolumn{1}{c|}{\textbf{B}} & \multicolumn{3}{l|}{\cellcolor[HTML]{D9D9D9}}                                                                                                                      							\\ \cline{14-14}
                                                                                                                     &                                                                                        					&                                                                 & \multicolumn{1}{c|}{}                                                      	& \multicolumn{1}{c|}{}                                                          								& \multicolumn{1}{c|}{}                                                      	& \multicolumn{1}{c|}{}                                                        									& \multicolumn{1}{c|}{}                                                      & \multicolumn{1}{c|}{}                                                        								& \multicolumn{1}{c|}{}                                                      	&                                                         									& \multicolumn{1}{c|}{}                                                       &                                                         								& \multicolumn{1}{c|}{\textbf{C}} & \multicolumn{3}{l|}{\cellcolor[HTML]{D9D9D9}}                                                                                                                      							\\ \cline{14-14}
                                                                                                                     &                                                                                        					& \multirow{-3}{*}{\textbf{S}}                                    & \multicolumn{1}{c|}{\multirow{-3}{*}{{\rotatebox[origin=c]{75}{$.810$}}}} 	& \multicolumn{1}{c|}{\multirow{-3}{*}{{\rotatebox[origin=c]{75}{\color[HTML]{0000FF}{\boldmath{$<.001$}}}}}}   & \multicolumn{1}{c|}{\multirow{-3}{*}{{\rotatebox[origin=c]{75}{$.867$}}}} 	& \multicolumn{1}{c|}{\multirow{-3}{*}{{\rotatebox[origin=c]{75}{\color[HTML]{0000FF}{\boldmath{$.005$}}}}}}   	& \multicolumn{1}{c|}{\multirow{-3}{*}{{\rotatebox[origin=c]{75}{$.807$}}}} & \multicolumn{1}{c|}{\multirow{-3}{*}{{\rotatebox[origin=c]{75}{\color[HTML]{0000FF}{\boldmath{$<.001$}}}}}} 	& \multicolumn{1}{c|}{\multirow{-3}{*}{{\rotatebox[origin=c]{75}{$.867$}}}} 	& \multirow{-3}{*}{{\rotatebox[origin=c]{75}{\color[HTML]{0000FF}{\boldmath{$<.001$}}}}} 	& \multicolumn{1}{c|}{\multirow{-3}{*}{{\rotatebox[origin=c]{75}{$4.950$}}}}  & \multirow{-3}{*}{{\rotatebox[origin=c]{75}{$.176$}}}   									& \multicolumn{1}{c|}{\textbf{D}} & \multicolumn{3}{l|}{\cellcolor[HTML]{D9D9D9}}                                                                                                                      							\\ \cline{3-14} 
                                                                                                                     &                                                                                       					&                                                                 & \multicolumn{1}{c|}{}                                                      	& \multicolumn{1}{c|}{}                                                          								& \multicolumn{1}{c|}{}                                                      	& \multicolumn{1}{c|}{}                                                        									& \multicolumn{1}{c|}{}                                                      & \multicolumn{1}{c|}{}                                                        								& \multicolumn{1}{c|}{}                                                      	&                                                         									& \multicolumn{1}{c|}{}                                                       &                                                         								& \multicolumn{1}{c|}{\textbf{B}} & \multicolumn{3}{l|}{\cellcolor[HTML]{D9D9D9}}                                                                                                                      							\\ \cline{14-14}
                                                                                                                     &                                                                                        					&                                                                 & \multicolumn{1}{c|}{}                                                      	& \multicolumn{1}{c|}{}                                                          								& \multicolumn{1}{c|}{}                                                      	& \multicolumn{1}{c|}{}                                                        									& \multicolumn{1}{c|}{}                                                      & \multicolumn{1}{c|}{}                                                        								& \multicolumn{1}{c|}{}                                                      	&                                                         									& \multicolumn{1}{c|}{}                                                       &                                                         								& \multicolumn{1}{c|}{\textbf{C}} & \multicolumn{3}{l|}{\cellcolor[HTML]{D9D9D9}}                                                                                                                      							\\ \cline{14-14}
                                                                                                                     & \multirow{-6}{*}{\textbf{{\rotatebox[origin=c]{90}{Midair}}}}                          					& \multirow{-3}{*}{\textbf{E}}                                    & \multicolumn{1}{c|}{\multirow{-3}{*}{{\rotatebox[origin=c]{75}{$.831$}}}} 	& \multicolumn{1}{c|}{\multirow{-3}{*}{{\rotatebox[origin=c]{75}{\color[HTML]{0000FF}{\boldmath{$<.001$}}}}}}   & \multicolumn{1}{c|}{\multirow{-3}{*}{{\rotatebox[origin=c]{75}{$.826$}}}} 	& \multicolumn{1}{c|}{\multirow{-3}{*}{{\rotatebox[origin=c]{75}{\color[HTML]{0000FF}{\boldmath{$<.001$}}}}}} 	& \multicolumn{1}{c|}{\multirow{-3}{*}{{\rotatebox[origin=c]{75}{$.816$}}}} & \multicolumn{1}{c|}{\multirow{-3}{*}{{\rotatebox[origin=c]{75}{\color[HTML]{0000FF}{\boldmath{$<.001$}}}}}} 	& \multicolumn{1}{c|}{\multirow{-3}{*}{{\rotatebox[origin=c]{75}{$.882$}}}} 	& \multirow{-3}{*}{{\rotatebox[origin=c]{75}{\color[HTML]{0000FF}{\boldmath{$.009$}}}}}   	& \multicolumn{1}{c|}{\multirow{-3}{*}{{\rotatebox[origin=c]{75}{$5.900$}}}}  & \multirow{-3}{*}{{\rotatebox[origin=c]{75}{$.117$}}}   									& \multicolumn{1}{c|}{\textbf{D}} & \multicolumn{3}{l|}{\cellcolor[HTML]{D9D9D9}}                                                                                                                      							\\ \cline{2-14} 
                                                                                                                     &                                                                                        					&                                                                 & \multicolumn{1}{c|}{}                                                      	& \multicolumn{1}{c|}{}                                                          								& \multicolumn{1}{c|}{}                                                      	& \multicolumn{1}{c|}{}                                                        									& \multicolumn{1}{c|}{}                                                      & \multicolumn{1}{c|}{}                                                        								& \multicolumn{1}{c|}{}                                                      	&                                                         									& \multicolumn{1}{c|}{}                                                       &                                                         								& \multicolumn{1}{c|}{\textbf{B}} & \multicolumn{3}{l|}{\cellcolor[HTML]{D9D9D9}}                                                                                                                      							\\ \cline{14-14}
                                                                                                                     &                                                                                        					&                                                                 & \multicolumn{1}{c|}{}                                                      	& \multicolumn{1}{c|}{}                                                          								& \multicolumn{1}{c|}{}                                                      	& \multicolumn{1}{c|}{}                                                        									& \multicolumn{1}{c|}{}                                                      & \multicolumn{1}{c|}{}                                                        								& \multicolumn{1}{c|}{}                                                      	&                                                         									& \multicolumn{1}{c|}{}                                                       &                                                         								& \multicolumn{1}{c|}{\textbf{C}} & \multicolumn{3}{l|}{\cellcolor[HTML]{D9D9D9}}                                                                                                                      							\\ \cline{14-14}
                                                                                                                     &                                                                                        					& \multirow{-3}{*}{\textbf{S}}                                    & \multicolumn{1}{c|}{\multirow{-3}{*}{{\rotatebox[origin=c]{75}{$.820$}}}} 	& \multicolumn{1}{c|}{\multirow{-3}{*}{{\rotatebox[origin=c]{75}{\color[HTML]{0000FF}{\boldmath{$<.001$}}}}}}   & \multicolumn{1}{c|}{\multirow{-3}{*}{{\rotatebox[origin=c]{75}{$.826$}}}} 	& \multicolumn{1}{c|}{\multirow{-3}{*}{{\rotatebox[origin=c]{75}{\color[HTML]{0000FF}{\boldmath{$<.001$}}}}}} 	& \multicolumn{1}{c|}{\multirow{-3}{*}{{\rotatebox[origin=c]{75}{$.836$}}}} & \multicolumn{1}{c|}{\multirow{-3}{*}{{\rotatebox[origin=c]{75}{\color[HTML]{0000FF}{\boldmath{$.001$}}}}}}   	& \multicolumn{1}{c|}{\multirow{-3}{*}{{\rotatebox[origin=c]{75}{$.851$}}}} 	& \multirow{-3}{*}{{\rotatebox[origin=c]{75}{\color[HTML]{0000FF}{\boldmath{$.001$}}}}}   	& \multicolumn{1}{c|}{\multirow{-3}{*}{{\rotatebox[origin=c]{75}{$5.750$}}}}  & \multirow{-3}{*}{{\rotatebox[origin=c]{75}{$.124$}}}   								    & \multicolumn{1}{c|}{\textbf{D}} & \multicolumn{3}{l|}{\multirow{-9}{*}{\cellcolor[HTML]{D9D9D9}}}                                                                                                    							\\ \cline{3-17} 
                                                                                                                     &                                                                                        					&                                                                 & \multicolumn{1}{c|}{}                                                      	& \multicolumn{1}{c|}{}                                                          								& \multicolumn{1}{c|}{}                                                      	& \multicolumn{1}{c|}{}                                                        									& \multicolumn{1}{c|}{}                                                      & \multicolumn{1}{c|}{}                                                        								& \multicolumn{1}{c|}{}                                                      	&                                                         									& \multicolumn{1}{c|}{}                                                       &                                                         								& \multicolumn{1}{c|}{\textbf{B}} & \multicolumn{1}{c|}{$1.000$}                                     	& \multicolumn{2}{c|}{\cellcolor[HTML]{D9D9D9}}																			\\ \cline{14-16} 
                                                                                                                     &                                                                                        					&                                                                 & \multicolumn{1}{c|}{}                                                      	& \multicolumn{1}{c|}{}                                                          								& \multicolumn{1}{c|}{}                                                      	& \multicolumn{1}{c|}{}                                                        									& \multicolumn{1}{c|}{}                                                      & \multicolumn{1}{c|}{}                                                        								& \multicolumn{1}{c|}{}                                                      	&                                                         									& \multicolumn{1}{c|}{}                                                       &                                                         								& \multicolumn{1}{c|}{\textbf{C}} & \multicolumn{1}{c|}{$.190$}                                     	& \multicolumn{1}{c|}{{\color[HTML]{0000FF} \boldmath{$.007$}}}     & \multicolumn{1}{c|}{\cellcolor[HTML]{D9D9D9}}		\\ \cline{14-17} 
    \multirow{-12}{*}{\rotatebox{90}{\textbf{\begin{tabular}[c]{@{}c@{}}Location Preference\\Ratings\end{tabular}}}} & \multirow{-6}{*}{\textbf{{\rotatebox[origin=c]{90}{Onskin}}}}                          					& \multirow{-3}{*}{\textbf{E}}                                    & \multicolumn{1}{c|}{\multirow{-3}{*}{{\rotatebox[origin=c]{75}{$.837$}}}} 	& \multicolumn{1}{c|}{\multirow{-3}{*}{{\rotatebox[origin=c]{75}{\color[HTML]{0000FF}{\boldmath{$.001$}}}}}}    & \multicolumn{1}{c|}{\multirow{-3}{*}{{\rotatebox[origin=c]{75}{$.792$}}}} 	& \multicolumn{1}{c|}{\multirow{-3}{*}{{\rotatebox[origin=c]{75}{\color[HTML]{0000FF}{\boldmath{$<.001$}}}}}} 	& \multicolumn{1}{c|}{\multirow{-3}{*}{{\rotatebox[origin=c]{75}{$.793$}}}} & \multicolumn{1}{c|}{\multirow{-3}{*}{{\rotatebox[origin=c]{75}{\color[HTML]{0000FF}{\boldmath{$<.001$}}}}}} 	& \multicolumn{1}{c|}{\multirow{-3}{*}{{\rotatebox[origin=c]{75}{$.845$}}}} 	& \multirow{-3}{*}{{\rotatebox[origin=c]{75}{\color[HTML]{0000FF}{\boldmath{$.001$}}}}}   	& \multicolumn{1}{c|}{\multirow{-3}{*}{{\rotatebox[origin=c]{75}{$15.850$}}}} & \multirow{-3}{*}{{\rotatebox[origin=c]{75}{\color[HTML]{0000FF}{\boldmath{$.001$}}}}}   & \multicolumn{1}{c|}{\textbf{D}} & \multicolumn{1}{c|}{$.367$}                                     	& \multicolumn{1}{c|}{{\color[HTML]{0000FF} \boldmath{$.006$}}}     & \multicolumn{1}{c|}{$1.000$}      				\\ \hline
    \end{tabular}%
    }
    \vspace{.15cm} 
    \captionof{table}{Statistical comparison of swipe performance metrics across regions in the 4-region distribution. Figure~\ref{fig:GestureMetrics_4R2R_RQ3} highlights statistically significant pairwise differences.}  
    \Description{Statistical comparison of swipe performance metrics across regions in the 4-region distribution. Figure~\ref{fig:GestureMetrics_4R2R_RQ3} highlights statistically significant pairwise differences.}
    \label{tab:GestureMetrics_4R2R_RQ3}  
\end{table}

\endgroup
}

This subsection examines how swipe starting and ending regions influence performance in midair and onskin interaction spaces, focusing on the 4- and 6-region segmentations identified as effective in Section~\ref{subsection:Results_RQ1_RQ2}. 
The 8-region layout is excluded due to limited effectiveness at higher region densities. 
We analyze both region-level effects of starting and ending region choice and shape-level trends across individual swipe trajectories, highlighting how sensitivity to region selection increases as region density grows and region span narrows.

\subsection{4-Region Midair Segmentation}
\label{section:Results_RQ3_4R2R_Midair}
{
In the 4-region midair condition (Figure~\ref{fig:GestureMetrics_4R2R_RQ3}), both objective performance and subjective feedback revealed systematic differences among regions. 
Gesture accuracy (\textbf{DV1}) was significantly influenced by the \textbf{starting region} (Friedman $\chi^2(3)=10.235$, $p=.017$), but not by the ending region. 
Swipes initiated from the horizontally aligned regions \textbf{C} (ear/nape, $M=98.958\%$) and \textbf{A} (nose/eye, $M=95.833\%$) achieved the highest accuracy overall, with region \textbf{C} significantly outperforming ($p=.027$) region \textbf{B} (temple), which exhibited the lowest accuracy ($M=88.542\%$). 
Although accuracy differences across ending regions were not statistically significant, region \textbf{A} exhibited slightly lower ending region accuracy ($M=89.583\%$), suggesting that swipes terminating near the eye-adjacent region may introduce greater endpoint variability.

Angular displacement (\textbf{DV4}) showed a complementary pattern.
Region choice significantly affected angular skew for both swipe initiation (Friedman $\chi^2(3)=10.012$, $p=.018$) and termination (Friedman $\chi^2(3)=19.312$, $p<.001$). 
Swipe segments starting from or ending at region \textbf{A} exhibited significantly greater angular skew than those associated with all other regions (A$\leftrightarrow$B: starting $p=.022$, ending $p<.001$; A$\leftrightarrow$C: starting $p=.004$, ending $p<.001$; A$\leftrightarrow$D: starting $p=.013$, ending $p=.002$), with trajectories consistently drifting toward region \textbf{B}. 
In contrast, swipess performed near the ear and cheek regions (\textbf{B}, \textbf{C} and \textbf{D}) showed substantially smaller angular deviations, indicating more stable midair trajectories.

Subjective \textbf{location preference ratings} aligned with these performance trends. 
Participants reported lower perceived muscle strain and higher comfort for the horizontally aligned regions \textbf{C} and \textbf{A} compared to the vertically oriented regions \textbf{B} (temple) and \textbf{D} (chin), although these differences were not statistically significant. 
Participants further noted that downward swipes from region \textbf{B} felt more comfortable than upward swipes from region \textbf{D}, reinforcing a general preference for horizontal articulation and downward motion.

Subjective \textbf{location preference ratings} did not differ significantly across regions for either initiation ($\chi^2(3)=4.950$, $p=.176$) or termination ($\chi^2(3)=5.900$, $p=.117$). 
Nevertheless, descriptive trends suggested slightly higher comfort ratings for regions \textbf{C} and \textbf{A} compared to \textbf{B} and \textbf{D}.
Participants further noted that downward swipes from region \textbf{B} felt more comfortable than upward swipes from region \textbf{D}, reinforcing a general preference for horizontal swipes and downward motion.

Taken together, these findings reveal a clear phase-dependent ranking of regions (Figure~\ref{fig:GestureMetrics_4R2R_RQ3}). 
For swipe initiation, regions ranked as \textbf{C > A $\approx$ D > B}, with region \textbf{C} providing the highest accuracy and stability and region \textbf{A} supporting accurate starts but with greater angular skew. 
For swipe termination, regions ranked as \textbf{C > D > B > A}, with region \textbf{C} again yielding the most stable endpoints and region \textbf{A} performing worst due to increased angular deviation and reduced endpoint control. 
This asymmetry highlights region \textbf{C} as the most robust choice for both starting and ending midair swipes, and region \textbf{A} as better suited for initiation than termination.

Analysis of individual midair swipe types (Figure~\ref{fig:GestureMetrics_4R2R_RQ3_Individual_MidairSwipes}) further revealed shape-dependent performance differences outlined in Table~\ref{tab:4_region_gesture_ranking_multimetric}. 
Overall, unidirectional and U-shaped swipes outperformed most angular L-shaped swipes, particularly in the nose–eye region (\textbf{AB}, \textbf{BA}) and cheek–nose region (\textbf{AD}, \textbf{DA}), which showed the weakest performance among angular variants.
Horizontal front-to-back \textbf{AC} swipes along the anterior–posterior axis (from nose to ear) and vertical down-to-up \textbf{DB} swipes (from jaw to temple) showed strong overall performance, with \textbf{DB} demonstrating notably low angular displacement and high user ratings.
Despite receiving above-average ratings, \textbf{CA} showed significant angular drift at swipe termination, resulting in a lower performance tier.
Among U-shaped swipes, the best performance was observered over the ear (\textbf{CC}), followed by the cheek ~/~jawline (\textbf{DD}) and temple (\textbf{BB}) regions.
Participants exhibited a preference for U-swipes executed above the nose or along the cheek (\textbf{CC}, \textbf{DD}), despite a slight increase in drift (\ie{}, angular displacement) near the nose.
Participant comments indicated that the spatial extent of regions \textbf{A} and \textbf{B} allowed for minor swipe drift while still remaining perceptually aligned with the intended target area. 
Visual references for all swipe types are presented in Figure~\ref{fig:GestureMetrics_4R2R_RQ3_Visual_Swipe_References}.

{
\begingroup
\setlength{\floatsep}{-1ex} 
\setlength{\intextsep}{-1ex} 
\setlength{\abovecaptionskip}{-1ex}
\setlength{\belowcaptionskip}{-1ex}
\setlength{\tabcolsep}{2pt}

\begin{table}[t]
    \centering
    \begin{subtable}{1.0\linewidth}
        \centering
        \resizebox{240pt}{!}{
            \begin{tabular}{lcc}
            \hline
            \textbf{Tier} & \textbf{Gestures} & \textbf{Justification} \\
            \hline
            
            1 (Best) & 
            \begin{tabular}{c}DB, CC\end{tabular}&
            \begin{tabular}{c}
            Perfect accuracy, strong ratings, and low drift.
            \end{tabular} \\\hline
            
            2 (Strong) & 
            \begin{tabular}{c}AC, BB\\CB, DD\end{tabular} &
            \begin{tabular}{c}
            High accuracy with low drift\\but slightly lower ratings.
            \end{tabular} \\\hline
            
            3 (Moderate) & 
            \begin{tabular}{c}BD, BC, DC\\CD, CA, AA\end{tabular}&
            \begin{tabular}{c}
            Good accuracy but moderate\\drift or ratings.
            \end{tabular} \\\hline
            
            4 (Weak) & 
            \begin{tabular}{c}AB, AD, BA\end{tabular} &
            \begin{tabular}{c}
            Lower ratings and/or reduced accuracy.
            \end{tabular} \\\hline
            
            5 (Worst) & DA &
            \begin{tabular}{c}
            High angular drift.
            \end{tabular} \\
            
            \hline
            \end{tabular}
        }
        \vspace{0.05cm}
        \subcaption{Midair}     
        \label{tab:4_region_midair_gesture_ranking_multimetric}
    \end{subtable}
    \begin{subtable}{1.0\linewidth}
        \centering
        \resizebox{240pt}{!}{
            \begin{tabular}{lcc}
            \hline
            \textbf{Tier} & \textbf{Gestures} & \textbf{Justification} \\
            \hline
            
            1 (Best) & 
            \begin{tabular}{c}BD, CC, DD\end{tabular}&
            \begin{tabular}{c}
            Perfect accuracy, minimal angular \\
            deviation, and highest user ratings.
            \end{tabular} \\\hline
            
            2 (Strong) & 
            \begin{tabular}{c}AC, CA\\DB, BB\end{tabular} &
            \begin{tabular}{c}
            High accuracy with moderate
            \\stability and strong ratings.
            \end{tabular} \\\hline
            
            3 (Moderate) & 
            \begin{tabular}{c}AD, CD\\DC, AA\end{tabular}&
            \begin{tabular}{c}
            Reliable but moderate ratings or higher drift.
            \end{tabular} \\\hline
            
            4 (Weak) & 
            \begin{tabular}{c}BC, CB\end{tabular} &
            \begin{tabular}{c}
            Lower ratings and slightly weaker \\
            performance despite stable trajectories.
            \end{tabular} \\\hline
            
            5 (Worst) & 
            \begin{tabular}{c}AB, BA\\DA\end{tabular} &
            \begin{tabular}{c}
            Lowest accuracy, highest drift, and lowest rating.
            \end{tabular} \\
            
            \hline
            \end{tabular}
        }
        \vspace{0.05cm}
        \subcaption{Onskin}    
        \label{tab:4_region_onskin_gesture_ranking_multimetric}
    \end{subtable}
    \vspace{-0.4cm}
    \caption{Multi-metric swipe ranking in 4-region segmentation based on accuracy, angular displacement, and user ratings.}
    \Description{Multi-metric swipe ranking in 4-region segmentation based on accuracy, angular displacement, and user ratings.}
    \label{tab:4_region_gesture_ranking_multimetric}
\end{table}

\endgroup
}

}

\subsection{4-Region Onskin Segmentation}
\label{section:Results_RQ3_4R2R_Onskin}
{
For the 4-region onskin segmentation (Figure~\ref{fig:GestureMetrics_4R2R_RQ3}), accuracy (\textbf{DV1}) remained consistently high and did not differ significantly for starting regions ($\chi^2(3)=4.400$, $p=.221$).
For ending reigons, however region choice shows a modest but significant effect ($\chi^2(3)=8.769$, $p=.032$), with region \textbf{B} exhibiting the highest termination accuracy. 
Despite this effect, overall accuracy remained above $93\%$ across all regions.
Participants reported practical constraints in ocular-adjacent areas - such as surface friction at the temple and hairline (region \textbf{B}) and discomfort or eyewear interference near the eyes (region \textbf{A}) - which were reflected in several low-performing swipe types involving these regions. 
Consistent with the midair condition, regions distal from the eyes were perceived as more reliable.

In contrast, angular displacement \textbf{(DV4)} varied significantly with facial location and closely mirrored midair patterns (Friedman: starting $\chi^2(3)=50.703$, ending $\chi^2(3)=50.703$; both $p<.001$). 
Gestures involving region \textbf{A} showed the largest angular deviation (starting $M=23.448^\circ$, ending $M=19.375^\circ$), whereas swipe segments performed near the ear and cheek regions (\textbf{C} and \textbf{D}) exhibited the smallest deviations (between $-3^\circ$ to $3^\circ$ degree from the central plane of the region).
Pairwise comparisons confirmed significantly greater deviation for region \textbf{A} relative to other regions (A$\leftrightarrow$B: starting $p=.003$, ending $p=.015$; A$\leftrightarrow$C and A$\leftrightarrow$D: starting/ending $p<.001$).
Region \textbf{B} exhibited similar but less severe constraints (B$\leftrightarrow$C: starting/ending $p<.001$; B$\leftrightarrow$D: ending $p=.005$).
These results indicate that onskin swipes near the occular area (\textbf{A} and \textbf{B}) introduce greater trajectory variability compared to midair swipes.

Subjective location preference ratings were consistent with onskin accuracy trends, and showed no significant differences for starting regions ($\chi^2(3)=5.750$, $p=.124$).  
However, ending-region ratings differed significantly (Friedman $\chi^2(3)=15.850$, $p=0.001$).
Region \textbf{B} (temple) received significantly lower ratings than regions \textbf{C} ($p=0.007$) and \textbf{D} ($p=0.006$), indicating reduced perceived comfort and usability near the temple area. 
Participants frequently reported eye-corner avoidance and hair friction during upward motion as contributing factors to these lower ratings.

Taken together, onskin regions show a clear ranking influenced by anatomical constraints and tactile interaction, with facial contact and ocular proximity imposing stronger limitations than midair swipes. Regions \textbf{C} and \textbf{D} serve as robust anchors, combining lower angular skew with higher comfort and control. 
In contrast, regions \textbf{A} and \textbf{B}, both near the ocular area, present different constraints. 
Region \textbf{B} offers a wider interaction area than \textbf{A}, allowing greater steering away from the eye—reflected in reduced skew and more stable accuracy. 
However, region \textbf{B} is constrained by hair friction and eyewear obstruction, while region \textbf{A} imposes stricter eye-safety constraints, leading to greater angular deviation. 
Overall, these factors establish our ranking for onskin swipe initiation and termination, as shown in Figure \ref{fig:GestureMetrics_4R2R_RQ3}. 
Regions \textbf{C} emerge as the most reliable anchors for both midair and onskin interactions, with caution advised for precision-critical onskin swipes near the eyes, particularly under eyewear.

Analysis of onskin swipe types reveals shape-dependent performance differences, summarized in Table~\ref{tab:4_region_onskin_gesture_ranking_multimetric} and illustrated in Figure~\ref{fig:GestureMetrics_4R2R_RQ3_Visual_Swipe_References}. Unidirectional and most U-shaped swipes exhibit consistently high recognition accuracy, with some achieving perfect accuracy across participants. 
Among unidirectional swipes, the horizontal temple-to-cheek \textbf{BD} swipe and the swipe between the nose (\textbf{A}) and ear (\textbf{C}) demonstrate reliable performance, with \textbf{BD} showcasing exceptional results, minimal angular displacement, and the highest user ratings.

Gestures near the eye corner display greater directional variability, increasing the risk of accidental ocular interaction. 
Several L-shaped swipes—specifically \textbf{AB}, \textbf{BA}, and \textbf{DA}—ranked among the lowest in performance due to swipe termination around the ocular area. 
Among U-shaped swipes, the best performance occurs in the ear (\textbf{C})  and the cheek~/~chin region (\textbf{D}) (\ie{}, \textbf{CC}, \textbf{DD}). 
Temple-centered U-swipes (BB) follow closely, displaying slightly more drift (\ie{}, angular displacement).

Compared to midair, onskin swipes generally provide better accuracy but exhibit slightly higher drift, affected by tactile contact and the need to avoid ocular interaction. 
Visual references for all swipe types are available in Figure~\ref{fig:GestureMetrics_4R2R_RQ3_Visual_Swipe_References}.

}

{
\begingroup
\setlength{\floatsep}{-1ex} 
\setlength{\intextsep}{-1ex} 
\setlength{\abovecaptionskip}{-1ex}
\setlength{\belowcaptionskip}{-1ex}
\setlength{\tabcolsep}{0.8pt}

\begin{figure*}[!b]
    \centering
    \includegraphics[width=500pt]{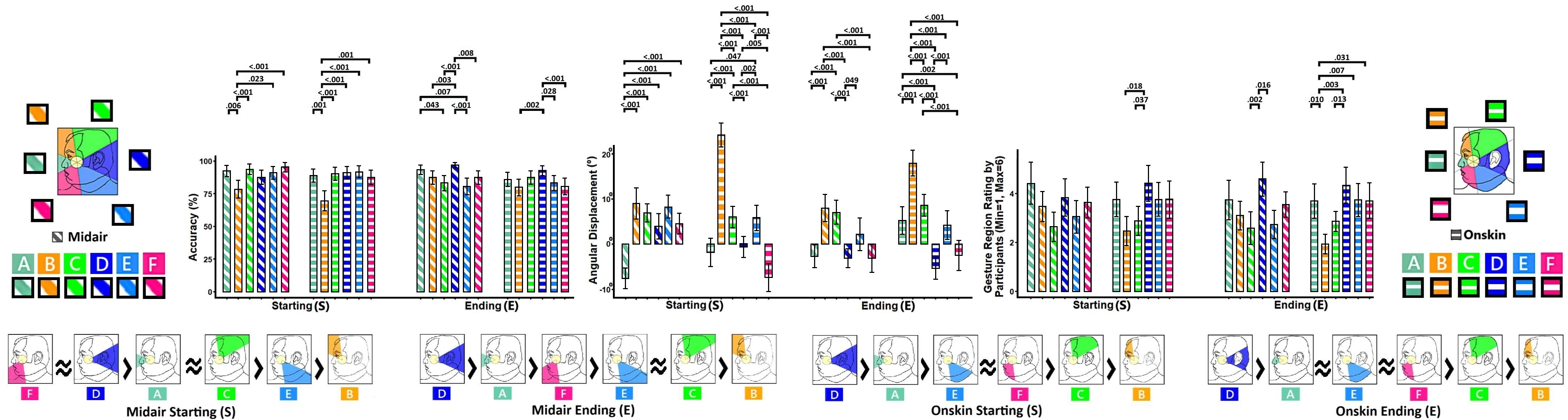}    
    \vspace{0.25cm} 
    \caption{Accuracy (DV1), average angular displacement (DV4) \& location preference ratings for staring and ending regions in 6-region segmentation.}  
    \Description{Three vertical bar charts illustrating accuracy, angular displacement, and relative region ratings across various starting and ending regions in a 6-region distribution around the cheekbone. Region A is around the nose, B around the eyes, C around the temple above ear, D around the ear, E under the ear and around the mandible angle, and F around the mouth and chin, all color-coded against a side image of the face. In the barcharts, slanted colored lines represent midair regions, while horizontal colored lines indicate onskin regions.}
    \label{fig:GestureMetrics_6R2R_RQ3}   
\end{figure*}

\begin{figure*}[!b]
    \centering
    \includegraphics[width=500pt]{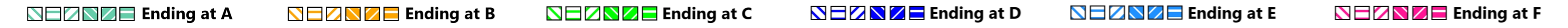}
    \begin{subfigure}[t]{0.355\linewidth}     
        \centering
        \captionsetup{justification=centering}      
        \includegraphics[width = 0.99\linewidth]{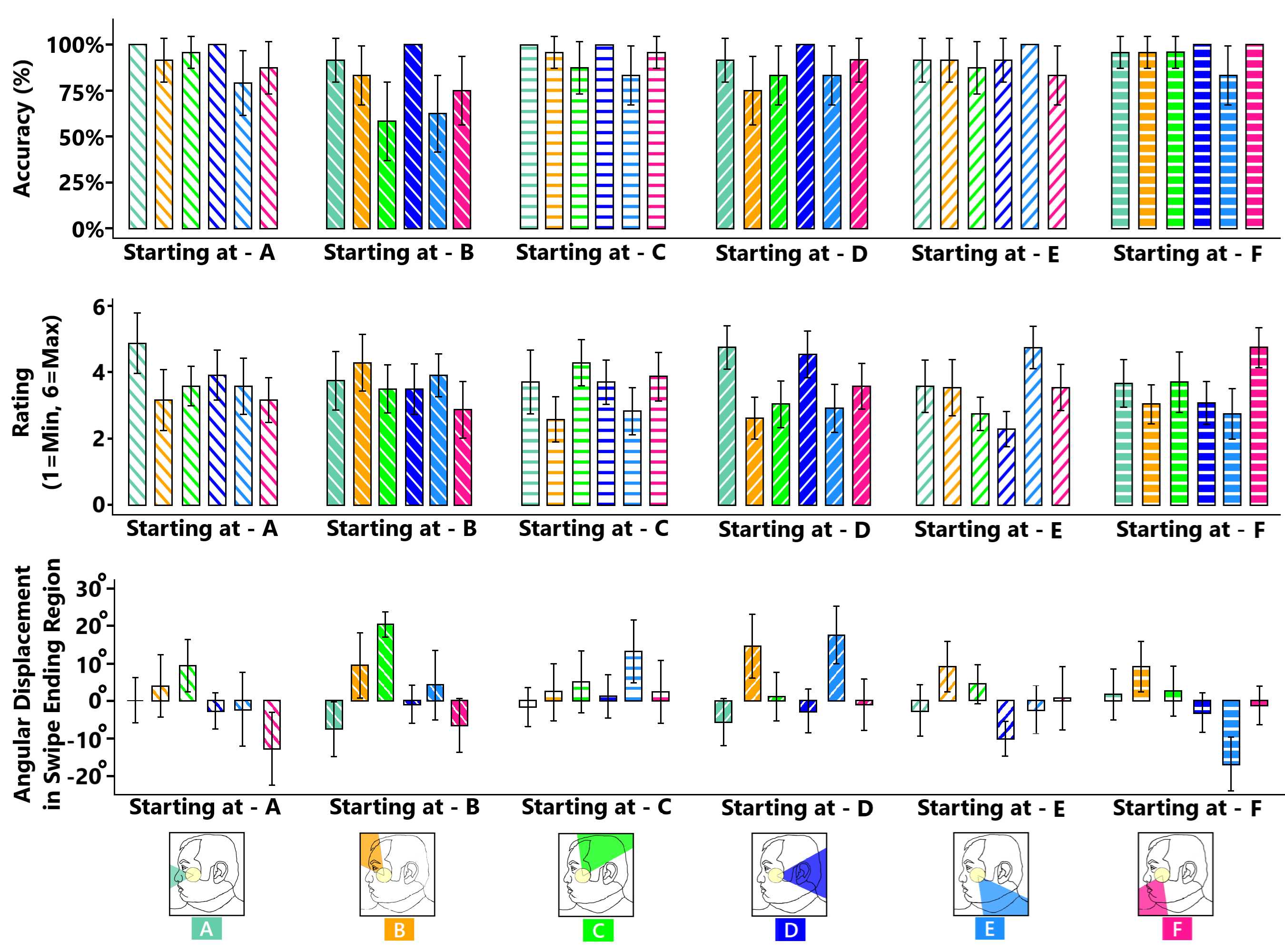}  
        \vspace{-0.5cm} 
        \caption{Midair swipes}  
        \label{fig:GestureMetrics_6R2R_RQ3_Individual_MidairSwipes}   
    \end{subfigure}
    \hfill
    \begin{subfigure}[t]{0.274\linewidth}     
        \centering
        \captionsetup{justification=centering}      
        \includegraphics[width = 0.9\linewidth]{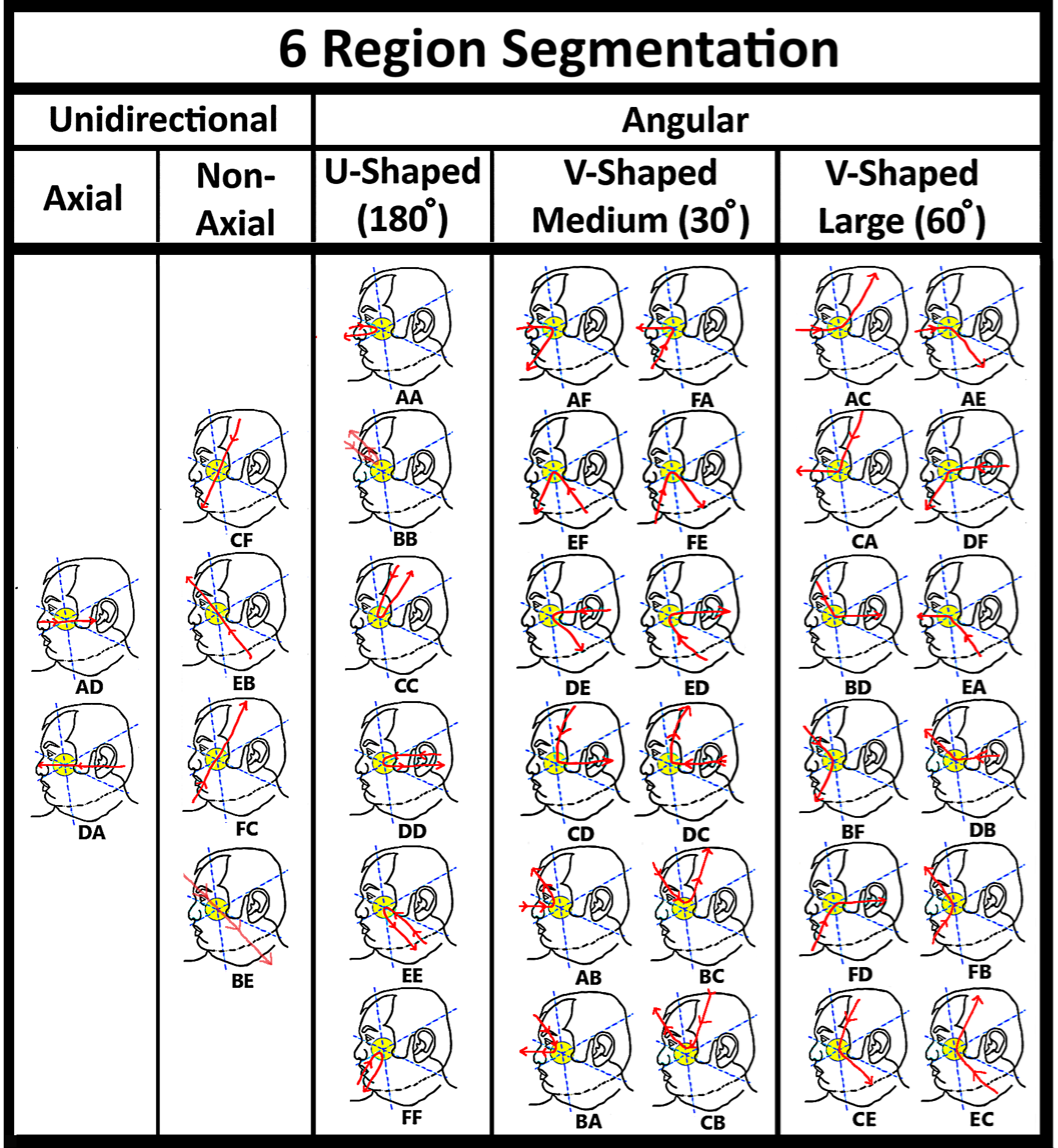}  
        \vspace{-0.15cm} 
        \caption{Visual swipe references}  
        \label{fig:GestureMetrics_6R2R_RQ3_Visual_Swipe_References}   
    \end{subfigure}
    \hfill
    \begin{subfigure}[t]{0.355\linewidth}     
        \centering
        \captionsetup{justification=centering}      
        \includegraphics[width = 0.99\linewidth]{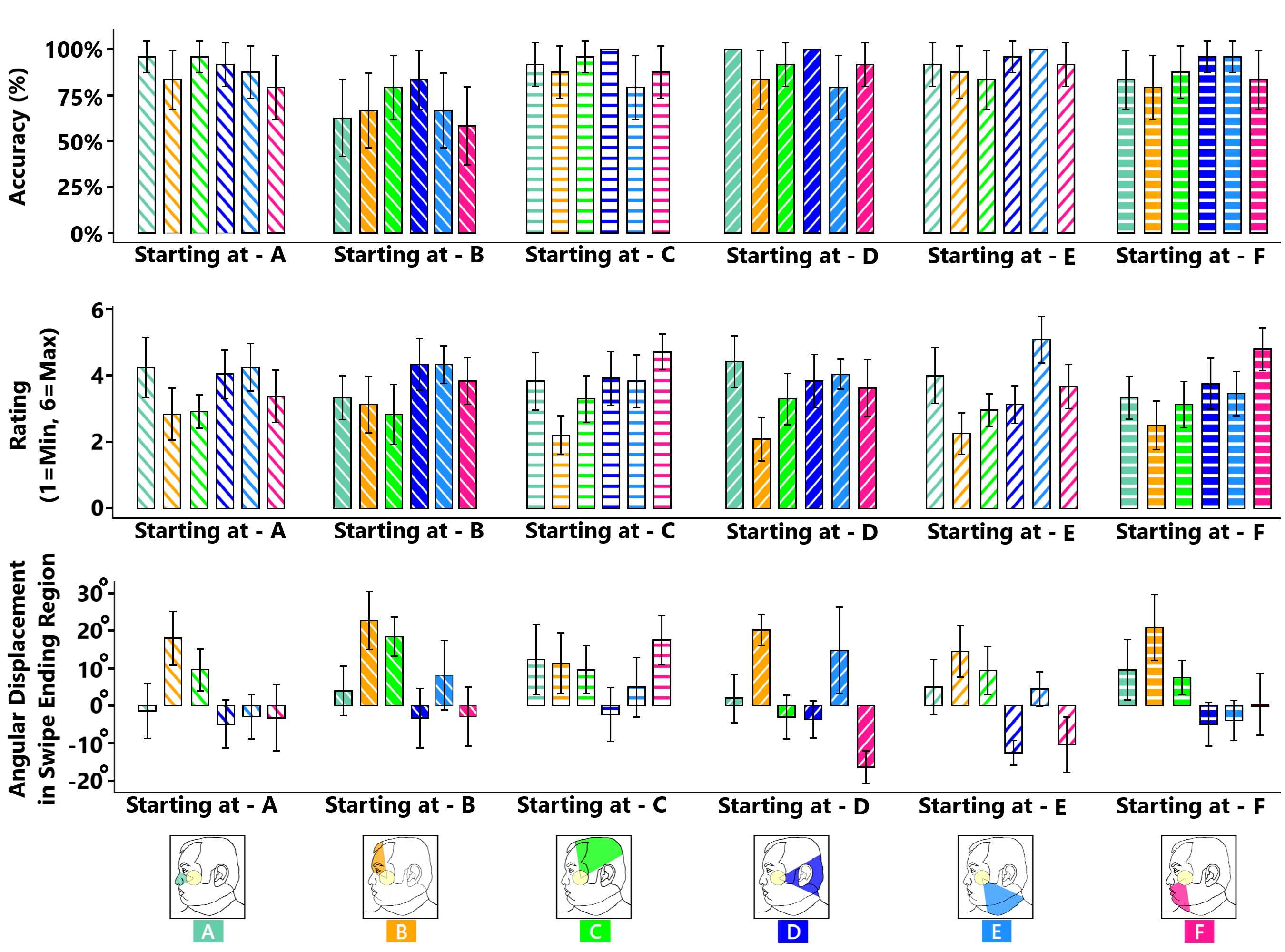}  
        \vspace{-0.5cm} 
        \caption{Onskin swipes}  
        \label{fig:GestureMetrics_6R2R_RQ3_Individual_OnskinSwipes}   
    \end{subfigure}
    \vspace{0.35cm} 
    \caption{Accuracy (DV1), average angular displacement (DV4) \& location preference ratings for individual swipe shapes in 6-region segmentation.}  
    \Description{Three vertical bar charts illustrating accuracy, angular displacement, and relative region ratings across various starting and ending regions in a 6-region distribution around the cheekbone. Region A is around the nose, B around the eyes, C around the temple above ear, D around the ear, E under the ear and around the mandible angle, and F around the mouth and chin, all color-coded against a side image of the face. In the barcharts, slanted colored lines represent midair regions, while horizontal colored lines indicate onskin regions.}
    \label{fig:GestureMetrics_6R2R_RQ3_Individual_Swipes}   
\end{figure*}

\begin{table}[!tb]
    \centering
    \resizebox{240pt}{!}
    {%
    \begin{tabular}{|c|c|c||cccccccccccc||cc||cccccc|}
    \hline
                                                                                                                     &                                                                                        				&                                     					& \multicolumn{12}{c||}{}                                                                                                                                                                                                                                                                                                                                                                                                                                                                                                                                                                                                                                                                                                                                                   																																																										& \multicolumn{2}{c||}{}                                                                                                                   						& \multicolumn{6}{c|}{}                                                                                                                                                                                                                                                                                                                         									\\
                                                                                                                     &                                                                                        				&                                     					& \multicolumn{12}{c||}{}                                                                                                                                                                                                                                                                                                                                                                                                                                                                                                                                                                                                                                                                                                                                                   																																																										& \multicolumn{2}{c||}{}                                                                                                                   						& \multicolumn{6}{c|}{}                                                                                                                                                                                                                                                                                                                         									\\
                                                                                                                     &                                                                                        				&                                     					& \multicolumn{12}{c||}{\multirow{-3}{*}{\textbf{Shapiro-Wilk Normality Test}}}                                                                                                                                                                                                                                                                                                                                                                                                                                                                                                                                                                                                                                                                                                                                                 																																													& \multicolumn{2}{c||}{}                                                                                                                   						& \multicolumn{6}{c|}{}                                                                                                                                                                                                                                                                                                                         									\\ \cline{4-15}
                                                                                                                     &                                                                                        				&                                     					& \multicolumn{2}{c||}{}                                                                                        																& \multicolumn{2}{c||}{}                                                                                        													& \multicolumn{2}{c||}{}                                                                                        												   & \multicolumn{2}{c||}{}                                                                                        														& \multicolumn{2}{c||}{}                                                                                        													& \multicolumn{2}{c||}{}                                                                                                                            & \multicolumn{2}{c||}{}                                                                                                                   						& \multicolumn{6}{c|}{}                                                                                                                                                                                                                                                                                                                         									\\
                                                                                                                     &                                                                                        				&                                     					& \multicolumn{2}{c||}{}                                                                                        																& \multicolumn{2}{c||}{}                                                                                        													& \multicolumn{2}{c||}{}                                                                                        												   & \multicolumn{2}{c||}{}                                                                                        														& \multicolumn{2}{c||}{}                                                                                        													& \multicolumn{2}{c||}{}                                                                                                                            & \multicolumn{2}{c||}{}                                                                                                                   						& \multicolumn{6}{c|}{}                                                                                                                                                                                                                                                                                                                         									\\
                                                                                                                     &                                                                                        				&                                     					& \multicolumn{2}{c||}{}                                                                                        																& \multicolumn{2}{c||}{}                                                                                        													& \multicolumn{2}{c||}{}                                                                                        												   & \multicolumn{2}{c||}{}                                                                                        														& \multicolumn{2}{c||}{}                                                                                        													& \multicolumn{2}{c||}{}                                                                                                                            & \multicolumn{2}{c||}{}                                                                                                                   						& \multicolumn{6}{c|}{}                                                                                                                                                                                                                                                                                                                         									\\
                                                                                                                     &                                                                                        				&                                     					& \multicolumn{2}{c||}{}                                                                                        																& \multicolumn{2}{c||}{}                                                                                        													& \multicolumn{2}{c||}{}                                                                                        												   & \multicolumn{2}{c||}{}                                                                                        														& \multicolumn{2}{c||}{}                                                                                        													& \multicolumn{2}{c||}{}                                                                                                                            & \multicolumn{2}{c||}{}                                                                                                                   						& \multicolumn{6}{c|}{}                                                                                                                                                                                                                                                                                                                         									\\
                                                                                                                     &                                                                                        				&                                     					& \multicolumn{2}{c||}{\multirow{-5}{*}{\textbf{\rotatebox{90}{\begin{tabular}[c]{@{}c@{}}Region\\A\end{tabular}}}}}                                                            & \multicolumn{2}{c||}{\multirow{-5}{*}{\textbf{\rotatebox{90}{\begin{tabular}[c]{@{}c@{}}Region\\B\end{tabular}}}}}                                                & \multicolumn{2}{c||}{\multirow{-5}{*}{\textbf{\rotatebox{90}{\begin{tabular}[c]{@{}c@{}}Region\\C\end{tabular}}}}}                                               & \multicolumn{2}{c||}{\multirow{-5}{*}{\textbf{\rotatebox{90}{\begin{tabular}[c]{@{}c@{}}Region\\D\end{tabular}}}}}                                                 & \multicolumn{2}{c||}{\multirow{-5}{*}{\textbf{\rotatebox{90}{\begin{tabular}[c]{@{}c@{}}Region\\E\end{tabular}}}}}                                                & \multicolumn{2}{c||}{\multirow{-5}{*}{\textbf{\rotatebox{90}{\begin{tabular}[c]{@{}c@{}}Region\\F\end{tabular}}}}}                                & \multicolumn{2}{c||}{\multirow{-8}{*}{\textbf{\rotatebox{90}{\begin{tabular}[c]{@{}c@{}}Friedman Test\\(Within-Subjects)\end{tabular}}}}} 					& \multicolumn{6}{c|}{\multirow{-8}{*}{\textbf{\begin{tabular}[c]{@{}c@{}}Post-Hoc Wilcoxon\\ Test with Bonferroni\\ Correction\end{tabular}}}}                                                                                                                                                                                                 									\\ \cline{4-23}
    \multirow{-9}{*}{\rotatebox{90}{\textbf{\begin{tabular}[c]{@{}c@{}}Gesture Metrics\end{tabular}}}}               & \multirow{-9}{*}{\rotatebox{90}{\textbf{\begin{tabular}[c]{@{}c@{}}Interaction Space\end{tabular}}}} & \multirow{-9}{*}{\rotatebox{90}{\textbf{Position}}} 	& \multicolumn{1}{c|}{\textbf{W}}              								 & \multicolumn{1}{c||}{\textbf{p}}                                               					& \multicolumn{1}{c|}{\textbf{W}}                             	& \multicolumn{1}{c||}{\textbf{p}}                                                					& \multicolumn{1}{c|}{\textbf{W}}              					& \multicolumn{1}{c||}{\textbf{p}}                                                                 & \multicolumn{1}{c|}{\textbf{W}}                               & \multicolumn{1}{c||}{\textbf{p}}                                               					& \multicolumn{1}{c|}{\textbf{W}}              					& \multicolumn{1}{c||}{\textbf{p}}                                                					& \multicolumn{1}{c|}{\textbf{W}}              					& \multicolumn{1}{c||}{\textbf{p}}                                                	& \multicolumn{1}{c|}{\boldmath{$\chi^2(5)$}}                       & \textbf{p}                                                              					& \multicolumn{1}{c|}{}  & \multicolumn{1}{c|}{A}                                       	& \multicolumn{1}{c|}{B}                                         	& \multicolumn{1}{c|}{C}                                         	& \multicolumn{1}{c|}{D}                                         	& \multicolumn{1}{c|}{E}                                					\\ \hline
                                                                                                                     &                                                                                        				&                                     					& \multicolumn{1}{c|}{}                        				  				 & \multicolumn{1}{c|}{{\color[HTML]{0000FF} }}                                   					& \multicolumn{1}{c|}{}                                       	& \multicolumn{1}{c|}{{\color[HTML]{0000FF} }}                                   					& \multicolumn{1}{c|}{}                        					& \multicolumn{1}{c|}{{\color[HTML]{0000FF} }}                                                     & \multicolumn{1}{c|}{}                                         & \multicolumn{1}{c|}{{\color[HTML]{0000FF} }}                                   					& \multicolumn{1}{c|}{}                        					& \multicolumn{1}{c|}{{\color[HTML]{0000FF} }}                                    					& \multicolumn{1}{c|}{}                        					& {\color[HTML]{0000FF} }                                   						& \multicolumn{1}{c|}{}                                         	& {\color[HTML]{0000FF} }                                                 					& \multicolumn{1}{c|}{B} & \multicolumn{1}{c|}{{\color[HTML]{0000FF} \boldmath{$.006$}}}   	& \multicolumn{4}{c|}{\cellcolor[HTML]{D9D9D9}}                       																																																					\\ \cline{18-20}
                                                                                                                     &                                                                                        				&                                     					& \multicolumn{1}{c|}{}                        				  				 & \multicolumn{1}{c|}{{\color[HTML]{0000FF} }}                                   					& \multicolumn{1}{c|}{}                                       	& \multicolumn{1}{c|}{{\color[HTML]{0000FF} }}                                   					& \multicolumn{1}{c|}{}                        					& \multicolumn{1}{c|}{{\color[HTML]{0000FF} }}                                                     & \multicolumn{1}{c|}{}                                         & \multicolumn{1}{c|}{{\color[HTML]{0000FF} }}                                   					& \multicolumn{1}{c|}{}                        					& \multicolumn{1}{c|}{{\color[HTML]{0000FF} }}                                    					& \multicolumn{1}{c|}{}                        					& {\color[HTML]{0000FF} }                                   						& \multicolumn{1}{c|}{}                                         	& {\color[HTML]{0000FF} }                                                 					& \multicolumn{1}{c|}{C} & \multicolumn{1}{c|}{$1.000$}                                   	& \multicolumn{1}{c|}{{\color[HTML]{0000FF} \boldmath{$<.001$}}} 	& \multicolumn{3}{c|}{\cellcolor[HTML]{D9D9D9}}                       																																				\\ \cline{18-21}
                                                                                                                     &                                                                                        				&                                     					& \multicolumn{1}{c|}{}                        				  				 & \multicolumn{1}{c|}{{\color[HTML]{0000FF} }}                                   					& \multicolumn{1}{c|}{}                                       	& \multicolumn{1}{c|}{{\color[HTML]{0000FF} }}                                   					& \multicolumn{1}{c|}{}                        					& \multicolumn{1}{c|}{{\color[HTML]{0000FF} }}                                                     & \multicolumn{1}{c|}{}                                         & \multicolumn{1}{c|}{{\color[HTML]{0000FF} }}                                   					& \multicolumn{1}{c|}{}                        					& \multicolumn{1}{c|}{{\color[HTML]{0000FF} }}                                    					& \multicolumn{1}{c|}{}                        					& {\color[HTML]{0000FF} }                                   						& \multicolumn{1}{c|}{}                                         	& {\color[HTML]{0000FF} }                                                 					& \multicolumn{1}{c|}{D} & \multicolumn{1}{c|}{$1.000$}                                   	& \multicolumn{1}{c|}{$.572$}                                     	& \multicolumn{1}{c|}{$.778$}                                     	& \multicolumn{2}{c|}{\cellcolor[HTML]{D9D9D9}}                       																			\\ \cline{18-22}
                                                                                                                     &                                                                                        				&                                     					& \multicolumn{1}{c|}{}                        				  				 & \multicolumn{1}{c|}{{\color[HTML]{0000FF} }}                                   					& \multicolumn{1}{c|}{}                                       	& \multicolumn{1}{c|}{{\color[HTML]{0000FF} }}                                   					& \multicolumn{1}{c|}{}                        					& \multicolumn{1}{c|}{{\color[HTML]{0000FF} }}                                                     & \multicolumn{1}{c|}{}                                         & \multicolumn{1}{c|}{{\color[HTML]{0000FF} }}                                   					& \multicolumn{1}{c|}{}                        					& \multicolumn{1}{c|}{{\color[HTML]{0000FF} }}                                    					& \multicolumn{1}{c|}{}                        					& {\color[HTML]{0000FF} }                                   						& \multicolumn{1}{c|}{}                                         	& {\color[HTML]{0000FF} }                                                 					& \multicolumn{1}{c|}{E} & \multicolumn{1}{c|}{$1.000$}                                   	& \multicolumn{1}{c|}{{\color[HTML]{0000FF} \boldmath{$.023$}}}     & \multicolumn{1}{c|}{$1.000$}                                     	& \multicolumn{1}{c|}{$1.000$}                                     	& \multicolumn{1}{c|}{\cellcolor[HTML]{D9D9D9}}								\\ \cline{18-23} 
                                                                                                                     &                                                                                        				& \multirow{-5}{*}{\textbf{S}}        					& \multicolumn{1}{c|}{\multirow{-5}{*}{\rotatebox{90}{$.292$}}} 			 & \multicolumn{1}{c|}{\multirow{-5}{*}{{\color[HTML]{0000FF} \rotatebox{90}{\boldmath{$<.001$}}}}} & \multicolumn{1}{c|}{\multirow{-5}{*}{\rotatebox{90}{$.506$}}} & \multicolumn{1}{c|}{\multirow{-5}{*}{{\color[HTML]{0000FF} \rotatebox{90}{\boldmath{$<.001$}}}}} 	& \multicolumn{1}{c|}{\multirow{-5}{*}{\rotatebox{90}{$.258$}}} & \multicolumn{1}{c|}{\multirow{-5}{*}{{\color[HTML]{0000FF} \rotatebox{90}{\boldmath{$<.001$}}}}} & \multicolumn{1}{c|}{\multirow{-5}{*}{\rotatebox{90}{$.387$}}} & \multicolumn{1}{c|}{\multirow{-5}{*}{{\color[HTML]{0000FF} \rotatebox{90}{\boldmath{$<.001$}}}}} 	& \multicolumn{1}{c|}{\multirow{-5}{*}{\rotatebox{90}{$.322$}}} & \multicolumn{1}{c|}{\multirow{-5}{*}{{\color[HTML]{0000FF} \rotatebox{90}{\boldmath{$<.001$}}}}} 	& \multicolumn{1}{c|}{\multirow{-5}{*}{\rotatebox{90}{$.220$}}}	& \multirow{-5}{*}{{\color[HTML]{0000FF} \rotatebox{90}{\boldmath{$<.001$}}}} 		& \multicolumn{1}{c|}{\multirow{-5}{*}{\rotatebox{90}{$33.271$}}}	& \multirow{-5}{*}{{\color[HTML]{0000FF} \rotatebox{90}{\boldmath{$<.001$}}}}               & \multicolumn{1}{c|}{F} & \multicolumn{1}{c|}{$1.000$}                                   	& \multicolumn{1}{c|}{{\color[HTML]{0000FF} \boldmath{$<.001$}}} 	& \multicolumn{1}{c|}{$1.000$}                                     	& \multicolumn{1}{c|}{$.186$}                                     	& \multicolumn{1}{c|}{$1.000$}                       						\\ \cline{3-23} 
                                                                                                                     &                                                                                        				&                                     					& \multicolumn{1}{c|}{}                        				  				 & \multicolumn{1}{c|}{{\color[HTML]{0000FF} }}                                   					& \multicolumn{1}{c|}{}                                       	& \multicolumn{1}{c|}{{\color[HTML]{0000FF} }}                                   					& \multicolumn{1}{c|}{}                        					& \multicolumn{1}{c|}{{\color[HTML]{0000FF} }}                                                     & \multicolumn{1}{c|}{}                                         & \multicolumn{1}{c|}{{\color[HTML]{0000FF} }}                                   					& \multicolumn{1}{c|}{}                        					& \multicolumn{1}{c|}{{\color[HTML]{0000FF} }}                                    					& \multicolumn{1}{c|}{}                        					& {\color[HTML]{0000FF} }                                   						& \multicolumn{1}{c|}{}                                         	& {\color[HTML]{0000FF} }                                                 					& \multicolumn{1}{c|}{B} & \multicolumn{1}{c|}{$.329$}                                   	& \multicolumn{4}{c|}{\cellcolor[HTML]{D9D9D9}}                       																																																					\\ \cline{18-20}
                                                                                                                     &                                                                                        				&                                     					& \multicolumn{1}{c|}{}                        				  				 & \multicolumn{1}{c|}{{\color[HTML]{0000FF} }}                                   					& \multicolumn{1}{c|}{}                                       	& \multicolumn{1}{c|}{{\color[HTML]{0000FF} }}                                   					& \multicolumn{1}{c|}{}                        					& \multicolumn{1}{c|}{{\color[HTML]{0000FF} }}                                                     & \multicolumn{1}{c|}{}                                         & \multicolumn{1}{c|}{{\color[HTML]{0000FF} }}                                   					& \multicolumn{1}{c|}{}                        					& \multicolumn{1}{c|}{{\color[HTML]{0000FF} }}                                    					& \multicolumn{1}{c|}{}                        					& {\color[HTML]{0000FF} }                                   						& \multicolumn{1}{c|}{}                                         	& {\color[HTML]{0000FF} }                                                 					& \multicolumn{1}{c|}{C} & \multicolumn{1}{c|}{{\color[HTML]{0000FF} \boldmath{$.043$}}}   	& \multicolumn{1}{c|}{$1.000$}                                     	& \multicolumn{3}{c|}{\cellcolor[HTML]{D9D9D9}}                       																																				\\ \cline{18-21}
                                                                                                                     &                                                                                        				&                                     					& \multicolumn{1}{c|}{}                        				  				 & \multicolumn{1}{c|}{{\color[HTML]{0000FF} }}                                   					& \multicolumn{1}{c|}{}                                       	& \multicolumn{1}{c|}{{\color[HTML]{0000FF} }}                                   					& \multicolumn{1}{c|}{}                        					& \multicolumn{1}{c|}{{\color[HTML]{0000FF} }}                                                     & \multicolumn{1}{c|}{}                                         & \multicolumn{1}{c|}{{\color[HTML]{0000FF} }}                                   					& \multicolumn{1}{c|}{}                        					& \multicolumn{1}{c|}{{\color[HTML]{0000FF} }}                                    					& \multicolumn{1}{c|}{}                        					& {\color[HTML]{0000FF} }                                   						& \multicolumn{1}{c|}{}                                         	& {\color[HTML]{0000FF} }                                                 					& \multicolumn{1}{c|}{D} & \multicolumn{1}{c|}{$1.000$}                                   	& \multicolumn{1}{c|}{{\color[HTML]{0000FF} \boldmath{$.003$}}}     & \multicolumn{1}{c|}{{\color[HTML]{0000FF} \boldmath{$<.001$}}} 	& \multicolumn{2}{c|}{\cellcolor[HTML]{D9D9D9}}                       																			\\ \cline{18-22}
                                                                                                                     &                                                                                        				&                                     					& \multicolumn{1}{c|}{}                        				  				 & \multicolumn{1}{c|}{{\color[HTML]{0000FF} }}                                   					& \multicolumn{1}{c|}{}                                       	& \multicolumn{1}{c|}{{\color[HTML]{0000FF} }}                                   					& \multicolumn{1}{c|}{}                        					& \multicolumn{1}{c|}{{\color[HTML]{0000FF} }}                                                     & \multicolumn{1}{c|}{}                                         & \multicolumn{1}{c|}{{\color[HTML]{0000FF} }}                                   					& \multicolumn{1}{c|}{}                        					& \multicolumn{1}{c|}{{\color[HTML]{0000FF} }}                                    					& \multicolumn{1}{c|}{}                        					& {\color[HTML]{0000FF} }                                   						& \multicolumn{1}{c|}{}                                         	& {\color[HTML]{0000FF} }                                                 					& \multicolumn{1}{c|}{E} & \multicolumn{1}{c|}{{\color[HTML]{0000FF} \boldmath{$.007$}}}   	& \multicolumn{1}{c|}{$1.000$}                                     	& \multicolumn{1}{c|}{$1.000$}                                     	& \multicolumn{1}{c|}{{\color[HTML]{0000FF} \boldmath{$<.001$}}} 	& \multicolumn{1}{c|}{\cellcolor[HTML]{D9D9D9}}								\\ \cline{18-23} 
                                                                                                                     & \multirow{-10}{*}{\rotatebox{90}{\textbf{Midair}}}                                     				& \multirow{-5}{*}{\textbf{E}}        					& \multicolumn{1}{c|}{\multirow{-5}{*}{\rotatebox{90}{$.220$}}} 			 & \multicolumn{1}{c|}{\multirow{-5}{*}{{\color[HTML]{0000FF} \rotatebox{90}{\boldmath{$<.001$}}}}} & \multicolumn{1}{c|}{\multirow{-5}{*}{\rotatebox{90}{$.363$}}} & \multicolumn{1}{c|}{\multirow{-5}{*}{{\color[HTML]{0000FF} \rotatebox{90}{\boldmath{$<.001$}}}}}  & \multicolumn{1}{c|}{\multirow{-5}{*}{\rotatebox{90}{$.430$}}} & \multicolumn{1}{c|}{\multirow{-5}{*}{{\color[HTML]{0000FF} \rotatebox{90}{\boldmath{$<.001$}}}}} & \multicolumn{1}{c|}{\multirow{-5}{*}{\rotatebox{90}{$.094$}}} & \multicolumn{1}{c|}{\multirow{-5}{*}{{\color[HTML]{0000FF} \rotatebox{90}{\boldmath{$<.001$}}}}}   & \multicolumn{1}{c|}{\multirow{-5}{*}{\rotatebox{90}{$.467$}}} & \multicolumn{1}{c|}{\multirow{-5}{*}{{\color[HTML]{0000FF} \rotatebox{90}{\boldmath{$<.001$}}}}}  & \multicolumn{1}{c|}{\multirow{-5}{*}{\rotatebox{90}{$.363$}}} & \multirow{-5}{*}{{\color[HTML]{0000FF} \rotatebox{90}{\boldmath{$<.001$}}}} 		& \multicolumn{1}{c|}{\multirow{-5}{*}{\rotatebox{90}{$34.211$}}}   & \multirow{-5}{*}{{\color[HTML]{0000FF} \rotatebox{90}{\boldmath{$<.001$}}}}               & \multicolumn{1}{c|}{F} & \multicolumn{1}{c|}{$.778$}                                   	& \multicolumn{1}{c|}{$1.000$}                                     	& \multicolumn{1}{c|}{$1.000$}                                     	& \multicolumn{1}{c|}{{\color[HTML]{0000FF} \boldmath{$.008$}}}     & \multicolumn{1}{c|}{$.774$}                        						\\ \cline{2-23} 
                                                                                                                     &                                                                                        				&                                     					& \multicolumn{1}{c|}{}                        				  				 & \multicolumn{1}{c|}{{\color[HTML]{0000FF} }}                                   					& \multicolumn{1}{c|}{}                                       	& \multicolumn{1}{c|}{{\color[HTML]{0000FF} }}                                   					& \multicolumn{1}{c|}{}                        					& \multicolumn{1}{c|}{{\color[HTML]{0000FF} }}                                                     & \multicolumn{1}{c|}{}                                         & \multicolumn{1}{c|}{{\color[HTML]{0000FF} }}                                   					& \multicolumn{1}{c|}{}                        					& \multicolumn{1}{c|}{{\color[HTML]{0000FF} }}                                   					& \multicolumn{1}{c|}{}                        					& {\color[HTML]{0000FF} }                                   						& \multicolumn{1}{c|}{}                                         	& {\color[HTML]{0000FF} }                                                 					& \multicolumn{1}{c|}{B} & \multicolumn{1}{c|}{{\color[HTML]{0000FF} \boldmath{$.001$}}}   	& \multicolumn{4}{c|}{\cellcolor[HTML]{D9D9D9}}                       																																																					\\ \cline{18-20}
                                                                                                                     &                                                                                        				&                                     					& \multicolumn{1}{c|}{}                        				  				 & \multicolumn{1}{c|}{{\color[HTML]{0000FF} }}                                   					& \multicolumn{1}{c|}{}                                       	& \multicolumn{1}{c|}{{\color[HTML]{0000FF} }}                                   					& \multicolumn{1}{c|}{}                        					& \multicolumn{1}{c|}{{\color[HTML]{0000FF} }}                                                     & \multicolumn{1}{c|}{}                                         & \multicolumn{1}{c|}{{\color[HTML]{0000FF} }}                                   					& \multicolumn{1}{c|}{}                        					& \multicolumn{1}{c|}{{\color[HTML]{0000FF} }}                                   					& \multicolumn{1}{c|}{}                        					& {\color[HTML]{0000FF} }                                   						& \multicolumn{1}{c|}{}                                         	& {\color[HTML]{0000FF} }                                                 					& \multicolumn{1}{c|}{C} & \multicolumn{1}{c|}{$1.000$}                                   	& \multicolumn{1}{c|}{{\color[HTML]{0000FF} \boldmath{$<.001$}}} 	& \multicolumn{3}{c|}{\cellcolor[HTML]{D9D9D9}}                       																																				\\ \cline{18-21}
                                                                                                                     &                                                                                        				&                                     					& \multicolumn{1}{c|}{}                        				  				 & \multicolumn{1}{c|}{{\color[HTML]{0000FF} }}                                   					& \multicolumn{1}{c|}{}                                       	& \multicolumn{1}{c|}{{\color[HTML]{0000FF} }}                                   					& \multicolumn{1}{c|}{}                        					& \multicolumn{1}{c|}{{\color[HTML]{0000FF} }}                                                     & \multicolumn{1}{c|}{}                                         & \multicolumn{1}{c|}{{\color[HTML]{0000FF} }}                                   					& \multicolumn{1}{c|}{}                        					& \multicolumn{1}{c|}{{\color[HTML]{0000FF} }}                                   					& \multicolumn{1}{c|}{}                        					& {\color[HTML]{0000FF} }                                   						& \multicolumn{1}{c|}{}                                         	& {\color[HTML]{0000FF} }                                                 					& \multicolumn{1}{c|}{D} & \multicolumn{1}{c|}{$1.000$}                                   	& \multicolumn{1}{c|}{{\color[HTML]{0000FF} \boldmath{$<.001$}}} 	& \multicolumn{1}{c|}{$1.000$}                                     	& \multicolumn{2}{c|}{\cellcolor[HTML]{D9D9D9}}                       																			\\ \cline{18-22}
                                                                                                                     &                                                                                        				&                                     					& \multicolumn{1}{c|}{}                        				  				 & \multicolumn{1}{c|}{{\color[HTML]{0000FF} }}                                   					& \multicolumn{1}{c|}{}                                       	& \multicolumn{1}{c|}{{\color[HTML]{0000FF} }}                                   					& \multicolumn{1}{c|}{}                        					& \multicolumn{1}{c|}{{\color[HTML]{0000FF} }}                                                     & \multicolumn{1}{c|}{}                                         & \multicolumn{1}{c|}{{\color[HTML]{0000FF} }}                                   					& \multicolumn{1}{c|}{}                        					& \multicolumn{1}{c|}{{\color[HTML]{0000FF} }}                                   					& \multicolumn{1}{c|}{}                        					& {\color[HTML]{0000FF} }                                   						& \multicolumn{1}{c|}{}                                         	& {\color[HTML]{0000FF} }                                                 					& \multicolumn{1}{c|}{E} & \multicolumn{1}{c|}{$1.000$}                                   	& \multicolumn{1}{c|}{{\color[HTML]{0000FF} \boldmath{$<.001$}}} 	& \multicolumn{1}{c|}{$1.000$}                                     	& \multicolumn{1}{c|}{$1.000$}                                     	& \multicolumn{1}{c|}{\cellcolor[HTML]{D9D9D9}}								\\ \cline{18-23} 
                                                                                                                     &                                                                                        				& \multirow{-5}{*}{\textbf{S}}        					& \multicolumn{1}{c|}{\multirow{-5}{*}{\rotatebox{90}{$.363$}}} 			 & \multicolumn{1}{c|}{\multirow{-5}{*}{{\color[HTML]{0000FF} \rotatebox{90}{\boldmath{$<.001$}}}}} & \multicolumn{1}{c|}{\multirow{-5}{*}{\rotatebox{90}{$.579$}}} & \multicolumn{1}{c|}{\multirow{-5}{*}{{\color[HTML]{0000FF} \rotatebox{90}{\boldmath{$<.001$}}}}}  & \multicolumn{1}{c|}{\multirow{-5}{*}{\rotatebox{90}{$.337$}}} & \multicolumn{1}{c|}{\multirow{-5}{*}{{\color[HTML]{0000FF} \rotatebox{90}{\boldmath{$<.001$}}}}} & \multicolumn{1}{c|}{\multirow{-5}{*}{\rotatebox{90}{$.323$}}} & \multicolumn{1}{c|}{\multirow{-5}{*}{{\color[HTML]{0000FF} \rotatebox{90}{\boldmath{$<.001$}}}}}   & \multicolumn{1}{c|}{\multirow{-5}{*}{\rotatebox{90}{$.308$}}} & \multicolumn{1}{c|}{\multirow{-5}{*}{{\color[HTML]{0000FF} \rotatebox{90}{\boldmath{$<.001$}}}}} 	& \multicolumn{1}{c|}{\multirow{-5}{*}{\rotatebox{90}{$.387$}}} & \multirow{-5}{*}{{\color[HTML]{0000FF} \rotatebox{90}{\boldmath{$<.001$}}}} 		& \multicolumn{1}{c|}{\multirow{-5}{*}{\rotatebox{90}{$52.981$}}}   & \multirow{-5}{*}{{\color[HTML]{0000FF} \rotatebox{90}{\boldmath{$<.001$}}}}               & \multicolumn{1}{c|}{F} & \multicolumn{1}{c|}{$1.000$}                                   	& \multicolumn{1}{c|}{{\color[HTML]{0000FF} \boldmath{$.001$}}}     & \multicolumn{1}{c|}{$1.000$}                                     	& \multicolumn{1}{c|}{$1.000$}                                     	& \multicolumn{1}{c|}{$1.000$}                   	 						\\ \cline{3-23} 
                                                                                                                     &                                                                                        				&                                     					& \multicolumn{1}{c|}{}                        				  				 & \multicolumn{1}{c|}{{\color[HTML]{0000FF} }}                                   					& \multicolumn{1}{c|}{}                                       	& \multicolumn{1}{c|}{{\color[HTML]{0000FF} }}                                   					& \multicolumn{1}{c|}{}                        					& \multicolumn{1}{c|}{{\color[HTML]{0000FF} }}                                                     & \multicolumn{1}{c|}{}                                         & \multicolumn{1}{c|}{{\color[HTML]{0000FF} }}                                   					& \multicolumn{1}{c|}{}                        					& \multicolumn{1}{c|}{{\color[HTML]{0000FF} }}                                   					& \multicolumn{1}{c|}{}                        					& {\color[HTML]{0000FF} }                                   						& \multicolumn{1}{c|}{}                                         	& {\color[HTML]{0000FF} }                                                 					& \multicolumn{1}{c|}{B} & \multicolumn{1}{c|}{$.778$}                                   	& \multicolumn{4}{c|}{\cellcolor[HTML]{D9D9D9}}                       																																																					\\ \cline{18-20}
                                                                                                                     &                                                                                        				&                                     					& \multicolumn{1}{c|}{}                        				  				 & \multicolumn{1}{c|}{{\color[HTML]{0000FF} }}                                   					& \multicolumn{1}{c|}{}                                       	& \multicolumn{1}{c|}{{\color[HTML]{0000FF} }}                                   					& \multicolumn{1}{c|}{}                        					& \multicolumn{1}{c|}{{\color[HTML]{0000FF} }}                                                     & \multicolumn{1}{c|}{}                                         & \multicolumn{1}{c|}{{\color[HTML]{0000FF} }}                                   					& \multicolumn{1}{c|}{}                        					& \multicolumn{1}{c|}{{\color[HTML]{0000FF} }}                                   					& \multicolumn{1}{c|}{}                        					& {\color[HTML]{0000FF} }                                   						& \multicolumn{1}{c|}{}                                         	& {\color[HTML]{0000FF} }                                                 					& \multicolumn{1}{c|}{C} & \multicolumn{1}{c|}{$1.000$}                                   	& \multicolumn{1}{c|}{$.531$}                                     	& \multicolumn{3}{c|}{\cellcolor[HTML]{D9D9D9}}                       																																				\\ \cline{18-21}
                                                                                                                     &                                                                                        				&                                     					& \multicolumn{1}{c|}{}                        				  				 & \multicolumn{1}{c|}{{\color[HTML]{0000FF} }}                                   					& \multicolumn{1}{c|}{}                                       	& \multicolumn{1}{c|}{{\color[HTML]{0000FF} }}                                   					& \multicolumn{1}{c|}{}                        					& \multicolumn{1}{c|}{{\color[HTML]{0000FF} }}                                                     & \multicolumn{1}{c|}{}                                         & \multicolumn{1}{c|}{{\color[HTML]{0000FF} }}                                   					& \multicolumn{1}{c|}{}                        					& \multicolumn{1}{c|}{{\color[HTML]{0000FF} }}                                   					& \multicolumn{1}{c|}{}                        					& {\color[HTML]{0000FF} }                                   						& \multicolumn{1}{c|}{}                                         	& {\color[HTML]{0000FF} }                                                 					& \multicolumn{1}{c|}{D} & \multicolumn{1}{c|}{$.125$}                                   	& \multicolumn{1}{c|}{{\color[HTML]{0000FF} \boldmath{$.002$}}}     & \multicolumn{1}{c|}{$.353$}                                    	& \multicolumn{2}{c|}{\cellcolor[HTML]{D9D9D9}}                       																			\\ \cline{18-22}
                                                                                                                     &                                                                                        				&                                     					& \multicolumn{1}{c|}{}                        				  				 & \multicolumn{1}{c|}{{\color[HTML]{0000FF} }}                                   					& \multicolumn{1}{c|}{}                                       	& \multicolumn{1}{c|}{{\color[HTML]{0000FF} }}                                   					& \multicolumn{1}{c|}{}                        					& \multicolumn{1}{c|}{{\color[HTML]{0000FF} }}                                                     & \multicolumn{1}{c|}{}                                         & \multicolumn{1}{c|}{{\color[HTML]{0000FF} }}                                   					& \multicolumn{1}{c|}{}                        					& \multicolumn{1}{c|}{{\color[HTML]{0000FF} }}                                   					& \multicolumn{1}{c|}{}                        					& {\color[HTML]{0000FF} }                                   						& \multicolumn{1}{c|}{}                                         	& {\color[HTML]{0000FF} }                                                 					& \multicolumn{1}{c|}{E} & \multicolumn{1}{c|}{$1.000$}                                   	& \multicolumn{1}{c|}{$1.000$}                                     	& \multicolumn{1}{c|}{$1.000$}                                     	& \multicolumn{1}{c|}{{\color[HTML]{0000FF} \boldmath{$.028$}}}     & \multicolumn{1}{c|}{\cellcolor[HTML]{D9D9D9}}								\\ \cline{18-23} 
    \multirow{-20}{*}{\textbf{\rotatebox{90}{\begin{tabular}[c]{@{}c@{}}Accuracy (DV1)\end{tabular}}}}               & \multirow{-10}{*}{\rotatebox{90}{\textbf{Onskin}}}                                     				& \multirow{-5}{*}{\textbf{E}}        					& \multicolumn{1}{c|}{\multirow{-5}{*}{\rotatebox{90}{$.387$}}} 			 & \multicolumn{1}{c|}{\multirow{-5}{*}{{\color[HTML]{0000FF} \rotatebox{90}{\boldmath{$<.001$}}}}} & \multicolumn{1}{c|}{\multirow{-5}{*}{\rotatebox{90}{$.475$}}} & \multicolumn{1}{c|}{\multirow{-5}{*}{{\color[HTML]{0000FF} \rotatebox{90}{\boldmath{$<.001$}}}}}  & \multicolumn{1}{c|}{\multirow{-5}{*}{\rotatebox{90}{$.363$}}} & \multicolumn{1}{c|}{\multirow{-5}{*}{{\color[HTML]{0000FF} \rotatebox{90}{\boldmath{$<.001$}}}}} & \multicolumn{1}{c|}{\multirow{-5}{*}{\rotatebox{90}{$.240$}}} & \multicolumn{1}{c|}{\multirow{-5}{*}{{\color[HTML]{0000FF} \rotatebox{90}{\boldmath{$<.001$}}}}}   & \multicolumn{1}{c|}{\multirow{-5}{*}{\rotatebox{90}{$.430$}}} & \multicolumn{1}{c|}{\multirow{-5}{*}{{\color[HTML]{0000FF} \rotatebox{90}{\boldmath{$<.001$}}}}}  & \multicolumn{1}{c|}{\multirow{-5}{*}{\rotatebox{90}{$.467$}}} & \multirow{-5}{*}{{\color[HTML]{0000FF} \rotatebox{90}{\boldmath{$<.001$}}}} 		& \multicolumn{1}{c|}{\multirow{-5}{*}{\rotatebox{90}{$23.073$}}}   & \multirow{-5}{*}{{\color[HTML]{0000FF} \rotatebox{90}{\boldmath{$<.001$}}}}               & \multicolumn{1}{c|}{F} & \multicolumn{1}{c|}{$1.000$}                                   	& \multicolumn{1}{c|}{$1.000$}                                     	& \multicolumn{1}{c|}{$.643$}                                     	& \multicolumn{1}{c|}{{\color[HTML]{0000FF} \boldmath{$<.001$}}} 	& \multicolumn{1}{c|}{$1.000$}                        						\\ \hline
                                                                                                                     &                                                                                        				&                                     					& \multicolumn{1}{c|}{}                        				  				 & \multicolumn{1}{c|}{{\color[HTML]{0000FF} }}                                   					& \multicolumn{1}{c|}{}                                       	& \multicolumn{1}{c|}{{\color[HTML]{0000FF} }}                                   					& \multicolumn{1}{c|}{}                        					& \multicolumn{1}{c|}{{\color[HTML]{0000FF} }}                                                     & \multicolumn{1}{c|}{}                                         & \multicolumn{1}{c|}{{\color[HTML]{0000FF} }}                                   					& \multicolumn{1}{c|}{}                        					& \multicolumn{1}{c|}{{\color[HTML]{0000FF} }}                                   					& \multicolumn{1}{c|}{}                        					& {\color[HTML]{0000FF} }                                   						& \multicolumn{1}{c|}{}                                         	& {\color[HTML]{0000FF} }                                                 					& \multicolumn{1}{c|}{B} & \multicolumn{1}{c|}{{\color[HTML]{0000FF} \boldmath{$<.001$}}} 	& \multicolumn{4}{c|}{\cellcolor[HTML]{D9D9D9}}                       																																																					\\ \cline{18-20}
                                                                                                                     &                                                                                        				&                                     					& \multicolumn{1}{c|}{}                        				  				 & \multicolumn{1}{c|}{{\color[HTML]{0000FF} }}                                   					& \multicolumn{1}{c|}{}                                       	& \multicolumn{1}{c|}{{\color[HTML]{0000FF} }}                                   					& \multicolumn{1}{c|}{}                        					& \multicolumn{1}{c|}{{\color[HTML]{0000FF} }}                                                     & \multicolumn{1}{c|}{}                                         & \multicolumn{1}{c|}{{\color[HTML]{0000FF} }}                                   					& \multicolumn{1}{c|}{}                        					& \multicolumn{1}{c|}{{\color[HTML]{0000FF} }}                                   					& \multicolumn{1}{c|}{}                        					& {\color[HTML]{0000FF} }                                   						& \multicolumn{1}{c|}{}                                         	& {\color[HTML]{0000FF} }                                                 					& \multicolumn{1}{c|}{C} & \multicolumn{1}{c|}{{\color[HTML]{0000FF} \boldmath{$<.001$}}} 	& \multicolumn{1}{c|}{$1.000$}                                     	& \multicolumn{3}{c|}{\cellcolor[HTML]{D9D9D9}}                       																																				\\ \cline{18-21}
                                                                                                                     &                                                                                        				&                                     					& \multicolumn{1}{c|}{}                        				  				 & \multicolumn{1}{c|}{{\color[HTML]{0000FF} }}                                   					& \multicolumn{1}{c|}{}                                       	& \multicolumn{1}{c|}{{\color[HTML]{0000FF} }}                                   					& \multicolumn{1}{c|}{}                        					& \multicolumn{1}{c|}{{\color[HTML]{0000FF} }}                                                     & \multicolumn{1}{c|}{}                                         & \multicolumn{1}{c|}{{\color[HTML]{0000FF} }}                                   					& \multicolumn{1}{c|}{}                        					& \multicolumn{1}{c|}{{\color[HTML]{0000FF} }}                                   					& \multicolumn{1}{c|}{}                        					& {\color[HTML]{0000FF} }                                   						& \multicolumn{1}{c|}{}                                         	& {\color[HTML]{0000FF} }                                                 					& \multicolumn{1}{c|}{D} & \multicolumn{1}{c|}{{\color[HTML]{0000FF} \boldmath{$<.001$}}} 	& \multicolumn{1}{c|}{$.570$}                                     	& \multicolumn{1}{c|}{$.910$}                                     	& \multicolumn{2}{c|}{\cellcolor[HTML]{D9D9D9}}                       																			\\ \cline{18-22}
                                                                                                                     &                                                                                        				&                                     					& \multicolumn{1}{c|}{}                        				  				 & \multicolumn{1}{c|}{{\color[HTML]{0000FF} }}                                   					& \multicolumn{1}{c|}{}                                       	& \multicolumn{1}{c|}{{\color[HTML]{0000FF} }}                                   					& \multicolumn{1}{c|}{}                        					& \multicolumn{1}{c|}{{\color[HTML]{0000FF} }}                                                     & \multicolumn{1}{c|}{}                                         & \multicolumn{1}{c|}{{\color[HTML]{0000FF} }}                                   					& \multicolumn{1}{c|}{}                        					& \multicolumn{1}{c|}{{\color[HTML]{0000FF} }}                                   					& \multicolumn{1}{c|}{}                        					& {\color[HTML]{0000FF} }                                   						& \multicolumn{1}{c|}{}                                         	& {\color[HTML]{0000FF} }                                                 					& \multicolumn{1}{c|}{E} & \multicolumn{1}{c|}{{\color[HTML]{0000FF} \boldmath{$<.001$}}} 	& \multicolumn{1}{c|}{$1.000$}                                     	& \multicolumn{1}{c|}{$1.000$}                                     	& \multicolumn{1}{c|}{$.260$}                                     	& \multicolumn{1}{c|}{\cellcolor[HTML]{D9D9D9}}								\\ \cline{18-23} 
                                                                                                                     &                                                                                        				& \multirow{-5}{*}{\textbf{S}}        					& \multicolumn{1}{c|}{\multirow{-5}{*}{\rotatebox{90}{$.948$}}} 			 & \multicolumn{1}{c|}{\multirow{-5}{*}{{\color[HTML]{0000FF} \rotatebox{90}{\boldmath{$<.001$}}}}} & \multicolumn{1}{c|}{\multirow{-5}{*}{\rotatebox{90}{$.960$}}} & \multicolumn{1}{c|}{\multirow{-5}{*}{{\color[HTML]{0000FF} \rotatebox{90}{\boldmath{$<.001$}}}}}  & \multicolumn{1}{c|}{\multirow{-5}{*}{\rotatebox{90}{$.957$}}} & \multicolumn{1}{c|}{\multirow{-5}{*}{{\color[HTML]{0000FF} \rotatebox{90}{\boldmath{$<.001$}}}}} & \multicolumn{1}{c|}{\multirow{-5}{*}{\rotatebox{90}{$.975$}}} & \multicolumn{1}{c|}{\multirow{-5}{*}{{\color[HTML]{0000FF} \rotatebox{90}{\boldmath{$.011$}}}}}    & \multicolumn{1}{c|}{\multirow{-5}{*}{\rotatebox{90}{$.962$}}} & \multicolumn{1}{c|}{\multirow{-5}{*}{{\color[HTML]{0000FF} \rotatebox{90}{\boldmath{$<.001$}}}}}  & \multicolumn{1}{c|}{\multirow{-5}{*}{\rotatebox{90}{$.925$}}} & \multirow{-5}{*}{{\color[HTML]{0000FF} \rotatebox{90}{\boldmath{$<.001$}}}} 		& \multicolumn{1}{c|}{\multirow{-5}{*}{\rotatebox{90}{$96.654$}}}   & \multirow{-5}{*}{{\color[HTML]{0000FF} \rotatebox{90}{\boldmath{$<.001$}}}}               & \multicolumn{1}{c|}{F} & \multicolumn{1}{c|}{{\color[HTML]{0000FF} \boldmath{$<.001$}}} 	& \multicolumn{1}{c|}{$.510$}                                     	& \multicolumn{1}{c|}{$1.000$}                                     	& \multicolumn{1}{c|}{$1.000$}                                     	& \multicolumn{1}{c|}{$1.000$}                       						\\ \cline{3-23} 
                                                                                                                     &                                                                                        				&                                     					& \multicolumn{1}{c|}{}                        				  				 & \multicolumn{1}{c|}{{\color[HTML]{0000FF} }}                                   					& \multicolumn{1}{c|}{}                                       	& \multicolumn{1}{c|}{{\color[HTML]{0000FF} }}                                   					& \multicolumn{1}{c|}{}                        					& \multicolumn{1}{c|}{{\color[HTML]{0000FF} }}                                                     & \multicolumn{1}{c|}{}                                         & \multicolumn{1}{c|}{{\color[HTML]{0000FF} }}                                   					& \multicolumn{1}{c|}{}                        					& \multicolumn{1}{c|}{{\color[HTML]{0000FF} }}                                   					& \multicolumn{1}{c|}{}                        					& {\color[HTML]{0000FF} }                                   						& \multicolumn{1}{c|}{}                                         	& {\color[HTML]{0000FF} }                                                 					& \multicolumn{1}{c|}{B} & \multicolumn{1}{c|}{{\color[HTML]{0000FF} \boldmath{$<.001$}}} 	& \multicolumn{4}{c|}{\cellcolor[HTML]{D9D9D9}}                       																																																					\\ \cline{18-20}
                                                                                                                     &                                                                                        				&                                     					& \multicolumn{1}{c|}{}                        				  				 & \multicolumn{1}{c|}{{\color[HTML]{0000FF} }}                                   					& \multicolumn{1}{c|}{}                                       	& \multicolumn{1}{c|}{{\color[HTML]{0000FF} }}                                   					& \multicolumn{1}{c|}{}                        					& \multicolumn{1}{c|}{{\color[HTML]{0000FF} }}                                                     & \multicolumn{1}{c|}{}                                         & \multicolumn{1}{c|}{{\color[HTML]{0000FF} }}                                   					& \multicolumn{1}{c|}{}                        					& \multicolumn{1}{c|}{{\color[HTML]{0000FF} }}                                   					& \multicolumn{1}{c|}{}                        					& {\color[HTML]{0000FF} }                                   						& \multicolumn{1}{c|}{}                                         	& {\color[HTML]{0000FF} }                                                 					& \multicolumn{1}{c|}{C} & \multicolumn{1}{c|}{{\color[HTML]{0000FF} \boldmath{$<.001$}}} 	& \multicolumn{1}{c|}{$1.000$}                                     	& \multicolumn{3}{c|}{\cellcolor[HTML]{D9D9D9}}                       																																				\\ \cline{18-21}
                                                                                                                     &                                                                                        				&                                     					& \multicolumn{1}{c|}{}                        				  				 & \multicolumn{1}{c|}{{\color[HTML]{0000FF} }}                                   					& \multicolumn{1}{c|}{}                                       	& \multicolumn{1}{c|}{{\color[HTML]{0000FF} }}                                   					& \multicolumn{1}{c|}{}                        					& \multicolumn{1}{c|}{{\color[HTML]{0000FF} }}                                                     & \multicolumn{1}{c|}{}                                         & \multicolumn{1}{c|}{{\color[HTML]{0000FF} }}                                   					& \multicolumn{1}{c|}{}                        					& \multicolumn{1}{c|}{{\color[HTML]{0000FF} }}                                   					& \multicolumn{1}{c|}{}                        					& {\color[HTML]{0000FF} }                                   						& \multicolumn{1}{c|}{}                                         	& {\color[HTML]{0000FF} }                                                 					& \multicolumn{1}{c|}{D} & \multicolumn{1}{c|}{$1.000$}                                   	& \multicolumn{1}{c|}{{\color[HTML]{0000FF} \boldmath{$<.001$}}} 	& \multicolumn{1}{c|}{{\color[HTML]{0000FF} \boldmath{$<.001$}}} 	& \multicolumn{2}{c|}{\cellcolor[HTML]{D9D9D9}}                       																			\\ \cline{18-22}
                                                                                                                     &                                                                                        				&                                     					& \multicolumn{1}{c|}{}                        				  				 & \multicolumn{1}{c|}{{\color[HTML]{0000FF} }}                                   					& \multicolumn{1}{c|}{}                                       	& \multicolumn{1}{c|}{{\color[HTML]{0000FF} }}                                   					& \multicolumn{1}{c|}{}                        					& \multicolumn{1}{c|}{{\color[HTML]{0000FF} }}                                                     & \multicolumn{1}{c|}{}                                         & \multicolumn{1}{c|}{{\color[HTML]{0000FF} }}                                   					& \multicolumn{1}{c|}{}                        					& \multicolumn{1}{c|}{{\color[HTML]{0000FF} }}                                   					& \multicolumn{1}{c|}{}                        					& {\color[HTML]{0000FF} }                                   						& \multicolumn{1}{c|}{}                                         	& {\color[HTML]{0000FF} }                                                 					& \multicolumn{1}{c|}{E} & \multicolumn{1}{c|}{$.330$}                                   	& \multicolumn{1}{c|}{$.196$}                                     	& \multicolumn{1}{c|}{$.181$}                                     	& \multicolumn{1}{c|}{{\color[HTML]{0000FF} \boldmath{$.049$}}}     & \multicolumn{1}{c|}{\cellcolor[HTML]{D9D9D9}}								\\ \cline{18-23} 
                                                                                                                     & \multirow{-10}{*}{\rotatebox{90}{\textbf{Midair}}}                                     				& \multirow{-5}{*}{\textbf{E}}        					& \multicolumn{1}{c|}{\multirow{-5}{*}{\rotatebox{90}{$.963$}}} 			 & \multicolumn{1}{c|}{\multirow{-5}{*}{{\color[HTML]{0000FF} \rotatebox{90}{\boldmath{$<.001$}}}}} & \multicolumn{1}{c|}{\multirow{-5}{*}{\rotatebox{90}{$.964$}}} & \multicolumn{1}{c|}{\multirow{-5}{*}{{\color[HTML]{0000FF} \rotatebox{90}{\boldmath{$<.001$}}}}}  & \multicolumn{1}{c|}{\multirow{-5}{*}{\rotatebox{90}{$.968$}}} & \multicolumn{1}{c|}{\multirow{-5}{*}{{\color[HTML]{0000FF} \rotatebox{90}{\boldmath{$.002$}}}}}  & \multicolumn{1}{c|}{\multirow{-5}{*}{\rotatebox{90}{$.924$}}} & \multicolumn{1}{c|}{\multirow{-5}{*}{{\color[HTML]{0000FF} \rotatebox{90}{\boldmath{$<.001$}}}}} 	& \multicolumn{1}{c|}{\multirow{-5}{*}{\rotatebox{90}{$.970$}}} & \multicolumn{1}{c|}{\multirow{-5}{*}{{\color[HTML]{0000FF} \rotatebox{90}{\boldmath{$.003$}}}}}   & \multicolumn{1}{c|}{\multirow{-5}{*}{\rotatebox{90}{$.962$}}} & \multirow{-5}{*}{{\color[HTML]{0000FF} \rotatebox{90}{\boldmath{$<.001$}}}} 		& \multicolumn{1}{c|}{\multirow{-5}{*}{\rotatebox{90}{$61.455$}}}   & \multirow{-5}{*}{{\color[HTML]{0000FF} \rotatebox{90}{\boldmath{$<.001$}}}}               & \multicolumn{1}{c|}{F} & \multicolumn{1}{c|}{$1.000$}                                   	& \multicolumn{1}{c|}{{\color[HTML]{0000FF} \boldmath{$<.001$}}} 	& \multicolumn{1}{c|}{{\color[HTML]{0000FF} \boldmath{$<.001$}}} 	& \multicolumn{1}{c|}{$1.000$}                                     	& \multicolumn{1}{c|}{$.776$}                        						\\ \cline{2-23} 
                                                                                                                     &                                                                                        				&                                     					& \multicolumn{1}{c|}{}                        				  				 & \multicolumn{1}{c|}{{\color[HTML]{0000FF} }}                                   					& \multicolumn{1}{c|}{}                                       	& \multicolumn{1}{c|}{{\color[HTML]{0000FF} }}                                   					& \multicolumn{1}{c|}{}                        					& \multicolumn{1}{c|}{{\color[HTML]{0000FF} }}                                                     & \multicolumn{1}{c|}{}                                         & \multicolumn{1}{c|}{{\color[HTML]{0000FF} }}                                   					& \multicolumn{1}{c|}{}                        					& \multicolumn{1}{c|}{{\color[HTML]{0000FF} }}                                   					& \multicolumn{1}{c|}{}                        					& {\color[HTML]{0000FF} }                                   						& \multicolumn{1}{c|}{}                                         	& {\color[HTML]{0000FF} }                                                					& \multicolumn{1}{c|}{B} & \multicolumn{1}{c|}{{\color[HTML]{0000FF} \boldmath{$<.001$}}} 	& \multicolumn{4}{c|}{\cellcolor[HTML]{D9D9D9}}                       																																																					\\ \cline{18-20}
                                                                                                                     &                                                                                        				&                                     					& \multicolumn{1}{c|}{}                        				  				 & \multicolumn{1}{c|}{{\color[HTML]{0000FF} }}                                   					& \multicolumn{1}{c|}{}                                       	& \multicolumn{1}{c|}{{\color[HTML]{0000FF} }}                                   					& \multicolumn{1}{c|}{}                        					& \multicolumn{1}{c|}{{\color[HTML]{0000FF} }}                                                     & \multicolumn{1}{c|}{}                                         & \multicolumn{1}{c|}{{\color[HTML]{0000FF} }}                                   					& \multicolumn{1}{c|}{}                        					& \multicolumn{1}{c|}{{\color[HTML]{0000FF} }}                                   					& \multicolumn{1}{c|}{}                        					& {\color[HTML]{0000FF} }                                   						& \multicolumn{1}{c|}{}                                         	& {\color[HTML]{0000FF} }                                                 					& \multicolumn{1}{c|}{C} & \multicolumn{1}{c|}{{\color[HTML]{0000FF} \boldmath{$<.001$}}} 	& \multicolumn{1}{c|}{{\color[HTML]{0000FF} \boldmath{$<.001$}}} 	& \multicolumn{3}{c|}{\cellcolor[HTML]{D9D9D9}}                       																																				\\ \cline{18-21}
                                                                                                                     &                                                                                        				&                                     					& \multicolumn{1}{c|}{}                        				  				 & \multicolumn{1}{c|}{{\color[HTML]{0000FF} }}                                   					& \multicolumn{1}{c|}{}                                       	& \multicolumn{1}{c|}{{\color[HTML]{0000FF} }}                                   					& \multicolumn{1}{c|}{}                        					& \multicolumn{1}{c|}{{\color[HTML]{0000FF} }}                                                     & \multicolumn{1}{c|}{}                                         & \multicolumn{1}{c|}{{\color[HTML]{0000FF} }}                                   					& \multicolumn{1}{c|}{}                        					& \multicolumn{1}{c|}{{\color[HTML]{0000FF} }}                                   					& \multicolumn{1}{c|}{}                        					& {\color[HTML]{0000FF} }                                   						& \multicolumn{1}{c|}{}                                         	& {\color[HTML]{0000FF} }                                                 					& \multicolumn{1}{c|}{D} & \multicolumn{1}{c|}{$1.000$}                                   	& \multicolumn{1}{c|}{{\color[HTML]{0000FF} \boldmath{$<.001$}}} 	& \multicolumn{1}{c|}{{\color[HTML]{0000FF} \boldmath{$<.001$}}} 	& \multicolumn{2}{c|}{\cellcolor[HTML]{D9D9D9}}                       																			\\ \cline{18-22}
                                                                                                                     &                                                                                        				&                                     					& \multicolumn{1}{c|}{}                        				  				 & \multicolumn{1}{c|}{{\color[HTML]{0000FF} }}                                   					& \multicolumn{1}{c|}{}                                       	& \multicolumn{1}{c|}{{\color[HTML]{0000FF} }}                                   					& \multicolumn{1}{c|}{}                        					& \multicolumn{1}{c|}{{\color[HTML]{0000FF} }}                                                     & \multicolumn{1}{c|}{}                                         & \multicolumn{1}{c|}{{\color[HTML]{0000FF} }}                                   					& \multicolumn{1}{c|}{}                        					& \multicolumn{1}{c|}{{\color[HTML]{0000FF} }}                                   					& \multicolumn{1}{c|}{}                        					& {\color[HTML]{0000FF} }                                   						& \multicolumn{1}{c|}{}                                         	& {\color[HTML]{0000FF} }                                                 					& \multicolumn{1}{c|}{E} & \multicolumn{1}{c|}{{\color[HTML]{0000FF} \boldmath{$.047$}}}   	& \multicolumn{1}{c|}{{\color[HTML]{0000FF} \boldmath{$<.001$}}} 	& \multicolumn{1}{c|}{$1.000$}                                     	& \multicolumn{1}{c|}{{\color[HTML]{0000FF} \boldmath{$.002$}}}     & \multicolumn{1}{c|}{\cellcolor[HTML]{D9D9D9}}								\\ \cline{18-23} 
                                                                                                                     &                                                                                        				& \multirow{-5}{*}{\textbf{S}}        					& \multicolumn{1}{c|}{\multirow{-5}{*}{\rotatebox{90}{$.948$}}}              & \multicolumn{1}{c|}{\multirow{-5}{*}{{\color[HTML]{0000FF} \rotatebox{90}{\boldmath{$<.001$}}}}} & \multicolumn{1}{c|}{\multirow{-5}{*}{\rotatebox{90}{$.888$}}} & \multicolumn{1}{c|}{\multirow{-5}{*}{{\color[HTML]{0000FF} \rotatebox{90}{\boldmath{$<.001$}}}}}  & \multicolumn{1}{c|}{\multirow{-5}{*}{\rotatebox{90}{$.954$}}} & \multicolumn{1}{c|}{\multirow{-5}{*}{{\color[HTML]{0000FF} \rotatebox{90}{\boldmath{$<.001$}}}}} & \multicolumn{1}{c|}{\multirow{-5}{*}{\rotatebox{90}{$.917$}}} & \multicolumn{1}{c|}{\multirow{-5}{*}{{\color[HTML]{0000FF} \rotatebox{90}{\boldmath{$<.001$}}}}} 	& \multicolumn{1}{c|}{\multirow{-5}{*}{\rotatebox{90}{$.867$}}} & \multicolumn{1}{c|}{\multirow{-5}{*}{{\color[HTML]{0000FF} \rotatebox{90}{\boldmath{$<.001$}}}}}  & \multicolumn{1}{c|}{\multirow{-5}{*}{\rotatebox{90}{$.963$}}} & \multirow{-5}{*}{{\color[HTML]{0000FF} \rotatebox{90}{\boldmath{$<.001$}}}} 		& \multicolumn{1}{c|}{\multirow{-5}{*}{\rotatebox{90}{$212.310$}}}  & \multirow{-5}{*}{{\color[HTML]{0000FF} \rotatebox{90}{\boldmath{$<.001$}}}}               & \multicolumn{1}{c|}{F} & \multicolumn{1}{c|}{$.239$}                                   	& \multicolumn{1}{c|}{{\color[HTML]{0000FF} \boldmath{$<.001$}}} 	& \multicolumn{1}{c|}{{\color[HTML]{0000FF} \boldmath{$<.001$}}} 	& \multicolumn{1}{c|}{{\color[HTML]{0000FF} \boldmath{$.005$}}}     & {\color[HTML]{0000FF} \boldmath{$<.001$}}          						\\ \cline{3-23} 
                                                                                                                     &                                                                                        				&                                     					& \multicolumn{1}{c|}{}                        				  				 & \multicolumn{1}{c|}{{\color[HTML]{0000FF} }}                                   					& \multicolumn{1}{c|}{}                                       	& \multicolumn{1}{c|}{{\color[HTML]{0000FF} }}                                   					& \multicolumn{1}{c|}{}                        					& \multicolumn{1}{c|}{{\color[HTML]{0000FF} }}                                                     & \multicolumn{1}{c|}{}                                         & \multicolumn{1}{c|}{{\color[HTML]{0000FF} }}                                   					& \multicolumn{1}{c|}{}                        					& \multicolumn{1}{c|}{{\color[HTML]{0000FF} }}                                   					& \multicolumn{1}{c|}{}                        					& {\color[HTML]{0000FF} }                                   						& \multicolumn{1}{c|}{}                                         	& {\color[HTML]{0000FF} }                                                 					& \multicolumn{1}{c|}{B} & \multicolumn{1}{c|}{{\color[HTML]{0000FF} \boldmath{$<.001$}}} 	& \multicolumn{4}{c|}{\cellcolor[HTML]{D9D9D9}}                       																																																					\\ \cline{18-20}
                                                                                                                     &                                                                                        				&                                     					& \multicolumn{1}{c|}{}                        				  				 & \multicolumn{1}{c|}{{\color[HTML]{0000FF} }}                                   					& \multicolumn{1}{c|}{}                                       	& \multicolumn{1}{c|}{{\color[HTML]{0000FF} }}                                   					& \multicolumn{1}{c|}{}                        					& \multicolumn{1}{c|}{{\color[HTML]{0000FF} }}                                                     & \multicolumn{1}{c|}{}                                         & \multicolumn{1}{c|}{{\color[HTML]{0000FF} }}                                   					& \multicolumn{1}{c|}{}                        					& \multicolumn{1}{c|}{{\color[HTML]{0000FF} }}                                   					& \multicolumn{1}{c|}{}                        					& {\color[HTML]{0000FF} }                                   						& \multicolumn{1}{c|}{}                                         	& {\color[HTML]{0000FF} }                                                 					& \multicolumn{1}{c|}{C} & \multicolumn{1}{c|}{$.297$}                                   	& \multicolumn{1}{c|}{{\color[HTML]{0000FF} \boldmath{$<.001$}}} 	& \multicolumn{3}{c|}{\cellcolor[HTML]{D9D9D9}}                       																																				\\ \cline{18-21}
                                                                                                                     &                                                                                        				&                                     					& \multicolumn{1}{c|}{}                        				  				 & \multicolumn{1}{c|}{{\color[HTML]{0000FF} }}                                   					& \multicolumn{1}{c|}{}                                       	& \multicolumn{1}{c|}{{\color[HTML]{0000FF} }}                                   					& \multicolumn{1}{c|}{}                        					& \multicolumn{1}{c|}{{\color[HTML]{0000FF} }}                                                     & \multicolumn{1}{c|}{}                                         & \multicolumn{1}{c|}{{\color[HTML]{0000FF} }}                                   					& \multicolumn{1}{c|}{}                        					& \multicolumn{1}{c|}{{\color[HTML]{0000FF} }}                                   					& \multicolumn{1}{c|}{}                        					& {\color[HTML]{0000FF} }                                   						& \multicolumn{1}{c|}{}                                         	& {\color[HTML]{0000FF} }                                                 					& \multicolumn{1}{c|}{D} & \multicolumn{1}{c|}{{\color[HTML]{0000FF} \boldmath{$<.001$}}} 	& \multicolumn{1}{c|}{{\color[HTML]{0000FF} \boldmath{$<.001$}}} 	& \multicolumn{1}{c|}{{\color[HTML]{0000FF} \boldmath{$<.001$}}} 	& \multicolumn{2}{c|}{\cellcolor[HTML]{D9D9D9}}                       																			\\ \cline{18-22}
                                                                                                                     &                                                                                        				&                                     					& \multicolumn{1}{c|}{}                        				  				 & \multicolumn{1}{c|}{{\color[HTML]{0000FF} }}                                   					& \multicolumn{1}{c|}{}                                       	& \multicolumn{1}{c|}{{\color[HTML]{0000FF} }}                                   					& \multicolumn{1}{c|}{}                        					& \multicolumn{1}{c|}{{\color[HTML]{0000FF} }}                                                     & \multicolumn{1}{c|}{}                                         & \multicolumn{1}{c|}{{\color[HTML]{0000FF} }}                                   					& \multicolumn{1}{c|}{}                        					& \multicolumn{1}{c|}{{\color[HTML]{0000FF} }}                                   					& \multicolumn{1}{c|}{}                        					& {\color[HTML]{0000FF} }                                   						& \multicolumn{1}{c|}{}                                         	& {\color[HTML]{0000FF} }                                                 					& \multicolumn{1}{c|}{E} & \multicolumn{1}{c|}{$1.000$}                                   	& \multicolumn{1}{c|}{{\color[HTML]{0000FF} \boldmath{$<.001$}}} 	& \multicolumn{1}{c|}{$.076$}                                     	& \multicolumn{1}{c|}{{\color[HTML]{0000FF} \boldmath{$<.001$}}} 	& \multicolumn{1}{c|}{\cellcolor[HTML]{D9D9D9}}								\\ \cline{18-23} 
    \multirow{-20}{*}{\rotatebox{90}{\textbf{\begin{tabular}[c]{@{}c@{}}Angular Displacement (DV4)\end{tabular}}}}   & \multirow{-10}{*}{\rotatebox{90}{\textbf{Onskin}}}                                     				& \multirow{-5}{*}{\textbf{E}}        					& \multicolumn{1}{c|}{\multirow{-5}{*}{\rotatebox{90}{$.963$}}} 			 & \multicolumn{1}{c|}{\multirow{-5}{*}{{\color[HTML]{0000FF} \rotatebox{90}{\boldmath{$<.001$}}}}} & \multicolumn{1}{c|}{\multirow{-5}{*}{\rotatebox{90}{$.964$}}} & \multicolumn{1}{c|}{\multirow{-5}{*}{{\color[HTML]{0000FF} \rotatebox{90}{\boldmath{$<.001$}}}}}  & \multicolumn{1}{c|}{\multirow{-5}{*}{\rotatebox{90}{$.967$}}}	& \multicolumn{1}{c|}{\multirow{-5}{*}{{\color[HTML]{0000FF} \rotatebox{90}{\boldmath{$.002$}}}}}  & \multicolumn{1}{c|}{\multirow{-5}{*}{\rotatebox{90}{$.926$}}} & \multicolumn{1}{c|}{\multirow{-5}{*}{{\color[HTML]{0000FF} \rotatebox{90}{\boldmath{$<.001$}}}}} 	& \multicolumn{1}{c|}{\multirow{-5}{*}{\rotatebox{90}{$.970$}}} & \multicolumn{1}{c|}{\multirow{-5}{*}{{\color[HTML]{0000FF} \rotatebox{90}{\boldmath{$.003$}}}}}   & \multicolumn{1}{c|}{\multirow{-5}{*}{\rotatebox{90}{$.962$}}} & \multirow{-5}{*}{{\color[HTML]{0000FF} \rotatebox{90}{\boldmath{$<.001$}}}} 		& \multicolumn{1}{c|}{\multirow{-5}{*}{\rotatebox{90}{$147.550$}}}  & \multirow{-5}{*}{{\color[HTML]{0000FF} \rotatebox{90}{\boldmath{$<.001$}}}}               & \multicolumn{1}{c|}{F} & \multicolumn{1}{c|}{{\color[HTML]{0000FF} \boldmath{$.002$}}}   	& \multicolumn{1}{c|}{{\color[HTML]{0000FF} \boldmath{$<.001$}}} 	& \multicolumn{1}{c|}{{\color[HTML]{0000FF} \boldmath{$<.001$}}} 	& \multicolumn{1}{c|}{$1.000$}                                     	& \multicolumn{1}{c|}{$.111$}                        						\\ \hline
                                                                                                                     &                                                                                        				&                                     					& \multicolumn{1}{c|}{}                        				  				 & \multicolumn{1}{c|}{{\color[HTML]{0000FF} }}                                   					& \multicolumn{1}{c|}{}                                       	& \multicolumn{1}{c|}{}                                                          					& \multicolumn{1}{c|}{}                        					& \multicolumn{1}{c|}{{\color[HTML]{0000FF} }}                                                     & \multicolumn{1}{c|}{}                                         & \multicolumn{1}{c|}{{\color[HTML]{0000FF} }}                                   					& \multicolumn{1}{c|}{}                        					& \multicolumn{1}{c|}{{\color[HTML]{0000FF} }}                                   					& \multicolumn{1}{c|}{}                        					&                                                           						& \multicolumn{1}{c|}{}                                         	& {\color[HTML]{0000FF} }                                                 					& \multicolumn{1}{c|}{B} & \multicolumn{1}{c|}{$.860$}                                   	& \multicolumn{4}{c|}{\cellcolor[HTML]{D9D9D9}}                       																																																					\\ \cline{18-20}
                                                                                                                     &                                                                                        				&                                     					& \multicolumn{1}{c|}{}                        				  				 & \multicolumn{1}{c|}{{\color[HTML]{0000FF} }}                                   					& \multicolumn{1}{c|}{}                                       	& \multicolumn{1}{c|}{}                                                          					& \multicolumn{1}{c|}{}                        					& \multicolumn{1}{c|}{{\color[HTML]{0000FF} }}                                                     & \multicolumn{1}{c|}{}                                         & \multicolumn{1}{c|}{{\color[HTML]{0000FF} }}                                   					& \multicolumn{1}{c|}{}                        					& \multicolumn{1}{c|}{{\color[HTML]{0000FF} }}                                   					& \multicolumn{1}{c|}{}                        					&                                                           						& \multicolumn{1}{c|}{}                                         	& {\color[HTML]{0000FF} }                                                 					& \multicolumn{1}{c|}{C} & \multicolumn{1}{c|}{$.120$}                                   	& \multicolumn{1}{c|}{$.940$}                                     	& \multicolumn{3}{c|}{\cellcolor[HTML]{D9D9D9}}                       																																				\\ \cline{18-21}
                                                                                                                     &                                                                                        				&                                     					& \multicolumn{1}{c|}{}                        				  				 & \multicolumn{1}{c|}{{\color[HTML]{0000FF} }}                                   					& \multicolumn{1}{c|}{}                                       	& \multicolumn{1}{c|}{}                                                          					& \multicolumn{1}{c|}{}                        					& \multicolumn{1}{c|}{{\color[HTML]{0000FF} }}                                                     & \multicolumn{1}{c|}{}                                         & \multicolumn{1}{c|}{{\color[HTML]{0000FF} }}                                   					& \multicolumn{1}{c|}{}                        					& \multicolumn{1}{c|}{{\color[HTML]{0000FF} }}                                   					& \multicolumn{1}{c|}{}                        					&                                                           						& \multicolumn{1}{c|}{}                                         	& {\color[HTML]{0000FF} }                                                 					& \multicolumn{1}{c|}{D} & \multicolumn{1}{c|}{$1.000$}                                   	& \multicolumn{1}{c|}{$1.000$}                                     	& \multicolumn{1}{c|}{$.200$}                                     	& \multicolumn{2}{c|}{\cellcolor[HTML]{D9D9D9}}                       																			\\ \cline{18-22}
                                                                                                                     &                                                                                        				&                                     					& \multicolumn{1}{c|}{}                        				  				 & \multicolumn{1}{c|}{{\color[HTML]{0000FF} }}                                   					& \multicolumn{1}{c|}{}                                       	& \multicolumn{1}{c|}{}                                                          					& \multicolumn{1}{c|}{}                        					& \multicolumn{1}{c|}{{\color[HTML]{0000FF} }}                                                     & \multicolumn{1}{c|}{}                                         & \multicolumn{1}{c|}{{\color[HTML]{0000FF} }}                                   					& \multicolumn{1}{c|}{}                        					& \multicolumn{1}{c|}{{\color[HTML]{0000FF} }}                                   					& \multicolumn{1}{c|}{}                        					&                                                           						& \multicolumn{1}{c|}{}                                         	& {\color[HTML]{0000FF} }                                                 					& \multicolumn{1}{c|}{E} & \multicolumn{1}{c|}{$.690$}                                   	& \multicolumn{1}{c|}{$1.000$}                                     	& \multicolumn{1}{c|}{$1.000$}                                     	& \multicolumn{1}{c|}{$1.000$}                                     	& \multicolumn{1}{c|}{\cellcolor[HTML]{D9D9D9}}								\\ \cline{18-23} 
                                                                                                                     &                                                                                        				& \multirow{-5}{*}{\textbf{S}}        					& \multicolumn{1}{c|}{\multirow{-5}{*}{\rotatebox{90}{$.734$}}} 			 & \multicolumn{1}{c|}{\multirow{-5}{*}{{\color[HTML]{0000FF} \rotatebox{90}{\boldmath{$<.001$}}}}} & \multicolumn{1}{c|}{\multirow{-5}{*}{\rotatebox{90}{$.926$}}} & \multicolumn{1}{c|}{\multirow{-5}{*}{\rotatebox{90}{$.077$}}}                                     & \multicolumn{1}{c|}{\multirow{-5}{*}{\rotatebox{90}{$.864$}}} & \multicolumn{1}{c|}{\multirow{-5}{*}{{\color[HTML]{0000FF} \rotatebox{90}{\boldmath{$.004$}}}}}  & \multicolumn{1}{c|}{\multirow{-5}{*}{\rotatebox{90}{$.879$}}} & \multicolumn{1}{c|}{\multirow{-5}{*}{{\color[HTML]{0000FF} \rotatebox{90}{\boldmath{$.008$}}}}}   	& \multicolumn{1}{c|}{\multirow{-5}{*}{\rotatebox{90}{$.840$}}} & \multicolumn{1}{c|}{\multirow{-5}{*}{{\color[HTML]{0000FF} \rotatebox{90}{\boldmath{$.001$}}}}}   & \multicolumn{1}{c|}{\multirow{-5}{*}{\rotatebox{90}{$.937$}}} & \multirow{-5}{*}{\rotatebox{90}{$.135$}}                    						& \multicolumn{1}{c|}{\multirow{-5}{*}{\rotatebox{90}{$13.333$}}}   & \multirow{-5}{*}{{\color[HTML]{0000FF} \rotatebox{90}{\boldmath{$.020$}}}}                & \multicolumn{1}{c|}{F} & \multicolumn{1}{c|}{$1.000$}                                   	& \multicolumn{1}{c|}{$1.000$}                                     	& \multicolumn{1}{c|}{$1.000$}                                     	& \multicolumn{1}{c|}{$1.000$}                                     	& \multicolumn{1}{c|}{$1.000$}                       						\\ \cline{3-23} 
                                                                                                                     &                                                                                        				&                                     					& \multicolumn{1}{c|}{}                        				  				 & \multicolumn{1}{c|}{{\color[HTML]{0000FF} }}                                   					& \multicolumn{1}{c|}{}                                       	& \multicolumn{1}{c|}{}                                                          					& \multicolumn{1}{c|}{}                        					& \multicolumn{1}{c|}{{\color[HTML]{0000FF} }}                                                     & \multicolumn{1}{c|}{}                                         & \multicolumn{1}{c|}{{\color[HTML]{0000FF} }}                                   					& \multicolumn{1}{c|}{}                        					& \multicolumn{1}{c|}{{\color[HTML]{0000FF} }}                                   					& \multicolumn{1}{c|}{}                        					&                                                           						& \multicolumn{1}{c|}{}                                         	& {\color[HTML]{0000FF} }                                                 					& \multicolumn{1}{c|}{B} & \multicolumn{1}{c|}{$1.000$}                                   	& \multicolumn{4}{c|}{\cellcolor[HTML]{D9D9D9}}                       																																																					\\ \cline{18-20}
                                                                                                                     &                                                                                        				&                                     					& \multicolumn{1}{c|}{}                        				  				 & \multicolumn{1}{c|}{{\color[HTML]{0000FF} }}                                   					& \multicolumn{1}{c|}{}                                       	& \multicolumn{1}{c|}{}                                                          					& \multicolumn{1}{c|}{}                        					& \multicolumn{1}{c|}{{\color[HTML]{0000FF} }}                                                     & \multicolumn{1}{c|}{}                                         & \multicolumn{1}{c|}{{\color[HTML]{0000FF} }}                                   					& \multicolumn{1}{c|}{}                        					& \multicolumn{1}{c|}{{\color[HTML]{0000FF} }}                                   					& \multicolumn{1}{c|}{}                        					&                                                           						& \multicolumn{1}{c|}{}                                         	& {\color[HTML]{0000FF} }                                                 					& \multicolumn{1}{c|}{C} & \multicolumn{1}{c|}{$1.000$}                                   	& \multicolumn{1}{c|}{$1.000$}                                     	& \multicolumn{3}{c|}{\cellcolor[HTML]{D9D9D9}}                       																																				\\ \cline{18-21}
                                                                                                                     &                                                                                        				&                                     					& \multicolumn{1}{c|}{}                        				  				 & \multicolumn{1}{c|}{{\color[HTML]{0000FF} }}                                   					& \multicolumn{1}{c|}{}                                       	& \multicolumn{1}{c|}{}                                                          					& \multicolumn{1}{c|}{}                        					& \multicolumn{1}{c|}{{\color[HTML]{0000FF} }}                                                     & \multicolumn{1}{c|}{}                                         & \multicolumn{1}{c|}{{\color[HTML]{0000FF} }}                                   					& \multicolumn{1}{c|}{}                        					& \multicolumn{1}{c|}{{\color[HTML]{0000FF} }}                                   					& \multicolumn{1}{c|}{}                        					&                                                           						& \multicolumn{1}{c|}{}                                         	& {\color[HTML]{0000FF} }                                                 					& \multicolumn{1}{c|}{D} & \multicolumn{1}{c|}{$1.000$}                                   	& \multicolumn{1}{c|}{$.295$}                                     	& \multicolumn{1}{c|}{{\color[HTML]{0000FF} \boldmath{$.002$}}}     & \multicolumn{2}{c|}{\cellcolor[HTML]{D9D9D9}}                       																			\\ \cline{18-22}
                                                                                                                     &                                                                                        				&                                     					& \multicolumn{1}{c|}{}                        				  				 & \multicolumn{1}{c|}{{\color[HTML]{0000FF} }}                                   					& \multicolumn{1}{c|}{}                                       	& \multicolumn{1}{c|}{}                                                          					& \multicolumn{1}{c|}{}                        					& \multicolumn{1}{c|}{{\color[HTML]{0000FF} }}                                                     & \multicolumn{1}{c|}{}                                         & \multicolumn{1}{c|}{{\color[HTML]{0000FF} }}                                   					& \multicolumn{1}{c|}{}                        					& \multicolumn{1}{c|}{{\color[HTML]{0000FF} }}                                   					& \multicolumn{1}{c|}{}                        					&                                                           						& \multicolumn{1}{c|}{}                                         	& {\color[HTML]{0000FF} }                                                 					& \multicolumn{1}{c|}{E} & \multicolumn{1}{c|}{$1.000$}                                   	& \multicolumn{1}{c|}{$1.000$}                                     	& \multicolumn{1}{c|}{$1.000$}                                     	& \multicolumn{1}{c|}{{\color[HTML]{0000FF} \boldmath{$.016$}}}     & \multicolumn{1}{c|}{\cellcolor[HTML]{D9D9D9}}								\\ \cline{18-23} 
                                                                                                                     & \multirow{-10}{*}{\rotatebox{90}{\textbf{Midair}}}                                     				& \multirow{-5}{*}{\textbf{E}}        					& \multicolumn{1}{c|}{\multirow{-5}{*}{\rotatebox{90}{$.832$}}}              & \multicolumn{1}{c|}{\multirow{-5}{*}{{\color[HTML]{0000FF} \rotatebox{90}{\boldmath{$.001$}}}}}  & \multicolumn{1}{c|}{\multirow{-5}{*}{\rotatebox{90}{$.932$}}} & \multicolumn{1}{c|}{\multirow{-5}{*}{\rotatebox{90}{$.106$}}}                                     & \multicolumn{1}{c|}{\multirow{-5}{*}{\rotatebox{90}{$.810$}}} & \multicolumn{1}{c|}{\multirow{-5}{*}{{\color[HTML]{0000FF} \rotatebox{90}{\boldmath{$<.001$}}}}} & \multicolumn{1}{c|}{\multirow{-5}{*}{\rotatebox{90}{$.736$}}} & \multicolumn{1}{c|}{\multirow{-5}{*}{{\color[HTML]{0000FF} \rotatebox{90}{\boldmath{$<.001$}}}}} 	& \multicolumn{1}{c|}{\multirow{-5}{*}{\rotatebox{90}{$.912$}}} & \multicolumn{1}{c|}{\multirow{-5}{*}{{\color[HTML]{0000FF} \rotatebox{90}{\boldmath{$.038$}}}}}   & \multicolumn{1}{c|}{\multirow{-5}{*}{\rotatebox{90}{$.948$}}} & \multirow{-5}{*}{\rotatebox{90}{$.240$}}                                   		& \multicolumn{1}{c|}{\multirow{-5}{*}{\rotatebox{90}{$20.262$}}}   & \multirow{-5}{*}{{\color[HTML]{0000FF} \rotatebox{90}{\boldmath{$.001$}}}}                & \multicolumn{1}{c|}{F} & \multicolumn{1}{c|}{$1.000$}                                   	& \multicolumn{1}{c|}{$1.000$}                                     	& \multicolumn{1}{c|}{$1.000$}                                     	& \multicolumn{1}{c|}{$.609$}                                     	& \multicolumn{1}{c|}{$1.000$}                       						\\ \cline{2-23} 
                                                                                                                     &                                                                                        				&                                     					& \multicolumn{1}{c|}{}                        				  				 & \multicolumn{1}{c|}{{\color[HTML]{0000FF} }}                                   					& \multicolumn{1}{c|}{}                                       	& \multicolumn{1}{c|}{{\color[HTML]{0000FF} }}                                                      & \multicolumn{1}{c|}{}                        					& \multicolumn{1}{c|}{{\color[HTML]{0000FF} }}                                                     & \multicolumn{1}{c|}{}                                         & \multicolumn{1}{c|}{{\color[HTML]{0000FF} }}                                   					& \multicolumn{1}{c|}{}                        					& \multicolumn{1}{c|}{}                                                          					& \multicolumn{1}{c|}{}                        					& {\color[HTML]{0000FF} }                                   						& \multicolumn{1}{c|}{}                                         	& {\color[HTML]{0000FF} }                                                 					& \multicolumn{1}{c|}{B} & \multicolumn{1}{c|}{$.223$}                                   	& \multicolumn{4}{c|}{\cellcolor[HTML]{D9D9D9}}                       																																																					\\ \cline{18-20}
                                                                                                                     &                                                                                        				&                                     					& \multicolumn{1}{c|}{}                        				  				 & \multicolumn{1}{c|}{{\color[HTML]{0000FF} }}                                   					& \multicolumn{1}{c|}{}                                       	& \multicolumn{1}{c|}{{\color[HTML]{0000FF} }}                                                      & \multicolumn{1}{c|}{}                        					& \multicolumn{1}{c|}{{\color[HTML]{0000FF} }}                                                     & \multicolumn{1}{c|}{}                                         & \multicolumn{1}{c|}{{\color[HTML]{0000FF} }}                                   					& \multicolumn{1}{c|}{}                        					& \multicolumn{1}{c|}{}                                                          					& \multicolumn{1}{c|}{}                        					& {\color[HTML]{0000FF} }                                   						& \multicolumn{1}{c|}{}                                         	& {\color[HTML]{0000FF} }                                                 					& \multicolumn{1}{c|}{C} & \multicolumn{1}{c|}{$1.000$}                                   	& \multicolumn{1}{c|}{$1.000$}                                     	& \multicolumn{3}{c|}{\cellcolor[HTML]{D9D9D9}}                       																																				\\ \cline{18-21}
                                                                                                                     &                                                                                        				&                                     					& \multicolumn{1}{c|}{}                        				  				 & \multicolumn{1}{c|}{{\color[HTML]{0000FF} }}                                   					& \multicolumn{1}{c|}{}                                       	& \multicolumn{1}{c|}{{\color[HTML]{0000FF} }}                                                      & \multicolumn{1}{c|}{}                        					& \multicolumn{1}{c|}{{\color[HTML]{0000FF} }}                                                     & \multicolumn{1}{c|}{}                                         & \multicolumn{1}{c|}{{\color[HTML]{0000FF} }}                                   					& \multicolumn{1}{c|}{}                        					& \multicolumn{1}{c|}{}                                                          					& \multicolumn{1}{c|}{}                        					& {\color[HTML]{0000FF} }                                   						& \multicolumn{1}{c|}{}                                         	& {\color[HTML]{0000FF} }                                                 					& \multicolumn{1}{c|}{D} & \multicolumn{1}{c|}{$1.000$}                                   	& \multicolumn{1}{c|}{{\color[HTML]{0000FF} \boldmath{$.018$}}}     & \multicolumn{1}{c|}{{\color[HTML]{0000FF} \boldmath{$.037$}}}     & \multicolumn{2}{c|}{\cellcolor[HTML]{D9D9D9}}                       																			\\ \cline{18-22}
                                                                                                                     &                                                                                        				&                                     					& \multicolumn{1}{c|}{}                        				  				 & \multicolumn{1}{c|}{{\color[HTML]{0000FF} }}                                   					& \multicolumn{1}{c|}{}                                       	& \multicolumn{1}{c|}{{\color[HTML]{0000FF} }}                                                      & \multicolumn{1}{c|}{}                        					& \multicolumn{1}{c|}{{\color[HTML]{0000FF} }}                                                     & \multicolumn{1}{c|}{}                                         & \multicolumn{1}{c|}{{\color[HTML]{0000FF} }}                                   					& \multicolumn{1}{c|}{}                        					& \multicolumn{1}{c|}{}                                                          					& \multicolumn{1}{c|}{}                        					& {\color[HTML]{0000FF} }                                   						& \multicolumn{1}{c|}{}                                         	& {\color[HTML]{0000FF} }                                                 					& \multicolumn{1}{c|}{E} & \multicolumn{1}{c|}{$1.000$}                                   	& \multicolumn{1}{c|}{.564}                                     	& \multicolumn{1}{c|}{$.816$}                                     	& \multicolumn{1}{c|}{$1.000$}                                     	& \multicolumn{1}{c|}{\cellcolor[HTML]{D9D9D9}}								\\ \cline{18-23} 
                                                                                                                     &                                                                                        				& \multirow{-5}{*}{\textbf{S}}        					& \multicolumn{1}{c|}{\multirow{-5}{*}{\rotatebox{90}{$.909$}}} 			 & \multicolumn{1}{c|}{\multirow{-5}{*}{{\color[HTML]{0000FF} \rotatebox{90}{\boldmath{$.033$}}}}}  & \multicolumn{1}{c|}{\multirow{-5}{*}{\rotatebox{90}{$.878$}}} & \multicolumn{1}{c|}{\multirow{-5}{*}{{\color[HTML]{0000FF} \rotatebox{90}{\boldmath{$.008$}}}}}   & \multicolumn{1}{c|}{\multirow{-5}{*}{\rotatebox{90}{$.914$}}} & \multicolumn{1}{c|}{\multirow{-5}{*}{{\color[HTML]{0000FF} \rotatebox{90}{\boldmath{$.042$}}}}}  & \multicolumn{1}{c|}{\multirow{-5}{*}{\rotatebox{90}{$.823$}}} & \multicolumn{1}{c|}{\multirow{-5}{*}{{\color[HTML]{0000FF} \rotatebox{90}{\boldmath{$<.001$}}}}} 	& \multicolumn{1}{c|}{\multirow{-5}{*}{\rotatebox{90}{$.919$}}} & \multicolumn{1}{c|}{\multirow{-5}{*}{\rotatebox{90}{$.054$}}}                                   	& \multicolumn{1}{c|}{\multirow{-5}{*}{\rotatebox{90}{$.882$}}} & \multirow{-5}{*}{{\color[HTML]{0000FF} \rotatebox{90}{\boldmath{$.009$}}}}   		& \multicolumn{1}{c|}{\multirow{-5}{*}{\rotatebox{90}{$17.167$}}}   & \multirow{-5}{*}{{\color[HTML]{0000FF} \rotatebox{90}{\boldmath{$.004$}}}}                & \multicolumn{1}{c|}{F} & \multicolumn{1}{c|}{$1.000$}                                   	& \multicolumn{1}{c|}{.411}                                     	& \multicolumn{1}{c|}{$1.000$}                                     	& \multicolumn{1}{c|}{$1.000$}                                     	& \multicolumn{1}{c|}{$1.000$}                       						\\ \cline{3-23} 
                                                                                                                     &                                                                                        				&                                     					& \multicolumn{1}{c|}{}                        				  				 & \multicolumn{1}{c|}{{\color[HTML]{0000FF} }}                                   					& \multicolumn{1}{c|}{}                                       	& \multicolumn{1}{c|}{{\color[HTML]{0000FF} }}                                                      & \multicolumn{1}{c|}{}                        					& \multicolumn{1}{c|}{{\color[HTML]{0000FF} }}                                                     & \multicolumn{1}{c|}{}                                         & \multicolumn{1}{c|}{{\color[HTML]{0000FF} }}                                   					& \multicolumn{1}{c|}{}                        					& \multicolumn{1}{c|}{{\color[HTML]{0000FF} }}                                   					& \multicolumn{1}{c|}{}                        					& {\color[HTML]{0000FF} }                                   						& \multicolumn{1}{c|}{}                                         	& {\color[HTML]{0000FF} }                                                 					& \multicolumn{1}{c|}{B} & \multicolumn{1}{c|}{{\color[HTML]{0000FF} \boldmath{$.010$}}}   	& \multicolumn{4}{c|}{\cellcolor[HTML]{D9D9D9}}                       																																																					\\ \cline{18-20}
                                                                                                                     &                                                                                        				&                                     					& \multicolumn{1}{c|}{}                        				  				 & \multicolumn{1}{c|}{{\color[HTML]{0000FF} }}                                   					& \multicolumn{1}{c|}{}                                       	& \multicolumn{1}{c|}{{\color[HTML]{0000FF} }}                                                      & \multicolumn{1}{c|}{}                        					& \multicolumn{1}{c|}{{\color[HTML]{0000FF} }}                                                     & \multicolumn{1}{c|}{}                                         & \multicolumn{1}{c|}{{\color[HTML]{0000FF} }}                                   					& \multicolumn{1}{c|}{}                        					& \multicolumn{1}{c|}{{\color[HTML]{0000FF} }}                                   					& \multicolumn{1}{c|}{}                        					& {\color[HTML]{0000FF} }                                   						& \multicolumn{1}{c|}{}                                         	& {\color[HTML]{0000FF} }                                                 					& \multicolumn{1}{c|}{C} & \multicolumn{1}{c|}{$1.000$}                                   	& \multicolumn{1}{c|}{.264}                                     	& \multicolumn{3}{c|}{\cellcolor[HTML]{D9D9D9}}                       																																				\\ \cline{18-21}
                                                                                                                     &                                                                                        				&                                     					& \multicolumn{1}{c|}{}                        				  				 & \multicolumn{1}{c|}{{\color[HTML]{0000FF} }}                                   					& \multicolumn{1}{c|}{}                                       	& \multicolumn{1}{c|}{{\color[HTML]{0000FF} }}                                                      & \multicolumn{1}{c|}{}                        					& \multicolumn{1}{c|}{{\color[HTML]{0000FF} }}                                                     & \multicolumn{1}{c|}{}                                         & \multicolumn{1}{c|}{{\color[HTML]{0000FF} }}                                   					& \multicolumn{1}{c|}{}                        					& \multicolumn{1}{c|}{{\color[HTML]{0000FF} }}                                   					& \multicolumn{1}{c|}{}                        					& {\color[HTML]{0000FF} }                                   						& \multicolumn{1}{c|}{}                                         	& {\color[HTML]{0000FF} }                                                 					& \multicolumn{1}{c|}{D} & \multicolumn{1}{c|}{$1.000$}                                   	& \multicolumn{1}{c|}{{\color[HTML]{0000FF} \boldmath{$.003$}}}     & \multicolumn{1}{c|}{{\color[HTML]{0000FF} \boldmath{$.013$}}}     & \multicolumn{2}{c|}{\cellcolor[HTML]{D9D9D9}}                       																			\\ \cline{18-22}
                                                                                                                     &                                                                                        				&                                     					& \multicolumn{1}{c|}{}                        				  				 & \multicolumn{1}{c|}{{\color[HTML]{0000FF} }}                                   					& \multicolumn{1}{c|}{}                                       	& \multicolumn{1}{c|}{{\color[HTML]{0000FF} }}                                                      & \multicolumn{1}{c|}{}                        					& \multicolumn{1}{c|}{{\color[HTML]{0000FF} }}                                                     & \multicolumn{1}{c|}{}                                         & \multicolumn{1}{c|}{{\color[HTML]{0000FF} }}                                   					& \multicolumn{1}{c|}{}                        					& \multicolumn{1}{c|}{{\color[HTML]{0000FF} }}                                   					& \multicolumn{1}{c|}{}                        					& {\color[HTML]{0000FF} }                                   						& \multicolumn{1}{c|}{}                                         	& {\color[HTML]{0000FF} }                                                 					& \multicolumn{1}{c|}{E} & \multicolumn{1}{c|}{$1.000$}                                   	& \multicolumn{1}{c|}{{\color[HTML]{0000FF} \boldmath{$.007$}}}     & \multicolumn{1}{c|}{$.576$}                                     	& \multicolumn{1}{c|}{$1.000$}                                     	& \multicolumn{1}{c|}{\cellcolor[HTML]{D9D9D9}}								\\ \cline{18-23} 
    \multirow{-20}{*}{\rotatebox{90}{\textbf{\begin{tabular}[c]{@{}c@{}}Location Preference Ratings\end{tabular}}}} & \multirow{-10}{*}{\rotatebox{90}{\textbf{Onskin}}}                                      				& \multirow{-5}{*}{\textbf{E}}        					& \multicolumn{1}{c|}{\multirow{-5}{*}{\rotatebox{90}{$.884$}}} 			 & \multicolumn{1}{c|}{\multirow{-5}{*}{{\color[HTML]{0000FF} \rotatebox{90}{\boldmath{$.010$}}}}}  & \multicolumn{1}{c|}{\multirow{-5}{*}{\rotatebox{90}{$.844$}}} & \multicolumn{1}{c|}{\multirow{-5}{*}{{\color[HTML]{0000FF} \rotatebox{90}{\boldmath{$.002$}}}}}   & \multicolumn{1}{c|}{\multirow{-5}{*}{\rotatebox{90}{$.914$}}} & \multicolumn{1}{c|}{\multirow{-5}{*}{{\color[HTML]{0000FF} \rotatebox{90}{\boldmath{$.043$}}}}}  & \multicolumn{1}{c|}{\multirow{-5}{*}{\rotatebox{90}{$.787$}}} & \multicolumn{1}{c|}{\multirow{-5}{*}{{\color[HTML]{0000FF} \rotatebox{90}{\boldmath{$<.001$}}}}} 	& \multicolumn{1}{c|}{\multirow{-5}{*}{\rotatebox{90}{$.873$}}} & \multicolumn{1}{c|}{\multirow{-5}{*}{{\color[HTML]{0000FF} \rotatebox{90}{\boldmath{$<.001$}}}}} 	& \multicolumn{1}{c|}{\multirow{-5}{*}{\rotatebox{90}{$.890$}}} & \multirow{-5}{*}{{\color[HTML]{0000FF} \rotatebox{90}{\boldmath{$.013$}}}}   		& \multicolumn{1}{c|}{\multirow{-5}{*}{\rotatebox{90}{$26.786$}}}   & \multirow{-5}{*}{{\color[HTML]{0000FF} \rotatebox{90}{\boldmath{$<.001$}}}}               & \multicolumn{1}{c|}{F} & \multicolumn{1}{c|}{$1.000$}                                   	& \multicolumn{1}{c|}{{\color[HTML]{0000FF} \boldmath{$.031$}}}     & \multicolumn{1}{c|}{$1.000$}                                     	& \multicolumn{1}{c|}{$1.000$}                                     	& \multicolumn{1}{c|}{$1.000$}                       						\\ \hline
    \end{tabular}%
    }
    \vspace{.15cm} 
    \captionof{table}{Statistical comparison of swipe performance metrics across regions in the 6-region distribution. Figure~\ref{fig:GestureMetrics_6R2R_RQ3} highlights statistically significant pairwise differences.}  
    \Description{Statistical comparison of swipe performance metrics across regions in the 6-region distribution. Figure~\ref{fig:GestureMetrics_6R2R_RQ3} highlights statistically significant pairwise differences.}
    \label{tab:GestureMetrics_6R2R_RQ3}  
\end{table}

\endgroup
}

\subsection{6-Region Midair Segmentation}
\label{section:Results_RQ3_6R2R_Midair}
{
Figure~\ref{fig:GestureMetrics_6R2R_RQ3} summarizes significant pairwise differences across metrics for both starting and ending regions in the 6-region midair segmentation.

For swipe initiation, starting region significantly affected accuracy (\textbf{DV1}; Friedman $\chi^2(5)=33.271$, $p<.001$), angular displacement (\textbf{DV4}; $\chi^2(5)=96.654$, $p<.001$), and subjective ratings ($\chi^2(5)=13.333$, $p=.020$). 
Regions \textbf{D} (above the ear) and \textbf{F} (above the chin) achieved the highest accuracy and lowest angular displacement, indicating stable trajectories during swipe initiation. 
Regions \textbf{A} (above the nose) and \textbf{C} (above the temple) followed, exhibiting high accuracy but mild directional bias toward nearby facial landmarks. 
Region \textbf{E} showed reduced performance, likely due to limited visual feedback outside the field of view. 
The eye-adjacent region \textbf{B} exhibited significantly lower accuracy than regions \textbf{A} ($p=.006$), \textbf{C} ($p<.001$), \textbf{E} ($p=.023$), and \textbf{F} ($p<.001$). 
Regions adjacent to eye corner (\textbf{B} and \textbf{C}) and region around the nape of neck (\textbf{E}) showed high angular displacement and ranked among the lowest. 
Together, accuracy, angular displacement, and subjective ratings revealed a consistent but multidimensional ordering of starting regions, with objective performance favoring \textbf{D$\approx$F > A$\approx$C > E > B} (Figure \ref{fig:GestureMetrics_6R2R_RQ3}).

For swipe ending regions in the 6-region midair segmentation (Figure~\ref{fig:GestureMetrics_6R2R_RQ3}), region choice significantly affected swipe accuracy (\textbf{DV1}; Friedman $\chi^2(5)=34.211$, $p<.001$), angular displacement (\textbf{DV4}; Friedman $\chi^2(5)=213.320$, $p<.001$), and subjective location preference ratings (Friedman $\chi^2(5)=26.786$, $p<.001$), indicating that spatial constraints continue to shape swipe trajectories beyond initiation. 
Region \textbf{D} was the most accurate endpoint, outperforming all other regions except \textbf{A} (D$\leftrightarrow$B $p=.003$, D$\leftrightarrow$C $p<.001$, D$\leftrightarrow$E $p<.001$, D$\leftrightarrow$F $p=.008$).
\textbf{D} also received the highest subjective ratings -- significantly higher than \textbf{C} ($p=0.002$) and \textbf{E} ($p=0.016$)—consistent with its low angular distortion and visually stable termination over the ear.
Region \textbf{A} followed closely, outperforming regions \textbf{C} ($p=.043$) and \textbf{E} ($p=.007$).
Although endings near the nose (\textbf{A}) skewed toward the chin around \textbf{F}, angular displacement for \textbf{A} remained comparable to regions \textbf{D}, \textbf{E}, and \textbf{F}, supporting both high accuracy and preference. 
Region \textbf{F} showed intermediate performance, with mild skew toward the jawline but sufficient visual feedback for reliable termination. 
In contrast, regions \textbf{C} and \textbf{E} produced the lowest accuracy and ratings, as endpoints frequently crossed into adjacent regions (C$\rightarrow$D, E$\rightarrow$F), reflecting poor spatial separability. 
Region \textbf{B} exhibited highest angular skew—directed toward the temple to avoid ocular interaction.
However, due to occular feedback \textbf{B} had slightly better accuracy and ratings than \textbf{C} and \textbf{E}, although not statistically significant.
Together, these measures revealed a clear relative ordering of ending regions (\textbf{D > A > F > C $\approx$ E > B}), shown visually in figure \ref{fig:GestureMetrics_6R2R_RQ3}.

Individual swipe-shape analysis (Figure~\ref{fig:GestureMetrics_6R2R_RQ3_Individual_MidairSwipes}) showed that swipe reliability and preference were strongly shaped by alignment with visually accessible regions exhibiting low angular displacement. 
Horizontal midair swipes (Figure \ref{fig:GestureMetrics_6R2R_RQ3_Visual_Swipe_References}) consistently achieved the highest accuracy and preference, followed by U-shaped swipes, indicating that motions aligned with the user’s horizontal field of awareness are more robust than vertical or angular trajectories. 
Non-axial swipes showed mixed performance: swipes directed toward lower-face regions (\eg{}, \textbf{CF}, \textbf{FC}) maintained relatively stable trajectories with moderate displacement, whereas swipes involving the eye-adjacent region B (\eg{}, \textbf{EB}, \textbf{BE}) exhibited greater angular deviation and lower ratings.

For U-shaped swipes, most regions (\ie{}, \textbf{AA}, \textbf{DD}, \textbf{EE}, and \textbf{FF}) remained consistently stable across accuracy and ratings.
In contrast, U-shaped swipes in the eye and temple regions (\textbf{B} and \textbf{C}) showed increased angular deviation and lower ratings, likely due to weaker spatial separation in midair, and reduced visual guidance in the upper-face region - specially for region \textbf{C} outside field of view.
Among V-shaped swipes, adjacent medium-angle swipes centered near the ear-adjacent region D (\eg{}, \textbf{CD}, \textbf{DC}, \textbf{FA}) maintained strong accuracy with moderate displacement, whereas swipes intersecting upper-face regions (\eg{}, \textbf{AB}, \textbf{BA}, \textbf{BC}, \textbf{CB}) showed greater angular deviation and reduced stability. 
Similarly, large V-shaped swipes spanning non-adjacent regions remained relatively robust when aligned along the posterior cheek axis (\eg{}, \textbf{CA}, \textbf{DF}, \textbf{EA}), but degraded when trajectories crossed upper-face regions (\eg{}, \textbf{BF}, \textbf{DB}, \textbf{CE}).
Midair swipes \textbf{CE} and \textbf{EC}, which span visually constrained regions, frequently drifted toward region \textbf{D}, resulting in lower accuracy and preference; these swipes therefore align more closely with \textbf{DD} than with distinct region-to-region inputs. 

Overall, midair swipe patterns show that executing swipes in upper-face regions \textbf{B} and \textbf{C} is challenging due to visual avoidance and spatial drift. 
Drift pattern observations suggest merging \textbf{BB}, \textbf{BC}, and \textbf{CC} into a U-shaped midair swipe (temple–cheekbone–temple), which lowers cognitive load while maintaining expressive interaction.
Additionally, the drift from \textbf{F} to \textbf{E} for \textbf{FE} swipe indicates that \textbf{FE} and \textbf{EF} could also combine into a single midair swipe, aligning with the U-shaped \textbf{EE} swipe. Furthermore, downward swipes from \textbf{B}~/~\textbf{C} to \textbf{E}~/~\textbf{F} (\ie{}, \textbf{CE}, \textbf{BF}, \textbf{BE}, \textbf{CF}) can be unified as a swipe-down action, while corresponding upward swipes (\textbf{EC}, \textbf{FB}, \textbf{EB}, \textbf{FC}) can be unified to create a complementary swipe-up action.
}

\subsection{6-Region Onskin Segmentation}
\label{section:Results_RQ3_6R2R_Onskin}
{
Figure~\ref{fig:GestureMetrics_6R2R_RQ3} summarizes all significant pairwise differences across metrics for both starting and ending regions in the 6-region onskin segmentation, with post-hoc $p$-values annotated.
For swipe initiation, region choice significantly influenced angular displacement (\textbf{DV4}; Friedman $\chi^2(5)=212.310$, $p<.001$), which in turn explained differences in accuracy (\textbf{DV1}; Friedman $\chi^2(5)=52.981$, $p<.001$) and starting-region self-ratings (Friedman $\chi^2(5)=17.167$, $p=.020$).
The eye-adjacent region \textbf{B} (around the eyebrow and outer eye corner) consistently performed worst, exhibiting significantly lower accuracy than all other regions ($p<.001$) and substantially higher angular displacement ($p<.001$).
Because region \textbf{B} occupies a small area tightly surrounding the eye in the 6-region layout, swipe segments frequently deflected into the temporal region \textbf{C}, increasing skew and retrials.
Region \textbf{C} (temple) showed improved accuracy relative to \textbf{B}, but remained less preferred, as reflected in its lower ratings.

In contrast, region \textbf{D} (cheekbone--tragus junction) emerged as the most effective and preferred starting location, receiving significantly higher ratings than regions \textbf{B} ($p=.018$) and \textbf{C} ($p=0.037$) and exhibiting among the lowest angular displacement values.
Region \textbf{A} (around the nose) showed comparable accuracy and angular stability to \textbf{D}, with no statistically significant differences between them.
Regions \textbf{A} and \textbf{F} (mouth~/~chin) benefited from combined haptic and visual feedback, supporting more controlled initiation and reduced overshoot.
Regions \textbf{E} (nape) and \textbf{F} showed similar objective performance, although \textbf{E} was rated slightly lower due to increased wrist bending.
Together, these results support the ordering \textbf{D > A > E $\approx$ F > C > B} for onskin swipe initiation (Figure~\ref{fig:GestureMetrics_6R2R_RQ3}).

For swipe termination, ending-region choice similarly affected accuracy (\textbf{DV1}; Friedman $\chi^2(5)=23.073$, $p<.001$), angular displacement (\textbf{DV4}; Friedman $\chi^2(5)=147.550$, $p<.001$), and ending-region self-ratings (Friedman $\chi^2(5)=26.786$, $p<.001$), indicating persistent region-dependent effects, although termination was more tolerant of brief traversal through constrained areas.
Region \textbf{D} again yielded the most reliable endpoints, achieving higher accuracy than regions \textbf{B} ($p=.002$), \textbf{E} ($p=.028$), and \textbf{F} ($p<.001$), and receiving the highest ratings, significantly exceeding those of regions \textbf{B} ($p=.003$) and \textbf{C} ($p=0.013$).
Region \textbf{B} ranked lowest overall, exhibiting significantly greater angular displacement than all other regions ($p<.001$) as endpoints deflected toward the temple to avoid ocular contact, resulting in the lowest accuracy and ratings.
Region \textbf{C} showed reduced skew and improved accuracy relative to \textbf{B} ($p<.001$), but remained suboptimal due to limited endpoint separability.
Regions \textbf{A} and \textbf{D} exhibited comparably low angular displacement and high accuracy, while region \textbf{F} showed intermediate performance; region \textbf{E} was consistently less preferred.
Across phases, the ordering \textbf{D > A $\approx$ E $\approx$ F > C > B} was preserved for onskin swipe termination (Figure~\ref{fig:GestureMetrics_6R2R_RQ3}).

Swipe-shape analysis (Figure~\ref{fig:GestureMetrics_6R2R_RQ3_Individual_OnskinSwipes}) revealed systematic  swipe stability patterns across the 5 swipe groups summarized in Table~\ref{tab:onskin_6region_swipe_shape_stability}. 
\textbf{Axial swipes} (\ie{}, \textbf{AD}, \textbf{DA}) were highly stable, achieving high accuracy, low angular displacement, and favorable ratings due to continuous tactile guidance along the cheek. 
In contrast, \textbf{non-axial swipes} showed mixed performance: swipes confined to cheek regions (\eg{}, \textbf{CF}, \textbf{FC}) maintained relatively high accuracy with moderate displacement, whereas swipes involving the eye-adjacent region \textbf{B} (\eg{}, \textbf{EB}, \textbf{BE}) exhibited greater angular deviation and lower ratings. 
For \textbf{U-shaped swipes}, regions \textbf{A}, \textbf{D}, \textbf{E}, and \textbf{F} remained consistently stable across DV1 and ratings and outperformed vertical non-axial swipes. 
In contrast, \textbf{BB} and \textbf{CC} showed increased angular deviation and lower ratings, reflecting upper-face interaction constraints: region \textbf{B} lies adjacent to the eye, where users tend to avoid contact with the eye corner and drift towards \textbf{C}, while region \textbf{C} may introduce additional variability due to hair-related friction. 

Among \textbf{V-shaped swipes}, adjacent medium-angle swipes avoiding region \textbf{B} (\eg{}, \textbf{CD}, \textbf{DC}, \textbf{EF}, \textbf{FE}, \textbf{FA}) maintained high accuracy with moderate displacement, whereas swipes intersecting the eye region (\eg{}, \textbf{AB}, \textbf{BA}, \textbf{BC}) showed greater displacement and lower ratings. 
Similarly, \textbf{V-shaped large swipes} spanning non-adjacent regions remained robust when anchored on the cheek (\eg{}, \textbf{DF}, \textbf{FD}), but degraded when involving region \textbf{B}. 
Regions \textbf{E} and \textbf{F} also exhibited distinct swipe behaviors, likely reflecting different tactile anchors along the mandible—region \textbf{E} near the mandibular angle and region \textbf{F} near the anterior jaw and chin—suggesting that these lower-face regions should remain separate.

Overall, onskin swipe drift patterns indicate that swipes originating or terminating in regions \textbf{B} and \textbf{C} remain challenging for users, particularly near the eye and temple areas. Merging \textbf{EC} with \textbf{FC} and \textbf{CE} with \textbf{CF} may therefore improve recognition robustness for vertical swipes while reducing reliance on the eye-adjacent region. 
More broadly, these patterns suggest that robust onskin swipe sets in the 6-region layout should prioritize horizontal, cheek-anchored trajectories while minimizing interaction near the eye and temple regions.

}

}

\end{appendices}

\end{document}